\documentclass[12pt]{article}
\usepackage[utf8]{inputenc}
\usepackage{graphicx} 
\usepackage[margin=1in]{geometry}
\usepackage{lscape}
\usepackage{pdflscape}
\usepackage{amsmath,amssymb}
\usepackage{tabularx}
\usepackage{booktabs}
\usepackage{dsfont}
\usepackage{ccaption}
\usepackage{dcolumn}
\usepackage{multirow}
\usepackage{nicefrac}
\usepackage{float}
\usepackage{subcaption}
\usepackage{comment}
\usepackage{xcolor}
\usepackage[singlespacing,nodisplayskipstretch]{setspace}
\usepackage{color}
\usepackage{url}
\usepackage{varioref}
\usepackage{eurosym}
\usepackage{mdframed}
\usepackage{rotating}
\usepackage{enumitem}
\usepackage{makecell}
\usepackage[normalem]{ulem} 
\usepackage{threeparttable}
\usepackage{appendix}
\usepackage{algorithm}

\usepackage{natbib}
\definecolor{navy}{rgb}{0,.3,.7}
\definecolor{myorange}{RGB}{238, 166, 63 }
\definecolor{myblue}{RGB}{29, 130, 231 }
\definecolor{green}{RGB}{49, 178, 54}
\definecolor{darkred}{RGB}{169,5,5}
\definecolor{darkgreen}{RGB}{0,152,0}
\usepackage[pdfpagemode=UseThumbs,%
     colorlinks=true,%
     urlcolor=navy,%
     citecolor=darkred,%citecolor=teal,
     linkcolor=darkred%linkcolor=teal
     ]{hyperref}
\usepackage{cleveref}
\graphicspath{{}}
\usepackage[font=bf]{caption} 
\newcolumntype{d}[1]{D{.}{.}{#1}}
\newcolumntype{e}{D{.}{.}{-1}}
\newcolumntype{.}{D{.}{.}{3}}
\newcommand{\bs}{\boldsymbol}

\begin{document}

%%%%%%%%%%%%%%%%%%%%%%%%%%%%%%%%%%%%%%%%%%%%%%%%%%%%%%%%%%%%%%%%%%%%
%  TITLE PAGE
%%%%%%%%%%%%%%%%%%%%%%%%%%%%%%%%%%%%%%%%%%%%%%%%%%%%%%%%%%%%%%%%%%%%

\title{\large A Game of Throngs:
How Spotlight Tourism Reshapes Local Economies \thanks{We thank Josep Amer-Mestre, Francesco Drago, Andrea Ichino, Jordi Jofre-Monseny, Fabrizia Mealli, and Alessandro Tarozzi, and seminar participants at the European University Institute for helpful comments. All errors are our own.}
\vspace{.8 cm}
}
\vspace{.8 cm}
\author{Mafalda Batalha \thanks{Department of Economics, European University Institute.} 
\and Javier Viviens \footnotemark[2] }

\date{\vspace{.8 cm} 16 November  2026 \\
}
\maketitle

\begin{abstract}
\noindent

In July 2017, San Juan de Gaztelugatxe --- a modest chapel in the Basque Country --- appeared as ``Dragonstone'' in \textit{Game of Thrones}, transforming a locally-known site with limited tourism activity into a global tourist destination overnight. We exploit this exogenous spotlight shock to estimate its causal effects on the local economy using a novel administrative dataset and a Synthetic Control approach. We find an abrupt increase in tourism activity, far exceeding the effects documented in the literature. Despite this surge, employment gains in tourism-related sectors were largely offset by losses elsewhere, resulting in no net effect on total employment. We also document signs of housing-market overheating, with increases in both property transactions and prices. We find no evidence that the shock displaced the local population. Instead, it led to a roughly twofold increase in arrivals from abroad. Taken together, these findings suggest that a substantial increase in tourism activity does not necessarily translate into broad-based local economic gains. Instead, tourism-led growth may entail significant sectoral reallocation and housing-market pressures, underscoring the importance of looking beyond tourism activity alone when assessing its wider economic effects.

\vspace{0.5 cm}

\vspace{0.2cm}
\end{abstract}

\thispagestyle{empty}

\newpage

\setcounter{page}{1}

\begin{spacing}{2}

\clearpage

%% =============================================================================
\section{Introduction}
\label{sec:intro}
%% =============================================================================

Tourism has become a central instrument of local economic policy, accounting for roughly 10\% of global GDP and employment \citep{world_travel_and_tourism_council_world_2026}. Cities, regions and small municipalities invest in place branding, cultural events, heritage promotion and destination marketing to attract visitors, expecting tourism to generate employment, support local businesses, and attract new investment. This strategy is especially appealing in places with limited industrial bases, where tourism may provide one of the few available sources of external demand. Yet the local consequences of tourism are not necessarily straightforward. While higher visitor demand may stimulate hospitality services, it may also shift economic activity away from higher-productivity sectors, increase pressure on local infrastructure and raise housing costs if residential units are converted into tourist accommodation.

Despite its policy importance, credible causal evidence on the local effects of tourism remains limited. The main identification challenge is that tourism exposure is endogenous as tourists self-select into destinations. Destinations that attract visitors tend to differ systematically from other places in their geography, natural amenities, cultural heritage, infrastructure and long-run economic trajectories. These persistent differences confound the effects of tourism with the characteristics that made destinations attractive to tourists in the first place. Moreover, quasi-experimental evidence is difficult to find because tourism exposure rarely changes abruptly for reasons unrelated to destination characteristics.

This paper addresses this identification challenge by exploiting a sudden increase in tourist visibility generated by the TV series \textit{Game of Thrones}. In July 2017, San Juan de Gaztelugatxe, a hermitage on a rocky islet in the municipality of Bakio in the Basque Country, appeared in Season 7 of the series as ``Dragonstone''.
The broadcast provided an exogenous shock to the site's tourism exposure, transforming a primarily regional destination into an internationally recognized site almost overnight. Before the episode aired, Gaztelugatxe received around 80{,}000 visitors per year, mostly from the Basque Country and the rest of Spain. After the broadcast, annual visitors rose to 727{,}995 in 2019 \citep{gomez-limon_elaboracion_2020}. The surge was large enough for the Basque government to introduce an online booking system in 2018 and later cap daily visitors at 1{,}500.

Using a synthetic control method \citep{abadie_economic_2003, abadie_synthetic_2010,ben-michael_augmented_2021}, we estimate the causal effect of this spotlight-tourism shock along four dimensions: tourism development, labor markets, housing markets and residential mobility. We find that the shock drastically increased tourism activity, with tourist accommodation increasing by 600\% and overnight stays by 50\% in the treated municipality.  This increase in tourism activity was accompanied by higher employment in tourism-related sectors. However, these gains were offset by declines in employment in other sectors, resulting in a not significant net effect on employment. The increase in tourism also affected local housing markets, both at the extensive and intensive margins. At the extensive margin, transactions show no significant immediate increase in the treated municipality, with effects materializing only in later periods, while at the intensive margin, the average transaction price increased more rapidly. Finally, we find no evidence of local population displacement, but a marked increase in immigration, with arrivals from abroad nearly doubling. These results suggest that the effects of spotlight tourism extend beyond the tourism sector.

Our findings speak directly to the policy debate on whether \textit{spotlight tourism} — visitor interest triggered by media exposure, viral content or cultural products — can serve as a development tool for smaller and more peripheral regions. This question is particularly policy-relevant because tourism remains highly concentrated within regions, leaving many smaller destinations struggling to capture the economic benefits associated with visitor demand.
%For example, Florence captures the bulk of Tuscan visitors while smaller historic towns in the region struggle to attract overnight stays; Lisbon and Porto absorb the majority of tourists in Portugal, while interior municipalities remain largely bypassed. 
Policymakers increasingly seek to redirect tourism flows towards these less-visited areas through initiatives such as heritage designations, social-media marketing campaigns, and subsidies to low-cost airlines aimed at improving accessibility and attracting new visitors. Yet credible evidence on the economic consequences of tourism for the local communities that host it is scarce. Our setting — a coastal village that went from local to global recognition overnight — provides quasi-experimental evidence on the magnitude, composition and persistence of the local economic response. 

The estimates presented in this paper have direct implications for the design of tourism-diversification policies in peripheral regions. Although spotlight tourism can generate substantial increases in tourism activity, we find no evidence that these gains translate into net employment growth or broadly shared local economic benefits. We document a reallocation of employment across sectors, but find no net increase in overall employment. We interpret this shift toward the tourism sector as a potential drawback of tourism-enhancing policies, since tourism activities tend to be associated with lower productivity and greater seasonality. We also find evidence that increased tourism activity puts pressure on local housing markets, increasing both transaction volumes and prices. These effects have asymmetric welfare implications: landlords and homeowners benefit from asset appreciation, while non-homeowners face declining housing affordability without sharing in the associated capital gains.

This paper contributes to three strands of literature, reviewed in Section~\ref{sec:lit}. The remainder of the paper is organized as follows. Section~\ref{sec:setting} describes the setting. Section~\ref{sec:data} presents the data. Section~\ref{sec:strategy} outlines the empirical strategy. Section~\ref{sec:results} presents main results. Section~\ref{sec:conclusion} concludes.

%% =============================================================================
\section{Literature Review}
\label{sec:lit}
%% =============================================================================

This paper contributes to three strands of the literature.

\paragraph{Economic impact of tourism.}
A growing body of work aims to estimate the causal effects of tourism on local economies. \cite{faber_tourism_2019} exploit the expansion of coastal resorts in Mexico to estimate the local effects of tourism development. They find positive effects on income and employment in nearby areas, although aggregate gains are more modest once general-equilibrium adjustments are taken into account. \cite{gonzalez_more_2025} use positive shocks to the tourism industry caused by terrorist attacks in foreign destinations that compete with Spain for tourists as exogenous variation in the demand for service jobs. They show that tourism increases employment in hospitality and local services, while potentially crowding out activity in tradeable sectors.   \citet{di_giacomo_tourism_2025} studies Italian municipalities between 2004 and 2019 and, in contrast to the positive local spillovers found by \citet{faber_tourism_2019} for Mexico, he finds that greater tourism exposure lowers employment-to-population and labour-force participation rates, reallocates employment from manufacturing and non-tourism services towards hospitality and entertainment, and reduces average per-capita and labour income while raising property income and house prices. He attributes these effects to an inflow of young, low-skilled migrants and to rising land costs that crowd out non-tourism activity, a Dutch-disease-type mechanism through which tourism may hinder long-run development. \cite{di_giacomo_service_2026} use a similar instrumental variable to study the educational response to tourism expansion in Italy. They find that an increase in tourism raises the demand for labor in service jobs and decreases college enrollment and completion in the year after the shock.  

Our contribution to this literature is a clean natural experiment that identifies the local, municipality-level causal effect of a sudden and unprecedented tourism demand shock in municipalities with very little prior tourism infrastructure.

\paragraph{Film and television as drivers of tourism.}
An empirical literature documents how media exposure shapes tourist destination choice. \cite{tkalec_effect_2017} show that the early seasons of \textit{Game of Thrones} increased tourist arrivals to filming locations in Croatia. \cite{contu_impact_2022} study the local development effects of film productions, while \cite{nocito_beam_2023} examine the broader relationship between media exposure and tourism demand. 

Existing work in this area has largely focused on changes in tourist arrivals or regional visibility. Much less is known about whether these shocks generate broader economic adjustments within affected localities. This paper extends that literature by studying not only whether tourism increased, but also how local economies responded to that increase. 

\paragraph{Economic effects of major events and cultural shocks.}

Finally, this paper relates to the literature on the economic effects of major cultural events and place-based visibility shocks. Economists have studied whether events such as the Olympics %, World Cups or European Capitals of Culture 
 or similar mass events generate lasting economic benefits for host locations. \cite{bronzini_economic_2020} assess the short- and long-term economic impact of the 2000 Great Jubilee in Italy, finding a significant effect on the employment rate, a shift of the local economy toward less productive sectors, such as construction and services, but no significant effect of urban development on housing prices. On the other hand,
\cite{mancini_all_2021} look at the 2006 Turin Winter Olympics and find a positive effect on municipal per capita debt, while value added per capita, employment rate, trade openness, and house prices show no significant effect. \cite{favero_next_2026} use synthetic difference-in-differences to evaluate the impact of Matera’s designation as European Capital of Culture. The paper finds that being nominated as European Capital of Culture increased firms and workers in the hospitality and construction sectors, as well as income and nightlight activity, but decreased overall unemployment. 

A recurring challenge in this literature is endogenous selection: host cities are often chosen precisely because they are already growing, investing or politically salient. Moreover, these events are typically accompanied by substantial public investment and long anticipation periods, making it difficult to separate the effect of increased visibility from the effect of concurrent infrastructure expansion. The \textit{Game of Thrones} shock studied in this paper differs from these settings in important ways. The filming-location decision was external to local economic trends, generated little anticipation and was not associated with major public investment before the shock occurred. The resulting increase in tourism therefore provides a cleaner setting in which to study how a sudden rise in global visibility affects small local economies.

%% =============================================================================
\section{Setting}
\label{sec:setting}
%% =============================================================================
The television series \textit{Game of Thrones} is widely regarded as one of the most popular shows in television history. Among its notable filming locations is San Juan de Gaztelugatxe, a historic chapel and coastal landmark in the Basque Country, which was featured in the series as the fictional location of Dragonstone. Its first appearance in the show was in Season 7, Episode 1, ``Dragonstone'', which premiered on July 16, 2017.

The Basque Autonomous Community, hereafter the Basque Country,\footnote{
We will use the term Basque Country to refer to the Basque Autonomous Community, also known in Basque as Euskadi. This should not be confused with the Basque Country region, Euskal Herria in Basque, which encompasses the Basque Autonomous Community, Navarre, Lower Navarre, Labourd and Soule. 
} is an autonomous community in northern Spain composed of three provinces: Biscay, whose provincial capital is Bilbao; Gipuzkoa, whose capital is Donostia-San Sebasti\'{a}n; and Araba/\'{A}lava, whose capital is Vitoria-Gasteiz. It is a well-established summer destination for domestic tourism. The main tourist destinations include the urban centers of Bilbao, Donostia-San Sebastián and Vitoria-Gasteiz; its coastline, with towns such as Zarautz, Hondarribia and Lekeitio attracting visitors for their beaches, coastal landscapes and surfing opportunities; and the scenic natural parks of the Basque interior, such as Gorbeia or Urkiola.

The filming location, Gaztelugatxe, is a 10th-century hermitage on a rocky islet off the Basque coast, connected to the mainland by a stone causeway. The site is located on the western Basque coast in Bizkaia, near the small village of Bakio, roughly 30 kilometers north of Bilbao. Bakio, which has a population of around 2,900, is known primarily for its beaches and belongs to the Mungialdea county, whereas Gaztelugatxe lies in the neighboring county of Busturialdea. Before 2017, Bakio recorded relatively few overnight stays compared with the Basque Country's most visited destinations. Gaztelugatxe itself attracted approximately 80{,}000 visitors annually, most of whom were Spanish or Basque visitors who came for the annual local festivities or as part of a scenic coastal hike.

In 2016, parts of Season 7 of \textit{Game of Thrones} were filmed at Gaztelugatxe. The islet was used as the setting for the fictional ``Dragonstone,'' which appeared in several episodes of the season premiered simultaneously across 170 countries on 16 July 2017. International media coverage linked ``Dragonstone'' to the Basque Country, while social-media posts featuring the location circulated widely. %Figure~\ref{fig:gtrends} plots worldwide Google search interest for ``Gaztelugatxe'': search activity was close to zero before 2017, spiked to the index maximum at the time of the premiere, and remained structurally higher thereafter.
The visitor response was both immediate and substantial. As shown in Table~\ref{tab:visitors}, annual visits to Gaztelugatxe increased by 478\% in 2017 alone and continued to rise in 2018 and 2019. The shock was also persistent. Years later, the site remains among the most visited landmarks in the Basque Country, with access managed through an online reservation system and a daily visitor cap. %of \(1{,}500\) introduced in 2021.
\begin{table}[H]
  \centering
  \caption{Visitors to San Juan de Gaztelugatxe}
  \label{tab:visitors}
  \begin{tabular}{lrr}
    \toprule
    Year & Visitors & Year-on-year growth \\
    \midrule
    2016 & 82{,}589  & ---     \\
    2017 & 477{,}249 & +478\%  \\
    2018 & 559{,}838 & +17\%   \\
    2019 & 727{,}995 & +30\%   \\
    \bottomrule
  \end{tabular}
  \begin{minipage}{\linewidth}
{\scriptsize Source: \cite{gomez-limon_elaboracion_2020}.
}
 \end{minipage}
\end{table}

The decision to film at Gaztelugatxe was made by the series production team and was based on the site's aesthetic qualities, rather than on any economic or policy criterion. No prior economic investment in the municipality by the production was made, and the Basque regional government did not direct public infrastructure spending toward the area in anticipation of the broadcast. 

%% =============================================================================
\section{Data}
\label{sec:data}
%% =============================================================================
We construct a panel dataset at the municipality level by combining several administrative data sources with varying coverage and frequency. This section describes the data sources and their construction.

To measure the local supply response to tourism demand, we use the REATE (\textit{Registro de Empresas y Actividades Turísticas de Euskadi}), the official Basque registry of tourist accommodation establishments. The registry contains the registration date, municipality and establishment type for every licensed tourist accommodation in the region, including hotels,
\textit{viviendas de uso turístico} (VUT, vacation rentals), rural
accommodation, campsites and other tourist establishments. We use registration dates to reconstruct a quarterly municipality-level panel of the stock of tourist accommodation establishments between 2010 and 2025. The main outcome is the number of tourist accommodation units per 1{,}000 residents. The
sample covers 238 out of the 252 Basque municipalities. The missing 14 municipalities are small, inland and have no registered tourist accommodation over 2010--2025; they are concentrated in the low-tourism Tolosaldea county (7 of its 29 municipalities), and their median 2016 population is 272, against 1{,}494 for the region as a whole.

To measure tourism demand, we use administrative data on the number of overnight stays recorded across all types of accommodation. The data are provided by EUSTAT (Basque Statistics Institute) and cover 78 Basque municipalities from 2011 to 2025 at a yearly frequency. The data are provided only for municipalities with sufficient registered tourist accommodation to meet the EUSTAT statistical confidentiality thresholds during the sample period. The missing municipalities are generally small and either have no registered tourist accommodation (and hence no registered overnight stays) or have too few establishments to meet the minimum requirements for statistical confidentiality. We use the number of overnight stays per 1,000 residents across different accommodation types as the outcome.

Labor-market outcomes are constructed from the administrative panel of Social Security affiliations (\textit{afiliados a la Seguridad Social}), obtained from the Spanish Treasury (\textit{Tesorería General de la Seguridad Social}, TGSS). These records contain the universe of workers registered with the Spanish
Social Security system and are available monthly for all Basque municipalities from January 2011 onward. The data include information on both contributory regime and sector of economic activity. The analysis focuses primarily on two regimes: the \textit{Régimen General}, covering salaried workers, and the
\textit{Régimen de Autónomos}, covering self-employed workers. Additional special regimes, including maritime, agricultural, domestic and mining workers, are recorded separately.

The TGSS data also report employment by 2-digit CNAE-2009 industry code, allowing us to construct sector-specific employment measures. %We define \textit{tourism-sector employment} as the sum of affiliations in accommodation, food and beverage services, travel agencies and tour operators, and arts, culture, and recreation. We group all remaining sectors, including manufacturing, construction, wholesale and retail trade, transport, professional services and public administration, into \textit{non-tourism employment}.
We use two distinct aggregations of the underlying records. The first, used to analyze employment by worker regime, aggregates the data to the municipality $\times$ month level across all CNAE codes while retaining the distinction between general-regime and self-employed workers. This panel is monthly and spans the period 2003--2025. The second, used for the sectoral analysis, aggregates to the municipality $\times$ CNAE-2digit level and retains only December observations for the period 2011--2025. The December restriction is imposed because municipality-level CNAE breakdowns prior to 2022 are only systematically available for the end-of-year reference month, making annual December snapshots the consistent unit of observation across the full sample period. For control and spillover municipalities, December tracks the annual average closely (ratios within ±5\%). For Bakio, December falls 4–16\% below the annual tourism employment average, suggesting seasonal concentration of tourism in summer months. Consequently, the sectoral estimates capture effects on sustained employment rather than temporary, fixed-term, or seasonal employment that may occur only during the summer months.

Another caveat of the TGSS data is that, to protect statistical confidentiality, the TGSS suppresses cells containing fewer than five affiliated workers, replacing the exact count with an interval $[1, 4]$. We adopt a conservative strategy and assign the lower bound of the interval, 1, to all suppressed cells. Sensitivity checks using the upper bound, 4, yield quantitatively similar results. Censoring is negligible in the regime-level panel for all outcomes analyzed, but it is more prevalent in the sector-disaggregated panel, where individual CNAE $\times$ regime cells are sparse. In this panel, approximately 62\% of municipality–sector–year cells report a lower bound for the total worker count, because at least one regime sub-component (typically self-employed) falls below the 5-worker disclosure threshold. We use the lower endpoint throughout, which assigns 1 to each suppressed sub-component. %Food service is the exception, given that both treated municipalities report exact counts throughout, while in accommodation and arts/culture the self-employed component is suppressed in virtually all pre-treatment cells for Bakio.

We use the Bank of Spain's Microdata on the Real Estate Registry Statistics (PROP) database \citep{banco_de_espana_microdata_2026}, which contains postal code- level information from property registries between 2013 and 2025. We restrict the sample to market sales of urban dwellings, excluding inheritances, donations, foreclosures and other non-market transfers and construct the number of transactions per 1{,}000 inhabitants and the mean transaction prices. 

To study residential mobility, we use the municipal tables of the Spanish Statistical Office (INE) on residential moves, from the \textit{Estad\'{i}stica de Variaciones Residenciales} (EVR). These tables report arrivals (registrations in the municipality) and departures (deregistrations) between 2011 and 2021 for every Basque municipality. We distinguish moves between Spanish municipalities from moves to and from abroad, and we also consider the total of the two.%\footnote{The EVR was replaced in 2022 by the \textit{Estad\'{i}stica de Migraciones y Cambios de Residencia} (EMCR), which is not comparable for international flows: in their overlap year, 2021, EMCR arrivals from abroad are about 17\% and departures abroad about 53\% above the EVR. We therefore restrict the analysis to 2011--2021, so that all outcomes share the same sample and the post-treatment period is 2017--2021.} 
As in the rest of the paper, outcomes are annual counts per 1,000 residents, using the 2016 population as baseline.

Population estimates are obtained from the EUSTAT and are available annually at the municipality level. We use these estimates to construct per-capita measures. We also use administrative data from the INE on overnight stays at the province level for all Spanish provinces, covering the period 1999--2025 at a monthly frequency. Finally, we compute driving-time distances within the Basque Country using the open-source OSRM routing engine. These distances are used to construct descriptive figures.

Table~\ref{tab:data} summarizes the panel structure across outcomes.

\begin{table}[H]
  \centering
  \caption{Data sources and panel structure}
  \label{tab:data}
  \begin{tabular}{llccc}
    \toprule
    Outcome & Source &  Municipalities & Frequency & Period \\
    \midrule
    Overnight stays        & EUSTAT & 78 & Yearly   & 2011--2025 \\
    Accommodation stock    & REATE  & 238 & Quarterly & 2010--2025 \\
    SS affiliations - regime         & TGSS   & 250 & Monthly & 2003--2025 \\
    SS affiliations - sector breakdown      & TGSS   & 250 & Yearly & 2011--2025 \\
    Housing prices/sales   & BdE    & 250 & Yearly    & 2013--2025 \\
    Residential Mobility & INE & 250 & Yearly & 2011--2021 \\
    Population & EUSTAT & 250 & Yearly & 2000--2024 \\
    Overnight stays (national) & INE & 47 provinces & Monthly & 1999--2025 \\
    \bottomrule
  \end{tabular}
  \vspace{-0.1cm}
\begin{minipage}{\linewidth}
{\scriptsize \setstretch{1}\textit{Notes:} The Basque Autonomous Community comprises 252 municipalities as of 2026. As explained in the text, EUSTAT publishes municipal overnight-stay counts only for 78 municipalities and REATE accommodation registry covers 238 municipalities.  The TGSS data cover every Basque municipality except Usansolo, incorporated in 2023, and Arratzu, which the Social Security affiliation series does not report separately (Arratzu was administratively merged with Gernika-Lumo between 1966 and 1994). Therefore, these two municipalities are dropped from the analysis. The national overnight-stays panel covers the 47 Spanish provinces outside the Basque Country, used only in Appendix~\ref{app:regional_shock}.}
 \end{minipage}

\end{table}

%% =============================================================================
\section{Empirical Strategy}
\label{sec:strategy}
%% =============================================================================

Our goal is to examine how the spotlight shock generated by \textit{Game of Thrones} affected smaller municipalities whose economies were not primarily based on tourism. We therefore designate Bakio, a small coastal town located approximately 2 kilometers from Gaztelugatxe,\footnote{Even if Bakio is the closest town to Gaztelugatxe, the hermitage officially belongs to the municipality of Bermeo, which is 11km away.} as the treated municipality. The main objective of our identification strategy is to determine what would have happened in Bakio had Gaztelugatxe never appeared in \textit{Game of Thrones}, i.e., to estimate the counterfactual outcomes for this municipality. To this end, we construct a Synthetic Control  \citep{abadie_economic_2003,abadie_synthetic_2010}, using the remaining municipalities in the Basque Country as potential control units.

\begin{figure}[H]
    \caption{Municipalities in the Basque Country by treatment group}
    \label{fig:map_groups}
    \centering
\includegraphics[width = \linewidth]{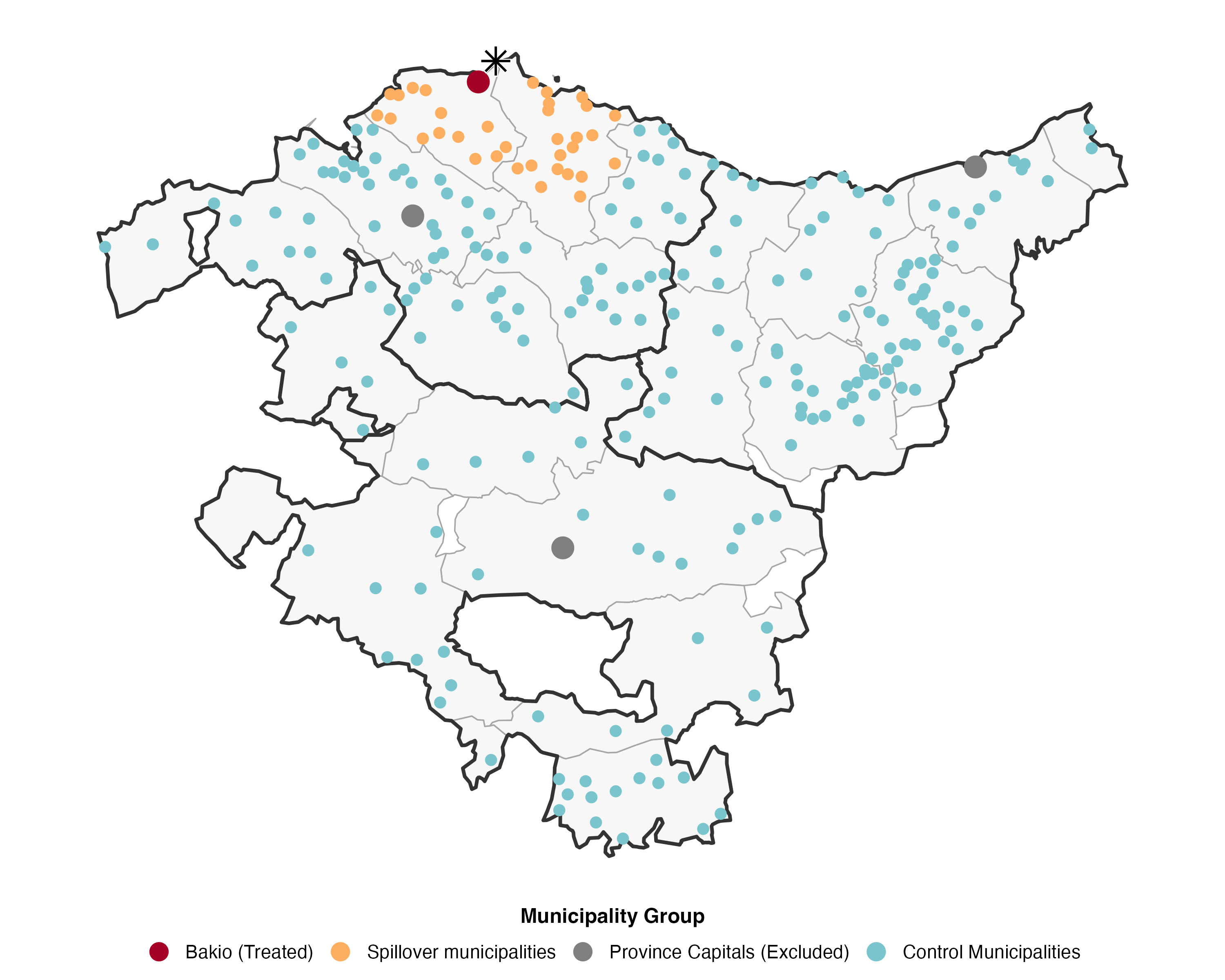}
\vspace{-0.2cm}
\begin{minipage}{\linewidth}
{\scriptsize \setstretch{1}\textit{Notes:} The star marks San Juan de Gaztelugatxe. Red: the treated municipality (Bakio). Orange: municipalities in the same counties as the treated unit and Gaztelugatxe, excluded from the donor pool and used to estimate spillover effects. Blue: the donor pool. Grey: The three provincial capitals, which are excluded from the analysis.}
 \end{minipage}
\end{figure}

We exclude two types of municipalities from the control group. First, we exclude the provincial capitals, which differ structurally from the treated municipalities along several dimensions \citep{abadie_using_2021}. Second, we exclude municipalities in the same counties as Gaztelugatxe and Bakio, which are geographically close to Gaztelugatxe and may therefore be exposed to the shock through spillovers. We do not expect these municipalities to be directly affected by the shock. Nevertheless, even if we believe these other municipalities are unlikely to experience a tourism response comparable to that of Bakio, their inclusion in the donor pool could contaminate the control group if the shock generated spillovers, for instance, through commuting workers. Although these municipalities are not suitable controls, we retain them as a separate group to investigate potential spillover effects. In Appendix \ref{app:spillover}, we use the control municipalities to construct synthetic controls for these excluded municipalities, allowing us to examine whether the shock affected nearby municipalities indirectly. Figure \ref{fig:map_groups} shows the location of the three groups of municipalities together with Gaztelugatxe: the treated municipality (Bakio, in red), the control municipalities (in blue), and the excluded municipalities that may be affected by spillovers (in orange). Figures \ref{fig:map_population} and \ref{fig:map_distance} present similar maps showing population and driving time from Gaztelugatxe, respectively. Finally, Table \ref{tab:summary_stats} reports summary statistics for the three groups. 

Our identification strategy for imputing the missing potential outcomes relies on the following assumptions. First, we maintain the Stable Unit Treatment Value Assumption (SUTVA) throughout the analysis, which rules out interference or spillover effects across units. Second, we assume no anticipation, such that the treatment does not affect outcomes prior to its onset. These assumptions are standard in the causal panel data literature \citep{arkhangelsky_causal_2024}.

\begin{table}[H]
\centering
\caption{\label{tab:summary_stats}Pre-treatment summary statistics }
\centering
\resizebox{\ifdim\width>\linewidth\linewidth\else\width\fi}{!}{
\fontsize{8}{10}\selectfont
\begin{tabular}[t]{lrrrrrrr}
\toprule
\multicolumn{1}{c}{ } & \multicolumn{1}{c}{Treated} & \multicolumn{3}{c}{Spillover municipalities} & \multicolumn{3}{c}{Control municipalities} \\
\cmidrule(l{3pt}r{3pt}){2-2} \cmidrule(l{3pt}r{3pt}){3-5} \cmidrule(l{3pt}r{3pt}){6-8}
Variable & Bakio & Spillover mean & Spillover SD & N & Control mean & Control SD & N\\
\midrule
\addlinespace[0.4em]
\multicolumn{8}{l}{\textbf{Panel A: Covariates}}\\
\hspace{1em}Driving time from Gaztelugatxe (min) & 10.4 & 37.2 & 10.1 & 33 & 85.7 & 25.3 & 210\\
\hspace{1em}Driving time from nearest capital (min) & 31.9 & 34.2 & 8.3 & 33 & 32.3 & 13.0 & 210\\
\hspace{1em}Population & 2,599.7 & 2,974.3 & 4,974.7 & 33 & 6,213.1 & 12,231.7 & 210\\
\addlinespace[0.4em]
\multicolumn{8}{l}{\textbf{Panel B: Tourism outcomes}}\\
\hspace{1em}Overnight stays per 1,000 residents & 2,105.7 & 4,392.8 & 4,399.8 & 11 & 4,659.4 & 6,645.2 & 63\\
\hspace{1em}Accommodation units per 1,000 residents & 2.89 & 1.73 & 2.12 & 34 & 1.50 & 2.38 & 197\\
\addlinespace[0.4em]
\multicolumn{8}{l}{\textbf{Panel C: Labor market outcomes (per 1,000 residents)}}\\
\hspace{1em}Total affiliated workers & 211.3 & 240.1 & 134.5 & 33 & 437.0 & 408.2 & 210\\
\textit{Regimen breakdown} \\
\hspace{1em}Salaried & 118.6 & 138.6 & 136.9 & 33 & 323.9 & 407.5 & 210\\
\hspace{1em}Self-employed & 83.3 & 86.6 & 18.0 & 33 & 98.8 & 41.5 & 210\\
\textit{Sector breakdown} \\
\hspace{1em}Accommodation & 1.0 & 2.1 & 4.1 & 33 & 3.1 & 11.5 & 210\\
\hspace{1em}Restaurants & 29.8 & 20.0 & 14.4 & 33 & 19.6 & 25.8 & 210\\
\hspace{1em}Administrative services & 14.5 & 3.3 & 3.8 & 33 & 12.9 & 26.6 & 210\\
\hspace{1em}Culture \& recreation & 1.7 & 1.9 & 2.5 & 33 & 3.0 & 4.7 & 210\\
\hspace{1em}Manufacturing & 2.8 & 57.1 & 114.7 & 33 & 174.9 & 250.9 & 210\\
\hspace{1em}Construction & 8.0 & 17.1 & 17.0 & 33 & 25.6 & 24.2 & 210\\
\hspace{1em}Wholesale & 20.0 & 24.2 & 16.7 & 33 & 56.8 & 111.6 & 210\\
\hspace{1em}Other services & 47.3 & 66.0 & 78.0 & 33 & 74.8 & 100.8 & 210\\
\addlinespace[0.4em]
\multicolumn{8}{l}{\textbf{Panel D: Housing market outcomes}}\\
\hspace{1em}Dwelling sales per 1,000 residents & 5.2 & 1.8 & 2.5 & 33 & 4.0 & 3.9 & 210\\
\hspace{1em}Mean sale price (EUR) & 180,844 & 215,713 & 49,724 & 16 & 162,757 & 55,706 & 180\\
\hspace{1em}Median sale price (EUR) & 181,267 & 204,101 & 49,885 & 16 & 153,615 & 52,849 & 180\\
\addlinespace[0.4em]
\multicolumn{8}{l}{\textbf{Panel E: Residential mobility (per 1,000 residents)}}\\
\textit{Arrivals} \\
\hspace{1em}Total arrivals & 59.0 & 50.1 & 10.2 & 34 & 45.1 & 14.1 & 210\\
\hspace{1em}Internal arrivals & 55.1 & 45.9 & 9.7 & 34 & 39.5 & 13.0 & 210\\
\hspace{1em}Arrivals from abroad & 3.9 & 4.3 & 2.4 & 34 & 5.6 & 4.1 & 210\\
\textit{Departures} \\
\hspace{1em}Total departures & 57.7 & 44.8 & 10.7 & 34 & 44.3 & 14.5 & 210\\
\hspace{1em}Internal departures & 52.5 & 41.1 & 10.4 & 34 & 39.5 & 13.6 & 210\\
\hspace{1em}Departures from abroad & 5.1 & 3.7 & 2.2 & 34 & 4.8 & 3.3 & 210\\
\bottomrule
\end{tabular}
}
\vspace{-0.1cm}
\begin{minipage}{1\linewidth \setstretch{0.75} } 
{\scriptsize Notes: All variables are averaged over the pre-treatment window. %: 2011--2016. 
Treated: Bakio. Spillover: all other municipalities in the counties of Mungialdea (Bakio's county) and Busturialdea (Gaztelugatxe's county). The difference between the number of municipalities reported in this table and in Table \ref{tab:data} arises because the three provincial capitals are excluded from the analysis. In addition, we exclude three municipalities: Gaztelu, Zierbena, and Ziortza-Bolibar, which enter the data only in 2013, 2010, and 2013, respectively. We exclude these municipalities to obtain a balanced panel, as required in the empirical strategy presented in Section \ref{sec:strategy}. In Panel D, N for price variables counts municipalities with at least one pre-treatment year with enough transactions to satisfy the disclosure rules; sales per 1,000 residents include municipalities with zero recorded sales.}
 \end{minipage}

\end{table}

%Our identification strategy for imputing the missing potential outcomes relies on the following assumptions. First, we maintain the Stable Unit Treatment Value Assumption (SUTVA) throughout the analysis, which rules out interference or spillover effects across units. Second, we assume no anticipation, such that the treatment does not affect outcomes prior to its onset. These assumptions are standard in the causal panel data literature \citep{arkhangelsky_causal_2024}.

SUTVA implies that each municipality's potential outcomes depend only on its own treatment status and are not affected by the treatment status of other municipalities. We consider this assumption plausible because our donor pool consists exclusively of municipalities located farther from Gaztelugatxe (and hence from the treated municipality, Bakio), and therefore less likely to have been directly exposed to the tourism shock. A potential threat to this assumption is that the spotlight shock was sufficiently large to affect tourism even in municipalities located farther from Gaztelugatxe. We consider this possibility unlikely because the Basque Country was already an established tourist destination prior to the appearance of Gaztelugatxe in \textit{Game of Thrones}, and the series did not generate a substantial increase in overall tourism in the region. Our identification strategy exploits variation in the geographic distribution of tourists within the Basque Country: the treatment primarily induced tourists to visit Bakio, a destination that was previously unlikely to be included in the typical tourist itinerary. Thus, the relevant shock is not a general increase in tourism throughout the Basque Country, but rather a geographically concentrated increase in tourist flows toward the area surrounding Gaztelugatxe. In Appendix \ref{app:regional_shock}, we provide evidence supporting this interpretation. Finally, in Appendix \ref{app:robust_bizkaia}, we show the results are robust if we drop all the municipalities that belong to the province of Bizkaia, leaving a larger buffer between Bakio and the closest municipality in the donor pool.

The no-anticipation assumption implies that the spotlight tourism shock did not affect outcomes before the first episode featuring Gaztelugatxe as Dragonstone was broadcast. Although the filming of \textit{Game of Thrones} at Gaztelugatxe was announced prior to the episode's release, we argue that this announcement generated limited public awareness and did not, by itself, substantially increase the site's attractiveness as a tourist destination. The spotlight effect emerged only once viewers could visually recognize Gaztelugatxe in the series and associate the location with \textit{Game of Thrones}. Thus, the episode's broadcast made the site salient to a broad audience and triggered tourists' interest in visiting it. 

Finally, following the standard formulation of the Synthetic Control method \citep{abadie_synthetic_2010,abadie_using_2021, ben-michael_augmented_2021}, we assume the following outcome model:
\begin{equation}
     Y_{it} = \tau_{it} D_{it} +  \boldsymbol{\lambda}_i' \mathbf{F}_t + \varepsilon_{it},
  \label{eq:gsc}
\end{equation}
where $D_{it}$ denotes the treatment indicator, equal to 1 if unit $i$ was treated at time $t$, $\tau_{it}$ captures the heterogeneous treatment effect for unit $i$ at time $t$, $\bs{F}_{t}$ is a vector of unobserved common factors, $\bs{\lambda}_{i}$ is a vector of latent factor loadings, and $\varepsilon_{it}$ is the idiosyncratic zero-mean error term, assumed to be strictly exogenous.\footnote{The model in equation \eqref{eq:gsc} is implicitly modeling potential outcomes as $Y_{it}(0) = \boldsymbol{\lambda}_i' \mathbf{F}_t + \varepsilon_{it} $ and $Y_{it}(1) = Y_{it}(0) + \tau_{it}$.} 

We estimate $\tau_{it}$ using the Ridge Augmented Synthetic Control (Ridge ASC) method proposed by \cite{ben-michael_augmented_2021}, which can be viewed as a bias-correction extension of the original Synthetic Control (SC) method \citep{abadie_economic_2003,abadie_synthetic_2010}. The original SC assigns weights to control units to construct a convex combination that approximates the treated unit's pre-treatment outcomes. Therefore, this approach relies on the existence of a convex combination of control units that can adequately reproduce the treated unit's potential outcome under control. When the synthetic control provides a poor pre-treatment fit, the resulting treatment effect estimates may be biased. Leaving the details to \cite{ben-michael_augmented_2021}, the Ridge ASC addresses this concern by first constructing a standard synthetic control and then using a ridge outcome model to estimate the bias arising from imperfect pre-treatment fit and correct the post-treatment estimates accordingly. When the pre-treatment fit is good, Ridge ASC and SC yield similar estimates. As the fit deteriorates, their estimates increasingly diverge. Another advantage of the Ridge ASC is that ridge regularization also spreads the weights more evenly across control units, reducing the reliance on a small number of donors. However, because Ridge ASC permits extrapolation beyond the convex hull of the donor pool and may assign negative weights to some control units, the resulting synthetic control loses interpretability. %In Appendix \ref{app:other_sc}, we show that our results are robust to using the SC estimator proposed by \cite{abadie_synthetic_2010}.

Regarding inference, we compute the standard error of $\hat \tau_{it}$ using a jackknife procedure, as implemented in the \texttt{augsynth} software of \citet{ben-michael_augmented_2021}. This procedure iteratively drops each control unit assigned a non-zero weight in the original fit, re-estimates the Ridge ASC on the remaining units, and uses the resulting leave-one-out estimates of $\hat \tau_{it}$ to construct a jackknife variance estimate.

As a robustness, we repeat the analysis using the standard Synthetic Control estimator \citep{abadie_synthetic_2010}, the Synthetic Difference-in-Differences estimator \citep{arkhangelsky_synthetic_2021} and the Two-Way Fixed Effect estimator. In Appendix \ref{app:robust_estimators}, we show that our results are robust to the estimator choice.
\begin{comment}
Regarding inference, because our data cover administrative records for the entire population of Basque municipalities rather than a sample drawn from some larger population, the relevant source of uncertainty is not sampling-based but design-based \citep{abadie_sampling-based_2020}. 
Our concern is not that we could have sampled different municipalities, but that the spotlight shock could, counterfactually, have landed on a different municipality than Bermeo and Bakio. We use randomization inference to test the sharp null hypothesis of no treatment effect \citep{fisher_design_1935,imbens_causal_2015}: 
\begin{equation} \label{eq:sharp_null}
    H_{0}: \tau_{it} = 0 \quad \forall i, \forall t \in T_{post},
\end{equation}
where $T_{post}$ denotes the set of post-treatment periods. Randomization or design based inference has the advantage of requiring neither normality assumptions nor asymptotic approximations, making it well suited to the small and finite number of municipalities in our setting. This is the standard mode of inference used in Synthetic Control literature \citep{abadie_synthetic_2010, abadie_using_2021, lei_inference_2025}. We compute p-values for the sharp null in \eqref{eq:sharp_null} the Leave-Two-Out placebo test proposed by \cite{lei_inference_2025}, which has desirable power properties in small samples.
\end{comment}

%% =============================================================================
\section{Results}
\label{sec:results}
%% =============================================================================
This section presents the results from the empirical strategy described in Section \ref{sec:strategy} for the treated municipality, Bakio. Appendix \ref{app:spillover} shows the same set of results for the spillover municipalities. We report both the treatment effect averaged over the post-treatment periods and the event-study estimates for each post-treatment period separately. The event-study plots also report estimates for the pre-treatment periods, allowing us to assess the quality of the synthetic control fit. The tables report standard errors for the estimated average treatment effects, while the event-study plots display 95\% confidence intervals.

Although some of our data are available at a finer temporal frequency than annually, we aggregate all outcomes to the yearly level to avoid fitting seasonal fluctuations rather than longer-term trajectories. This aggregation also facilitates comparisons across outcomes. Finally, we express labor market, accommodation, housing sales and number of in- and out-movers in per-capita terms, measured per 1,000 inhabitants. We use the 2016 population for all periods, so that changes in the resulting outcomes reflect changes in the underlying variables rather than changes in population.

\subsection{Tourism Development}
\label{subsec:res_tourism}

\begin{figure}[H]
    \caption{Event study estimates for tourism outcomes.}
    \label{fig:tourism}
    \centering
    \begin{tabular}{cc}
        \includegraphics[width=0.48\linewidth]{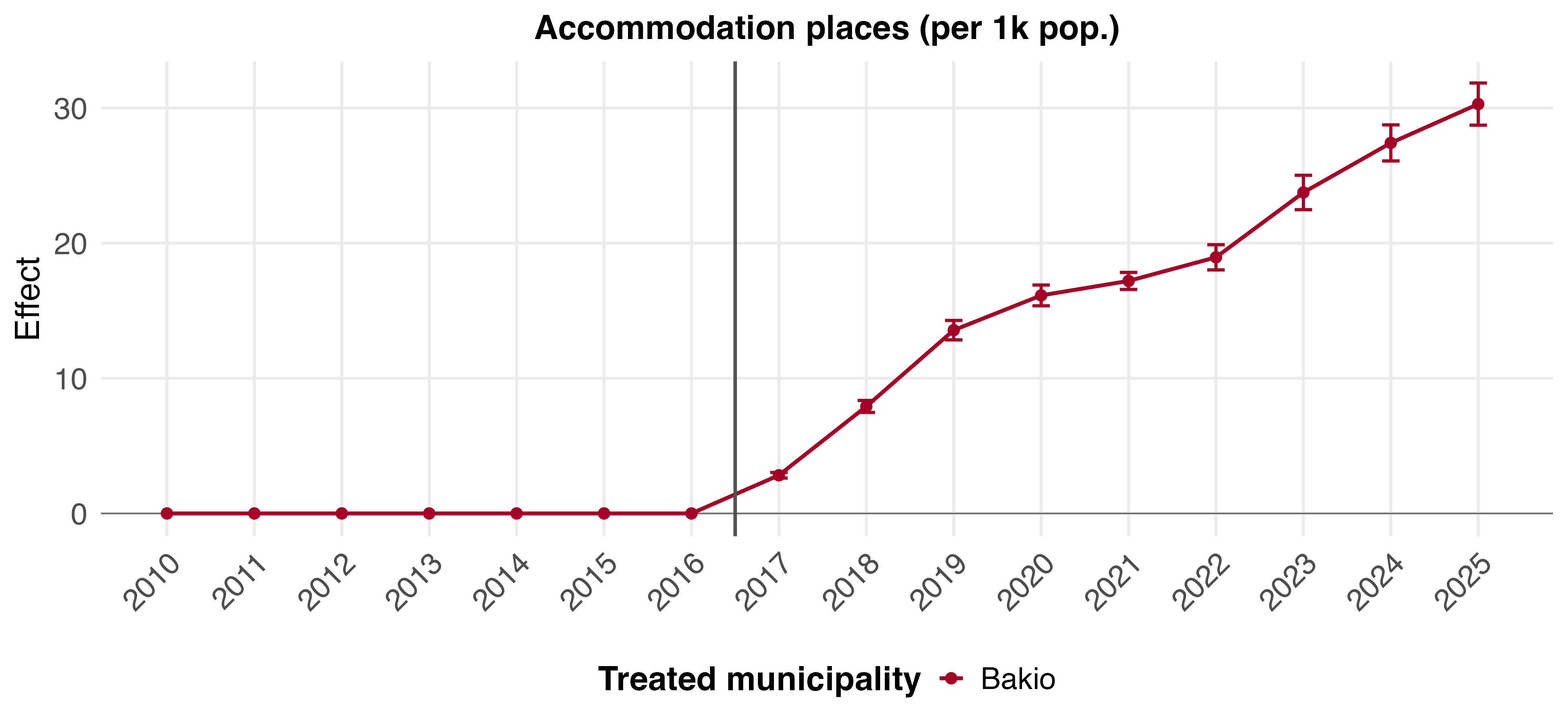}
        &
        \includegraphics[width=0.48\linewidth]{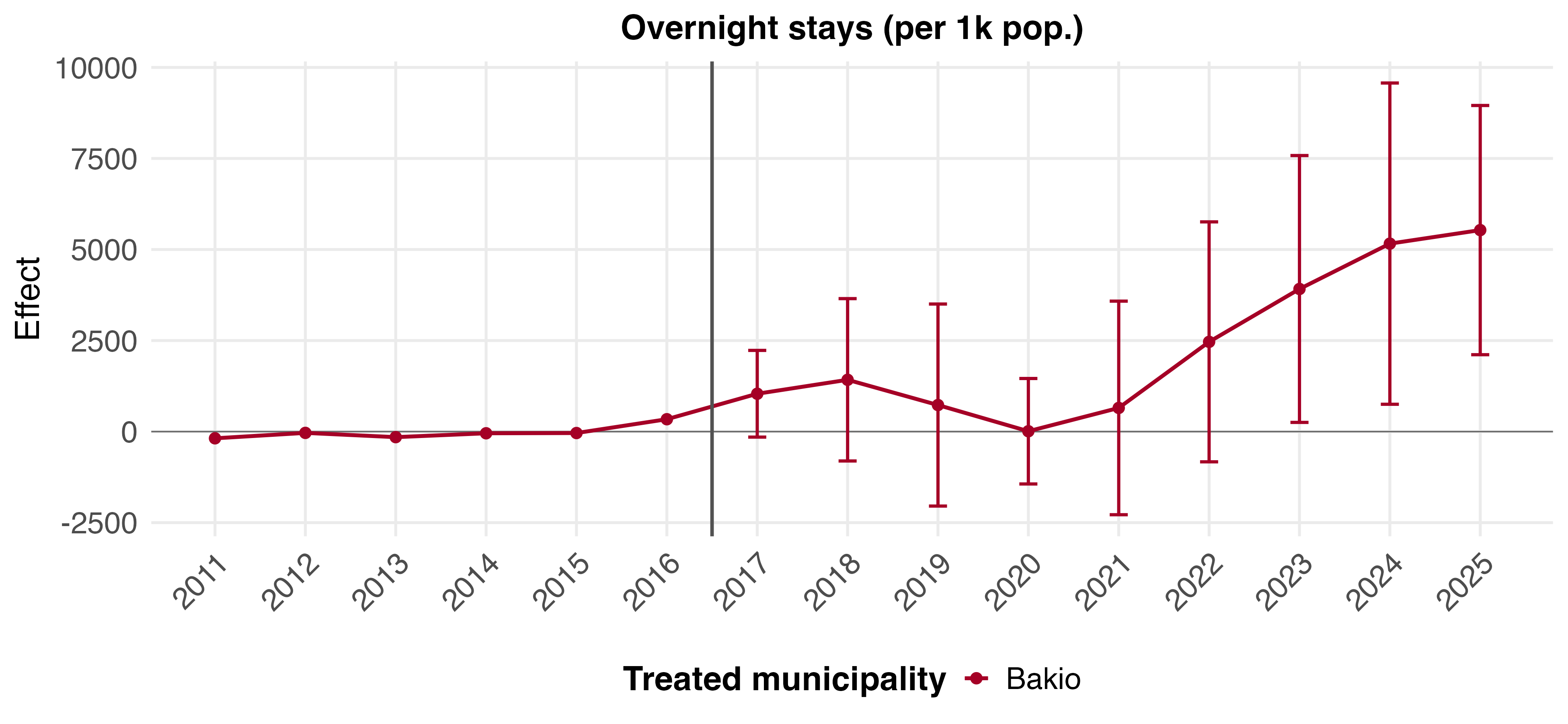}
    \end{tabular}

    \vspace{-0.2cm}
    \begin{minipage}{\linewidth}
        {\scriptsize \setstretch{1}\textit{Notes:} The plots show the difference between the observed outcome and its synthetic control, together with 95\% confidence intervals. Estimates are obtained using the Ridge Augmented Synthetic Control estimator \citep{ben-michael_augmented_2021}, estimated separately for each outcome and each treated municipality using the same donor pool of Basque municipalities. %Other accommodation includes hotels, hostels, camping sites and agroturismos. 
        Figure \ref{fig:tourism_trends} presents the same results in levels rather than as differences between the observed outcome and the synthetic control.}
    \end{minipage}
\end{figure}

Table~\ref{tab:tourism} reports the ATT for the two
main tourism outcomes. Figure~\ref{fig:tourism} shows the corresponding event-study plots while Figure~\ref{fig:tourism_trends} plots the observed outcomes and their estimated synthetic controls in levels over time. As descriptive evidence of the demand shock, recorded visits to Gaztelugatxe rose from roughly 80{,}000 per year before 2017 to 727{,}995 in 2019 --- a 478\% increase in 2017 alone (Table~\ref{tab:visitors}).

The estimated effect on tourist accommodation is large and grows over the post-treatment period. By the end of the sample, the shock is equivalent to roughly seven times the 2016 stock of accommodation in Bakio. Splitting the sample by accommodation type, we find that the positive effect is primarily driven by the conversion of residential properties into short-term tourist accommodation rather than by new hotel construction (see Appendix \ref{app:acco_types} for the corresponding estimates). Taken together, the results indicate a large and persistent expansion of tourist accommodation capacity.

The magnitude of these estimates is substantially greater than the accommodation effects reported in the existing literature. \cite{favero_next_2026} report an 81\% increase in tourist accommodation following Matera's designation as European Capital of Culture, a 2.25 increase versus the counterfactual; in our treated municipality, the \textit{Game of Thrones} broadcast raised accommodation from 10 units in 2016 to around 100 in 2025, against roughly 22 the synthetic control would predict, such that, by the end of the sample period, accommodation facilities are about 4.5 times the counterfactual. This contrast is informative about the nature of the shock we study. The existing literature primarily examines destinations that were already established tourist locations, or where investment in tourism development had been increasing gradually, such as Rome, Turin, Matera or Dubrovnik. Our setting instead captures the sudden exposure of a small municipality in which tourism had previously accounted for only a limited share of economic activity, being thrust into the global spotlight and undergoing a rapid transformation of its tourism structure.

Turning to overnight stays, the effect is large and persistent. Overnight stays start to increase soon after the shock, and the effect falls back to zero only during the pandemic years, 2020 and 2021. It then grows steadily: by 2025, overnight stays are about 5,500 per 1,000 residents above the synthetic control, an increase of roughly 150\% relative to Bakio's 2016 level of 3,713, and statistically significant at the 5\% level. This effect is again substantially larger than those documented in the existing literature. \cite{favero_next_2026} find a 34\% increase in overnight stays following Matera's designation as European Capital of Culture, while \cite{mancini_all_2021} report an increase of 17\% following the 2006 Winter Olympic Games in Turin. The much larger effect we document is consistent with the distinctive nature of our setting: rather than an incremental increase in tourism in an established destination, we capture the sudden emergence of mass tourism in a municipality where tourism had previously played a limited economic role.

\begin{table}[h!]
\centering
\caption{Tourism outcomes: augmented synthetic control estimates}
\small
\label{tab:tourism}
\begin{tabular}{lcc}
\toprule
 & \multicolumn{2}{c}{Bakio} \\
 \cmidrule(lr){2-3}
 & \multicolumn{1}{c}{Baseline (2016)} & \multicolumn{1}{c}{ATT} \\
\midrule
Accommodation places (per 1k pop.) & 3.852 & 17.564*** \\
 &  & (0.348) \\
[3pt]
Overnight stays (per 1k pop.) & 3,713 & 2,325 \\
 &  & (1,427) \\
[3pt]
\bottomrule
\end{tabular}
\vspace{-0.2cm}
\begin{minipage}{0.99\linewidth  \setstretch{0.75}}
{\scriptsize Notes: Estimates are obtained using the Ridge Augmented Synthetic Control estimator \citep{ben-michael_augmented_2021}. The baseline columns report the value of each outcome in 2016. The ATT columns report the average treatment effect over the post-treatment period (2017--2025), with standard errors in parentheses. $^{***}p<0.01$, $^{**}p<0.05$, $^{*}p<0.10$.}
 \end{minipage}
\end{table}

\subsection{Local Labor Markets}
\label{subsec:res_labour}

\begin{figure}[H]
    \caption{Event study estimates for total employment.}
    \label{fig:labor}
    \centering
\begin{tabular}{cc}
\includegraphics[width = 0.6\linewidth]{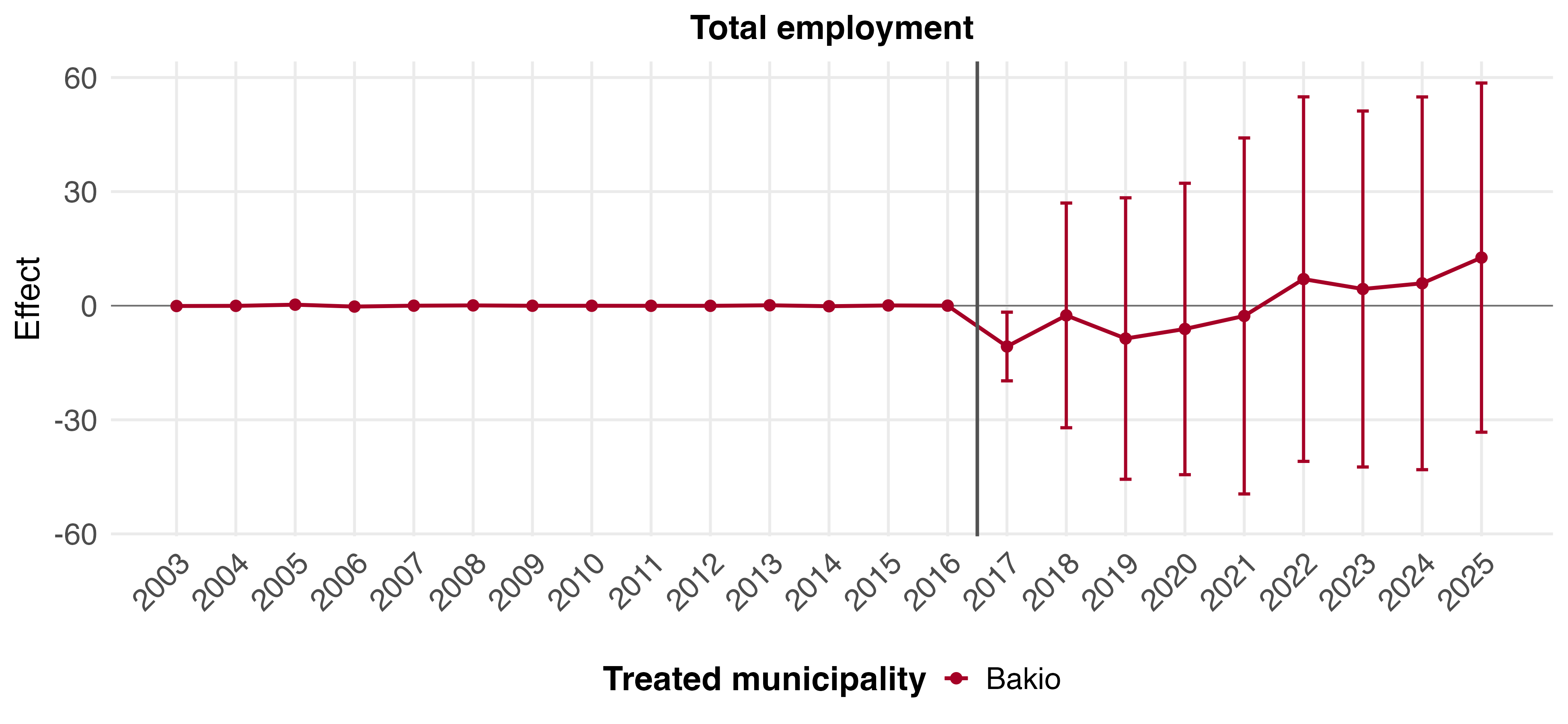}    
\end{tabular}
\vspace{-0.2cm}
\begin{minipage}{\linewidth}
{\scriptsize \setstretch{1}\textit{Notes:} The plots show the difference between the observed outcome and its synthetic control, together with 95\% confidence intervals. Estimates are obtained using the  Ridge Augmented Synthetic Control estimator \citep{ben-michael_augmented_2021}, estimated separately for each outcome and each treated municipality using the same donor pool of Basque municipalities. Figure \ref{fig:labor_trends} presents the same results in levels rather than as differences between the observed outcome and the synthetic control.}
 \end{minipage}
\end{figure}

\begin{figure}[H]
    \caption{Event study estimates for regime breakdown.}
    \label{fig:labor_reg}
    \centering
\begin{tabular}{cc}
\includegraphics[width = 8 cm]{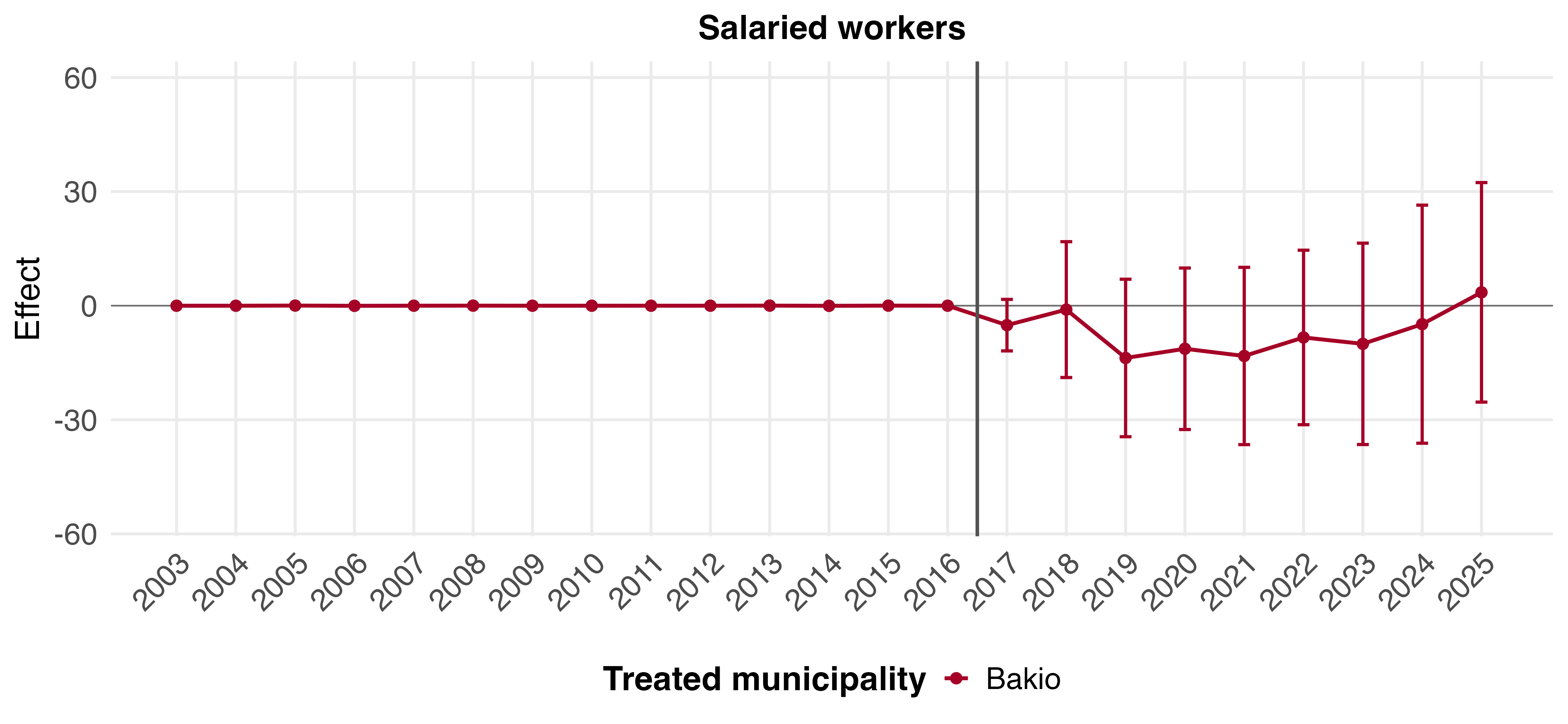}    
&  
\includegraphics[width = 8 cm]{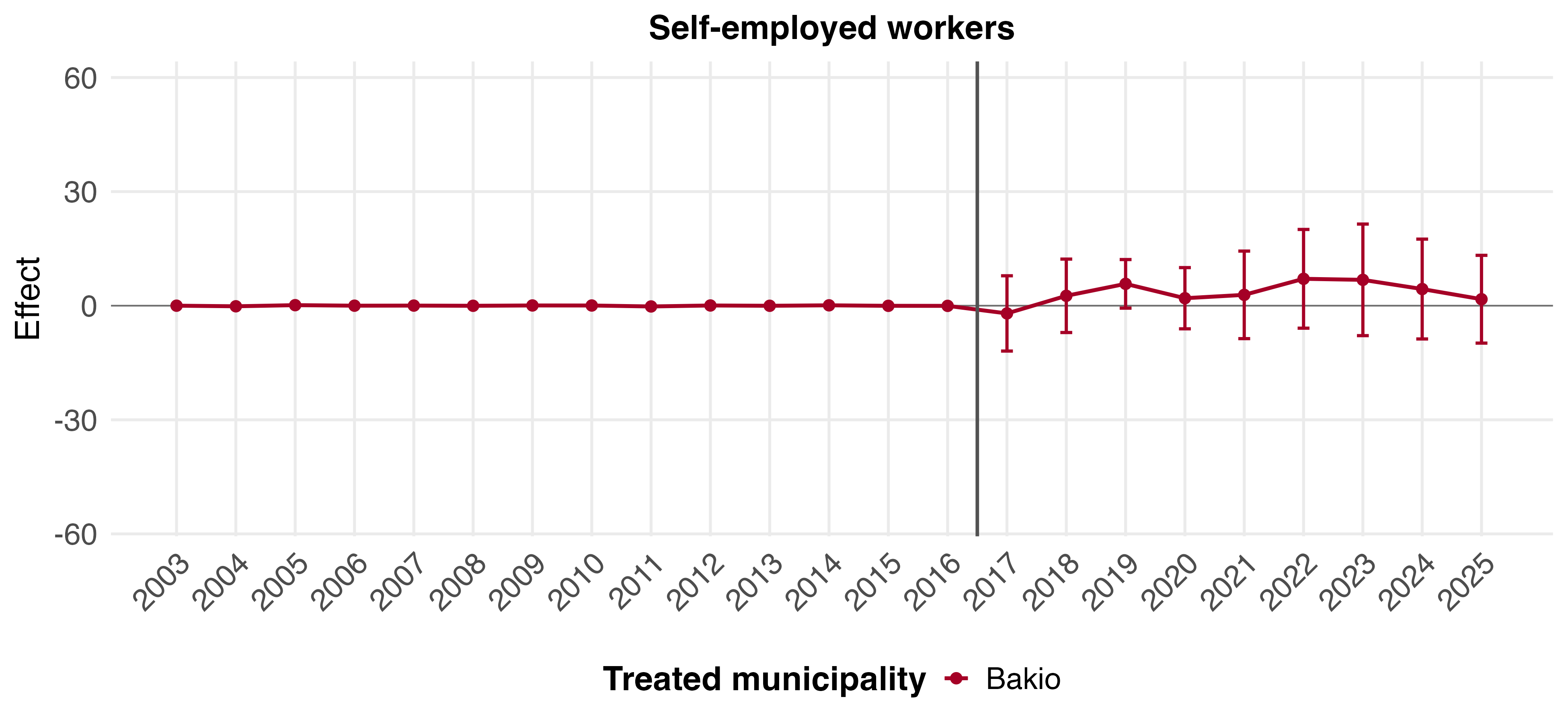} 
\end{tabular}
\vspace{-0.2cm}
\begin{minipage}{\linewidth}
{\scriptsize \setstretch{1}\textit{Notes:} The plots show the difference between the observed outcome and its synthetic control, together with 95\% confidence intervals. Estimates are obtained using the Ridge Augmented Synthetic Control estimator \citep{ben-michael_augmented_2021}, estimated separately for each outcome and each treated municipality using the same donor pool of Basque municipalities. Figure \ref{fig:labor_reg_trends} presents the same results in levels rather than as differences between the observed outcome and the synthetic control.}
 \end{minipage}
\end{figure}

\begin{figure}[H]
    \centering
    \caption{Event study estimates for sector breakdown.}
    \label{fig:labor_sector}

    \begin{tabular}{cc}
        \includegraphics[width=0.48\textwidth]{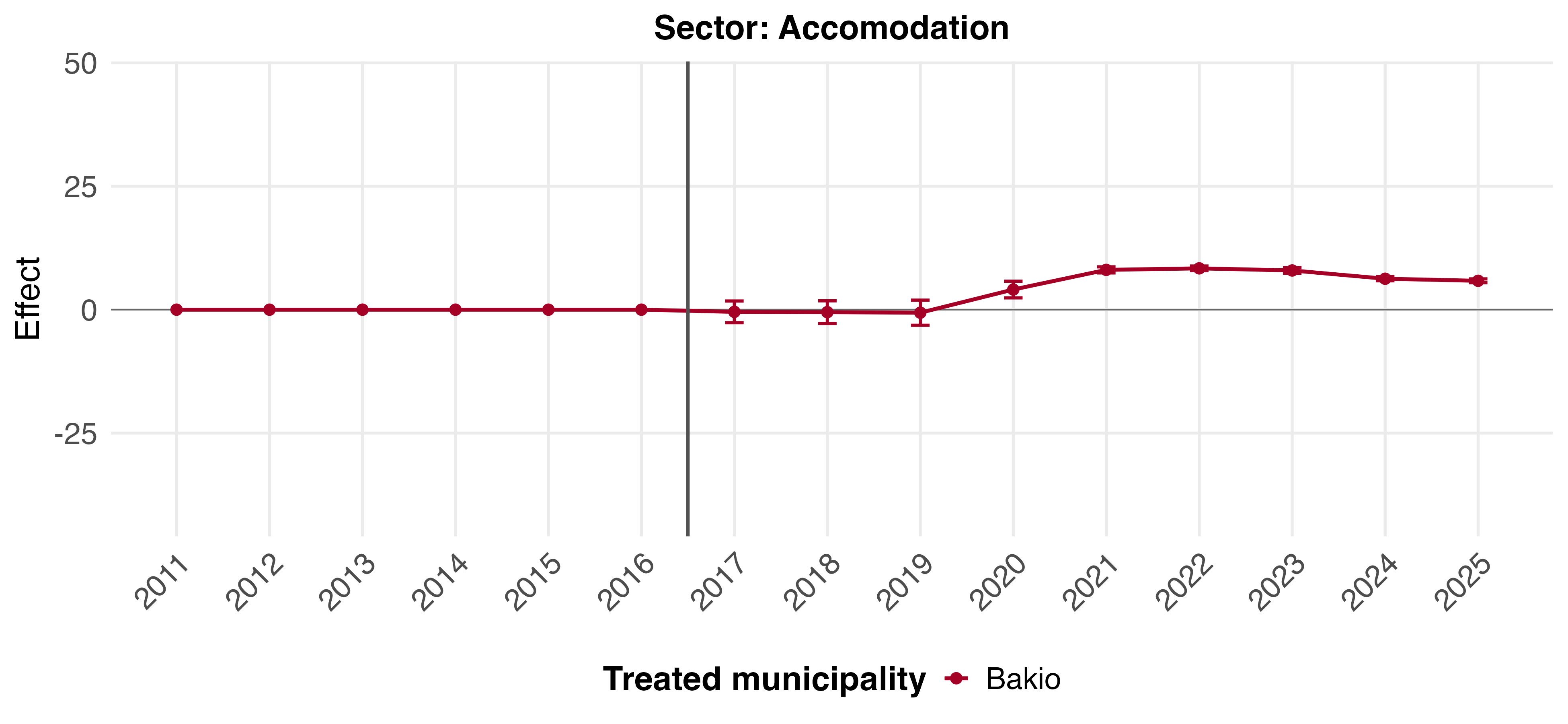} &
        \includegraphics[width=0.48\textwidth]{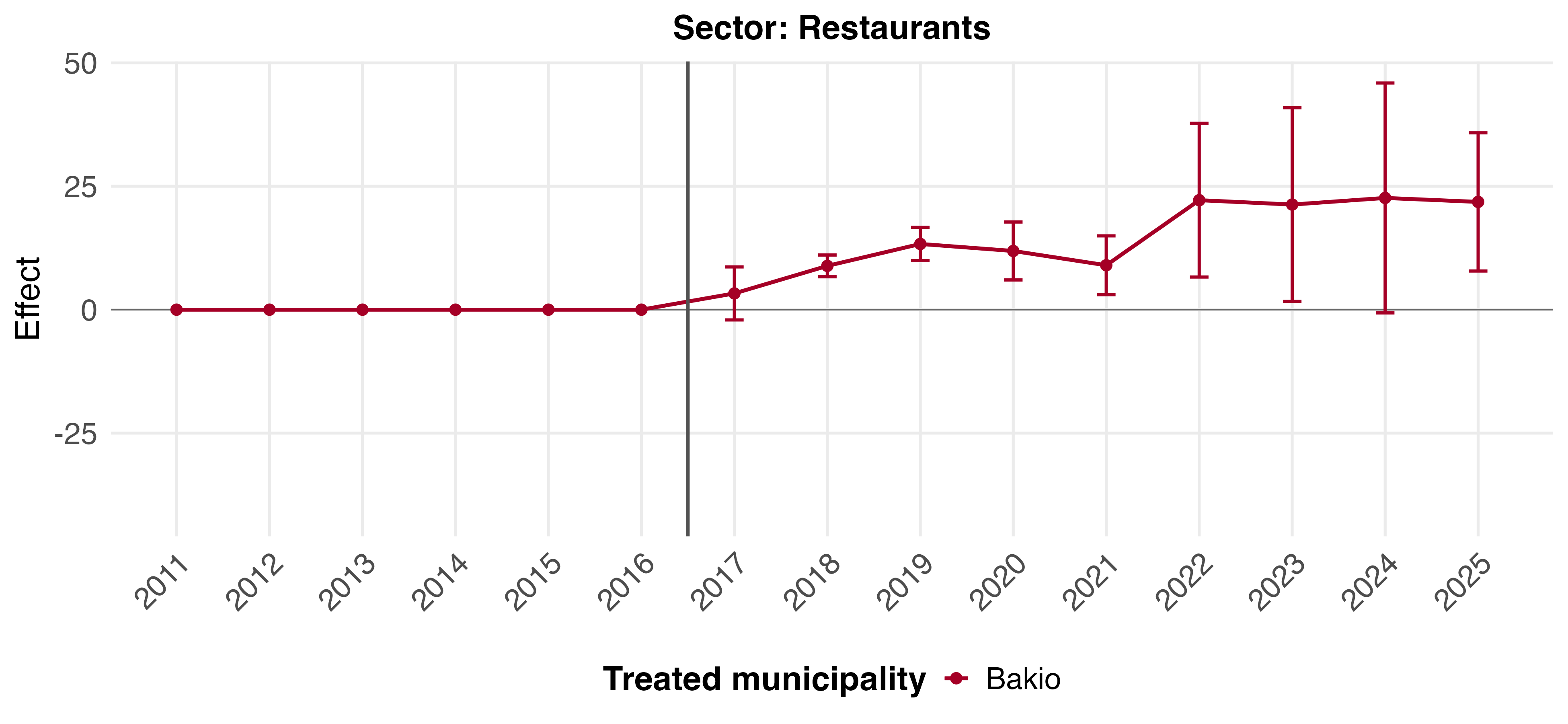} \\

        \includegraphics[width=0.48\textwidth]{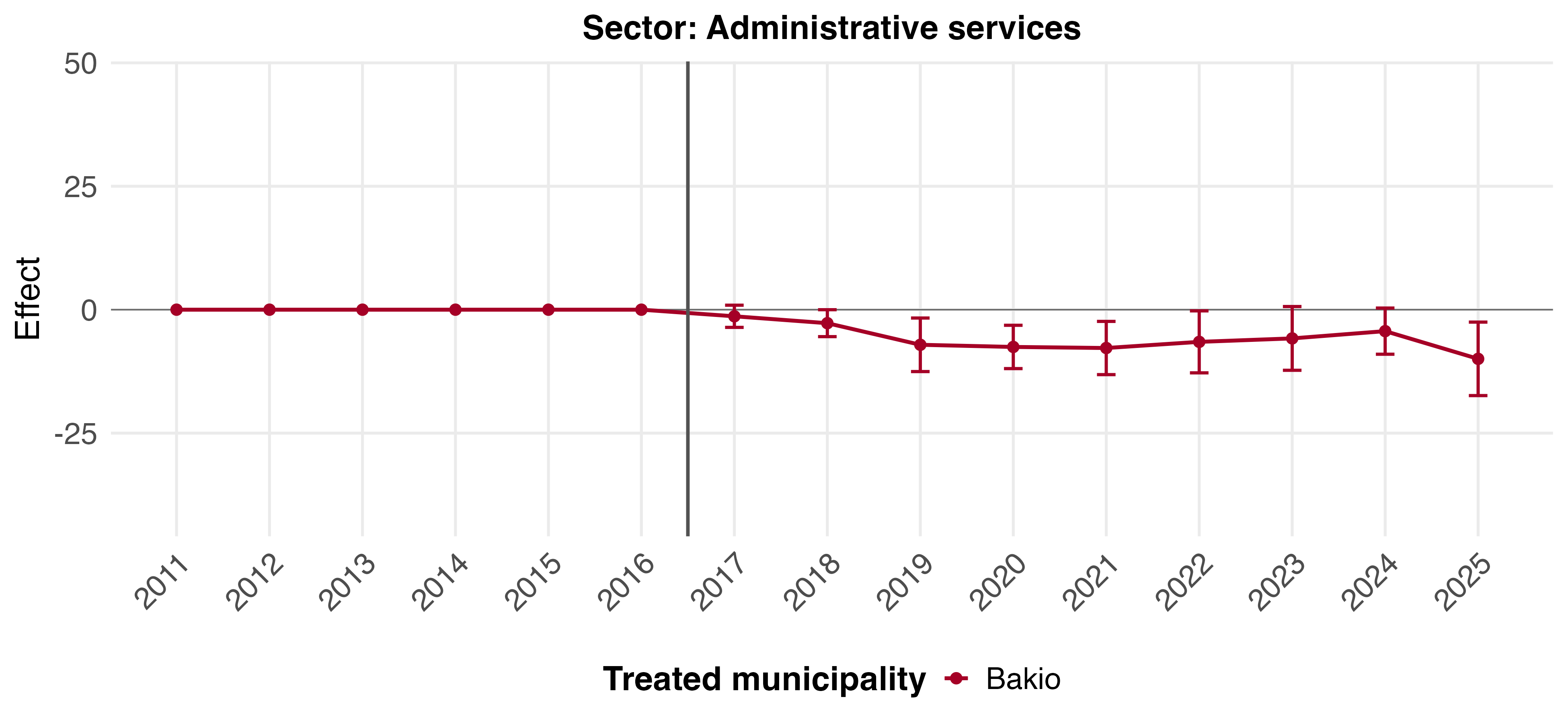} &
        \includegraphics[width=0.48\textwidth]{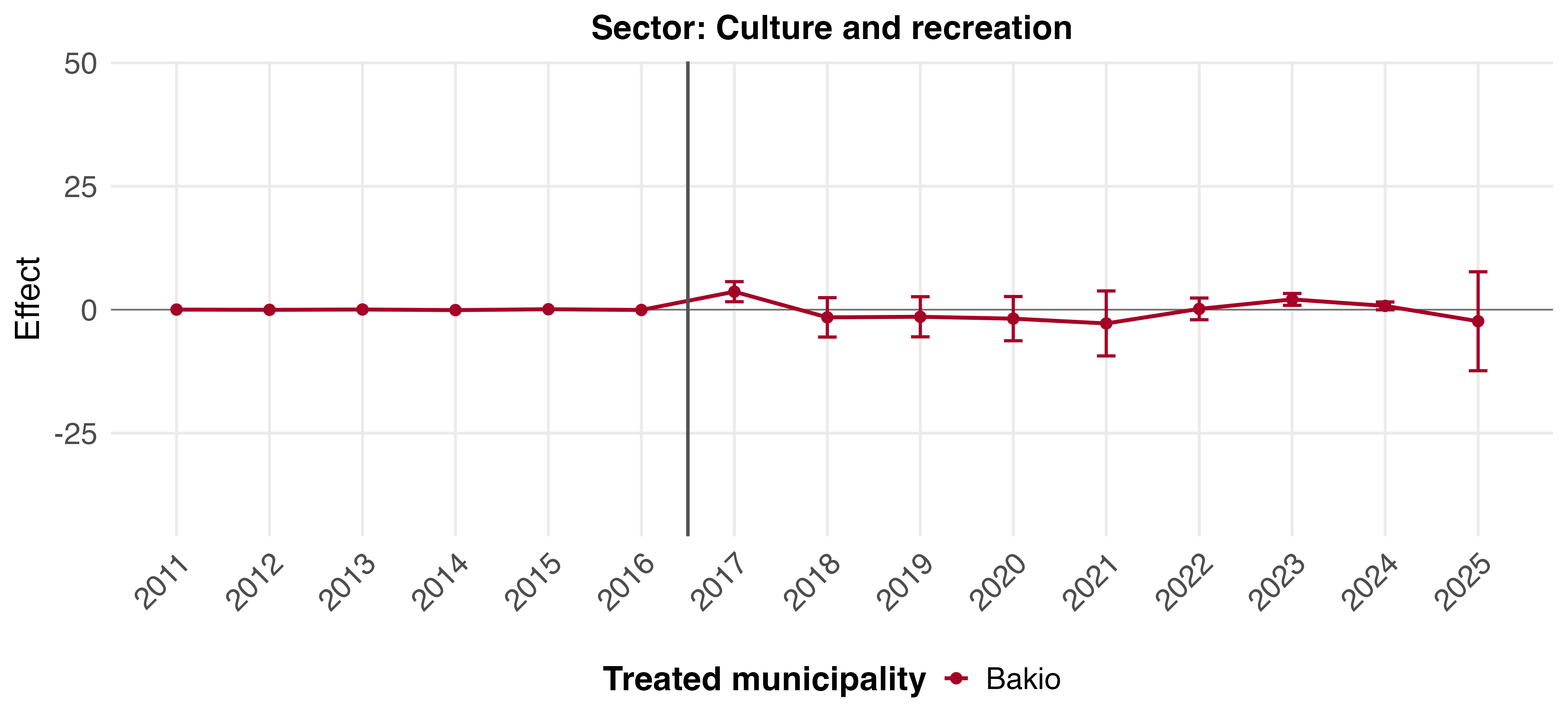} \\

        \includegraphics[width=0.48\textwidth]{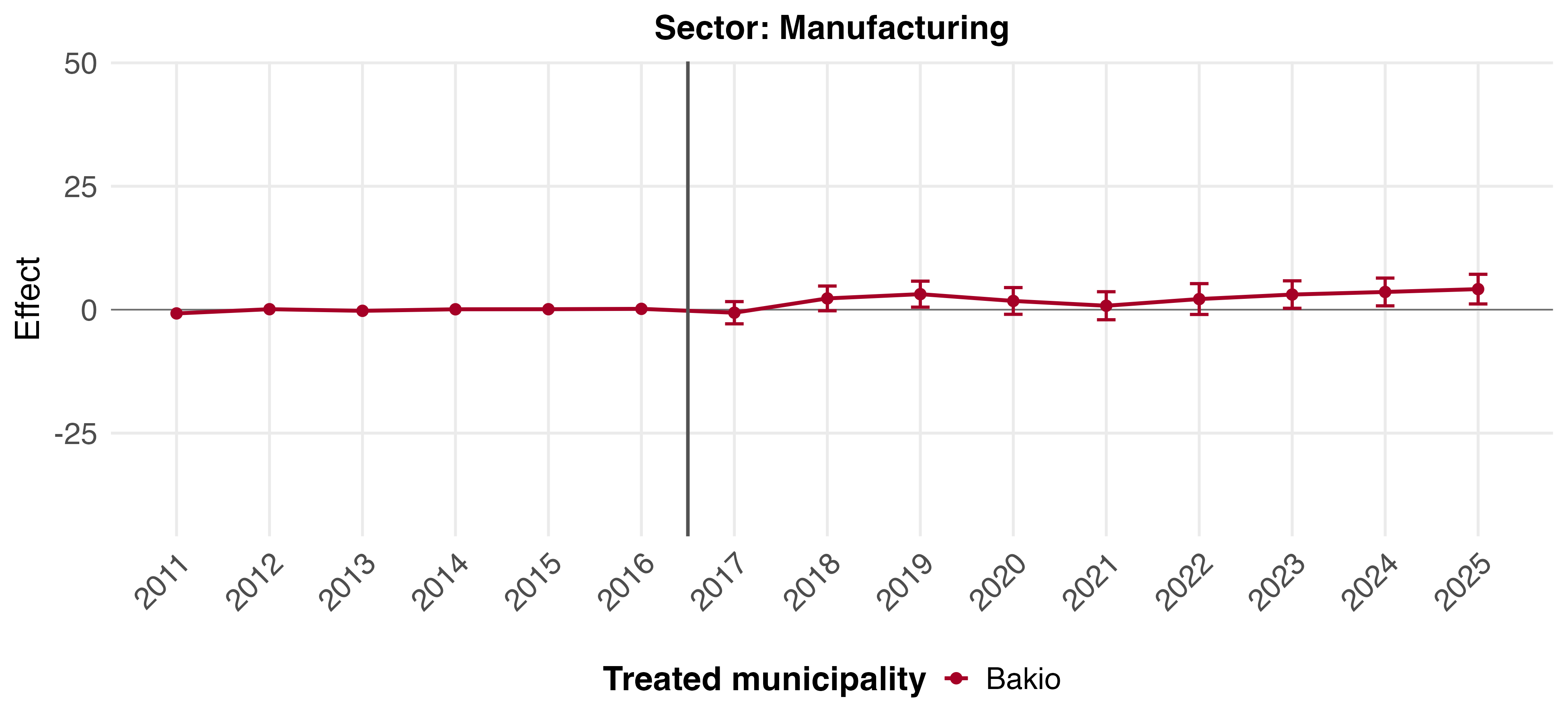} &
        \includegraphics[width=0.48\textwidth]{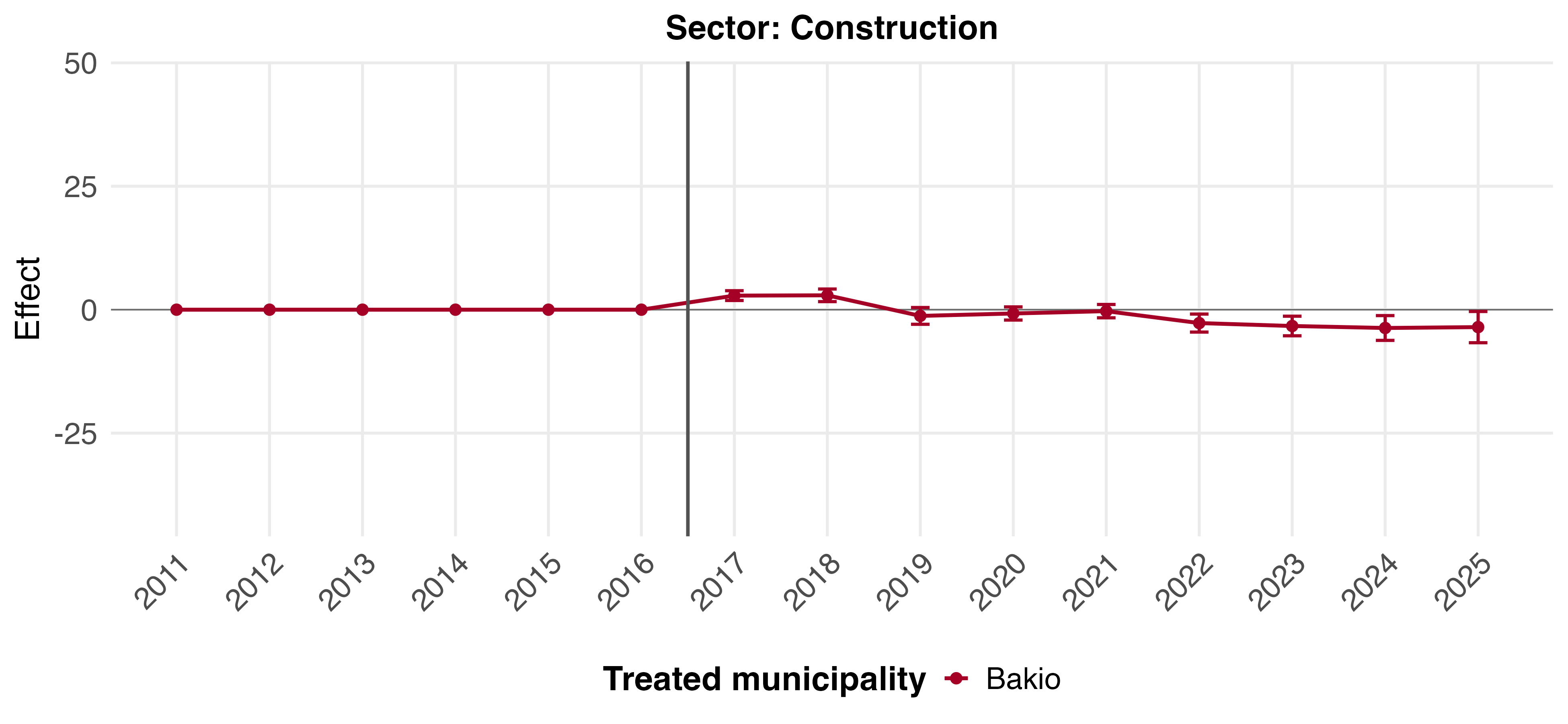} \\

        \includegraphics[width=0.48\textwidth]{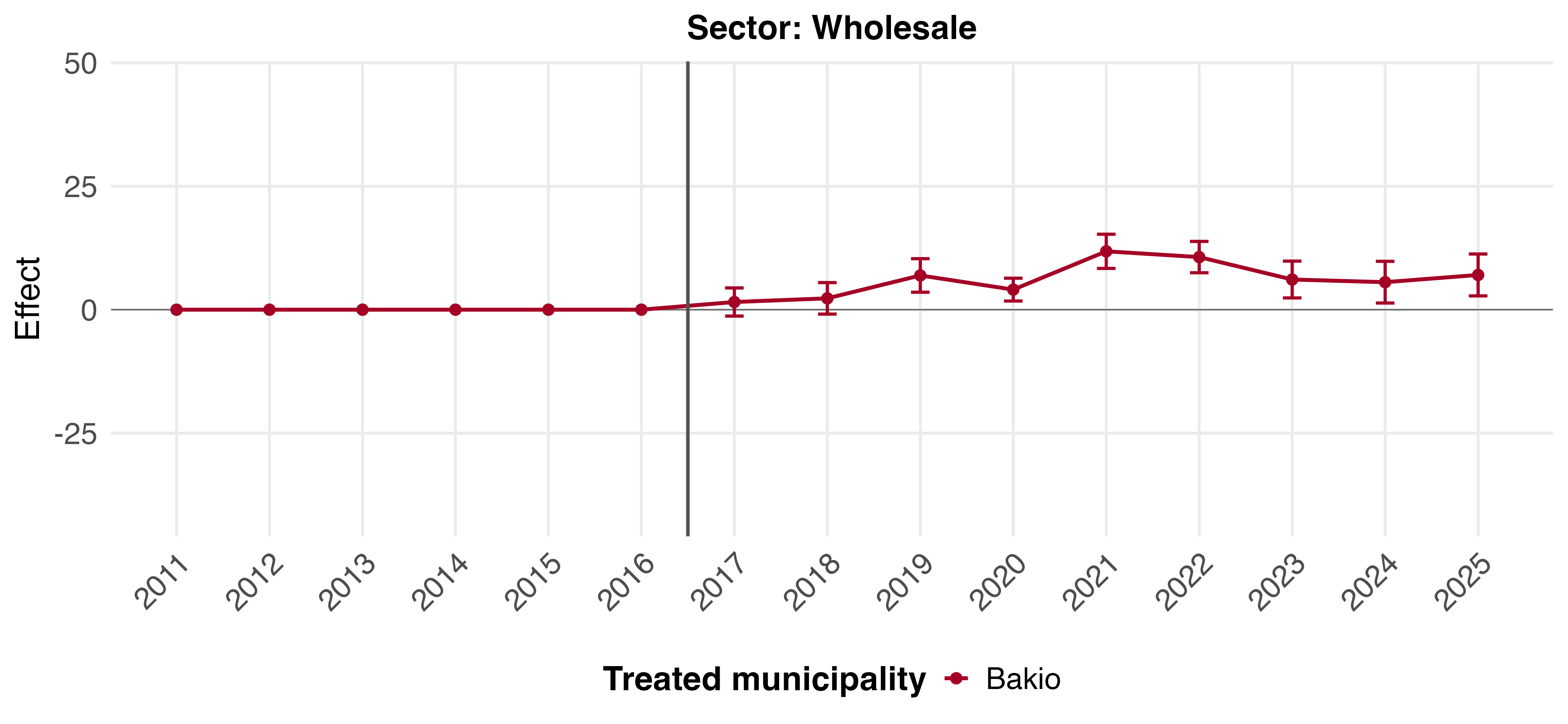} &
        \includegraphics[width=0.48\textwidth]{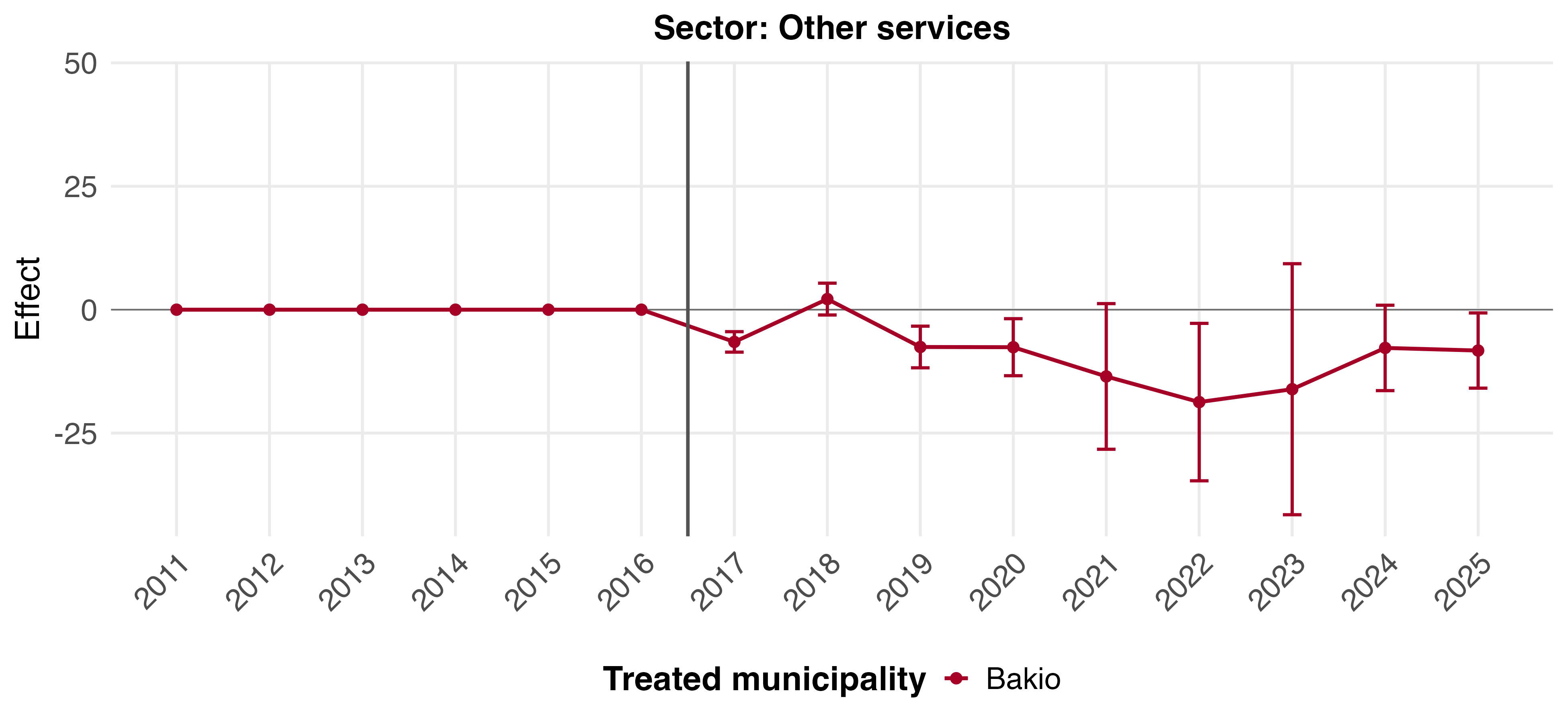}
    \end{tabular}

    \vspace{0.1cm}
    \begin{minipage}{\linewidth}
        {\scriptsize \setstretch{1}\textit{Notes:} The plots show the difference between the observed outcome and its synthetic control, together with 95\% confidence intervals. Estimates are obtained using the Ridge Augmented Synthetic Control estimator \citep{ben-michael_augmented_2021}, estimated separately for each outcome and each treated municipality using the same donor pool of Basque municipalities. Figure \ref{fig:labor_sector_trends} presents the same results in levels rather than as differences between the observed outcome and the synthetic control.}
    \end{minipage}
\end{figure}

The previous section shows that the spotlight shock generated a substantial increase in tourism activity. We next examine whether this increase in visitor demand translated into broader changes in local employment. While higher visitor demand may be expected to create jobs in accommodation, restaurants and leisure services, the extent to which other sectors may be affected is less clear. In this section, we look at the effect of the \textit{Game of Thrones} broadcast on total employment and sector-specific employment. Table \ref{tab:labor} presents the average effects over the post-treatment period, while Figures \ref{fig:labor}, \ref{fig:labor_reg} and \ref{fig:labor_sector} plot the event-study estimates. Figures \ref{fig:labor_trends}, \ref{fig:labor_reg_trends} and \ref{fig:labor_sector_trends} instead plot the observed outcomes and their synthetic controls in levels over time.

One caveat is that our sectoral employment data are December snapshots covering 2011--2025. This timing matters because employment in tourism-related sectors in coastal municipalities is likely to be seasonal, meaning that December observations may miss part of the employment response during the summer months. The sectoral estimates should therefore be interpreted as effects on persistent, year-round employment, abstracting from seasonal and temporary responses in the local labor market.

We find no effect on total employment. This aggregate effect masks substantial reallocation across sectors. Employment in tourism-related sectors, particularly accommodation and restaurants, increases substantially and persistently by 188\% and 44\%, respectively. These gains extend beyond tourism, with positive effects also emerging in wholesale and manufacturing.\footnote{Manufacturing is the only sector where the estimates obtained with the Augmented Synthetic Control do not coincided with the Synthetic Difference-in-Differences estimates, which are close to 0 and not significant. Therefore, we interpret the positive effect on manufacturing employment with caution.} At the same time, employment declines in other sectors, such as construction and services, offset these gains. Thus, while the spotlight shock generated substantial employment reallocation across sectors, the gains in some sectors were largely offset by losses in others, leaving total employment unchanged.

The null effect on total employment also applies to the regimen specific employment, as neither regime shows a significant effect: salaried employment falls by 7.1 per 1{,}000 and self-employment rises by 3.5 per 1{,}000, both well within the confidence interval, so the sectoral reallocation described above nets out across contract types as well as in the aggregate.

These results are consistent with the findings of \cite{mancini_all_2021} and \cite{gonzalez_more_2025}, who find no significant effects on employment. At the same time, they contrast with much of the existing literature, which documents positive employment effects of tourism both within tourism-related sectors and through spillovers to non-tourism sectors, ultimately resulting in a net increase in employment \citep{faber_tourism_2019, bronzini_economic_2020, nocito_beam_2023, favero_next_2026, di_giacomo_service_2026}. 

Taken together, these results show that the increase in tourism did not translate into improved local labor market outcomes. This finding has important policy implications, as it calls into question the assumption that increasing the tourism salience of peripheral areas and attracting more visitors will necessarily generate broad-based local economic benefits. %Policies aimed at promoting tourism in less developed or peripheral areas may result in counterproductive labor market outcomes in the long term.

\begin{table}[h!]
\centering
\caption{Labor market outcomes: augmented synthetic control estimates}
\footnotesize
\label{tab:labor}
\begin{tabular}{lcc}
\toprule
 & \multicolumn{2}{c}{Bakio} \\
 \cmidrule(lr){2-3}
 & \multicolumn{1}{c}{Baseline (2016)} & \multicolumn{1}{c}{ATT} \\
\midrule
\multicolumn{3}{l}{\textit{Panel A: Aggregate employment}} \\
\quad Total & 228.653 & -0.088 \\
 &  & (18.986) \\
[3pt]
\midrule
\multicolumn{3}{l}{\textit{Panel B: Regime breakdown}} \\
\quad Salaried & 130.265 & -7.128 \\
 &  & (9.826) \\
[3pt]
\quad Self-employed & 87.185 & 3.455 \\
 &  & (5.294) \\
[3pt]
\midrule
\multicolumn{3}{l}{\textit{Panel C: Sector breakdown}} \\
\quad Accommodation & 2.311 & 4.336*** \\
 &  & (0.609) \\
[3pt]
\quad Restaurants & 33.513 & 14.922*** \\
 &  & (4.820) \\
[3pt]
\quad Administrative services & 15.408 & -5.904*** \\
 &  & (2.222) \\
[3pt]
\quad Culture \& recreation & 1.156 & -0.365 \\
 &  & (1.822) \\
[3pt]
\quad Manufacturing & 2.696 & 2.262* \\
 &  & (1.342) \\
[3pt]
\quad Construction & 6.934 & -1.091** \\
 &  & (0.555) \\
[3pt]
\quad Wholesale & 24.653 & 6.227*** \\
 &  & (1.472) \\
[3pt]
\quad Other Services & 51.233 & -9.323** \\
 &  & (3.831) \\
[3pt]
\bottomrule
\end{tabular}
\vspace{-0.2cm}
\begin{minipage}{0.99\linewidth  \setstretch{0.75}}
{\scriptsize Notes: Estimates are obtained using the Ridge Augmented Synthetic Control estimator \citep{ben-michael_augmented_2021}. All outcomes are measured as the number of affiliated workers in the corresponding category per 1,000 inhabitants. The baseline columns report the value of each outcome in 2016. The ATT columns report the average treatment effect over the post-treatment period (2017--2025), with standard errors in parentheses. $^{***}p<0.01$, $^{**}p<0.05$, $^{*}p<0.10$.}
 \end{minipage}
\end{table}

%\clearpage 

% Housing Markets

\subsection{Housing Market}
\label{subsec:res_housing}

Tourism expansion may affect local economies not only through employment and business activity but also through the housing market. A growing literature finds that the conversion of residential dwellings into short-term tourist accommodation can raise rents and property values by reducing the housing available to residents and increasing the returns to tourist-oriented use \citep{garcia-lopez_short-term_2020,koster_short-term_2021,duso_airbnb_2024}. More broadly, tourism growth may increase a destination's attractiveness and generate new economic opportunities, stimulating residential property demand and market activity. This can put upward pressure on house prices while also encouraging more housing market transactions, as buyers and sellers respond to changing market conditions and expectations about future property values. We therefore examine both house prices and transactions to capture complementary dimensions of how tourism expansion reshapes local housing markets. 

To measure the effects on real estate markets, we use housing transaction records and restrict the sample to market sales of urban dwellings, excluding inheritances, donations, foreclosures and other transfers that do not reflect market transactions. We consider both the intensive and extensive margins of the housing market. On the extensive margin, we examine the number of transactions per 1{,}000 inhabitants. On the intensive margin, we consider the logarithm of the mean transaction price.

\begin{figure}[H]
    \caption{Event study estimates for housing prices and sales.}
    \label{fig:housing}
    \centering
    \begin{tabular}{cc}
    \includegraphics[width=8cm]{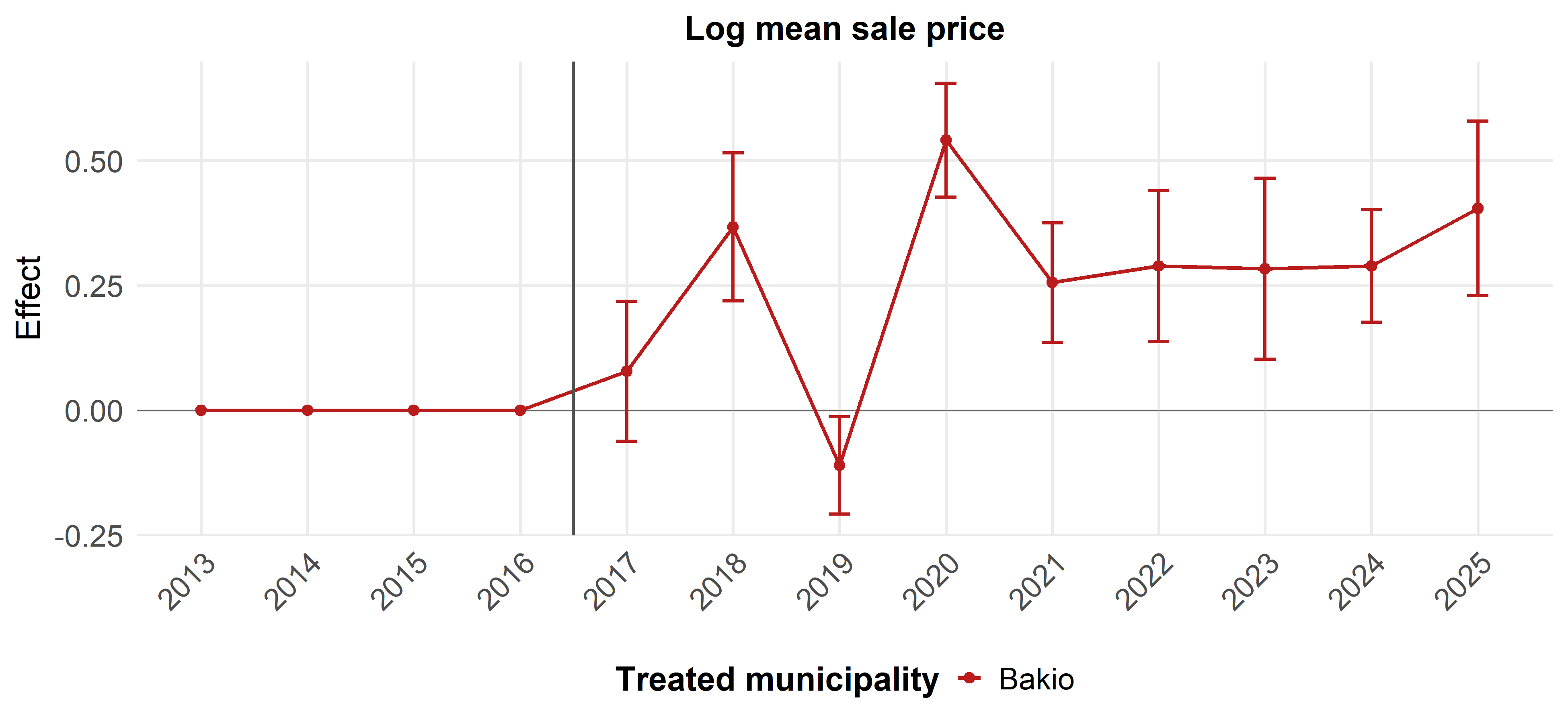} &
    \includegraphics[width=8cm]{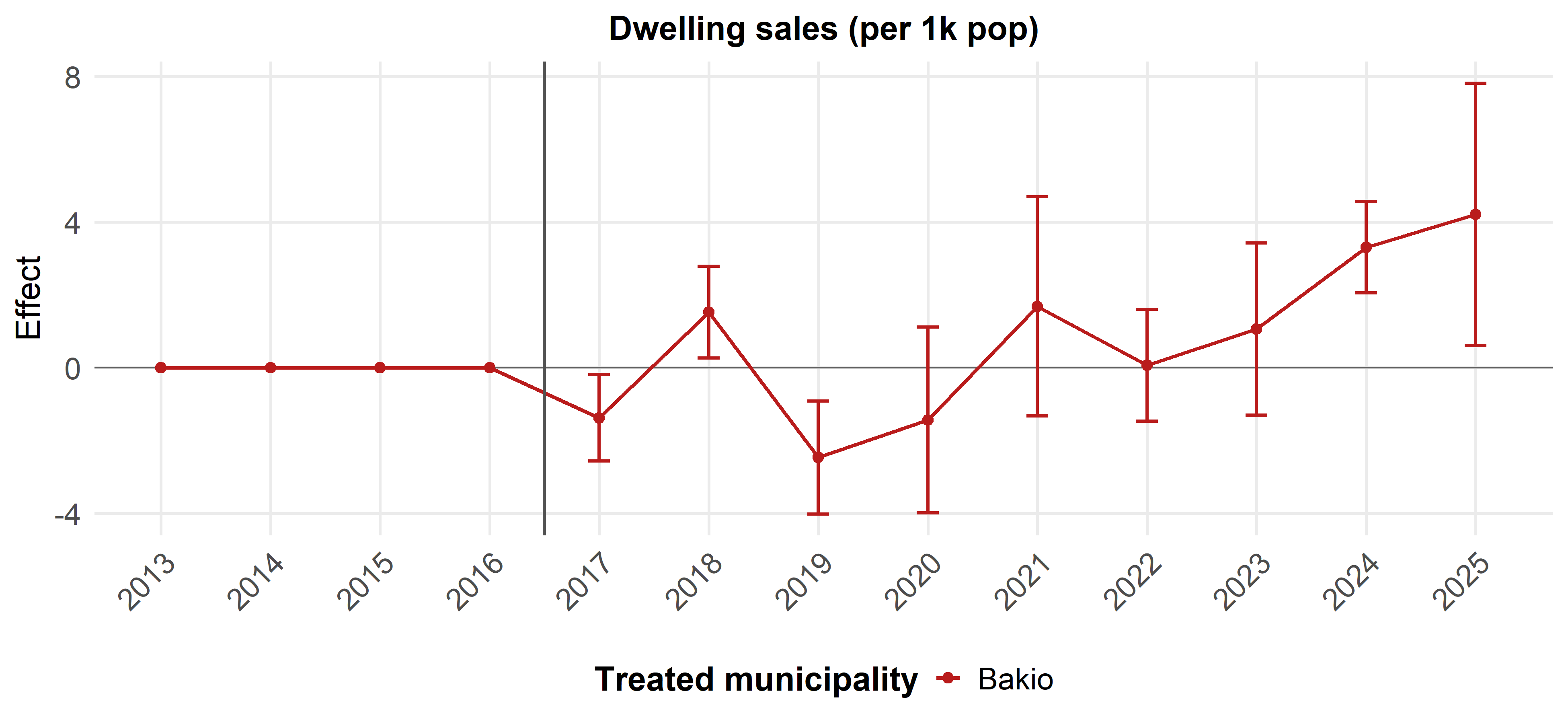} 
    \end{tabular}

    \vspace{-0.2cm}
    \begin{minipage}{\linewidth}
    {\scriptsize \setstretch{1}\textit{Notes:} The plots show the difference between the observed outcome and its synthetic control, together with 95\% confidence intervals. Estimates are obtained using the Ridge Augmented Synthetic Control estimator \citep{ben-michael_augmented_2021}, estimated separately for each outcome and each treated municipality using the same donor pool of Basque municipalities. Figure \ref{fig:housing_trends} presents the same results in levels rather than as differences between the observed outcome and the synthetic control.}
    \end{minipage}
\end{figure}

Table~\ref{tab:housing} reports the average post-treatment effects, while Figure~\ref{fig:housing} shows the corresponding event studies and Figure \ref{fig:housing_trends} displays the same results in levels, showing the observed outcomes alongside their synthetic-control counterparts rather than the differences between them.

\begin{table}[h!]
\centering
\caption{Housing market outcomes: augmented synthetic control estimates}
\footnotesize
\label{tab:housing}
\begin{tabular}{lcc}
\toprule
 & \multicolumn{2}{c}{Bakio} \\
 \cmidrule(lr){2-3}
 & \multicolumn{1}{c}{Baseline (2016)} & \multicolumn{1}{c}{ATT} \\
\midrule
\quad Dwelling sales (per 1k pop.) & 3.852 & 0.732 \\
 & & (0.637) \\
[3pt]
\quad Log mean sale price & 11.824 & 0.266*** \\
 & & (0.047) \\
[3pt]
\bottomrule
\end{tabular}
\vspace{-0.2cm}
\begin{minipage}{0.99\linewidth \setstretch{0.75}}
{\scriptsize Notes: Estimates are obtained using the Ridge Augmented Synthetic Control estimator \citep{ben-michael_augmented_2021}. Sales are counts per 1,000 inhabitants (2016 population). The ATT column reports the average treatment effect over the post-treatment period (2017--2025), with jackknife standard errors in parentheses. $^{***}p<0.01$, $^{**}p<0.05$, $^{*}p<0.10$.}
\end{minipage}
\end{table}

At the extensive margin, the evidence is consistent with a delayed increase in housing demand following the expansion of tourism. The number of transactions increases by $0.7$ per 1{,}000 inhabitants on average, although the estimate is not statistically significant. The event study shows that estimates alternate in sign during the first post-treatment years and become clearly positive only in 2024 and 2025. %The number of mortgages follows a similar pattern, with a positive effect of 3.5 additional mortgages per 1{,}000 inhabitants driven primarily by effects emerging in the post-pandemic years.
At the intensive margin, house prices exhibit a similarly delayed response. The average treatment effect on the log mean transaction price is $+0.27$, with the increase concentrated during and after the pandemic. %By contrast, we find no statistically significant effect on the average loan amount; none of the post-treatment coefficients differs significantly from zero. One possible explanation is a shift in the composition of buyers, with a growing share of investors entering the local housing market. If these investors finance a larger share of their purchases with equity rather than mortgage debt, this compositional shift could help explain why transaction prices increased without a corresponding increase in average loan amounts.

Taken together, the results are consistent with a tourism expansion that gradually increased the economic value of local housing. Tourist accommodation expanded substantially after 2017. The subsequent increase in both housing prices and, eventually, transactions is consistent with stronger demand for local property. The delayed response also suggests that the housing market did not adjust immediately to the initial increase in tourist arrivals.

These results contrast with the evidence from major cultural events, which typically finds no significant effect on local house prices \citep{bronzini_economic_2020,mancini_all_2021}. A potentially important distinction is the persistence of the underlying demand shock. While mega-events generate temporary increases in visitor demand, the \textit{Game of Thrones} shock produced a persistent rise in overnight stays that remained high years after the series ended. The delayed housing response in Bakio suggests that property markets respond not simply to an immediate influx of visitors, but to a lasting change in the expected returns to owning property in a tourist destination. In a municipality with limited housing supply, the persistence of tourism demand may therefore translate into higher property values as residential and tourism-related uses compete for available stock. 

\subsection{Residential Mobility} \label{subsec:res_mobility}

The previous sections show that the shock expanded tourist accommodation sharply, left total employment unchanged but had different sectoral responses, and raised local house prices. In this section, we ask whether the tourism shock affected residential mobility, measured by changes of registered residence into and out of the municipality, as well as total population registered.

\begin{figure}[H]
    \caption{Event study estimates for population.}
    \label{fig:population}
    \centering
\begin{tabular}{cc}
\includegraphics[width = 0.6\linewidth]{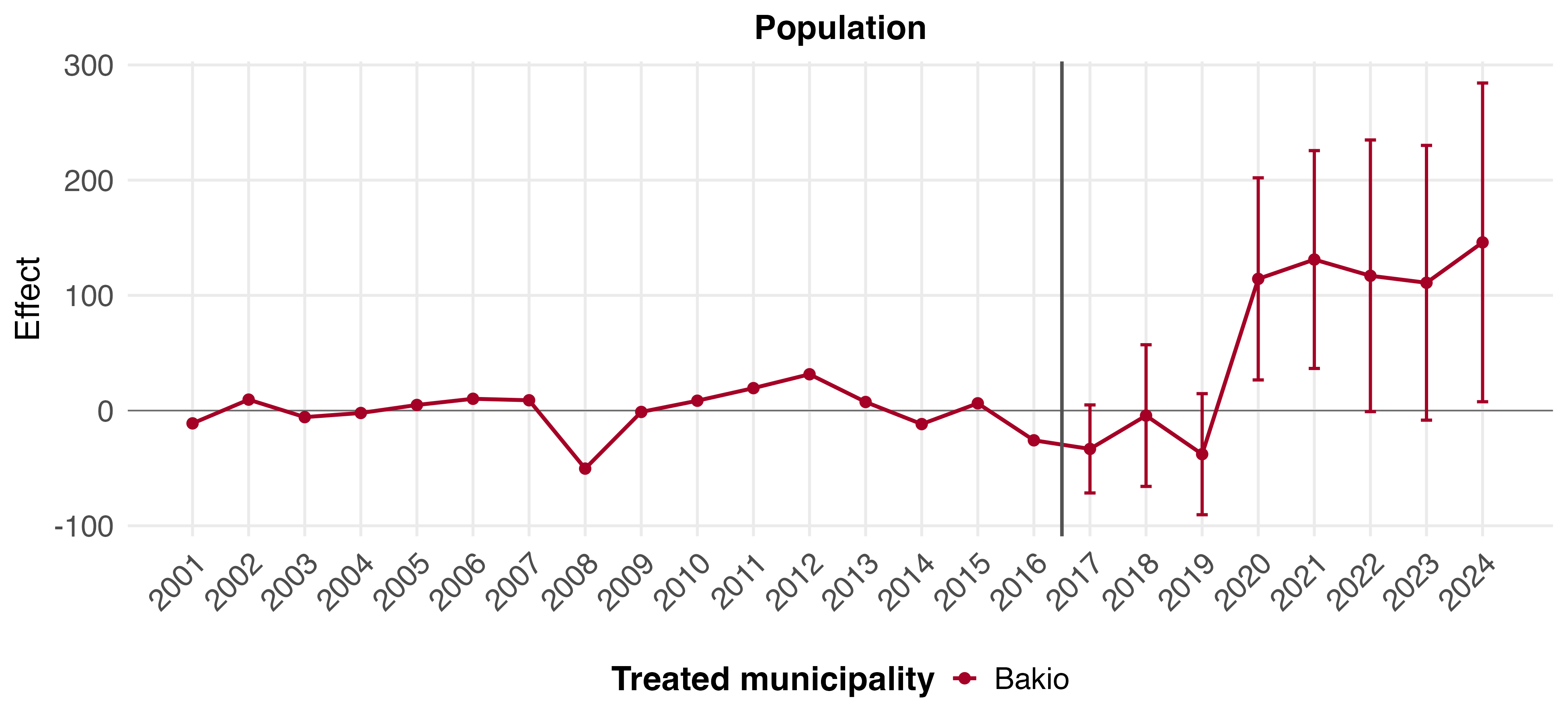}    
\end{tabular}
\vspace{-0.2cm}
\begin{minipage}{\linewidth}
{\scriptsize \setstretch{1}\textit{Notes:} The plots show the difference between the observed outcome and its synthetic control, together with 95\% confidence intervals. Estimates are obtained using the  Ridge Augmented Synthetic Control estimator \citep{ben-michael_augmented_2021}, estimated separately for each outcome and each treated municipality using the same donor pool of Basque municipalities. Figure \ref{fig:labor_trends} presents the same results in levels rather than as differences between the observed outcome and the synthetic control.}
 \end{minipage}
\end{figure}

\begin{figure}[H]
    \caption{Event study estimates for residential mobility.}
    \label{fig:mobility_es}
    \centering
    \setlength{\tabcolsep}{1pt}
    \begin{tabular}{cc}
    \includegraphics[width=0.40\linewidth]{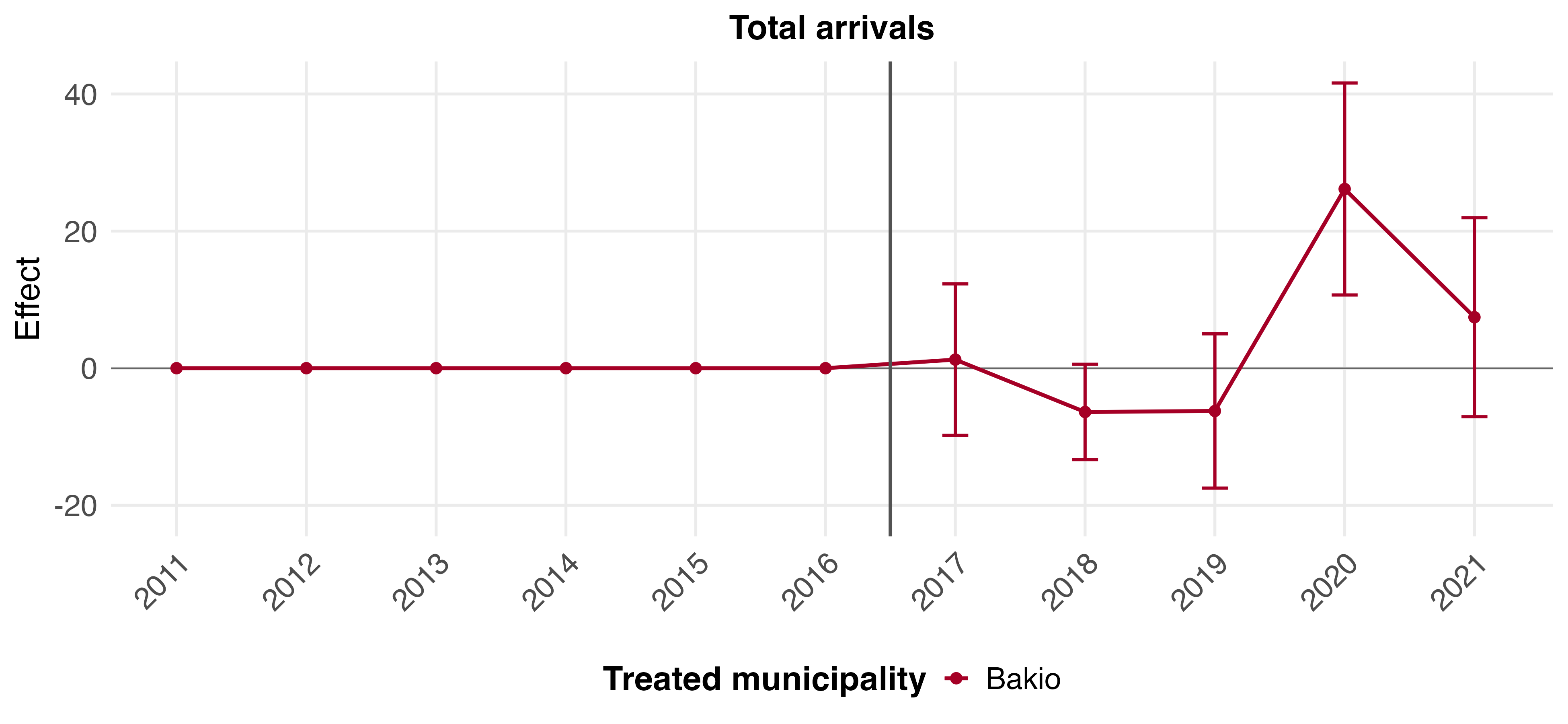} &
    \includegraphics[width=0.40\linewidth]{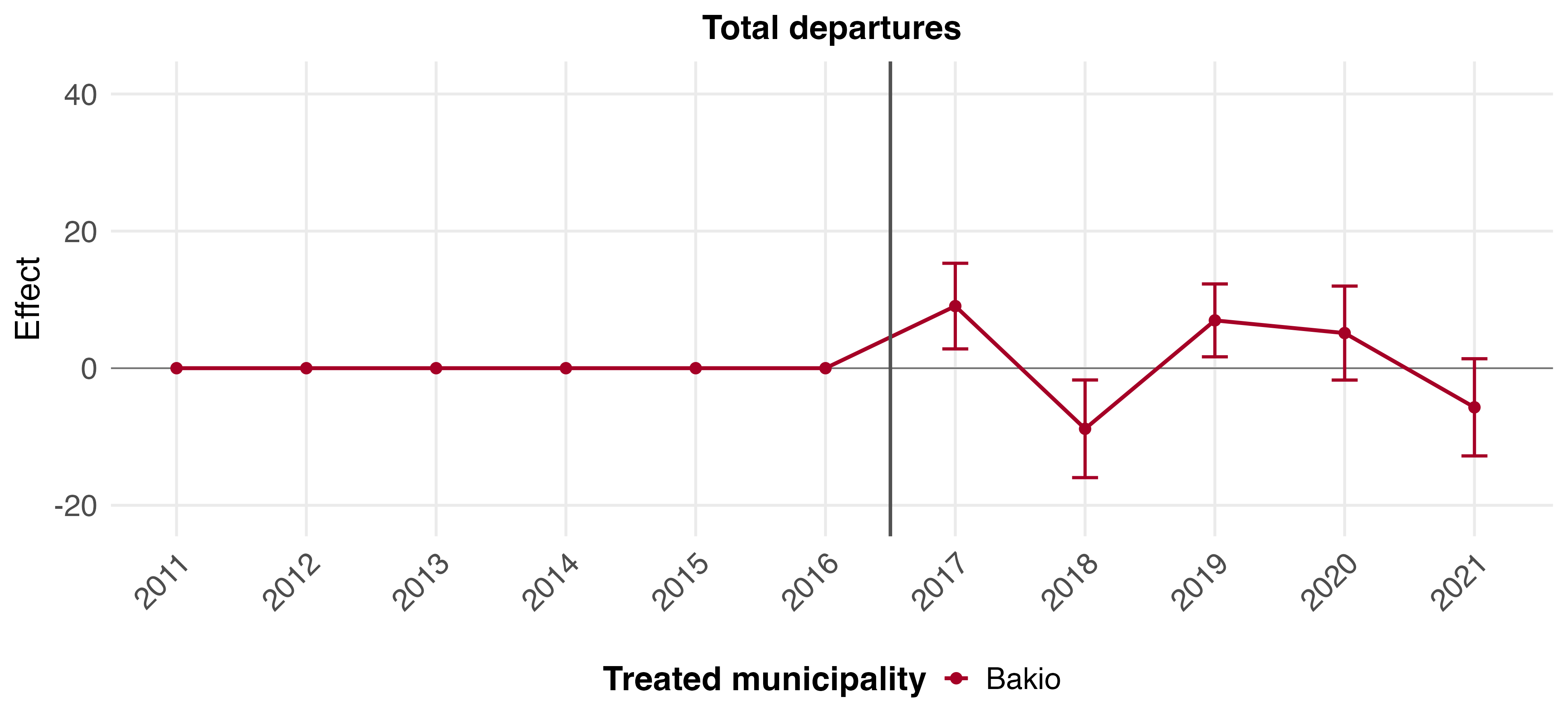} \\
    \includegraphics[width=0.40\linewidth]{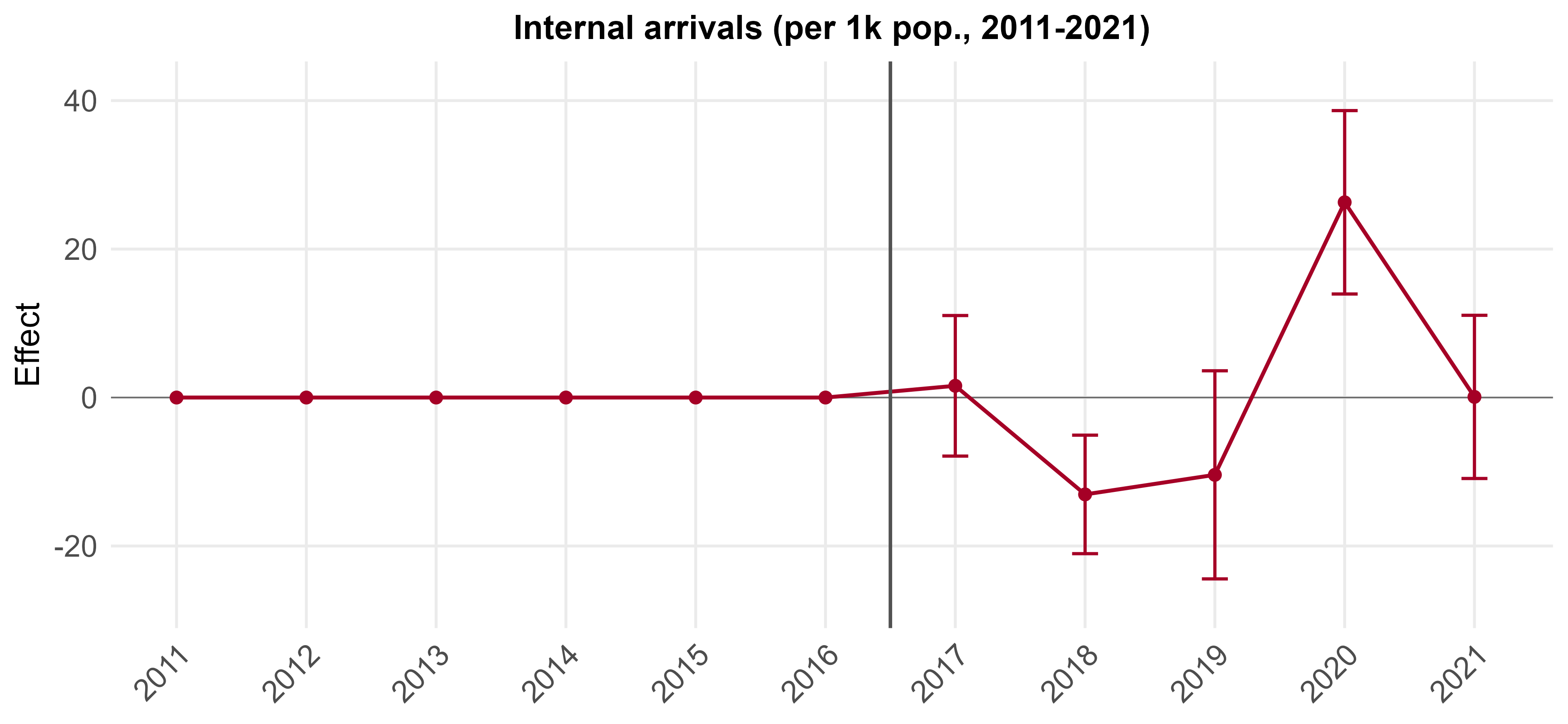} &
    \includegraphics[width=0.40\linewidth]{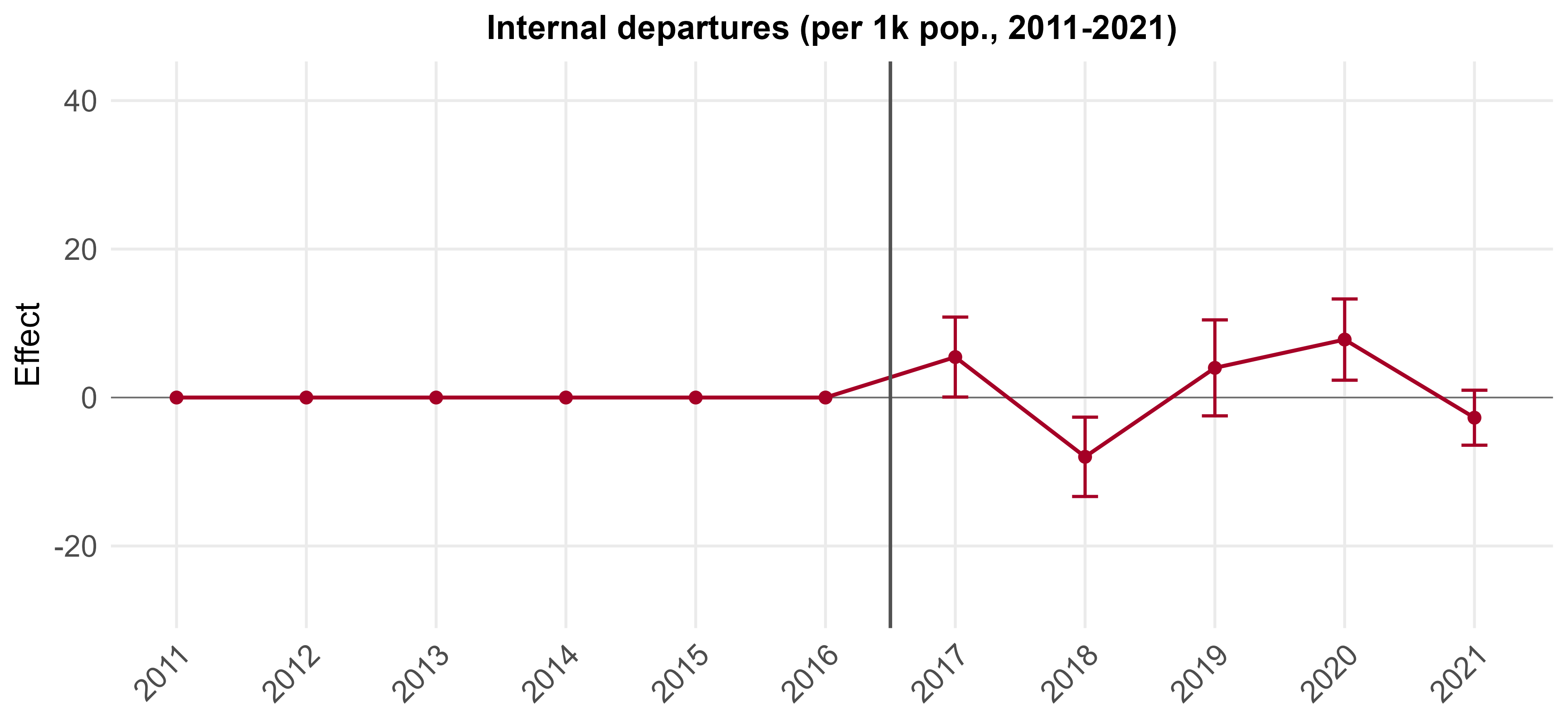} \\
    \includegraphics[width=0.40\linewidth]{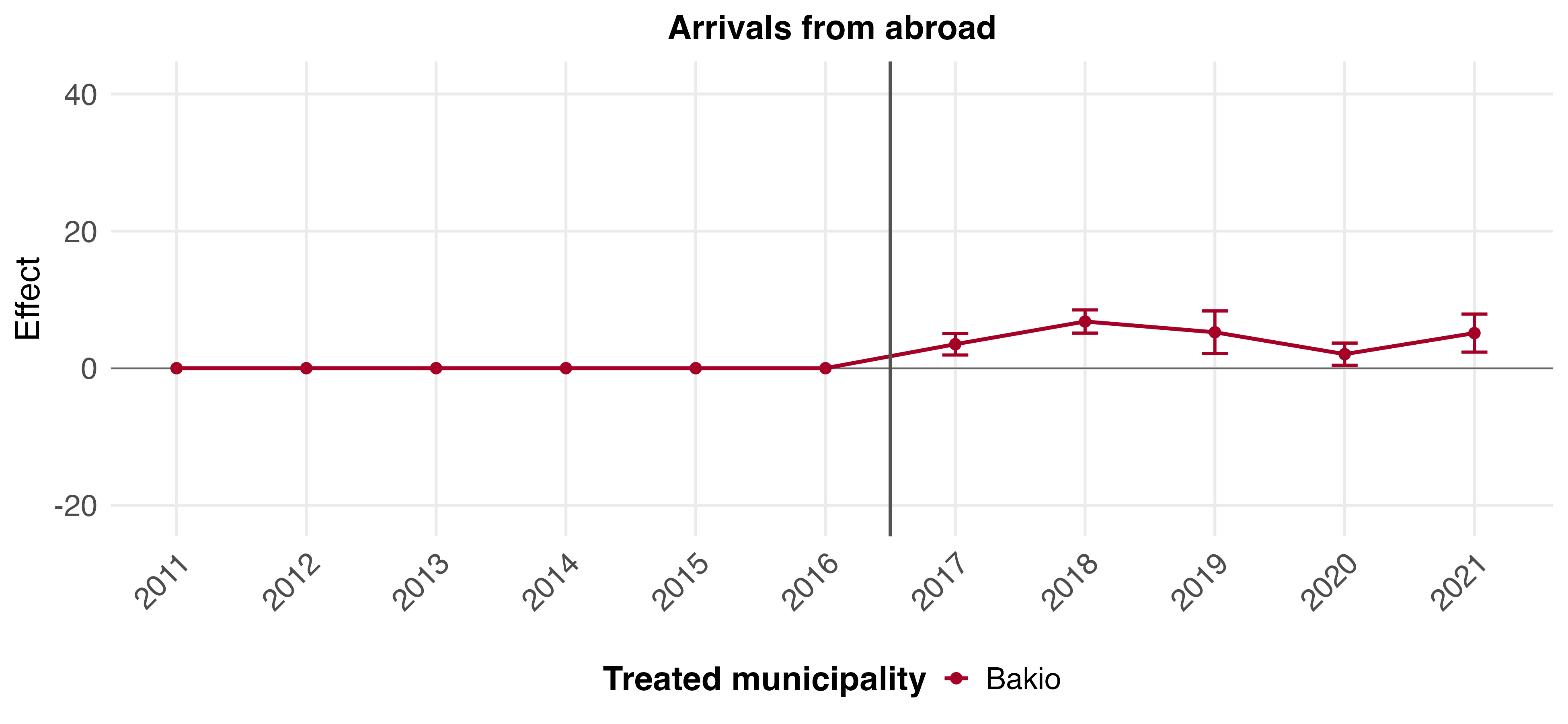} &
    \includegraphics[width=0.40\linewidth]{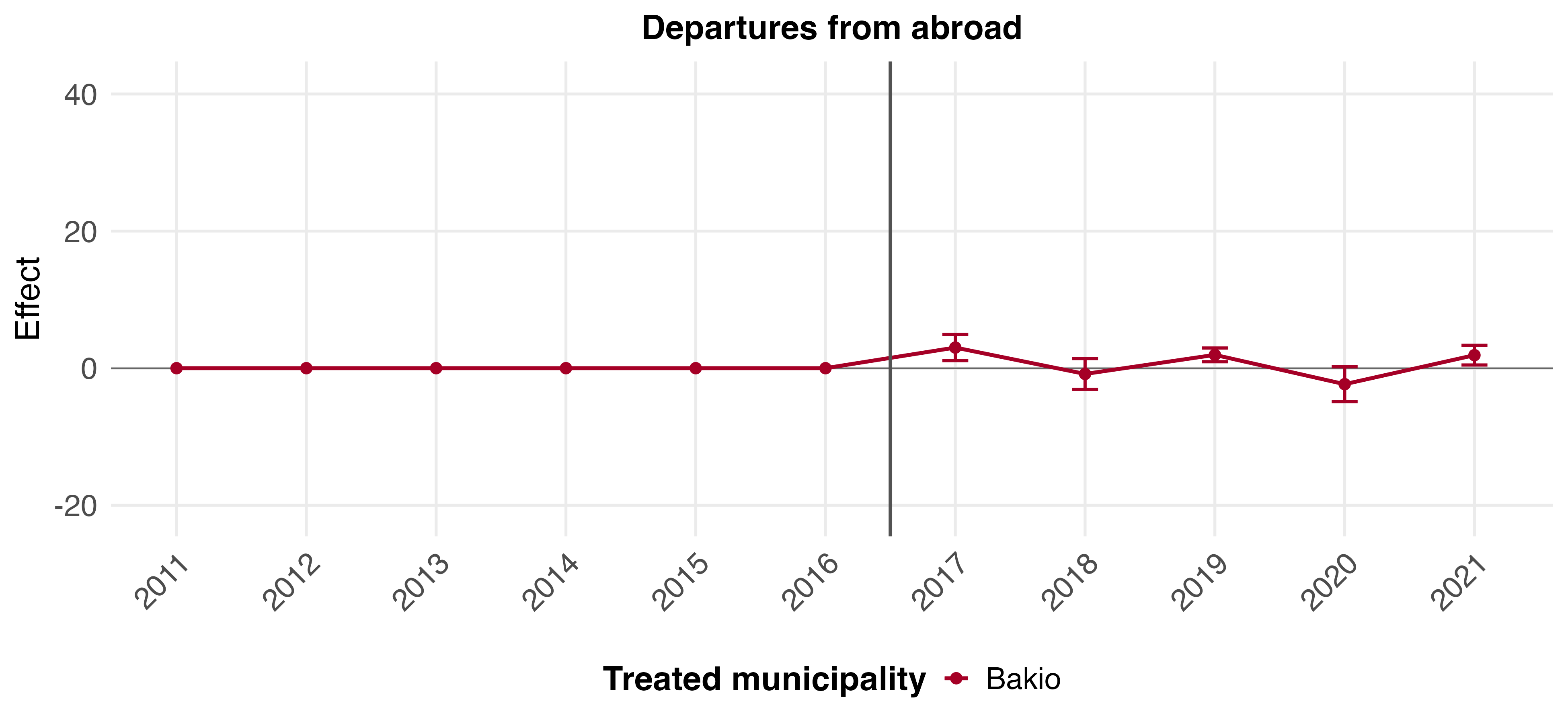}
    \end{tabular}
    \vspace{-0.2cm}
    \begin{minipage}{\linewidth}
    {\scriptsize \setstretch{1}\textit{Notes:} The plots show the difference between the observed outcome and its synthetic control for Bakio, together with 95\% confidence intervals. The left column shows arrivals and the right column departures; rows show all moves (abroad and internal), moves between Spanish municipalities and moves to and from abroad. Outcomes are annual moves per 1,000 residents (2016 population), 2011--2021, from the \textit{Estad\'{i}stica de Variaciones Residenciales} of the Spanish Statistical Office (INE). Estimates are obtained using the Ridge Augmented Synthetic Control estimator \citep{ben-michael_augmented_2021}, estimated separately for each outcome using the same donor pool of Basque municipalities. Within each row, arrivals and departures are drawn on the same vertical scale.}
    \end{minipage}
\end{figure}

Table \ref{tab:mobility_main} reports the average effects on arrivals (Panel A), departures (Panel B), and total population (Panel C), and Figures \ref{fig:population} and \ref{fig:mobility_es} shows the corresponding event studies. We consider all moves, moves between Spanish municipalities and moves to and from abroad. 

Overall, we find no evidence that the tourism shock displaced the local population. Instead, the clearest effect is a substantial increase in arrivals from abroad, which nearly double relative to their pre-shock level. This increase is accompanied by a moderate but significant effect on registered population. Taken together, the results suggest that the tourism shock changed the composition of population flows, increasing immigration without generating detectable domestic displacement.

%Overall, the shock had no significant effect on total arrivals or total departures: the average effect on arrivals is 4.4 per 1,000 residents and on departures 1.3, but neither is statistically significant. The same holds for moves between Spanish municipalities.

%The clearest effect is on arrivals from abroad. Arrivals from abroad rise by 4.5 per 1,000 residents on average, roughly doubling the 2016 level of 4.6, and the effect is significant at the 1\% level. The event study shows that the effect appears in 2017, peaks in 2018, dips in 2020 and recovers in 2021. Departures abroad increase by a smaller amount (0.7, against a baseline of 3.1, significant at the 10\% level), although the yearly estimates are uneven and change sign across years.

\begin{table}[h!]
\centering
\caption{Residential mobility: synthetic control estimates, Bakio}
\footnotesize
\label{tab:mobility_main}
\begin{tabular}{lcc}
\toprule
 & \multicolumn{2}{c}{Bakio} \\
 \cmidrule(lr){2-3}
 & \multicolumn{1}{c}{Baseline (2016)} & \multicolumn{1}{c}{ATT} \\
\midrule
\multicolumn{3}{l}{\textit{Panel A: Arrivals}} \\
Total arrivals & 68.952 & 4.438 \\
 &  & (3.820) \\
[3pt]
Internal arrivals & 64.330 & 0.250 \\
 &  & (3.655) \\
[3pt]
Arrivals from abroad & 4.622 & 4.543*** \\
 &  & (0.684) \\
[3pt]
\midrule
\multicolumn{3}{l}{\textit{Panel B: Departures}} \\
Total departures & 60.092 & 1.321 \\
 &  & (2.450) \\
[3pt]
Internal departures & 57.011 & 1.066 \\
 &  & (2.086) \\
[3pt]
Departures abroad & 3.082 & 0.735* \\
 &  & (0.395) \\
[3pt]
\midrule
\multicolumn{3}{l}{\textit{Panel C: Population}} \\
Population (residents) & 2,584 & 68 \\
 &  & (42) \\
[3pt]
\bottomrule
\end{tabular}
\vspace{-0.2cm}
\begin{minipage}{0.99\linewidth  \setstretch{0.75}}
{\scriptsize Notes: Estimates are obtained using the Ridge Augmented Synthetic Control estimator \citep{ben-michael_augmented_2021}, estimated separately for each outcome and treated municipality using the same donor pool of Basque municipalities. Outcomes are annual counts of residential moves (changes of registered residence) per 1,000 residents (population fixed at 2016), from the \textit{Estad\'{i}stica de Variaciones Residenciales} (EVR) of the Spanish Statistical Office (INE). Panels A and B use the same sample, 2011--2021, with post-treatment period 2017--2021. Panel C reports the registered population (number of residents) at the end of each year, from the municipal register (\textit{Padr\'{o}n}), with sample 2001--2024 and post-treatment period 2017--2024. Arrivals are registrations in the municipality and departures are deregistrations. Moves between Spanish municipalities are changes of residence from or to another Spanish municipality, and moves to and from abroad are registrations from or deregistrations to a foreign country. Standard errors in parentheses. $^{***}p<0.01$, $^{**}p<0.05$, $^{*}p<0.10$.}
 \end{minipage}
\end{table}

% \clearpage

\subsection{Spillover effects}

Appendix \ref{app:spillover} presents the corresponding results for the spillover municipalities. The spillover municipalities experience a rise in tourist accommodation, but not the other responses we find in Bakio. Accommodation increases significantly (by 2.7 units per 1,000 residents, against a 2016 level of 1.0), almost entirely through vacation rentals, but this is considerably smaller than in Bakio and is not accompanied by a significant increase in overnight stays. In the labor market, total employment does not change significantly. Unlike in Bakio, we find no significant employment gains in accommodation, restaurants or culture. In the housing market, dwelling sales increase significantly, while prices do not. The sales estimate may be inflated by a change in the coverage of the land registry (Appendix \ref{app:spillover}), so we do not interpret it as evidence of a housing-market response. Finally, residential mobility is largely unaffected. Arrivals from abroad increase significantly but by much less than in Bakio (0.6 per 1,000 residents, against 4.5), and we find no significant change in total arrivals or departures, in moves between Spanish municipalities, or in departures abroad.

%% =============================================================================
\section{Conclusion}
\label{sec:conclusion}
%% =============================================================================

This paper exploits the 2017 appearance of San Juan de Gaztelugatxe in \textit{Game of Thrones} as an exogenous shock to tourism demand in small coastal municipalities in the Basque Country. Combining administrative panel data with a Ridge Augmented Synthetic Control estimator, we provide a multidimensional causal assessment of a spotlight tourism shock, focusing on the tourism activity response and its consequences for local labor and housing markets, and population moves.

The shock led to a sustained increase in tourist arrivals and accommodation capacity. In the labor market, it generated employment gains in tourism-related sectors, which were offset by declines in other sectors, with no significant effect on total employment. The housing market exhibited a delayed response, with increases in transaction volumes and selling prices. Regarding residential mobility, we find a significant increase in arrivals from abroad, whith no significant effect on displacement of the local population. %As such, the shock seems to change the composition of new residents rather than the overall volume of moves. 
The main results are robust to expanding the buffer around Bakio by excluding all Bizkaia municipalities from the donor pool (Appendix \ref{app:robust_bizkaia}), and to using alternative estimators (Appendix \ref{app:robust_estimators}).

These results have important policy implications. Policymakers often seek to raise the profile of peripheral regions with limited tourism activity through spotlight events or media exposure, expecting the resulting increase in visitor demand to stimulate local economic development. Our findings suggest, however, that greater tourism activity does not necessarily translate into broad-based economic gains. The extent to which spotlight tourism generates local benefits may depend on the type of exposure, the tourists it attracts, and the channels through which their spending affects the local economy. More broadly, policymakers should assess whether increasing tourism visibility is indeed the most effective strategy for promoting economic development in peripheral regions.

Two caveats qualify the external validity of our estimates. First, the \textit{Game of Thrones} shock was extraordinarily large and abrupt: a globally popular series simultaneously broadcast across 170 countries. Smaller-scale spotlight events may produce different effects. Second, the type of tourists attracted by a spotlight shock may shape its economic consequences. In our setting, the shock was driven by the global popularity of a mainstream television series, which may have attracted visitors primarily because the municipalities were featured in \textit{Game of Thrones}, rather than because of an intrinsic interest in their cultural, environmental or other local attributes. This type of mass tourism may have different local economic effects from spotlight shocks that attract visitors who are more engaged with the destination itself and more likely to value or seek out its cultural and environmental assets. In particular, spotlight shocks associated with more sustainability-oriented or culturally motivated forms of tourism may generate different, and potentially more broadly distributed, economic effects. Our estimates should therefore not be interpreted as representative of the effects of all forms of tourism promotion or destination marketing, but rather of the type of mass tourism generated by spotlight shocks driven by television, social media, and viral content more broadly.
%% =============================================================================
%% References
%\clearpage
%\nocite{*}
\bibliography{references} 

\begin{thebibliography}{}

\bibitem[Abadie, 2021]{abadie_using_2021}
Abadie, A. (2021).
\newblock Using {Synthetic} {Controls}: {Feasibility}, {Data} {Requirements},
  and {Methodological} {Aspects}.
\newblock {\em Journal of Economic Literature}, 59(2):391--425.

\bibitem[Abadie et~al., 2010]{abadie_synthetic_2010}
Abadie, A., Diamond, A., and Hainmueller, J. (2010).
\newblock Synthetic {Control} {Methods} for {Comparative} {Case} {Studies}:
  {Estimating} the {Effect} of {California}’s {Tobacco} {Control} {Program}.
\newblock {\em Journal of the American Statistical Association},
  105(490):493--505.

\bibitem[Abadie and Gardeazabal, 2003]{abadie_economic_2003}
Abadie, A. and Gardeazabal, J. (2003).
\newblock The {Economic} {Costs} of {Conflict}: {A} {Case} {Study} of the
  {Basque} {Country}.
\newblock {\em American Economic Review}, 93(1):113--132.

\bibitem[Arkhangelsky et~al., 2021]{arkhangelsky_synthetic_2021}
Arkhangelsky, D., Athey, S., Hirshberg, D.~A., Imbens, G.~W., and Wager, S.
  (2021).
\newblock Synthetic {Difference}-in-{Differences}.
\newblock {\em American Economic Review}, 111(12):4088--4118.

\bibitem[Arkhangelsky and Imbens, 2024]{arkhangelsky_causal_2024}
Arkhangelsky, D. and Imbens, G. (2024).
\newblock Causal models for longitudinal and panel data: a survey.
\newblock {\em The Econometrics Journal}, 27(3):C1--C61.

\bibitem[{Banco de España}, 2026]{banco_de_espana_microdata_2026}
{Banco de España} (2026).
\newblock Microdata on the {Real} {Estate} {Registry} {Statistics} ({PROP}).
\newblock (Version 2.0) [Dataset]. Banco de España.
  https://doi.org/10.48719/BELAB.PROP1325\_02.

\bibitem[Ben-Michael et~al., 2021]{ben-michael_augmented_2021}
Ben-Michael, E., Feller, A., and Rothstein, J. (2021).
\newblock The {Augmented} {Synthetic} {Control} {Method}.
\newblock {\em Journal of the American Statistical Association},
  116(536):1789--1803.

\bibitem[Bronzini et~al., 2020]{bronzini_economic_2020}
Bronzini, R., Mocetti, S., and Mongardini, M. (2020).
\newblock The economic effects of big events: {Evidence} from the great jubilee
  2000 in {Rome}.
\newblock {\em Journal of Regional Science}, 60(4):801--822.

\bibitem[Contu and Pau, 2022]{contu_impact_2022}
Contu, G. and Pau, S. (2022).
\newblock The impact of {TV} series on tourism performance: the case of {Game}
  of {Thrones}.
\newblock {\em Empirical Economics}, 63(6):3313--3341.

\bibitem[Di~Giacomo, 2025]{di_giacomo_tourism_2025}
Di~Giacomo, G. (2025).
\newblock Tourism in {Industrialized} {Countries}: {Catalyst} for {Growth} or
  {Economic} {Burden}?
\newblock Working Paper.

\bibitem[Di~Giacomo and Lerch, 2026]{di_giacomo_service_2026}
Di~Giacomo, G. and Lerch, B. (2026).
\newblock Service {Jobs} and {Education}: {Evidence} from {Tourism} {Shocks} in
  {Italy}.
\newblock {\em CESifo Working Paper Series}.
\newblock Number: 12383.

\bibitem[Duso et~al., 2024]{duso_airbnb_2024}
Duso, T., Michelsen, C., Schaefer, M., and Tran, K.~D. (2024).
\newblock Airbnb and rental markets: {Evidence} from {Berlin}.
\newblock {\em Regional Science and Urban Economics}, 106:104007.

\bibitem[Faber and Gaubert, 2019]{faber_tourism_2019}
Faber, B. and Gaubert, C. (2019).
\newblock Tourism and {Economic} {Development}: {Evidence} from {Mexico}'s
  {Coastline}.
\newblock {\em American Economic Review}, 109(6):2245--2293.

\bibitem[Favero and Malisan, 2026]{favero_next_2026}
Favero, L. and Malisan, I. (2026).
\newblock Next job: {Matera}. {How} tourism and spotlight shape the local
  labour market.
\newblock {\em Regional Science and Urban Economics}, page 104289.

\bibitem[Garcia-López et~al., 2020]{garcia-lopez_short-term_2020}
Garcia-López, M.~A., Jofre-Monseny, J., Martínez-Mazza, R., and Segú, M.
  (2020).
\newblock Do short-term rental platforms affect housing markets? {Evidence}
  from {Airbnb} in {Barcelona}.
\newblock {\em Journal of Urban Economics}, 119:103278.

\bibitem[González and Surovtseva, 2025]{gonzalez_more_2025}
González, L. and Surovtseva, T. (2025).
\newblock Do {More} {Tourists} {Promote} {Local} {Employment}?
\newblock {\em Journal of Human Resources}.

\bibitem[Gómez-Limón and García~Ventura, 2020]{gomez-limon_elaboracion_2020}
Gómez-Limón, J. and García~Ventura, D. (2020).
\newblock Elaboración de un estudio sobre la capacidad de acogida de uso
  público en el {Biotopo} {Protegido} de {San} {Juan} de {Gaztelugatxe}
  ({Bizkaia}).
\newblock Technical report, Diputación Foral de Bizkaia.

\bibitem[Koster et~al., 2021]{koster_short-term_2021}
Koster, H. R.~A., van Ommeren, J., and Volkhausen, N. (2021).
\newblock Short-term rentals and the housing market: {Quasi}-experimental
  evidence from {Airbnb} in {Los} {Angeles}.
\newblock {\em Journal of Urban Economics}, 124:103356.

\bibitem[Mancini and Papini, 2021]{mancini_all_2021}
Mancini, A.~L. and Papini, G. (2021).
\newblock All that {Glitters} is not {Gold}. {An} {Economic} {Evaluation} of
  the {Turin} {Winter} {Olympics}.
\newblock Bank of Italy Temi di Discussione (Working Paper) No. 1355.

\bibitem[Nocito et~al., 2023]{nocito_beam_2023}
Nocito, S., Sartarelli, M., and Sobbrio, F. (2023).
\newblock A beam of light: {Media}, tourism and economic development.
\newblock {\em Journal of Urban Economics}, 137(C).

\bibitem[Tkalec et~al., 2017]{tkalec_effect_2017}
Tkalec, M., Zilic, I., and Recher, V. (2017).
\newblock The effect of film industry on tourism: \textit{{Game} of {Thrones}}
  and {Dubrovnik}.
\newblock {\em International Journal of Tourism Research}, 19(6):705--714.

\bibitem[{World Travel and Tourism Council},
  2026]{world_travel_and_tourism_council_world_2026}
{World Travel and Tourism Council} (2026).
\newblock World {Travel} \& {Tourism} {Economic} {Impact} {Research}.
\newblock Technical report.

\end{thebibliography}
\bibliographystyle{apalike}

%===================================================================
%===================================================================
%APPENDIX
%===================================================================
%===================================================================
\newpage

\appendix

\renewcommand{\theequation}{A\arabic{equation}}
\renewcommand{\thetable}{A\arabic{table}}
\renewcommand{\thefigure}{A\arabic{figure}}
\renewcommand{\thelemma}{A\arabic{lemma}}
\renewcommand{\theproposition}{A\arabic{proposition}}
\renewcommand{\thecorollary}{A\arabic{corollary}}
\setcounter{equation}{0}
\setcounter{table}{0}
\setcounter{figure}{0}
\setcounter{lemma}{0}
\linespread{1}

\newpage
\vspace{-0.4 cm}

\clearpage
%\begin{appendices}

\section{Additional results} \label{app:additional_results}

\subsection{Effect on touristic capacity by accommodation type} \label{app:acco_types}

This section splits the results on touristic accommodation by accommodation type. We show that most of the effect is driven by an increase in VUTs, i.e., residential properties used to accommodate tourists.

\begin{table}[h!]
\centering
\caption{Accommodation types: augmented synthetic control estimates}
\small
\label{tab:tourism_acco_type}
\begin{tabular}{lcc}
\toprule
 & \multicolumn{2}{c}{Bakio} \\
 \cmidrule(lr){2-3}
 & \multicolumn{1}{c}{Baseline (2016)} & \multicolumn{1}{c}{ATT} \\
\midrule
Accommodation places (per 1k pop.) & 3.852 & 17.564*** \\
 &  & (0.348) \\
[3pt]
Viviendas Uso Turístico (per 1k pop.) & 0.770 & 16.207*** \\
 &  & (0.224) \\
[3pt]
Other accommodation (per 1k pop.) & 3.082 & 0.884*** \\
 &  & (0.233) \\
[3pt]
\bottomrule
\end{tabular}
\vspace{-0.2cm}
\begin{minipage}{0.99\linewidth  \setstretch{0.75}}
{\scriptsize Notes: Estimates are obtained using the Ridge Augmented Synthetic Control estimator \citep{ben-michael_augmented_2021}. The baseline columns report the value of each outcome in 2016. The ATT columns report the average treatment effect over the post-treatment period (2017--2025), with standard errors in parentheses. $^{***}p<0.01$, $^{**}p<0.05$, $^{*}p<0.10$.}
 \end{minipage}
\end{table}

\begin{figure}[H]
    \caption{Event study estimates for accommodation types.}
    \label{fig:tourism_acco_type}
    \centering
    \begin{tabular}{cc}
    \includegraphics[width=0.48\linewidth]{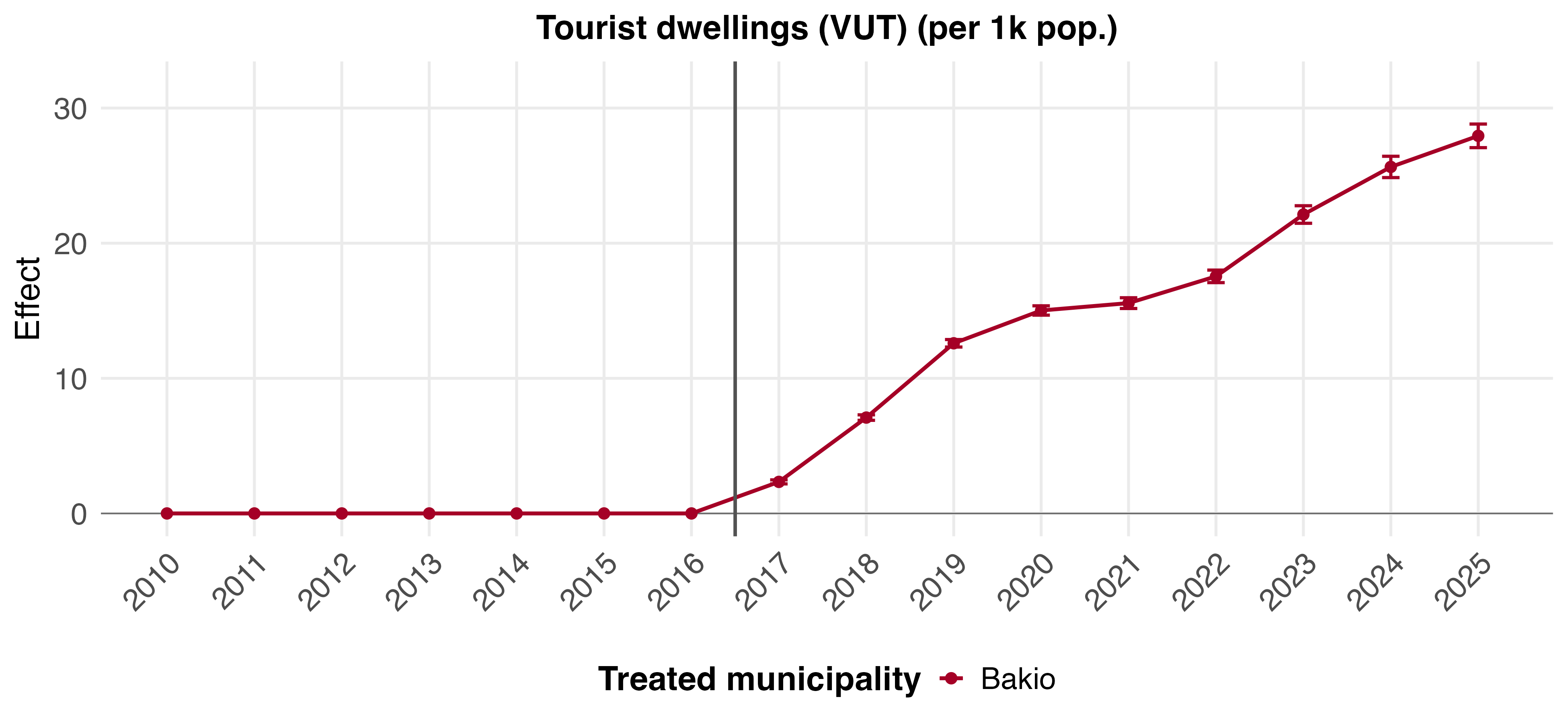}
&
    \includegraphics[width=0.48\linewidth]{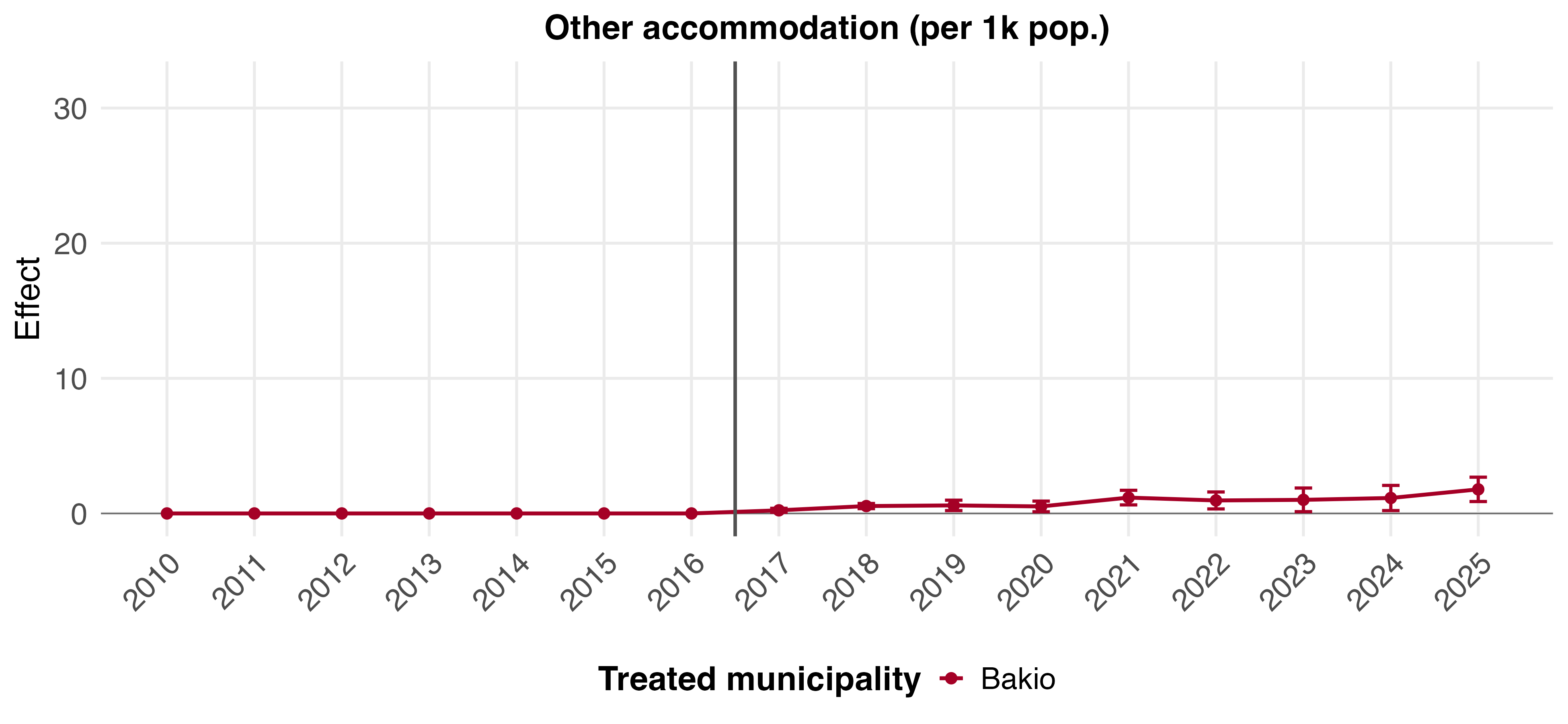}

    \end{tabular}

    \vspace{-0.2cm}
    \begin{minipage}{\linewidth}
        {\scriptsize \setstretch{1}\textit{Notes:} The plots show the difference between the observed outcome and its synthetic control, together with 95\% confidence intervals. Estimates are obtained using the Ridge Augmented Synthetic Control estimator \citep{ben-michael_augmented_2021}, estimated separately for each outcome and each treated municipality using the same donor pool of Basque municipalities. Other accommodation includes hotels, hostels, camping sites and agroturismos. Figure \ref{fig:tourism_trends} presents the same results in levels rather than as differences between the observed outcome and the synthetic control.}
    \end{minipage}
\end{figure}

\subsection{Spillover effects} \label{app:spillover}
This section extends the analysis to municipalities in the same counties as Bakio and Gaztelugatxe that were excluded from the donor pool to examine the spillover effects of the spotlight shock. Using the same donor pool as for Bakio, we construct a synthetic control for each municipality using the Ridge ASC estimator \citep{ben-michael_augmented_2021}. The tables in this section report the average effect across all post-treatment periods and across all spillover municipalities. The figures in this section plot the event-study estimates of the shock for the average spillover effect, together with the effects for Bakio discussed in Section \ref{sec:results}, to facilitate comparison of the results. %Some municipalities in the spillover group have very small populations, meaning that even modest changes in absolute terms can translate into large per-capita effects. The magnitude of these estimates should therefore be interpreted with caution.%, as large per-capita effects do not necessarily imply substantial changes in absolute terms.

\subsubsection{Tourism Development}

\begin{figure}[H]
    \caption{Event study estimates for tourism outcomes, spillover municipalities.}
    \label{fig:tourism_spillover}
    \centering
    \begin{tabular}{cc}
    \includegraphics[width = 8 cm]{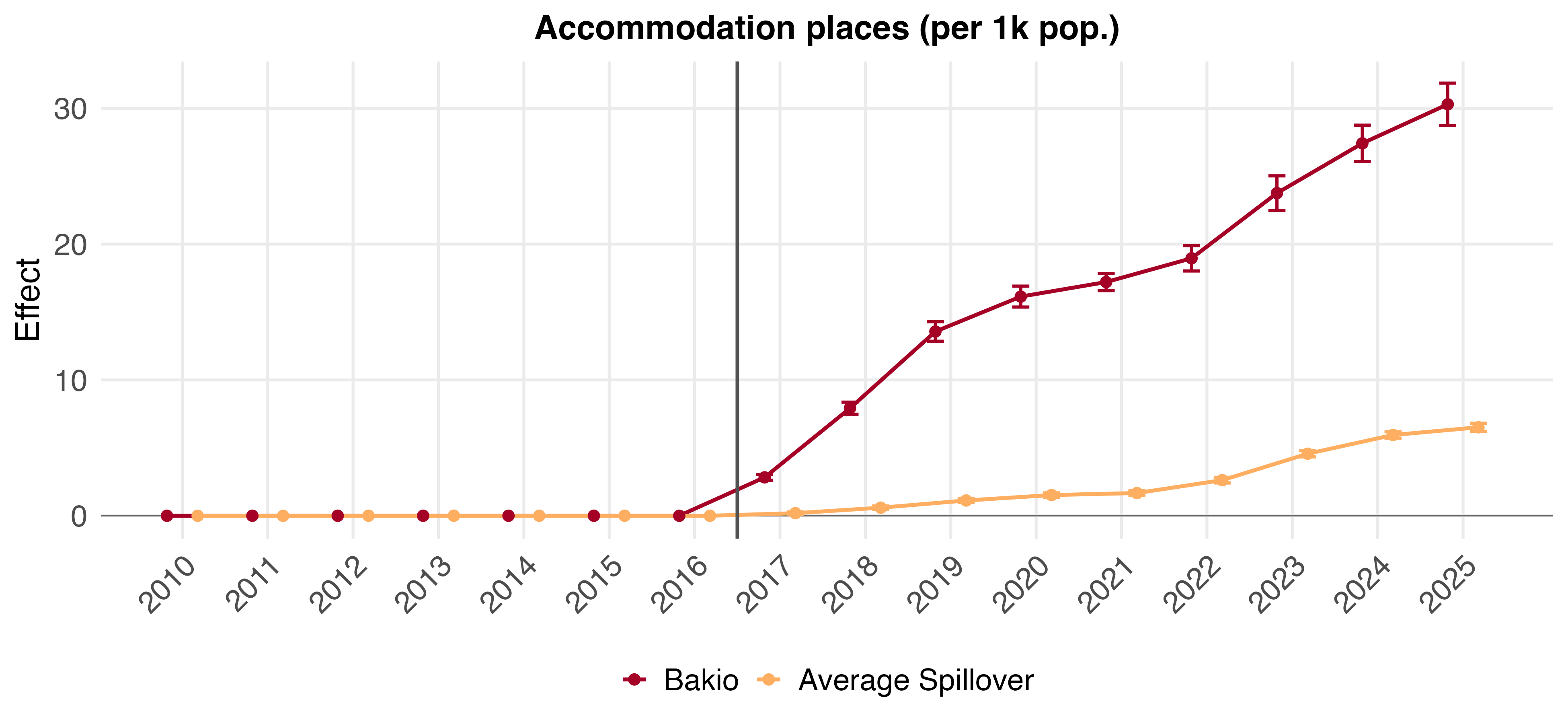} 
    &
    \includegraphics[width = 8 cm]{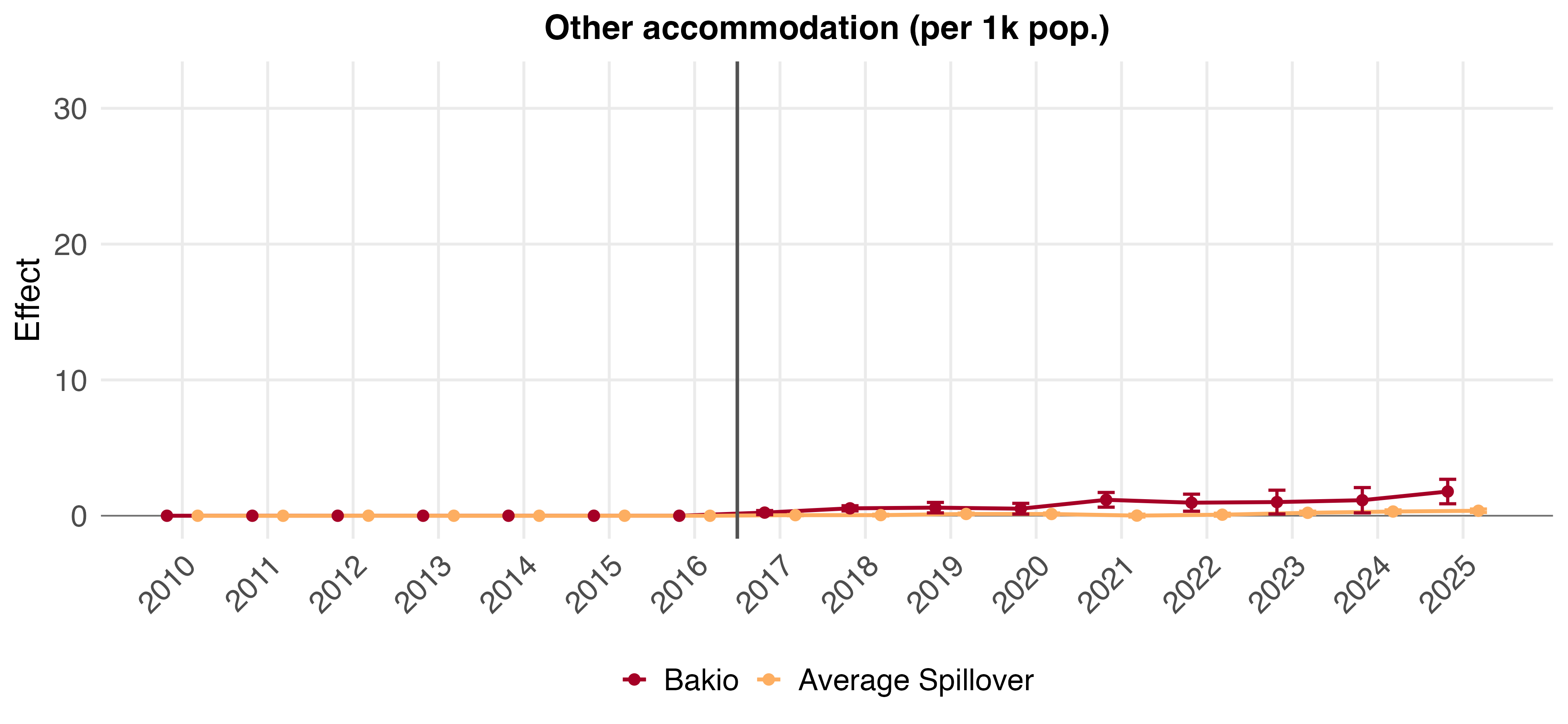}    \\
    \includegraphics[width = 8 cm]{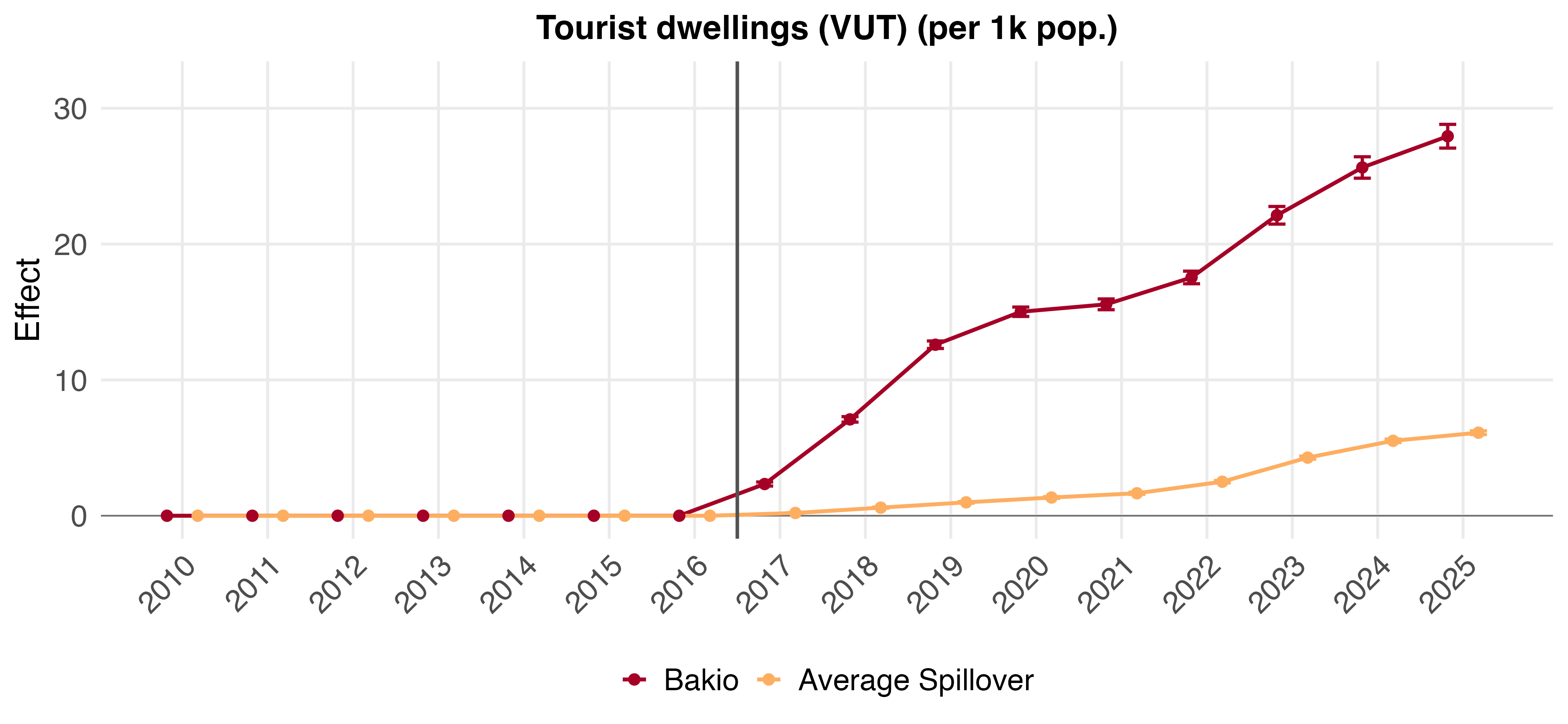} 
    &
    \includegraphics[width = 8 cm]{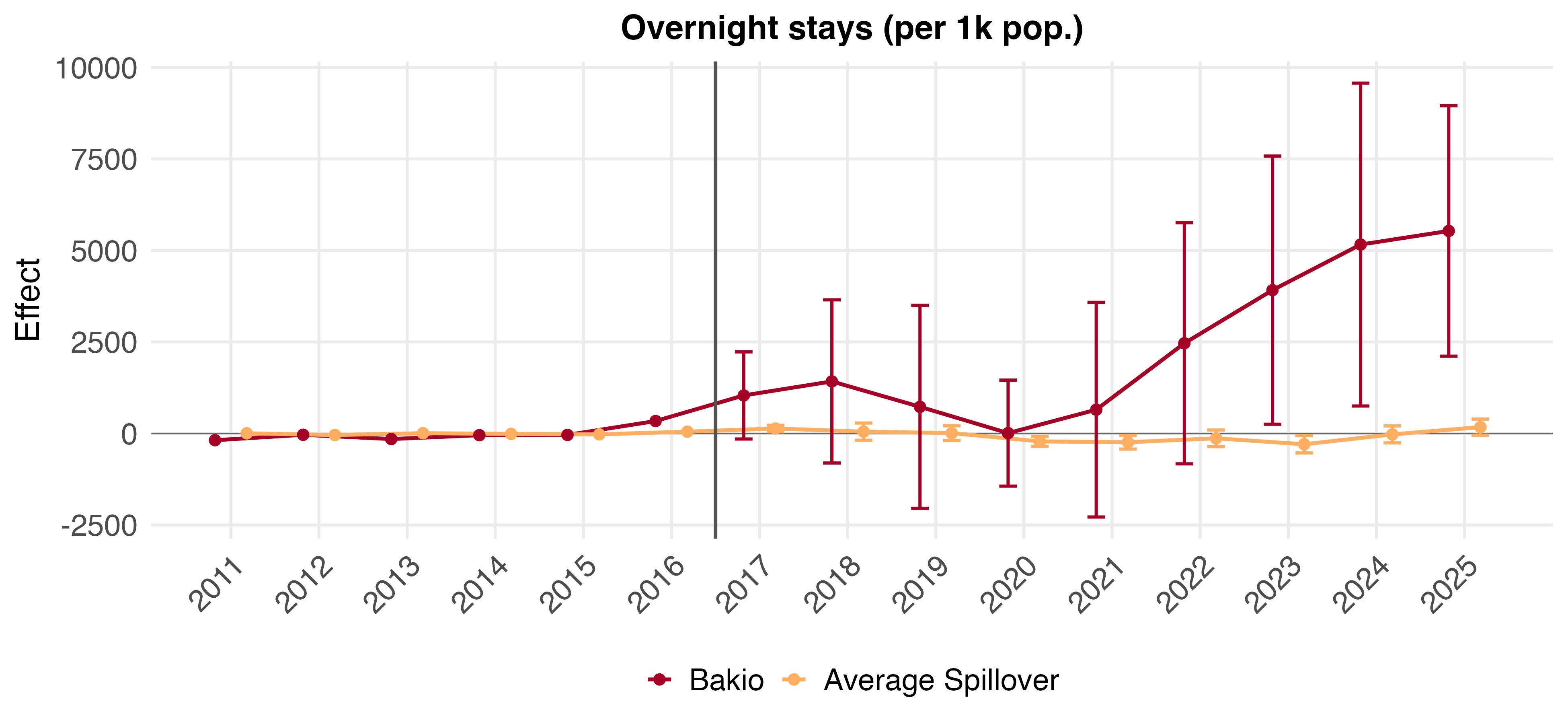} 
\end{tabular}
\vspace{-0.2cm}
\begin{minipage}{\linewidth}
   {\scriptsize \setstretch{1}\textit{Notes:} The plots show the difference between the observed outcome and its synthetic control, together with 95\% confidence intervals. The red line corresponds to the treated municipality, Bakio, and the orange line to the average spillover group (the municipalities that belong to Bakio's or Gaztelugatxe's counties). Estimates are obtained using the Ridge Augmented Synthetic Control estimator \citep{ben-michael_augmented_2021}, estimated separately for each outcome and each treated municipality using the same donor pool of Basque municipalities. }
    \end{minipage}
\end{figure}

Figure \ref{fig:tourism_spillover} shows a positive, statistically significant, and persistent spillover effect on the number of tourist accommodation units per 1,000 inhabitants. The magnitude is smaller in relative terms than the effects in the treated municipality.Regarding overnight stays, we find no statistically significant effect, suggesting that these municipalities were less directly exposed to the shock than Bakio. Nevertheless, the absence of an increase in overnight stays does not rule out the possibility that these municipalities were affected through other spillover channels.

\begin{table}[!h]
\centering
\caption{Tourism outcomes: average spillover effect (augmented synthetic control estimates)}
\small
\label{tab:tourism_spillover}
\begin{tabular}{lcc}
\toprule
 & \multicolumn{1}{c}{Baseline (2016)} & \multicolumn{1}{c}{ATT} \\
\midrule
Accommodation places (per 1k pop.) & 0.987 & 2.749*** \\
 &  & (0.082) \\
 Viviendas Uso Turístico (per 1k pop.) & 0.161 & 2.578*** \\
 &  & (0.039) \\
 Other accommodation (per 1k pop.) & 0.826 & 0.148*** \\
 &  & (0.040) \\
[3pt]
Overnight stays (per 1k pop.) & 1,872 & -60 \\
 &  & (86) \\
[3pt]
\bottomrule
\end{tabular}
\vspace{-0.2cm}
\begin{minipage}{0.99\linewidth  \setstretch{0.75}}
{\scriptsize Notes: Estimates are obtained using the Ridge Augmented Synthetic Control estimator \citep{ben-michael_augmented_2021}, estimated separately for each outcome and treated municipality using the same donor pool of Basque municipalities. ``Spillover" municipalities are all municipalities other than Bakio in the counties of Mungialdea (Bakio's county) and Busturialdea (Gaztelugatxe's county). The baseline column reports the 2016-population-weighted mean value of each outcome across spillover municipalities in 2016. The ATT column reports the 2016-population-weighted mean of the spillover municipalities' individual average treatment effects over the post-treatment period (2017--2025). $^{***}p<0.01$, $^{**}p<0.05$, $^{*}p<0.10$.}
 \end{minipage}
\end{table}

\subsubsection{Local Labor Markets}
\begin{figure}[H]
    \caption{Event study estimates for total employment.}
    \label{fig:labor_spillover}
    \centering
\begin{tabular}{cc}
\includegraphics[width = 0.6\linewidth]{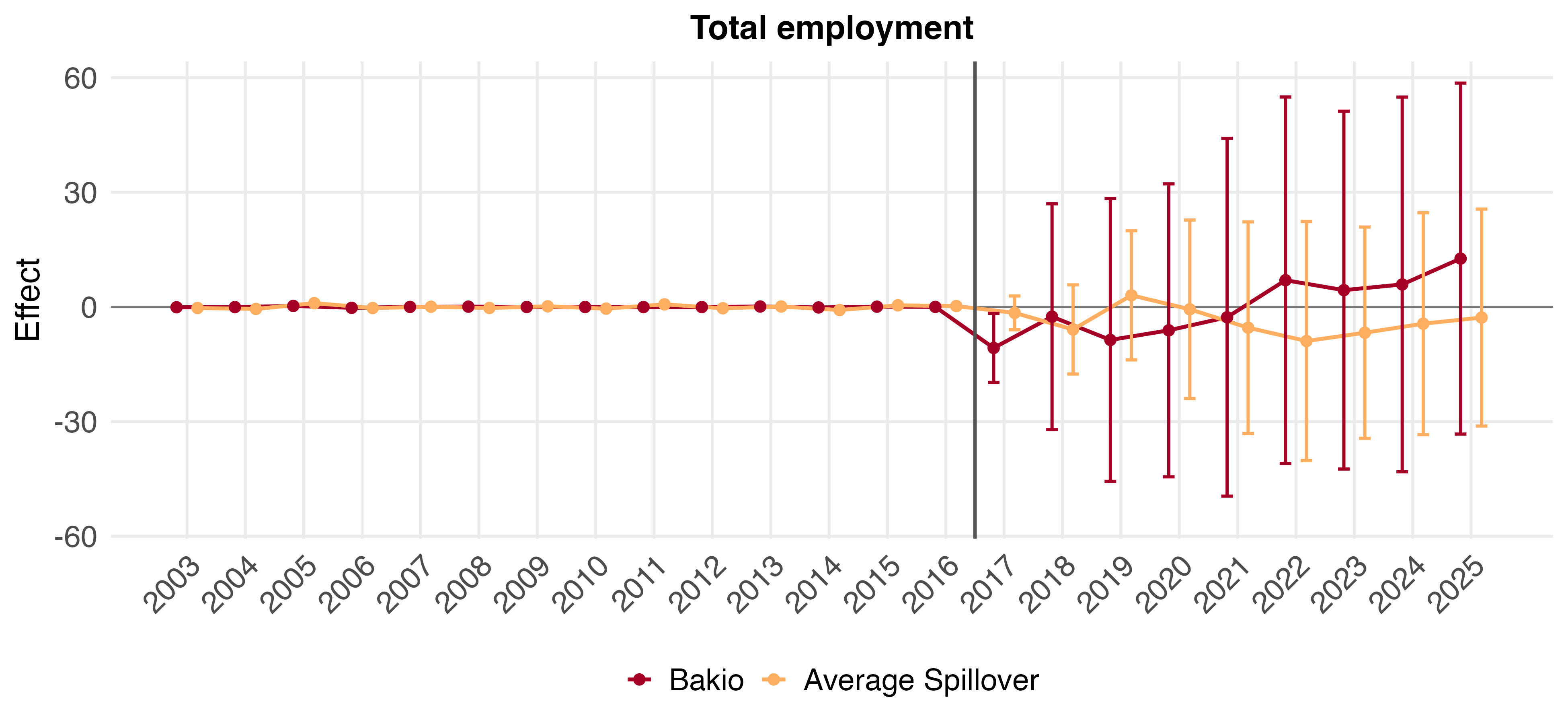}    
\end{tabular}
\vspace{-0.2cm}
\begin{minipage}{\linewidth}
   {\scriptsize \setstretch{1}\textit{Notes:} The plots show the difference between the observed outcome and its synthetic control, together with 95\% confidence intervals. The red line corresponds to the treated municipality, Bakio, and the orange line to the average spillover group (the municipalities that belong to Bakio's or Gaztelugatxe's counties). Estimates are obtained using the Ridge Augmented Synthetic Control estimator \citep{ben-michael_augmented_2021}, estimated separately for each outcome and each treated municipality using the same donor pool of Basque municipalities. }
    \end{minipage}
\end{figure}

\begin{figure}[H]
    \caption{Event study estimates for regime breakdown.}
    \label{fig:labor_reg_spillover}
    \centering
\begin{tabular}{cc}
\includegraphics[width = 8 cm]{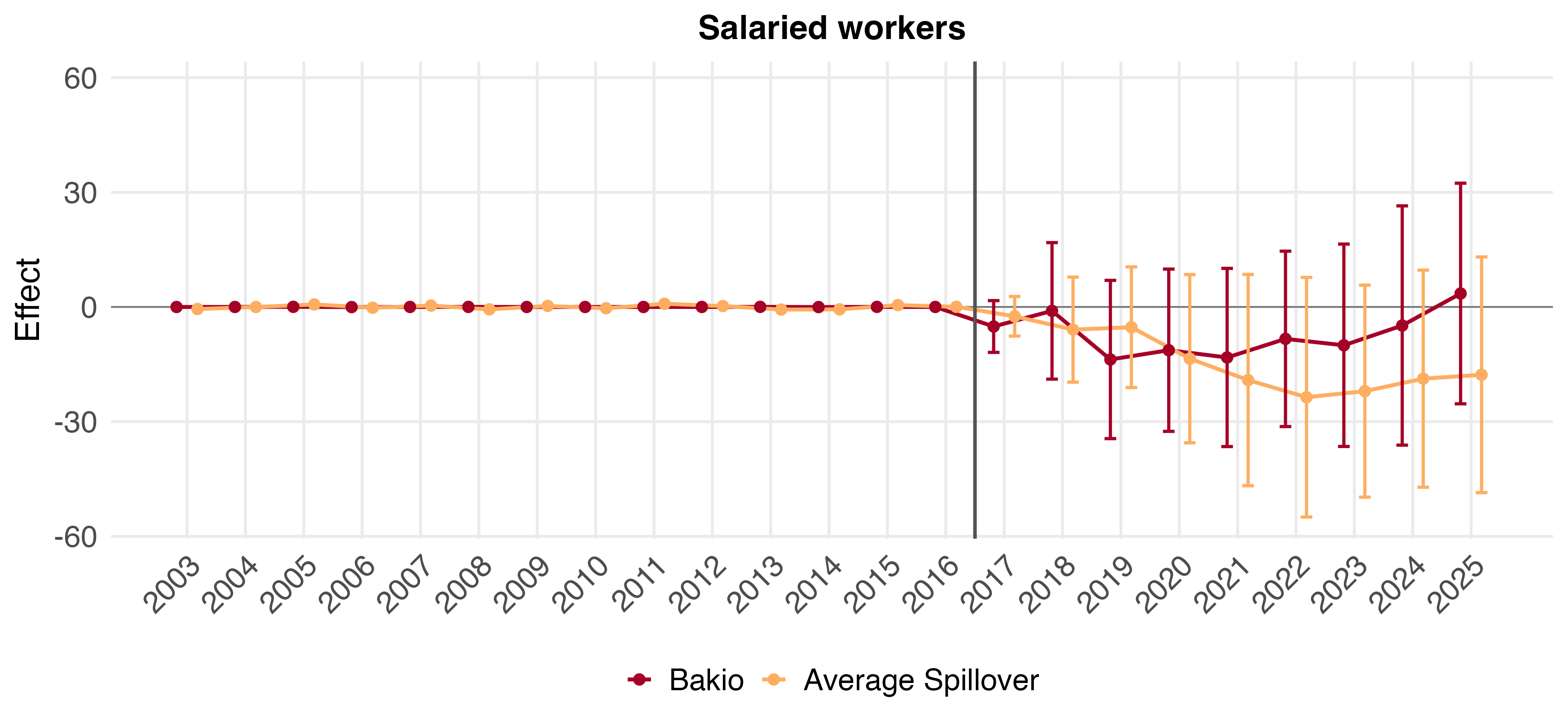}    
&  
\includegraphics[width = 8 cm]{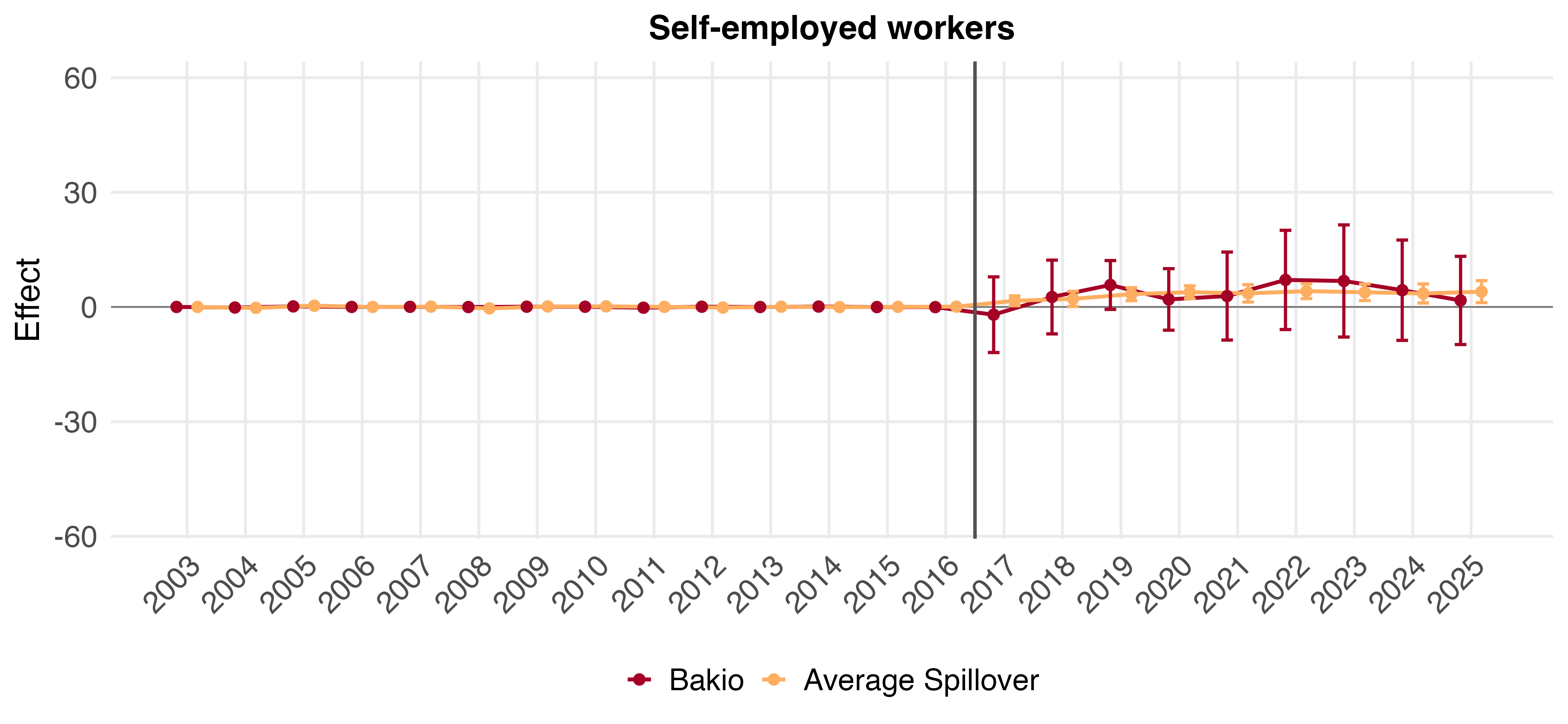} 
\end{tabular}
\vspace{-0.2cm}
\begin{minipage}{\linewidth}
   {\scriptsize \setstretch{1}\textit{Notes:} The plots show the difference between the observed outcome and its synthetic control, together with 95\% confidence intervals. The red line corresponds to the treated municipality, Bakio, and the orange line to the average spillover group (the municipalities that belong to Bakio's or Gaztelugatxe's counties). Estimates are obtained using the Ridge Augmented Synthetic Control estimator \citep{ben-michael_augmented_2021}, estimated separately for each outcome and each treated municipality using the same donor pool of Basque municipalities. }
    \end{minipage}
\end{figure}

\begin{figure}[H]
    \caption{Event study estimates for sector breakdown.}
    \label{fig:labor_sector_spillover}
    \centering
    \begin{tabular}{cc}
    \includegraphics[width=8cm]{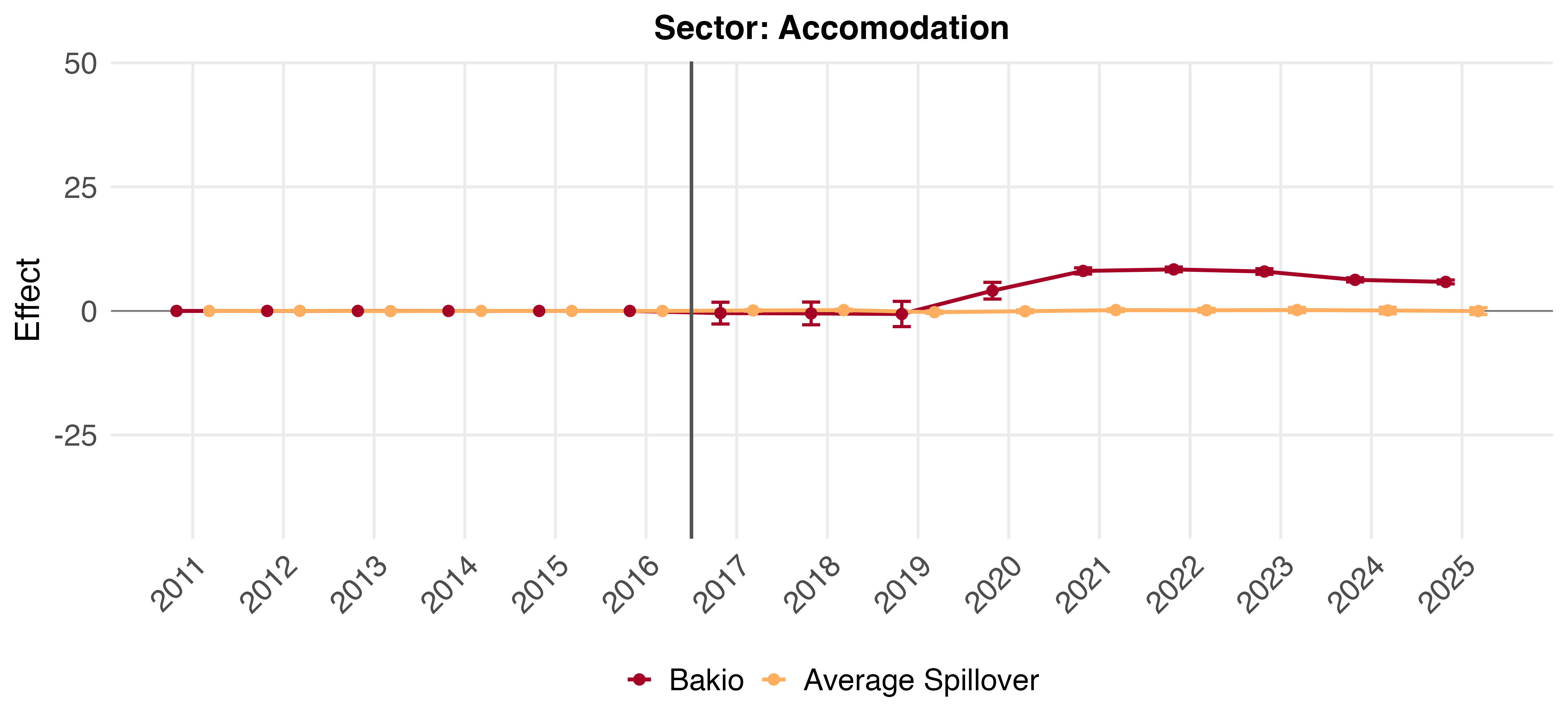} &
    \includegraphics[width=8cm]{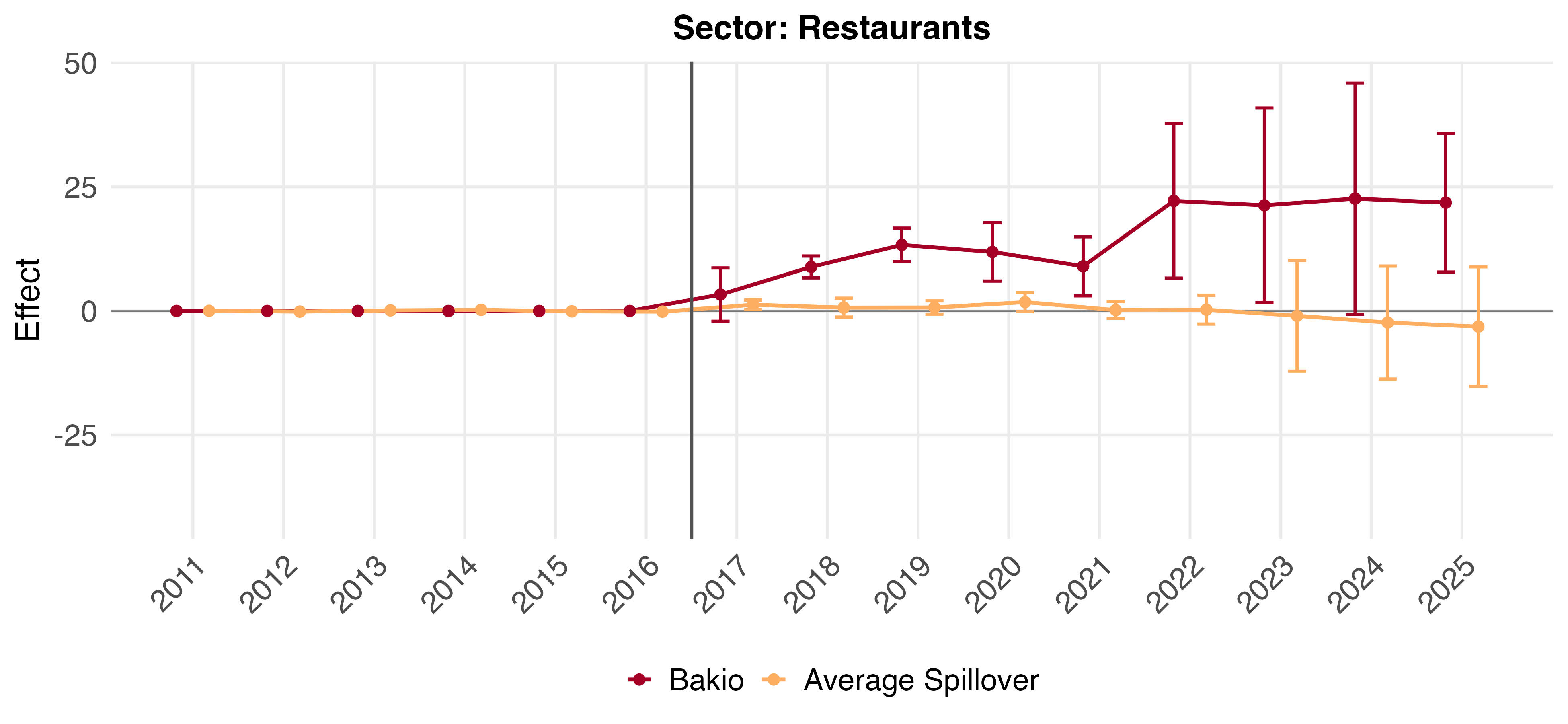}  \\
    \includegraphics[width=8cm]{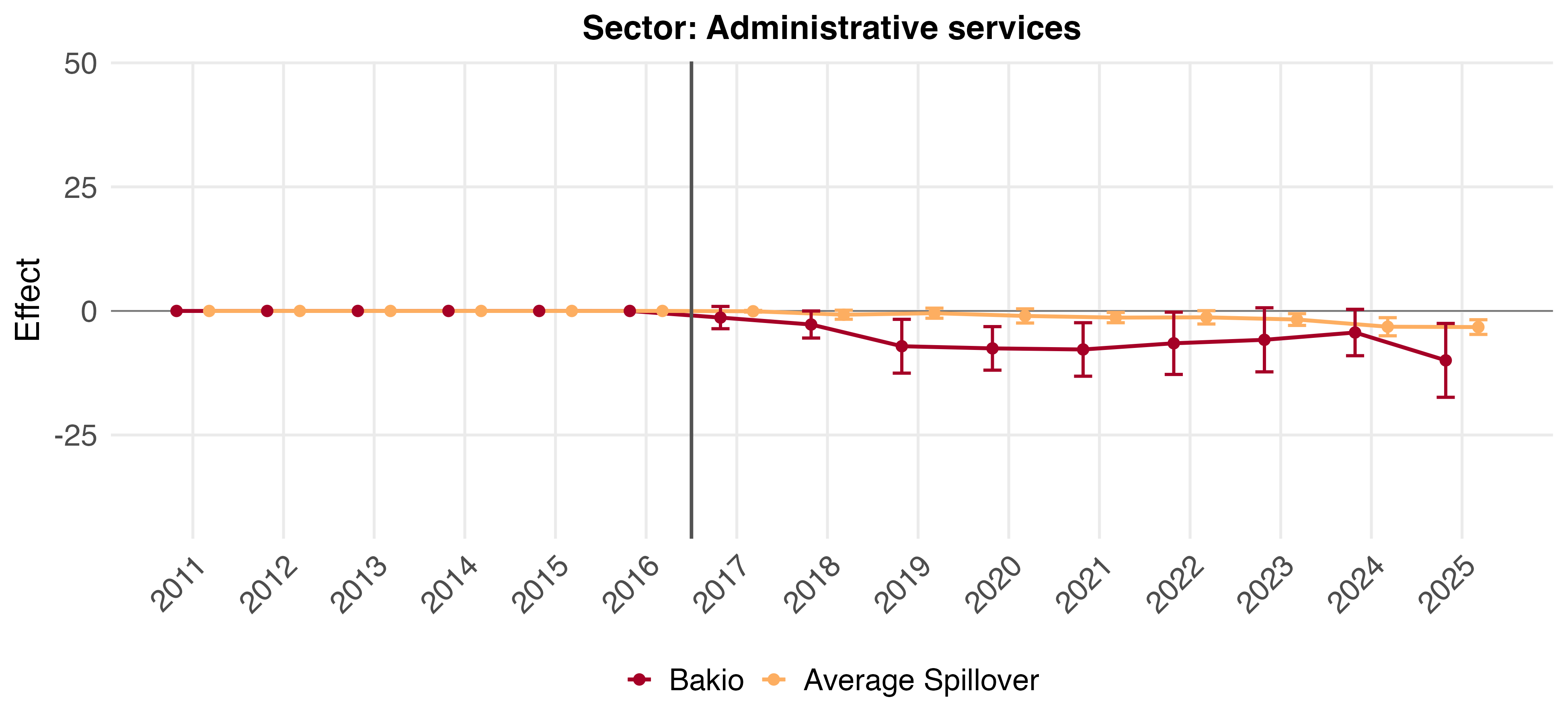} &
    \includegraphics[width=8cm]{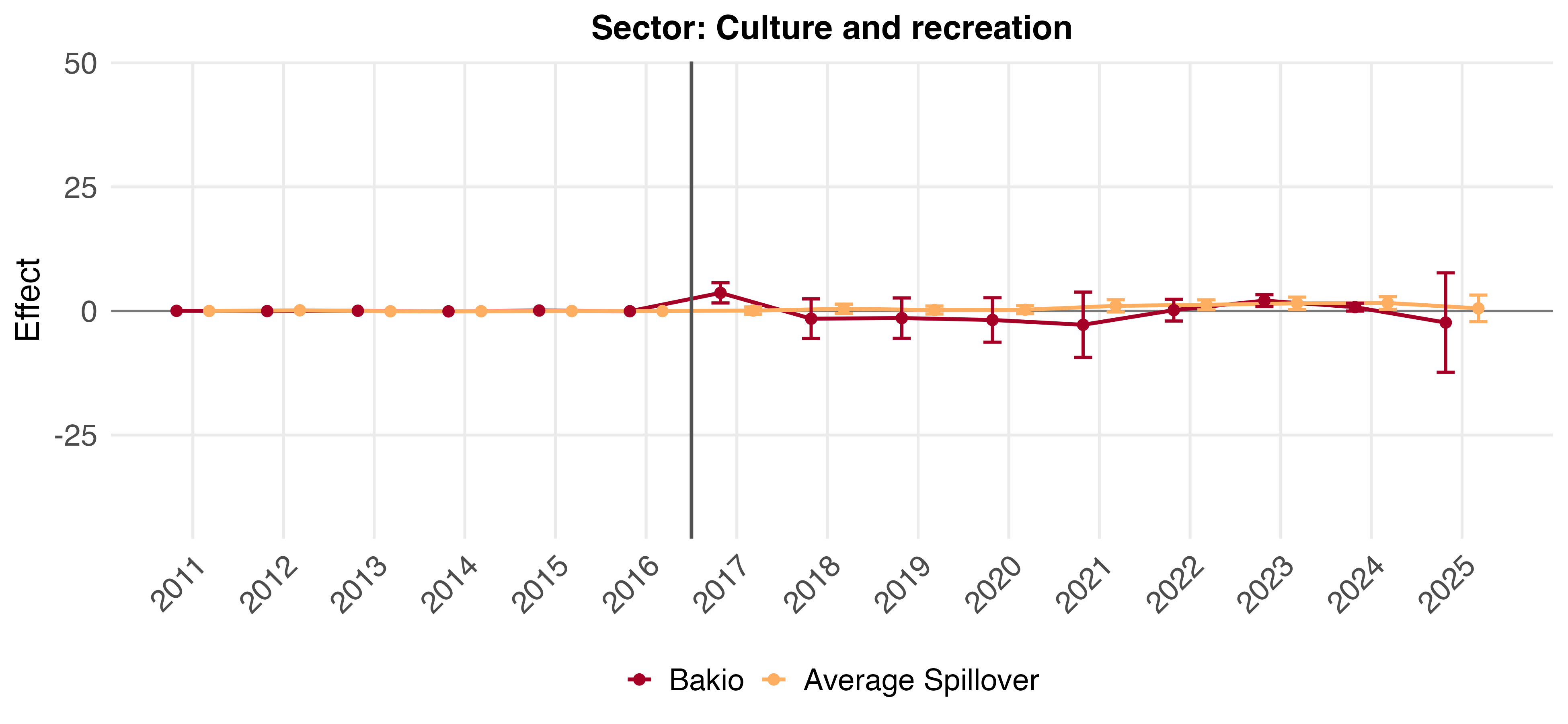} \\
    \includegraphics[width=8cm]{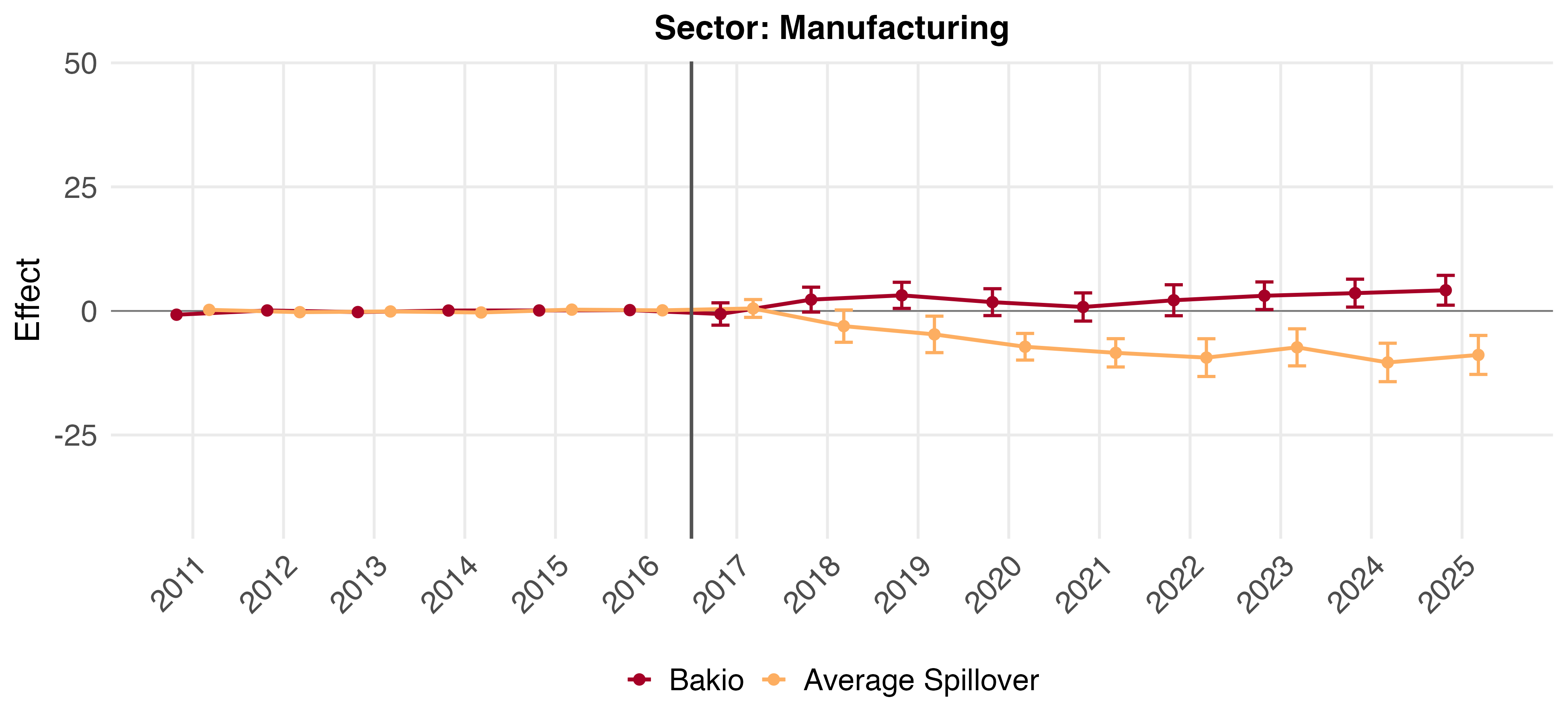} &
    \includegraphics[width=8cm]{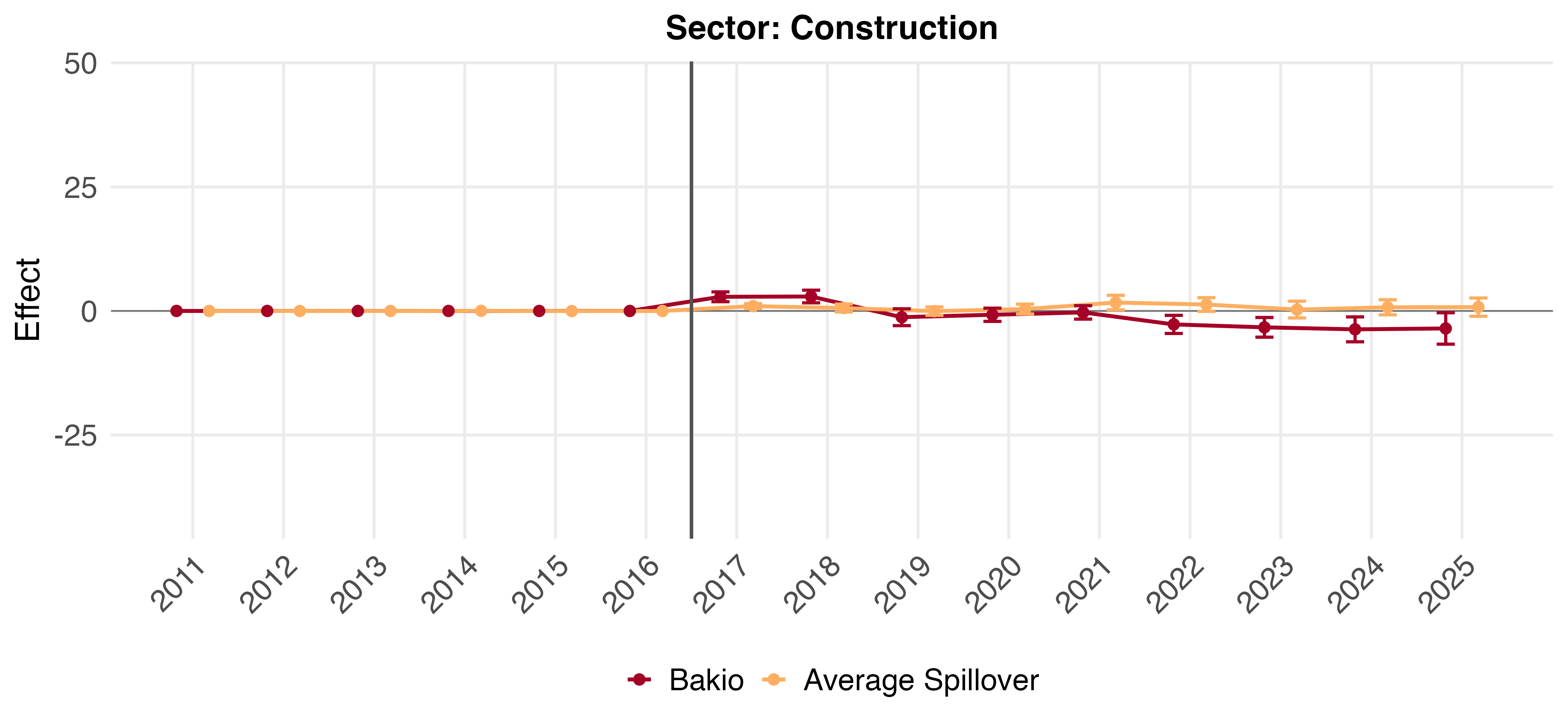} \\
    \includegraphics[width=8cm]{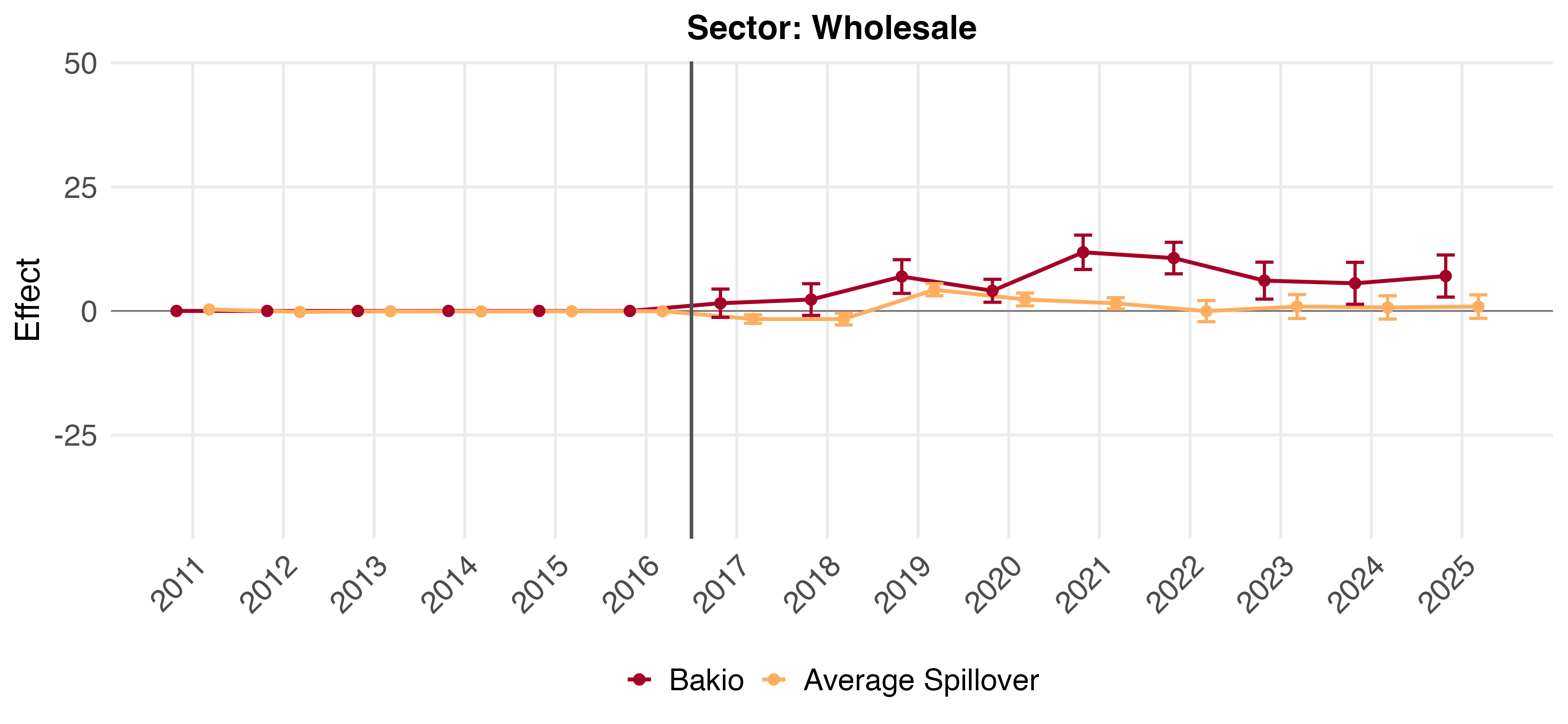} &
    \includegraphics[width=8cm]{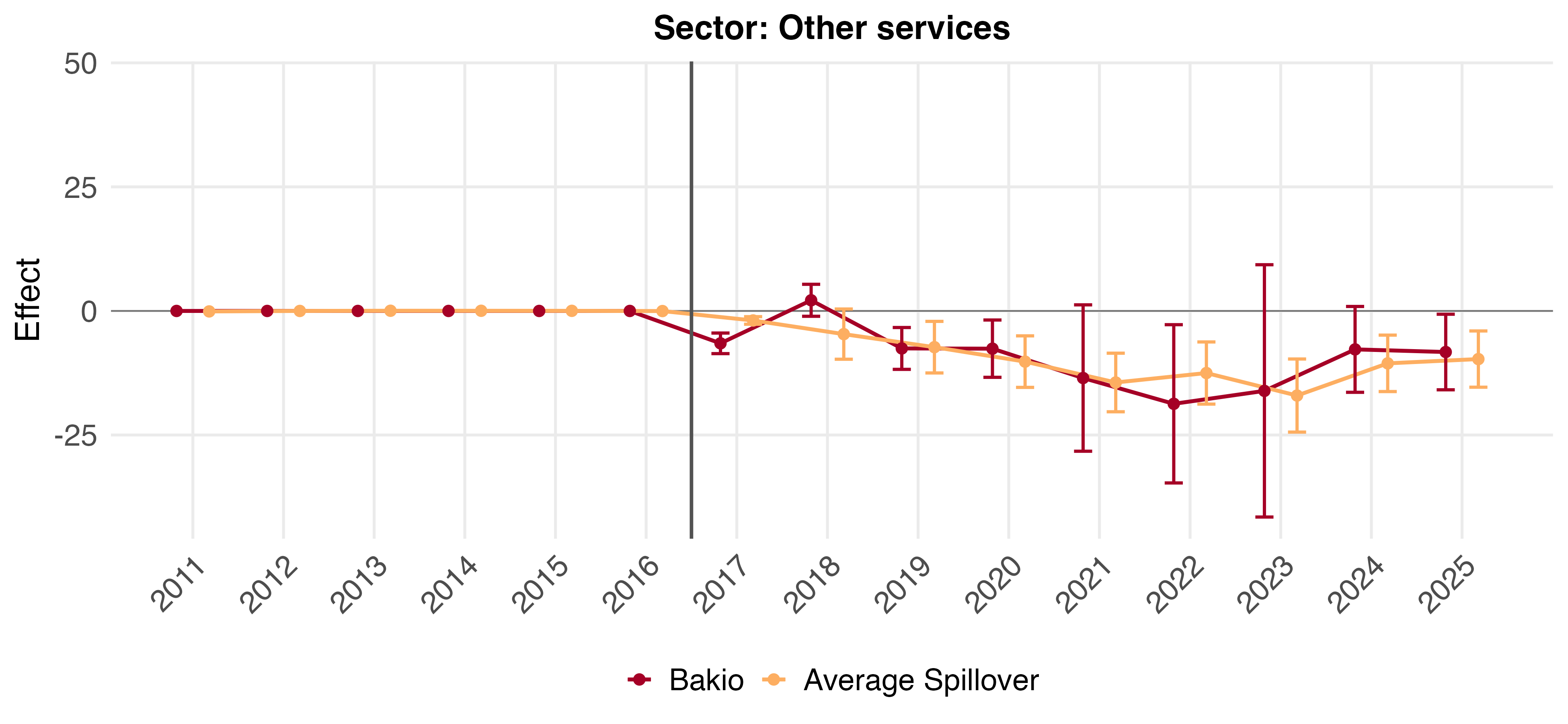} 
    \end{tabular}

    \vspace{-0.2cm}
    \begin{minipage}{\linewidth}
   {\scriptsize \setstretch{1}\textit{Notes:} The plots show the difference between the observed outcome and its synthetic control, together with 95\% confidence intervals. The red line corresponds to the treated municipality, Bakio, and the orange line to the average spillover group (the municipalities that belong to Bakio's or Gaztelugatxe's counties). Estimates are obtained using the Ridge Augmented Synthetic Control estimator \citep{ben-michael_augmented_2021}, estimated separately for each outcome and each treated municipality using the same donor pool of Basque municipalities. }
    \end{minipage}
\end{figure}

Table \ref{tab:labor_spillover} and Figures \ref{fig:labor_spillover}, \ref{fig:labor_reg_spillover}, and \ref{fig:labor_sector_spillover} report the average spillover effects on local labor markets. Unlike in Bakio, there is no evidence of employment gains in tourism-related sectors: the estimates for accommodation ($+0.06$ per 1,000 residents), restaurants ($-0.19$) and culture and recreation ($+0.77$) are small and not statistically significant. Employment declines significantly in manufacturing, other services and administrative services, while the effects on construction and wholesale trade are not significant.

Total employment does not change significantly. By regime (Table~\ref{tab:labor_spillover}, Panel B, and Figure~\ref{fig:labor_reg_spillover}), the point estimate for salaried employment is negative but not significant, whereas self-employment increases significantly ($+3.3$, against a 2016 level of 83.5). 

\begin{table}[H]
\centering
\caption{Labor market outcomes: average spillover effect (augmented synthetic control estimates)}
\footnotesize
\label{tab:labor_spillover}
\begin{tabular}{lcc}
\toprule
 & \multicolumn{1}{c}{Baseline (2016)} & \multicolumn{1}{c}{ATT} \\
\midrule
\multicolumn{3}{l}{\textit{Panel A: Aggregate employment}} \\
\quad Total & 263.765 & -3.695 \\
 &  & (10.584) \\
[3pt]
\midrule
\multicolumn{3}{l}{\textit{Panel B: Regime breakdown}} \\
\quad Salaried & 156.641 & -14.278 \\
 &  & (10.086) \\
[3pt]
\quad Self-employed & 83.477 & 3.331*** \\
 &  & (0.889) \\
[3pt]
\midrule
\multicolumn{3}{l}{\textit{Panel C: Sector breakdown}} \\
\quad Accommodation & 1.436 & 0.058 \\
 &  & (0.143) \\
[3pt]
\quad Restaurants & 21.468 & -0.187 \\
 &  & (2.282) \\
[3pt]
\quad Administrative services & 4.581 & -1.453** \\
 &  & (0.574) \\
[3pt]
\quad Culture \& recreation & 4.955 & 0.765 \\
 &  & (0.521) \\
[3pt]
\quad Manufacturing & 60.591 & -6.547*** \\
 &  & (1.435) \\
[3pt]
\quad Construction & 19.112 & 0.740 \\
 &  & (0.555) \\
[3pt]
\quad Wholesale & 41.378 & 0.808 \\
 &  & (0.603) \\
[3pt]
\quad Other Services & 78.105 & -9.815*** \\
 &  & (2.472) \\
[3pt]
\bottomrule
\end{tabular}
\vspace{-0.2cm}
\begin{minipage}{0.99\linewidth  \setstretch{0.75}}
{\scriptsize Notes: Estimates are obtained using the Ridge Augmented Synthetic Control estimator \citep{ben-michael_augmented_2021}, estimated separately for each outcome and treated municipality using the same donor pool of Basque municipalities. ``Spillover" municipalities are all municipalities other than Bakio in the counties of Mungialdea (Bakio's county) and Busturialdea (Gaztelugatxe's county). All outcomes are measured as the number of affiliated workers in the corresponding category per 1,000 inhabitants. The baseline column reports the 2016-population-weighted mean value of each outcome across spillover municipalities in 2016. The ATT column reports the 2016-population-weighted mean of the spillover municipalities' individual average treatment effects over the post-treatment period (2017--2025). $^{***}p<0.01$, $^{**}p<0.05$, $^{*}p<0.10$.}
 \end{minipage}
\end{table}

\subsubsection{Housing Market}
\begin{figure}[H]
    \caption{Event study estimates for housing prices and sales.}
    \label{fig:housing_reg_spillover}
    \centering
\begin{tabular}{cc}
\includegraphics[width = 8 cm]{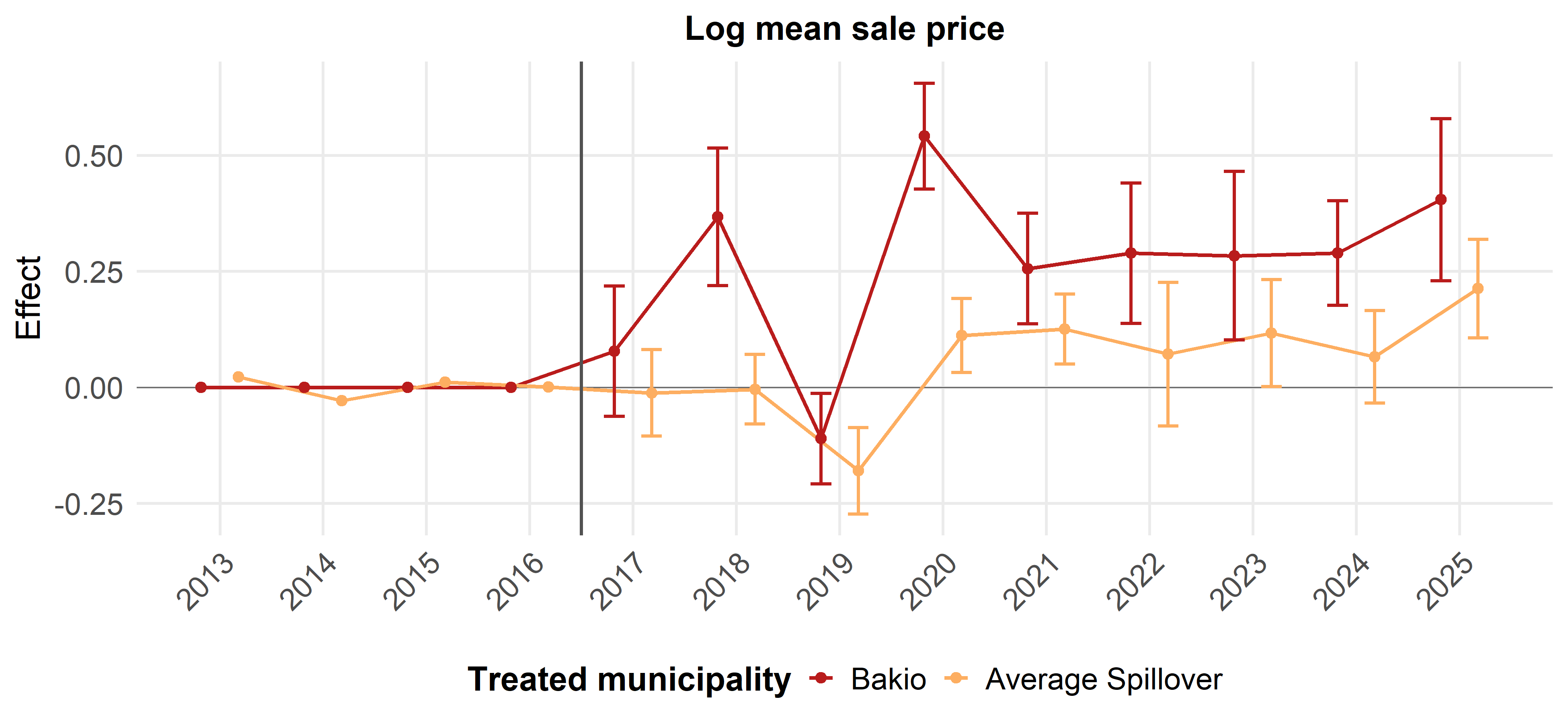}    
&  
\includegraphics[width = 8 cm]{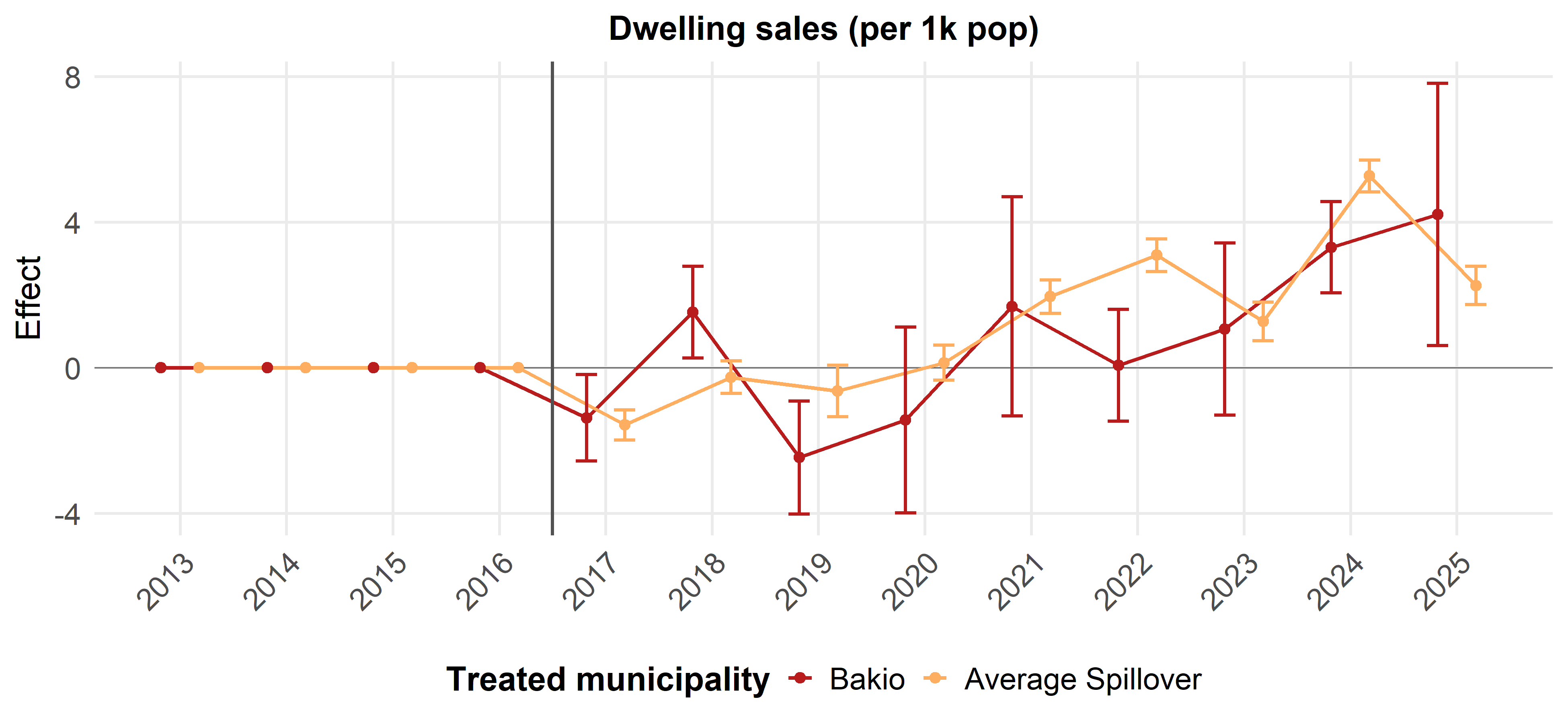} 
\end{tabular}
\vspace{-0.2cm}
\begin{minipage}{\linewidth}
   {\scriptsize \setstretch{1}\textit{Notes:} The plots show the difference between the observed outcome and its synthetic control, together with 95\% confidence intervals. The red line corresponds to the treated municipality, Bakio, and the orange line to the average spillover group (the municipalities that belong to Bakio's or Gaztelugatxe's counties). Estimates are obtained using the Ridge Augmented Synthetic Control estimator \citep{ben-michael_augmented_2021}, estimated separately for each outcome and treated municipality using the same donor pool of Basque municipalities. }
    \end{minipage}
\end{figure}

Table~\ref{tab:housing_spillover} and Figure~\ref{fig:housing_reg_spillover} report the spillover estimates for the housing market. Dwelling sales in the spillover municipalities rise by 1.28 per 1,000 inhabitants relative to their synthetic controls. Sale prices, in contrast, show no significant effect: the ATT on log mean prices is 0.06 but not statistically significant. We interpret the increase in sales cautiously because it may partly reflect a change in administrative data coverage.  According to Banco de Espa\~na, the Land Registry office serving most of the treated area (Bizkaia's Registry 10) changed its software in October 2020. Before that date, the registry reported only a subset of its transactions to the Banco de Espa\~na, and the missing records cannot be recovered. %This registry covers 27 of the 33 spillover municipalities.
In three of the largest municipalities in the spillover group, Bermeo, Gernika-Lumo, and Mungia, the number of recorded sales in 2021–2024 is roughly four to five times the level recorded in 2013–2019, compared with an increase of approximately 1.5 times in the typical Basque municipality.  Because these three towns account for most sales in the spillover group, the post-2020 increase in the estimated sales ATT may reflect both a genuine housing-market response and the change in registry coverage. Since the donor municipalities are unaffected by this reporting break, the synthetic control cannot account for it. Although average sale prices are not mechanically affected by the number of recorded transactions, changes in reporting coverage may also have altered the composition of observed sales. We therefore treat the estimated spillover effect on sales as potentially overstated and interpret the price estimates with the caveat that changes in transaction composition cannot be ruled out.

\begin{table}[h!]
\centering
\caption{Housing market outcomes: average spillover effect (augmented synthetic control estimates)}
\footnotesize
\label{tab:housing_spillover}
\begin{tabular}{lcc}
\toprule
 & \multicolumn{1}{c}{Baseline (2016)} & \multicolumn{1}{c}{ATT} \\
\midrule
\quad Dwelling sales (per 1k pop.) & 1.729 & 1.278*** \\
 & & (0.121) \\
\quad Log mean sale price & 12.195 & 0.060 \\
 & & (0.043) \\
[3pt]
\bottomrule
\end{tabular}
\vspace{-0.2cm}
\begin{minipage}{0.99\linewidth \setstretch{0.75}}
{\scriptsize Notes: Estimates are obtained using the Ridge Augmented Synthetic Control estimator \citep{ben-michael_augmented_2021}, estimated separately for each outcome and treated municipality using the same donor pool of Basque municipalities. ``Spillover'' municipalities are all municipalities other than Bakio in the treated counties. Sales are counts per 1,000 inhabitants (2016 population). The baseline column reports the mean value of each outcome across spillover municipalities in 2016. The ATT column reports the mean of the spillover municipalities' individual average treatment effects over the post-treatment period (2017--2025). $^{***}p<0.01$, $^{**}p<0.05$, $^{*}p<0.10$.}
\end{minipage}
\end{table}

%%%%%%%%%%%%%%%%%%%%%%%%%%%%%%%%%%%%
%%%%%%%%%%%%%%%%%%%%%%%%%%%%%%%%%%%%
%%%%%%%%%%%%%%%%%%%%%%%%%%%%%%%%%%%%

\subsubsection{Residential mobility}
\label{app:mobility_spillover}
\begin{figure}[H]
    \caption{Event study estimates for residential mobility, spillover municipalities.}
    \label{fig:mobility_es_spillover}
    \centering
    \setlength{\tabcolsep}{1pt}
    \begin{tabular}{cc}
    \includegraphics[width=0.45\linewidth]{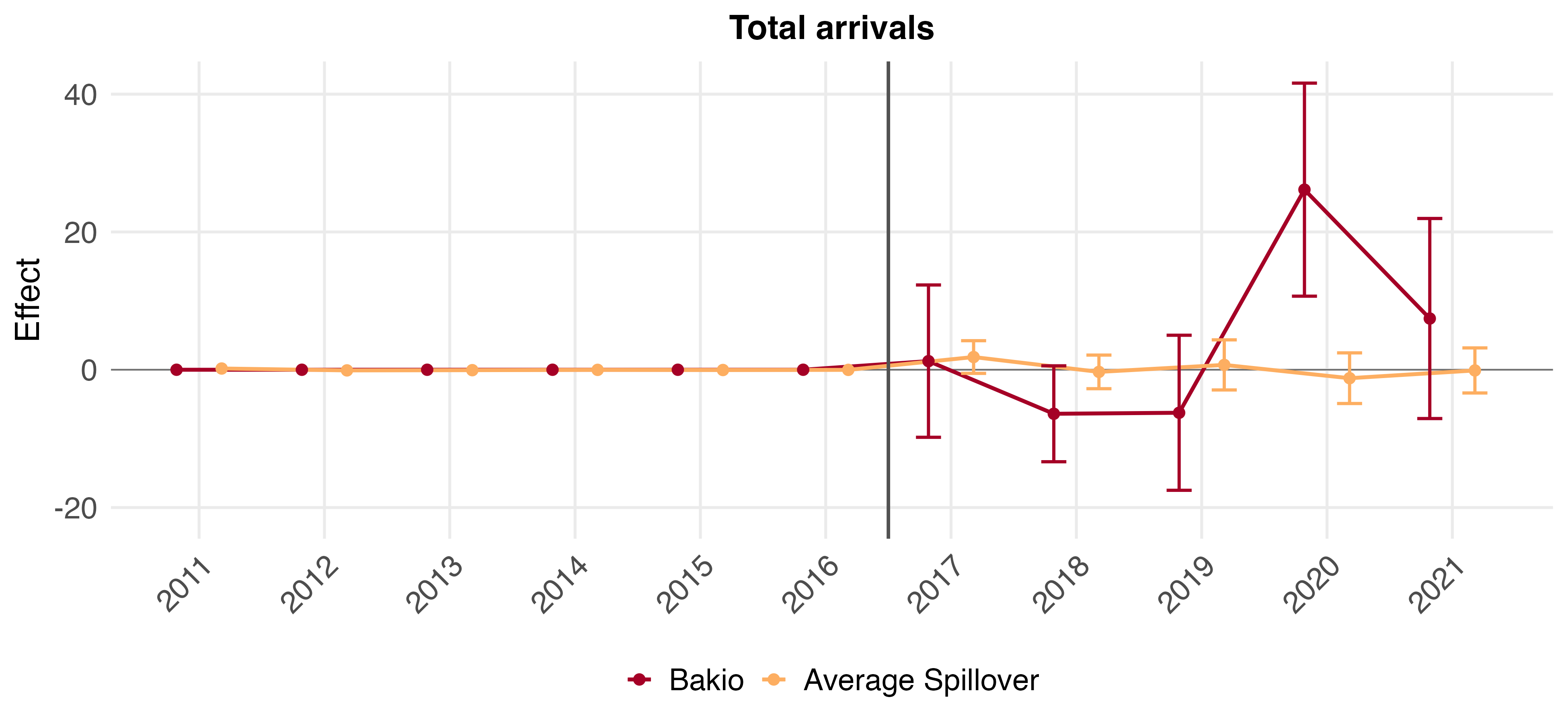} &
    \includegraphics[width=0.45\linewidth]{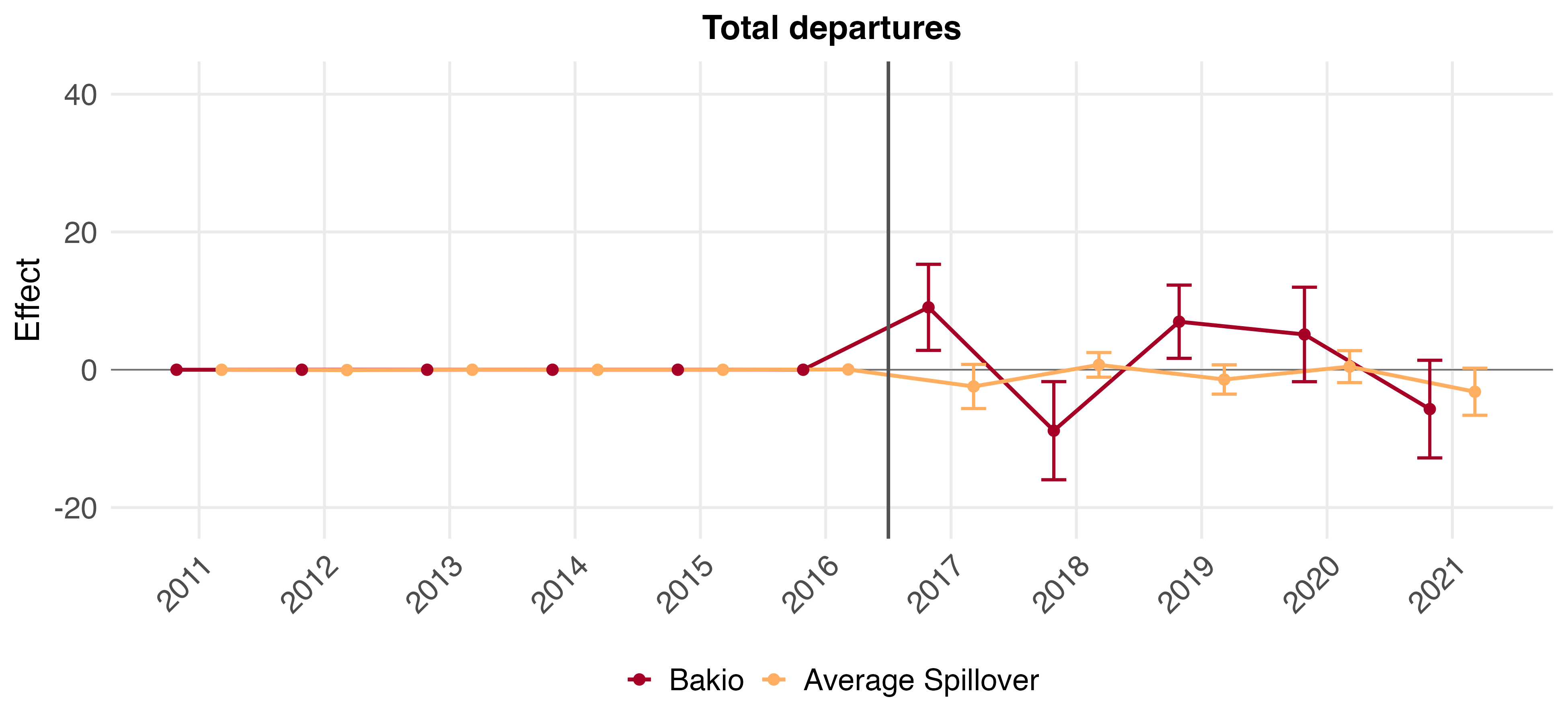} \\
    \includegraphics[width=0.45\linewidth]{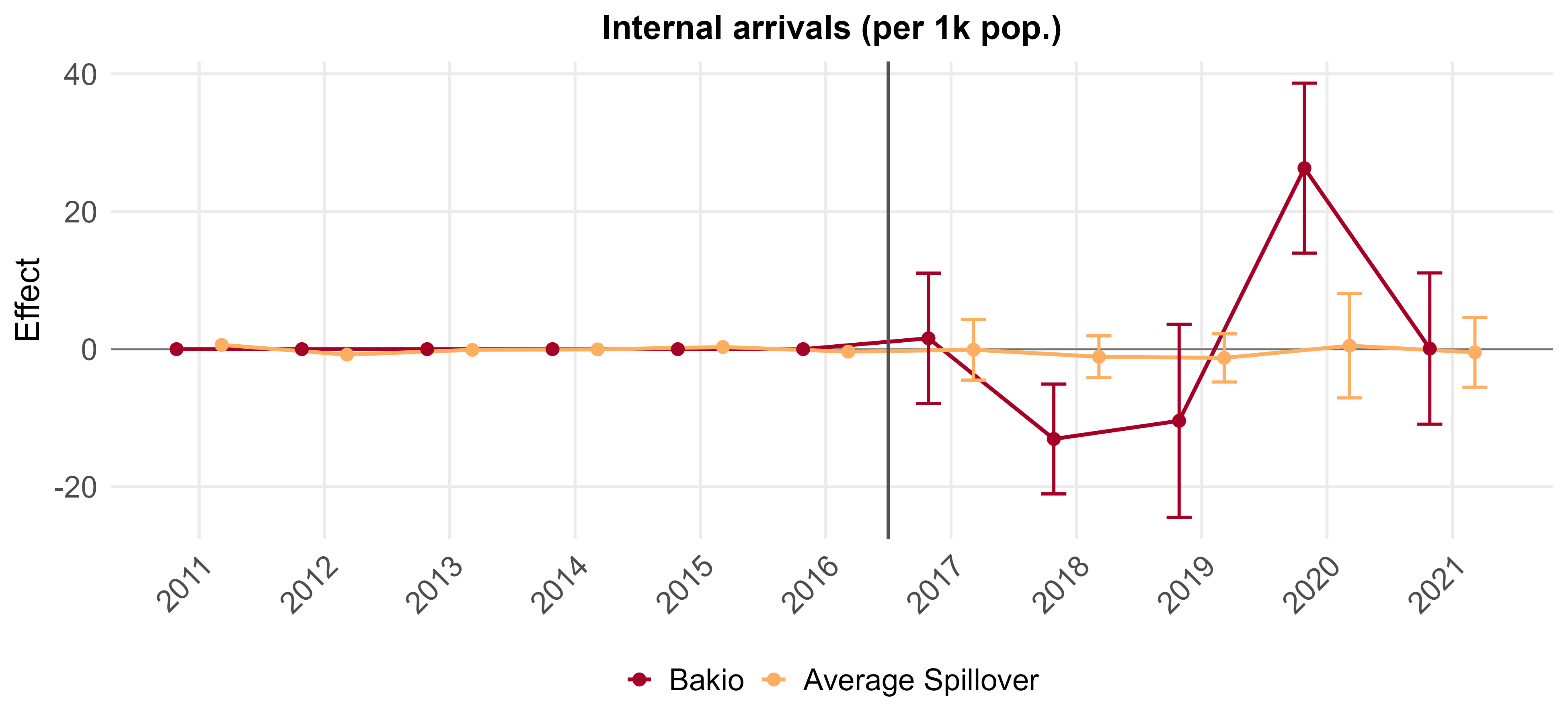} &
    \includegraphics[width=0.45\linewidth]{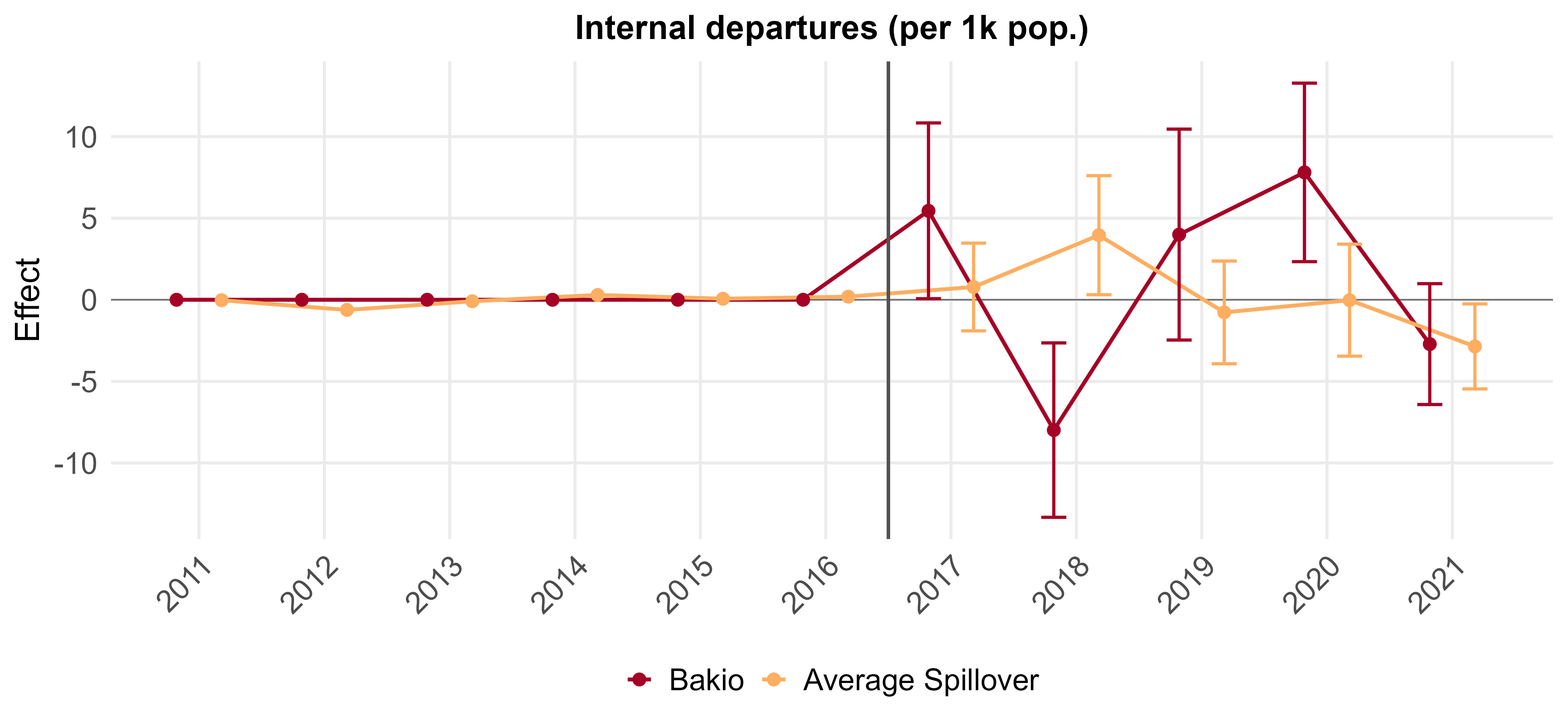} \\
    \includegraphics[width=0.45\linewidth]{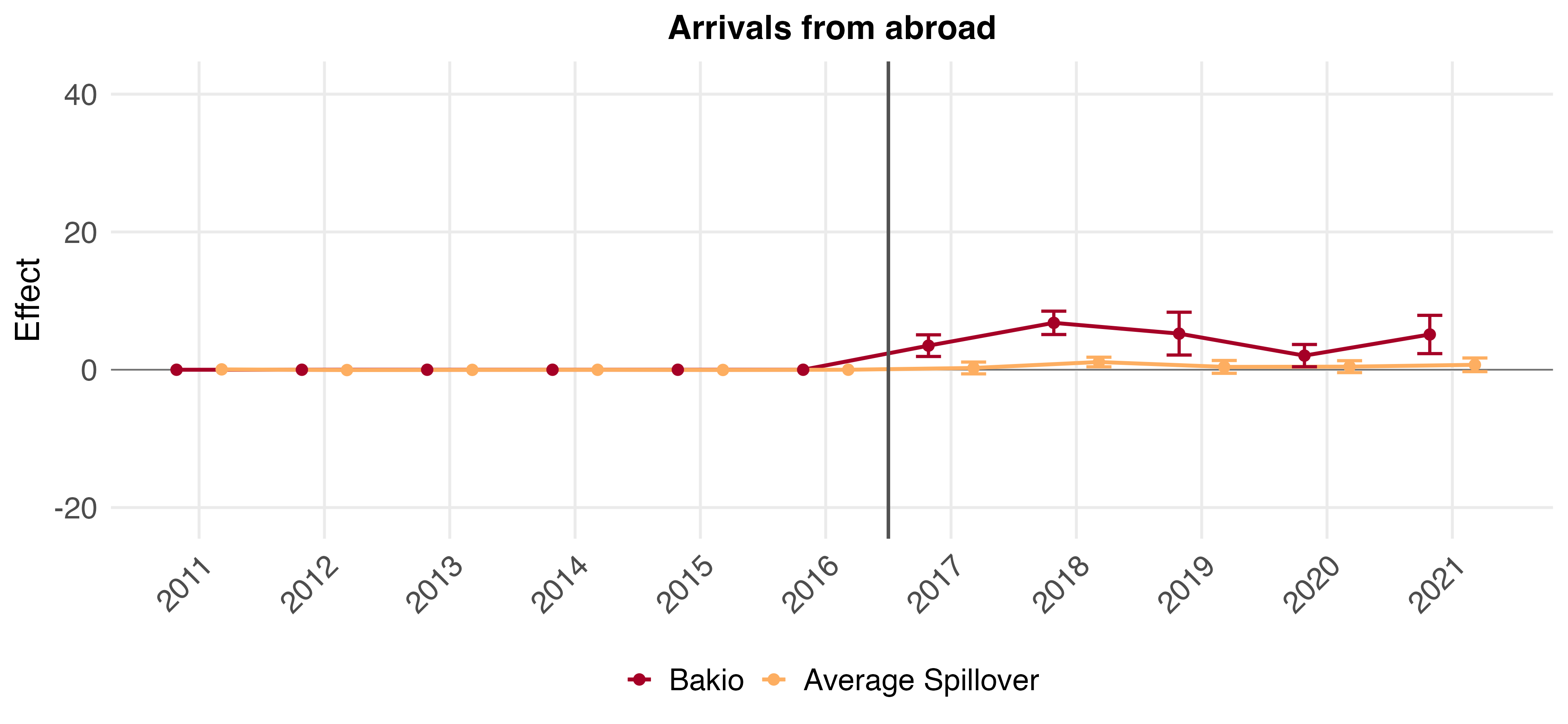} &
    \includegraphics[width=0.45\linewidth]{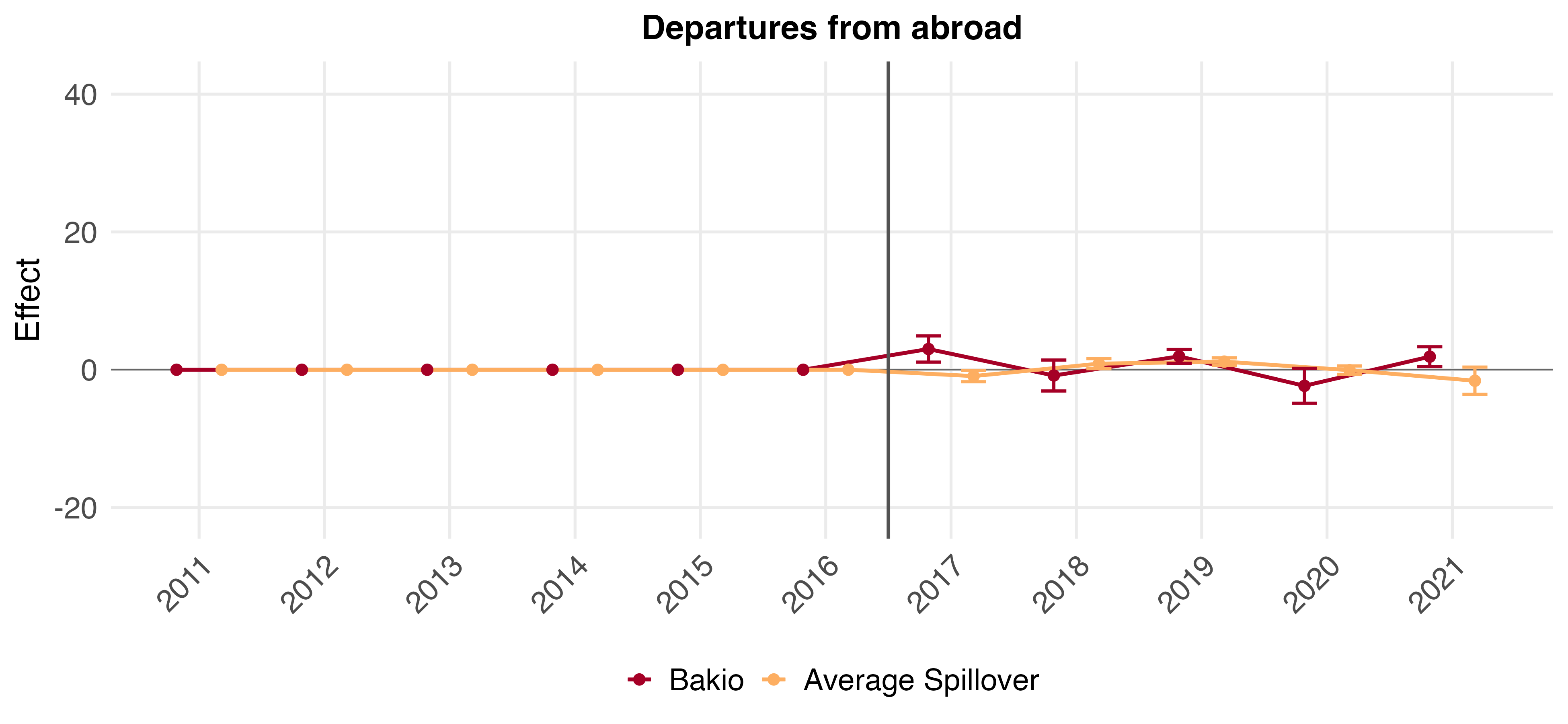}
    \end{tabular}
    \vspace{-0.2cm}
    \begin{minipage}{\linewidth}
    {\scriptsize \setstretch{1}\textit{Notes:} The plots show the difference between the observed outcome and its synthetic control, together with 95\% confidence intervals. The red line corresponds to the treated municipality, Bakio, and the orange line to the average spillover group (the municipalities that belong to Bakio's or Gaztelugatxe's counties). The left column shows arrivals and the right column departures; rows show all moves (abroad and internal), moves between Spanish municipalities and moves to and from abroad. Outcomes are annual moves per 1,000 residents (2016 population), 2011--2021, from the \textit{Estad\'{i}stica de Variaciones Residenciales} of the Spanish Statistical Office (INE). Estimates are obtained using the Ridge Augmented Synthetic Control estimator \citep{ben-michael_augmented_2021}, estimated separately for each outcome and each treated municipality using the same donor pool of Basque municipalities. Within each row, arrivals and departures are drawn on the same vertical scale.}
    \end{minipage}
\end{figure}

This section reports the residential mobility results for the spillover municipalities. Table \ref{tab:mobility_main_spillover} reports the average effect across spillover municipalities and Figure \ref{fig:mobility_es_spillover} plots the event studies for the average spillover effect together with those for Bakio.

The effect on arrivals from abroad is positive and significant, but much smaller than Bakio. Total arrivals and total departures do not change significantly, and neither do moves between Spanish municipalities, departures abroad, or total population registered. 

\begin{table}[!h]
\centering
\caption{Residential mobility: average spillover effects}
\footnotesize
\label{tab:mobility_main_spillover}
\begin{tabular}{lcc}
\toprule
 & \multicolumn{2}{c}{Average spillover} \\
 \cmidrule(lr){2-3}
 & \multicolumn{1}{c}{Baseline (2016)} & \multicolumn{1}{c}{ATT} \\
\midrule
\multicolumn{3}{l}{\textit{Panel A: Arrivals}} \\
Total arrivals & 48.077 & 0.182 \\
 &  & (1.074) \\
[3pt]
Internal arrivals & 40.816 & -0.375 \\
 &  & (0.742) \\
[3pt]
Arrivals from abroad & 7.261 & 0.592** \\
 &  & (0.249) \\
[3pt]
\midrule
\multicolumn{3}{l}{\textit{Panel B: Departures}} \\
Total departures & 42.548 & -1.175 \\
 &  & (0.991) \\
[3pt]
Internal departures & 36.959 & -0.071 \\
 &  & (0.689) \\
[3pt]
Departures abroad & 5.589 & -0.104 \\
 &  & (0.192) \\
[3pt]
\midrule
\multicolumn{3}{l}{\textit{Panel C: Population}} \\
Population (residents) & 11,073 & 31 \\
 &  & (140) \\
[3pt]
\bottomrule
\end{tabular}
\vspace{-0.2cm}
\begin{minipage}{0.99\linewidth  \setstretch{0.75}}
{\scriptsize Notes: Estimates are obtained using the Ridge Augmented Synthetic Control estimator \citep{ben-michael_augmented_2021}, estimated separately for each outcome and treated municipality using the same donor pool of Basque municipalities. Outcomes are annual counts of residential moves (changes of registered residence) per 1,000 residents (population fixed at 2016), from the \textit{Estad\'{i}stica de Variaciones Residenciales} (EVR) of the Spanish Statistical Office (INE). Panels A and B use the same sample, 2011--2021, with post-treatment period 2017--2021. Panel C reports the registered population (number of residents) at the end of each year, from the municipal register (\textit{Padr\'{o}n}), with sample 2001--2024 and post-treatment period 2017--2024. Arrivals are registrations in the municipality and departures are deregistrations. Moves between Spanish municipalities are changes of residence from or to another Spanish municipality, and moves to and from abroad are registrations from or deregistrations to a foreign country. ``Spillover" municipalities are all municipalities other than Bakio in the Busturia-Urdaibai and Uribe-Butroe counties. The baseline column reports the 2016-population-weighted mean of each outcome across spillover municipalities in 2016. The ATT column reports the 2016-population-weighted mean of the spillover municipalities' individual average treatment effects over the post-treatment period. $^{***}p<0.01$, $^{**}p<0.05$, $^{*}p<0.10$.}
 \end{minipage}
\end{table}

\begin{figure}[H]
    \caption{Event study estimates for population, spillover municipalities.}
    \label{fig:population_spillover}
    \centering
    
    \includegraphics[width=0.45\linewidth]{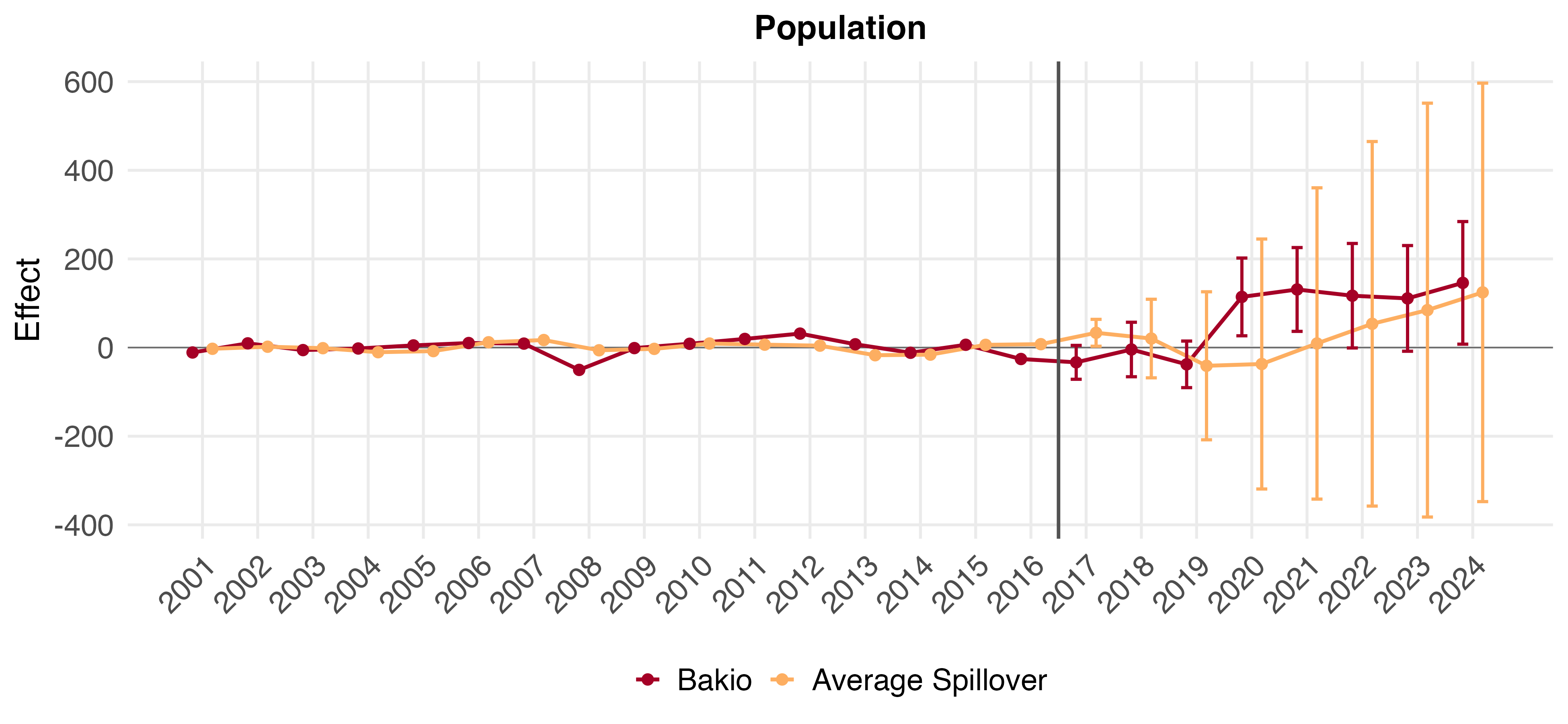} 
    \vspace{-0.2cm}
    \begin{minipage}{\linewidth}
    {\scriptsize \setstretch{1}\textit{Notes:} The plots show the difference between the observed outcome and its synthetic control, together with 95\% confidence intervals. The red line corresponds to the treated municipality, Bakio, and the orange line to the average spillover group (the municipalities that belong to Bakio's or Gaztelugatxe's counties). Estimates are obtained using the Ridge Augmented Synthetic Control estimator \citep{ben-michael_augmented_2021}, estimated separately for each outcome and each treated municipality using the same donor pool of Basque municipalities.}
    \end{minipage}
\end{figure}

\clearpage

\subsection{Regional vs local shock} \label{app:regional_shock}
A potential concern is that the \textit{Game of Thrones} shock increased tourism demand throughout the Basque Country, for example by raising awareness of the region as a tourist destination, rather than primarily affecting the area surrounding the filming location. Such a regional effect would contaminate our donor pool and undermine the identification strategy. To assess this possibility, we estimate the effect of the shock on total overnight stays in the three Basque provinces, using the remaining 47 Spanish provinces as the donor pool. We use annual province-level (NUTS 3) data on overnight stays from the INE covering 1999--2025 and apply the same Ridge Augmented Synthetic Control estimator \citep{ben-michael_augmented_2021} used in the main analysis. This province-level series is distinct from the thin municipal overnight-stays data discussed in Section~\ref{sec:data}; here it serves only as a test of whether the shock operated at the regional rather than the local level.

\begin{figure}[H]
    \caption{Event study estimates for overnight stays, Basque Country vs. synthetic control.}
    \label{fig:regional_vs_local}
    \centering
\begin{tabular}{cc}
\includegraphics[width = 0.8\linewidth]{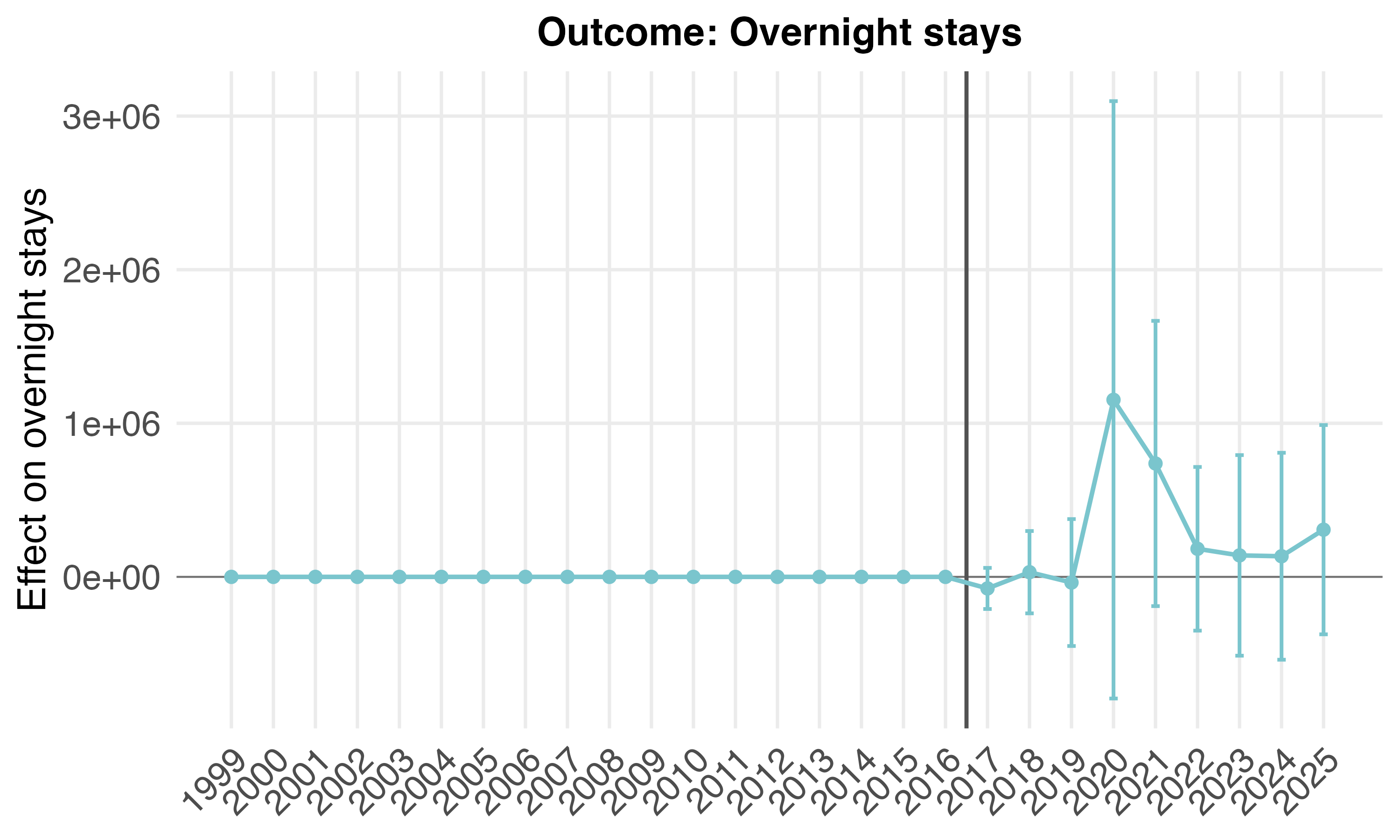}    
\end{tabular}
\vspace{-0.2cm}
\begin{minipage}{\linewidth}
{\scriptsize \setstretch{1}\textit{Notes:} The plots show the difference between the observed outcome for the Basque Country and its synthetic control, together with 95\% confidence intervals. The synthetic control is computed using the 47 Spanish provinces other than Araba, Bizkaia, and Gipuzkoa. Estimates are obtained using the Ridge Augmented Synthetic Control estimator \citep{ben-michael_augmented_2021}. }
 \end{minipage}
\end{figure}

Figure \ref{fig:regional_vs_local} plots the event-study estimates for the post-treatment period. We cannot reject the null hypothesis of no treatment effect in any post-treatment period. Excluding 2020 and 2021, which are particularly noisy due to the COVID-19 pandemic, the estimated average effect is 97{,}522 additional overnight stays, with a standard error of 201{,}229. The estimate is neither statistically significant nor economically large: it corresponds to an increase of approximately 5\% relative to the 2016 provincial level. We find no evidence that the shock raised tourism at the regional scale.

Hence, we find no evidence that the spotlight shock affected tourism throughout the Basque Country. This result is reassuring for our identification strategy, as it supports the assumption that the control municipalities were not exposed to the shock, and that the treatment reflects a geographically concentrated local tourism shock rather than a broader regional shock affecting the Basque Country as a whole.

\subsection{Expanding the buffer around treated municipality} \label{app:robust_bizkaia}
\begin{figure}[H]
    \caption{Municipalities in the Basque Country by treatment group, robust version}
    \label{fig:map_groups_robust}
    \centering
\includegraphics[width = \linewidth]{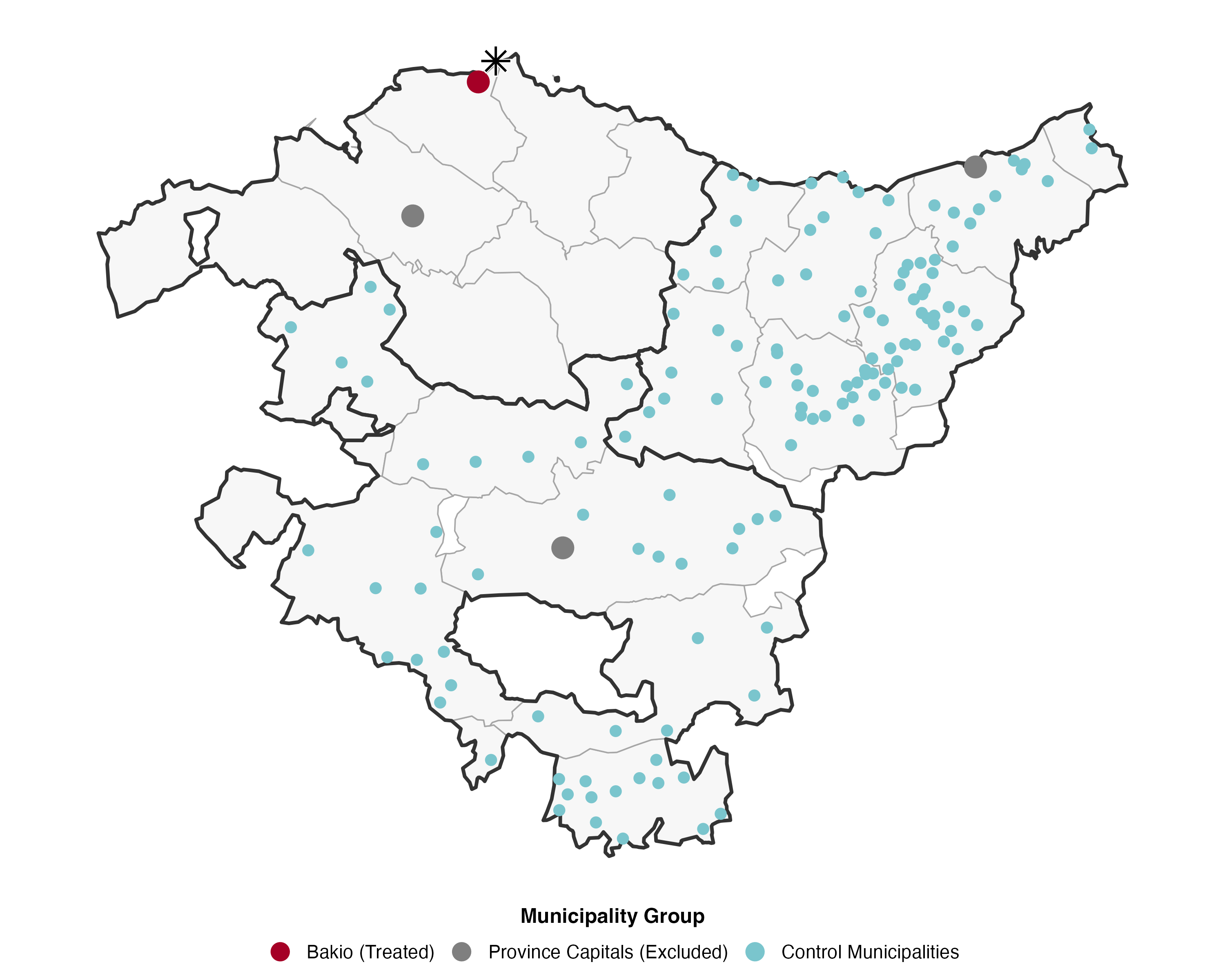}
\vspace{-0.2cm}
\begin{minipage}{\linewidth}
{\scriptsize \setstretch{1}\textit{Notes:} The star marks San Juan de Gaztelugatxe. Red: the treated municipality (Bakio). Orange: municipalities in the same counties as the treated unit and Gaztelugatxe, excluded from the donor pool and used to estimate spillover effects. Blue: the donor pool. Grey: The three provincial capitals, which are excluded from the analysis.}
 \end{minipage}
\end{figure}

\begin{figure}[H]
    \caption{Event study estimates for excluding all municipalities from Bizkaia province from the donor pool.}
    \label{fig:robustness_bizkaia}
    \centering
    \setlength{\tabcolsep}{1pt}
    \begin{tabular}{ccc}
    \includegraphics[width=0.3\linewidth]{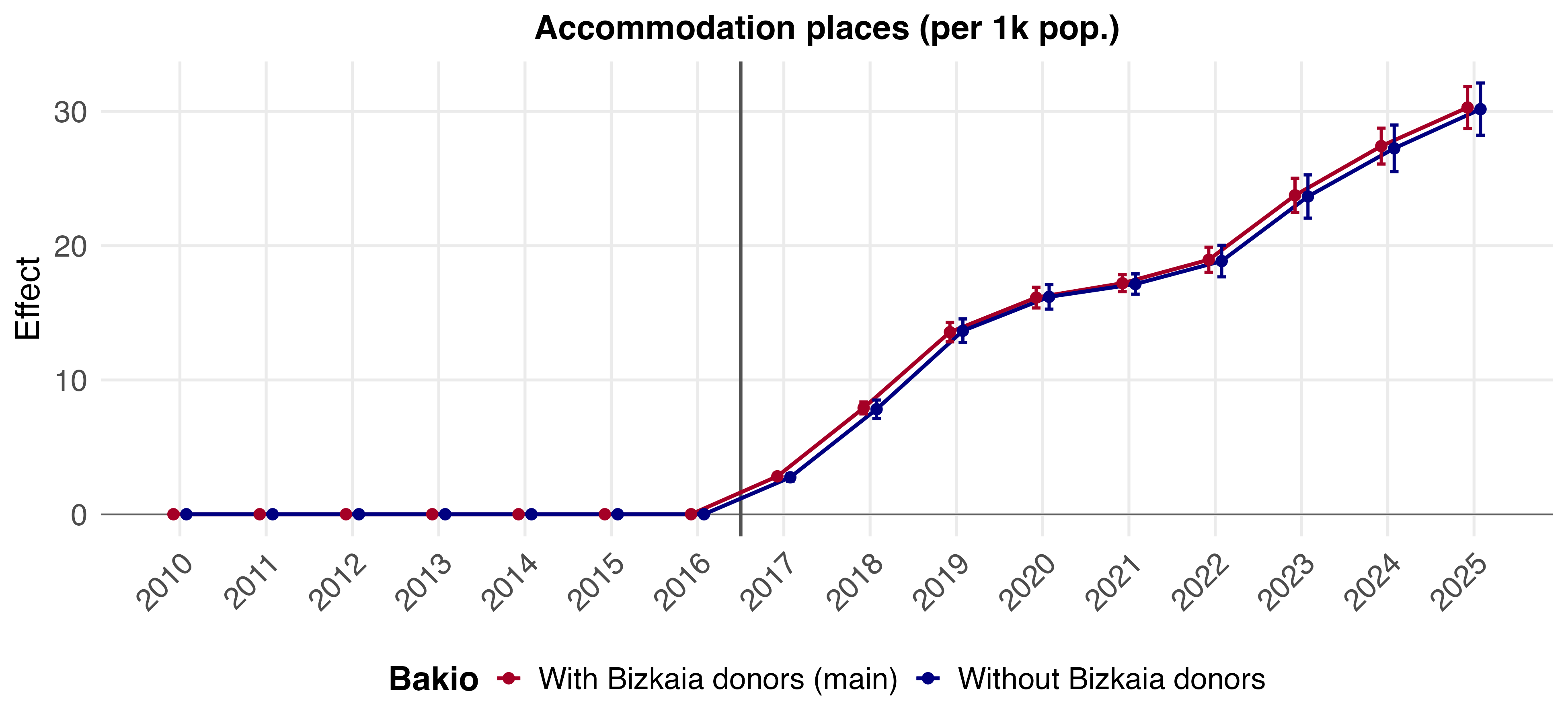} &
    \includegraphics[width=0.3\linewidth]{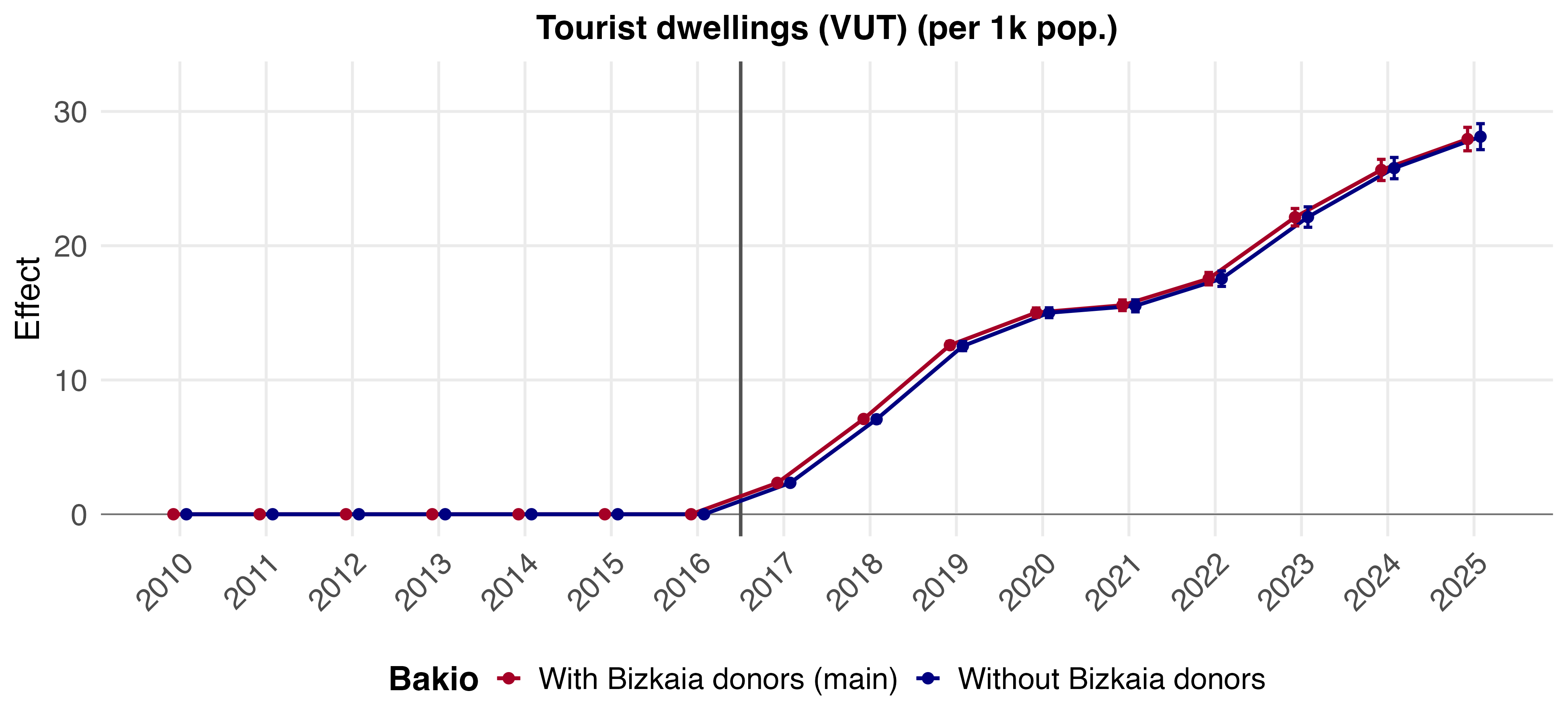} &
    \includegraphics[width=0.3\linewidth]{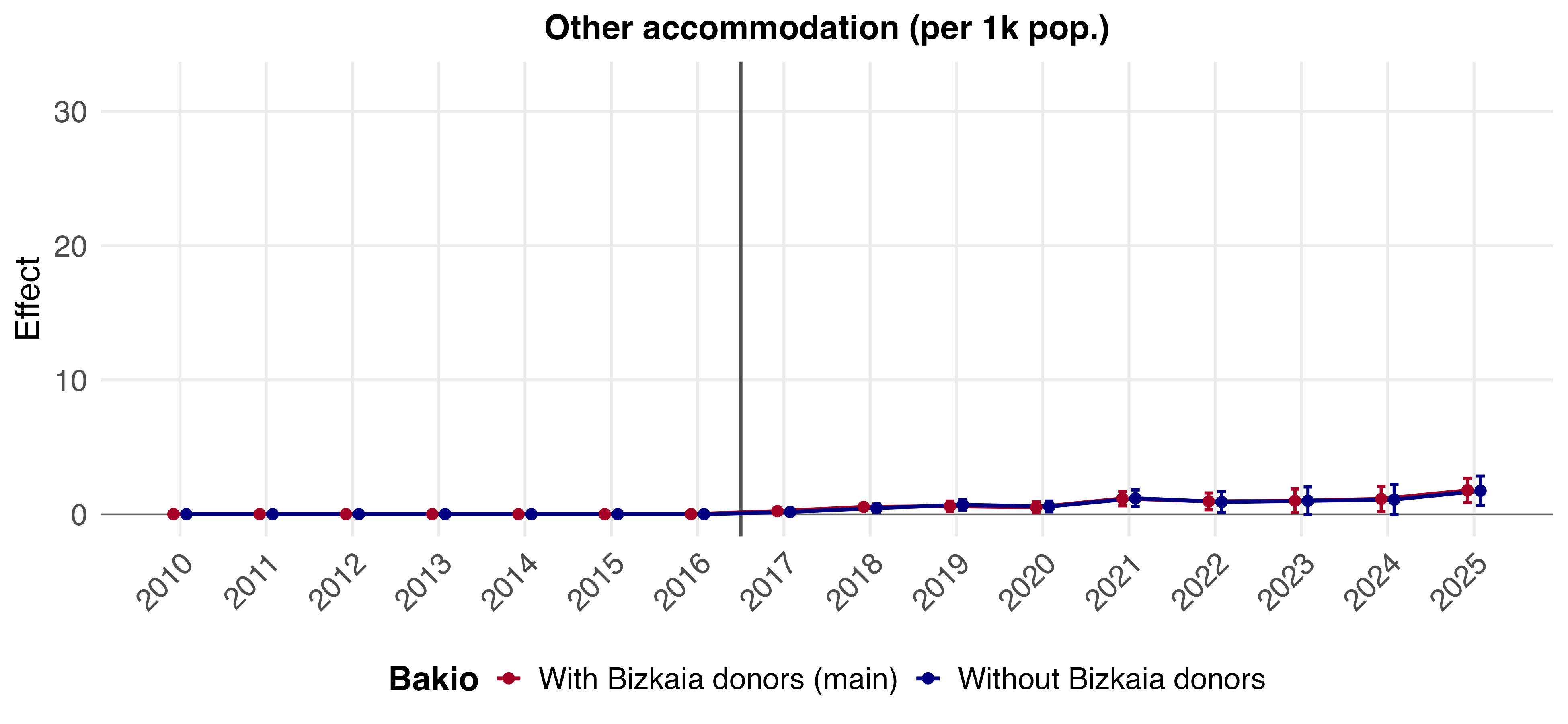} \\

    \includegraphics[width=0.3\linewidth]{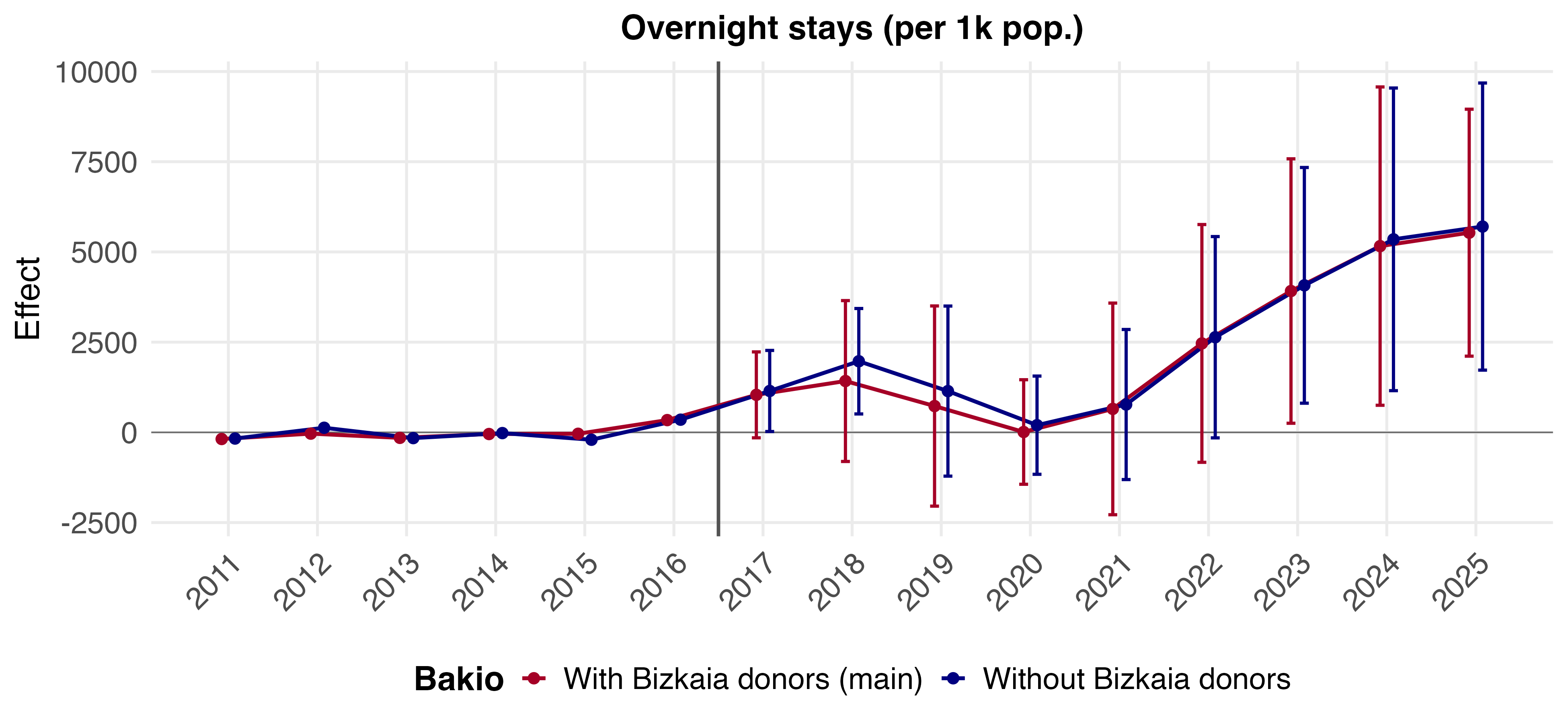} &
    \includegraphics[width=0.3\linewidth]{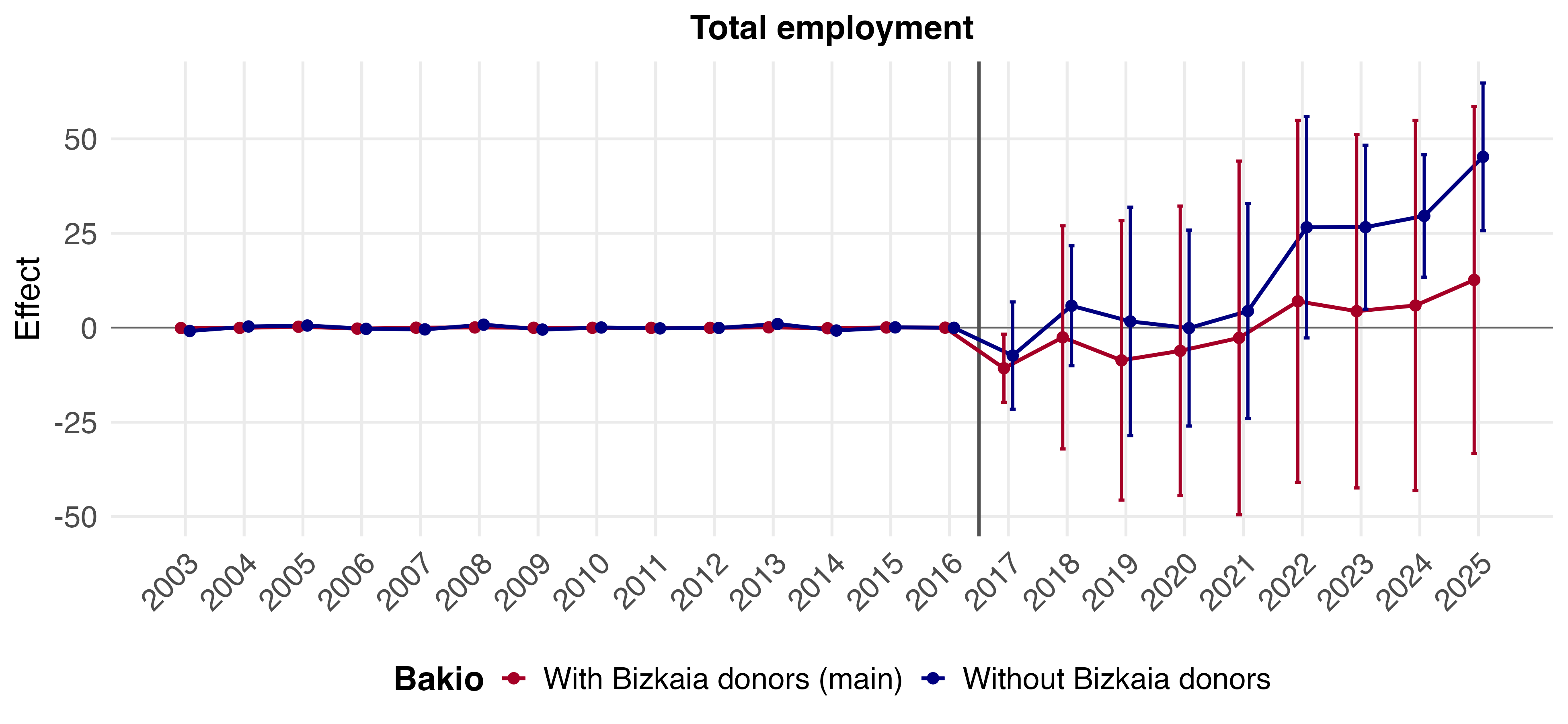} &
    \includegraphics[width=0.3\linewidth]{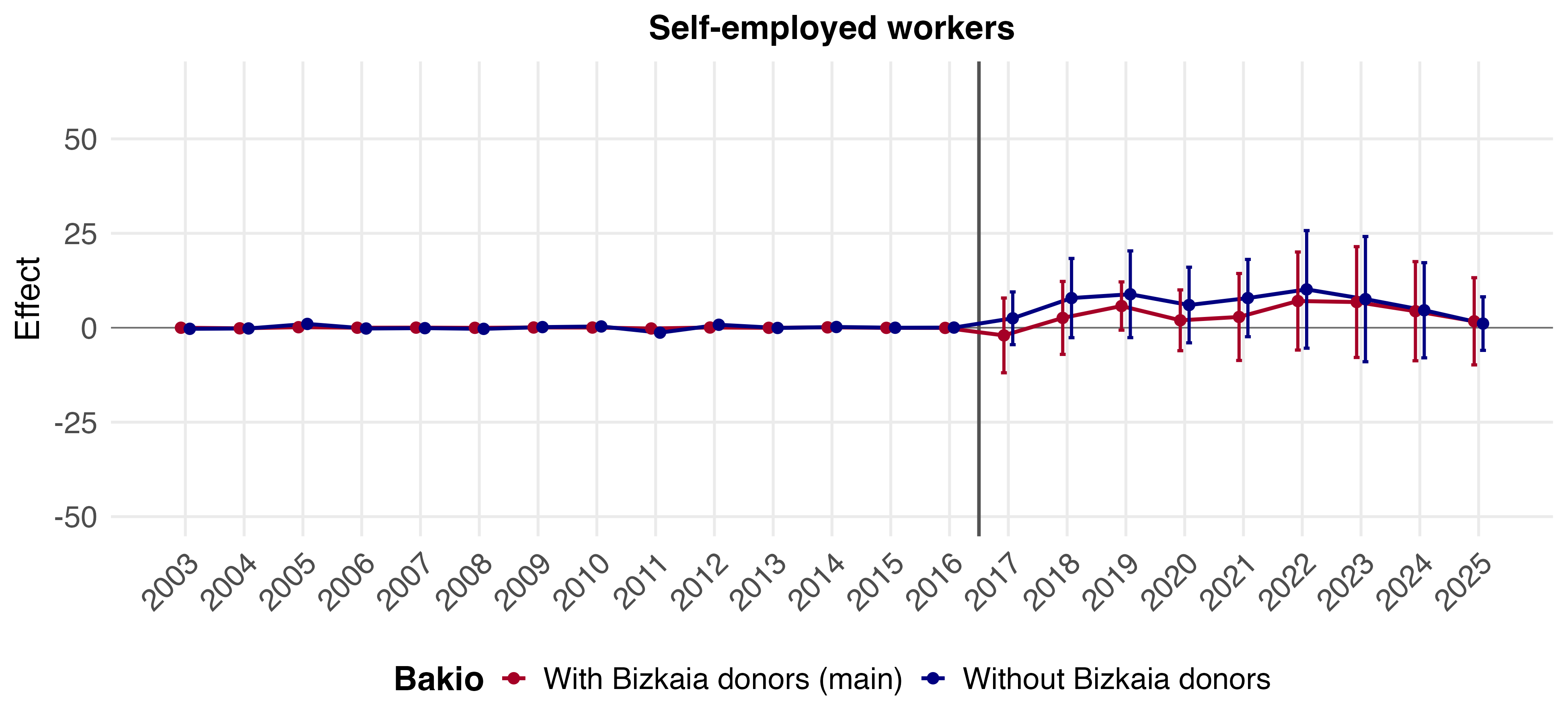} \\

    \includegraphics[width=0.3\linewidth]{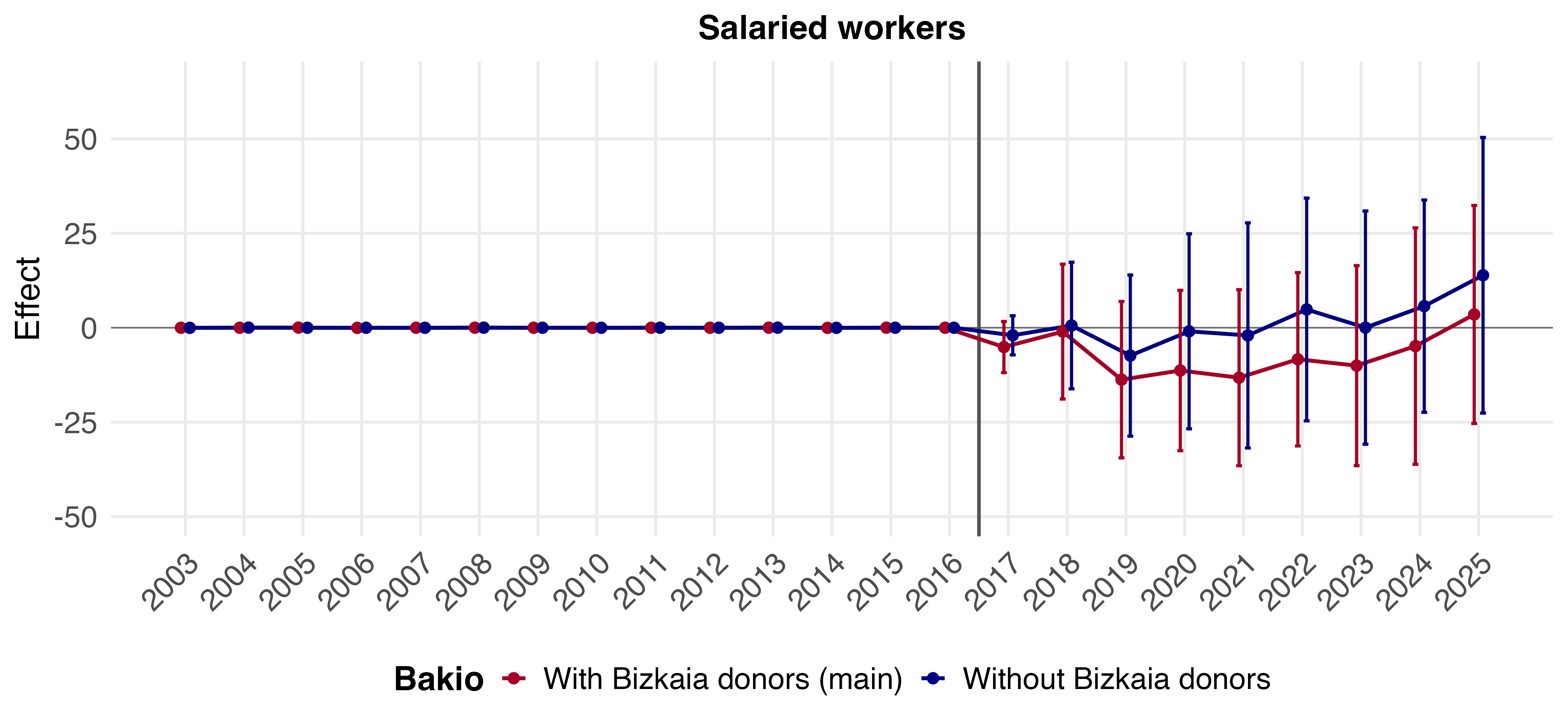} &
    \includegraphics[width=0.3\linewidth]{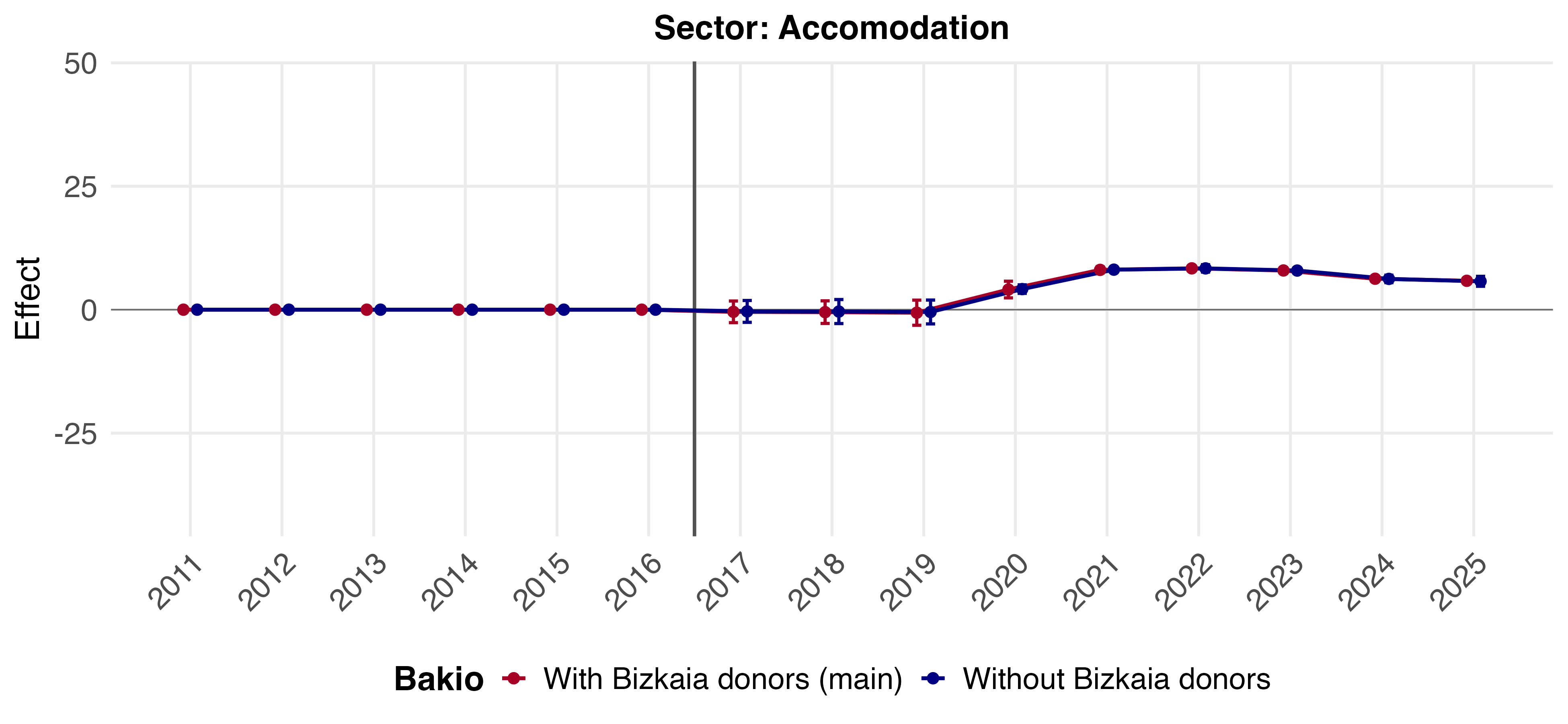} &
    \includegraphics[width=0.3\linewidth]{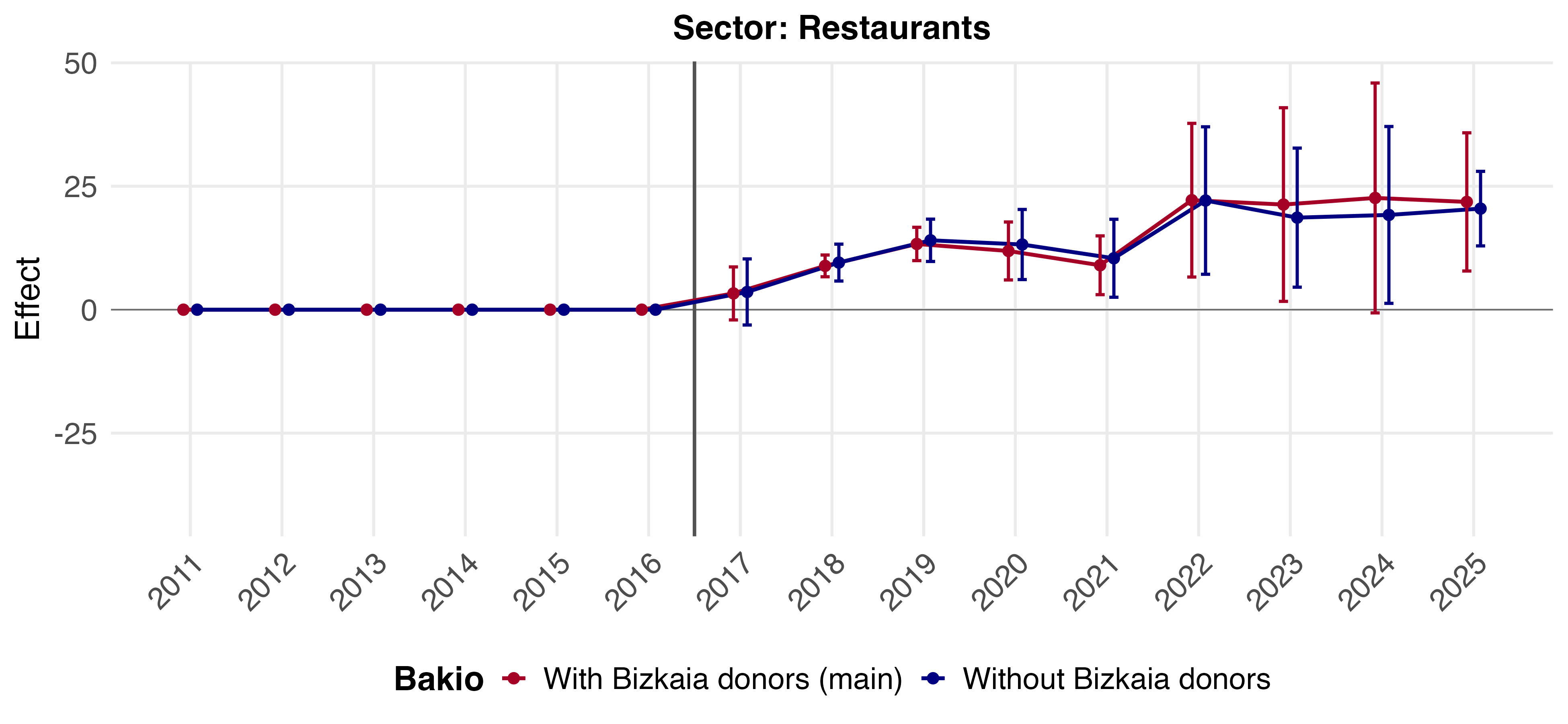} \\

    \includegraphics[width=0.3\linewidth]{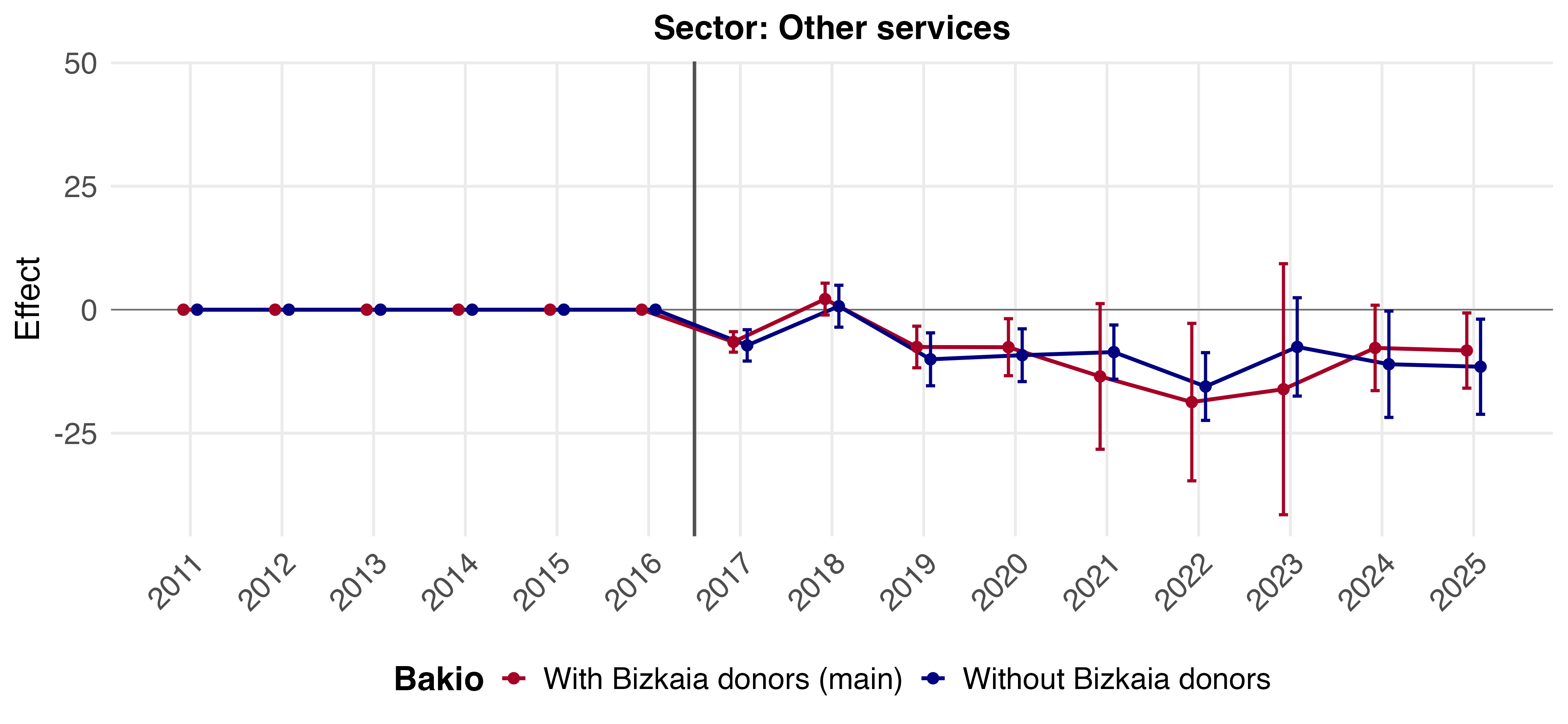} &
    \includegraphics[width=0.3\linewidth]{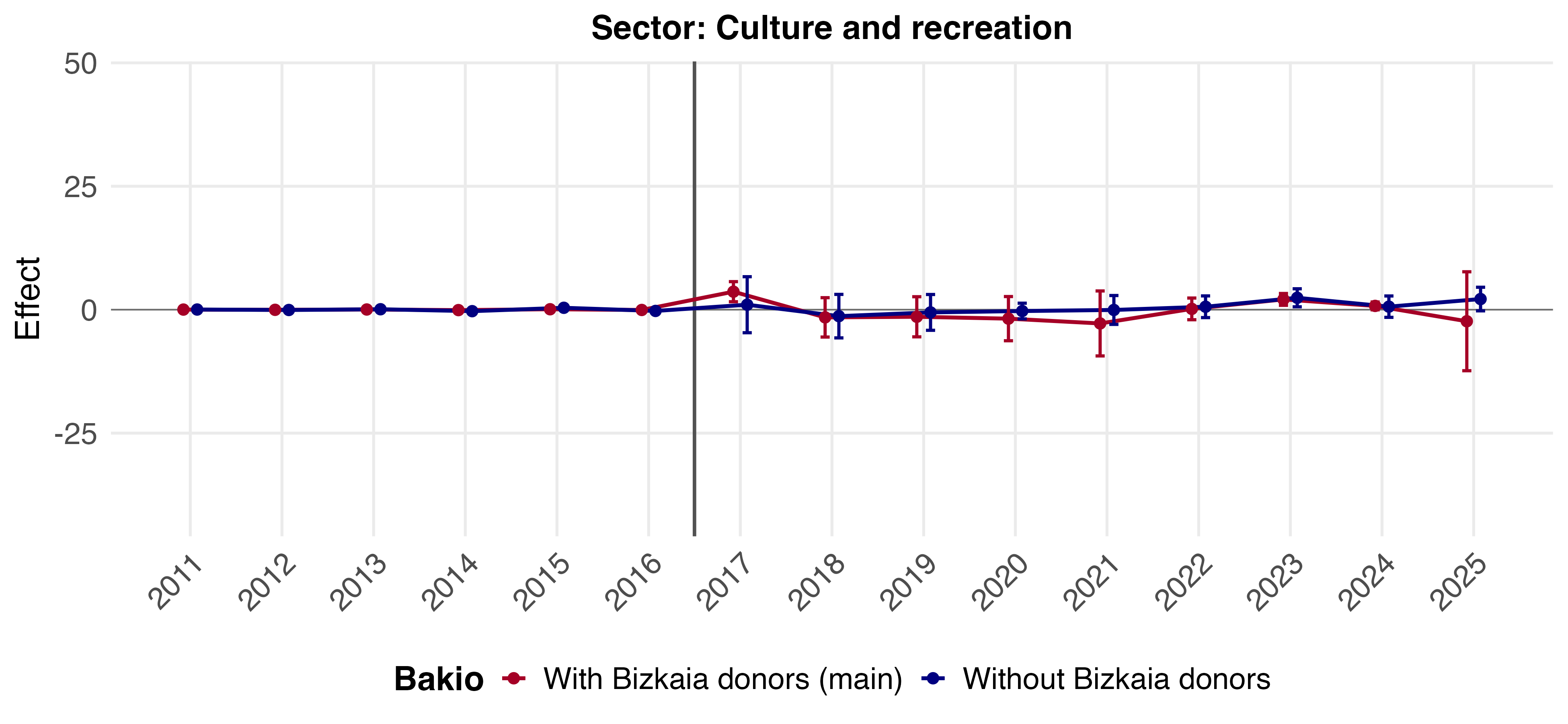} &
    \includegraphics[width=0.3\linewidth]{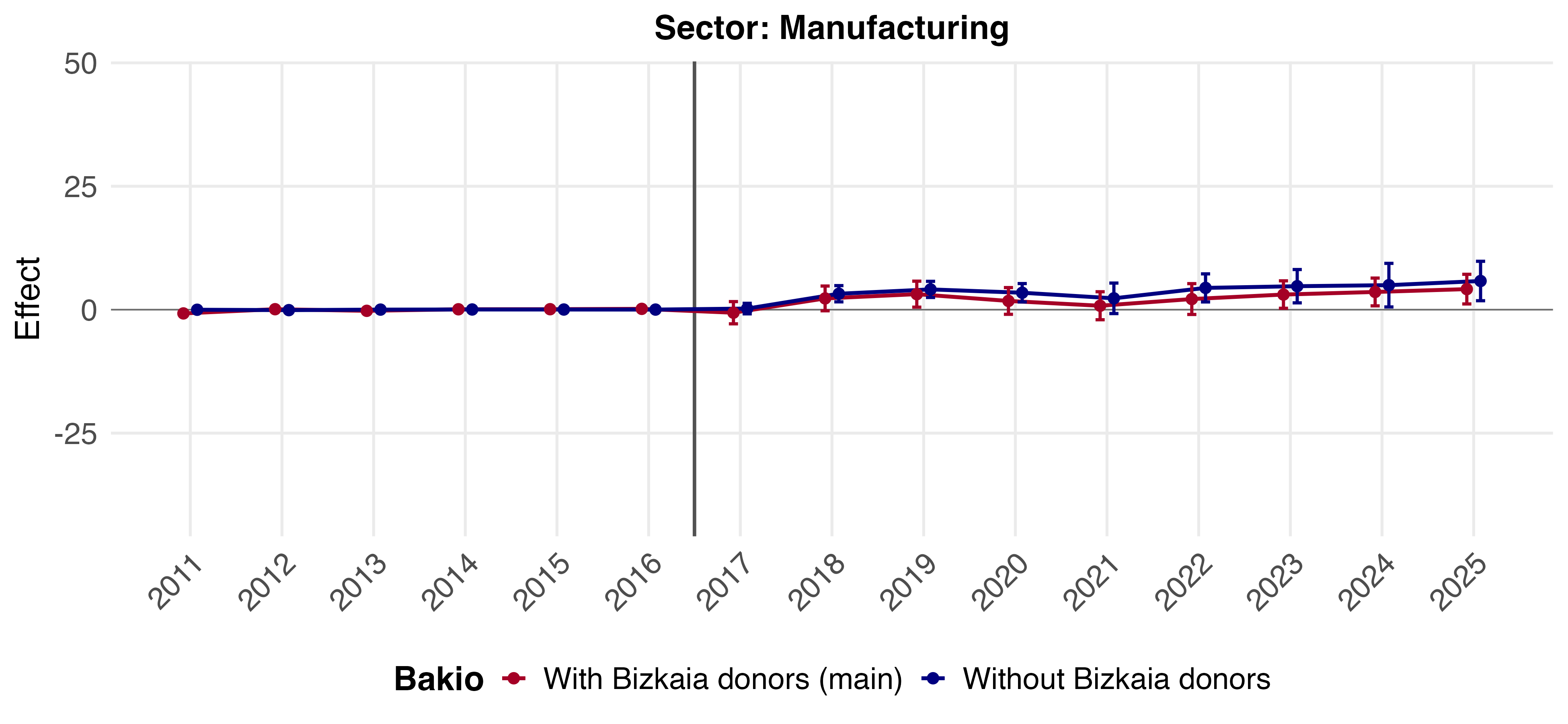} \\

    \includegraphics[width=0.3\linewidth]{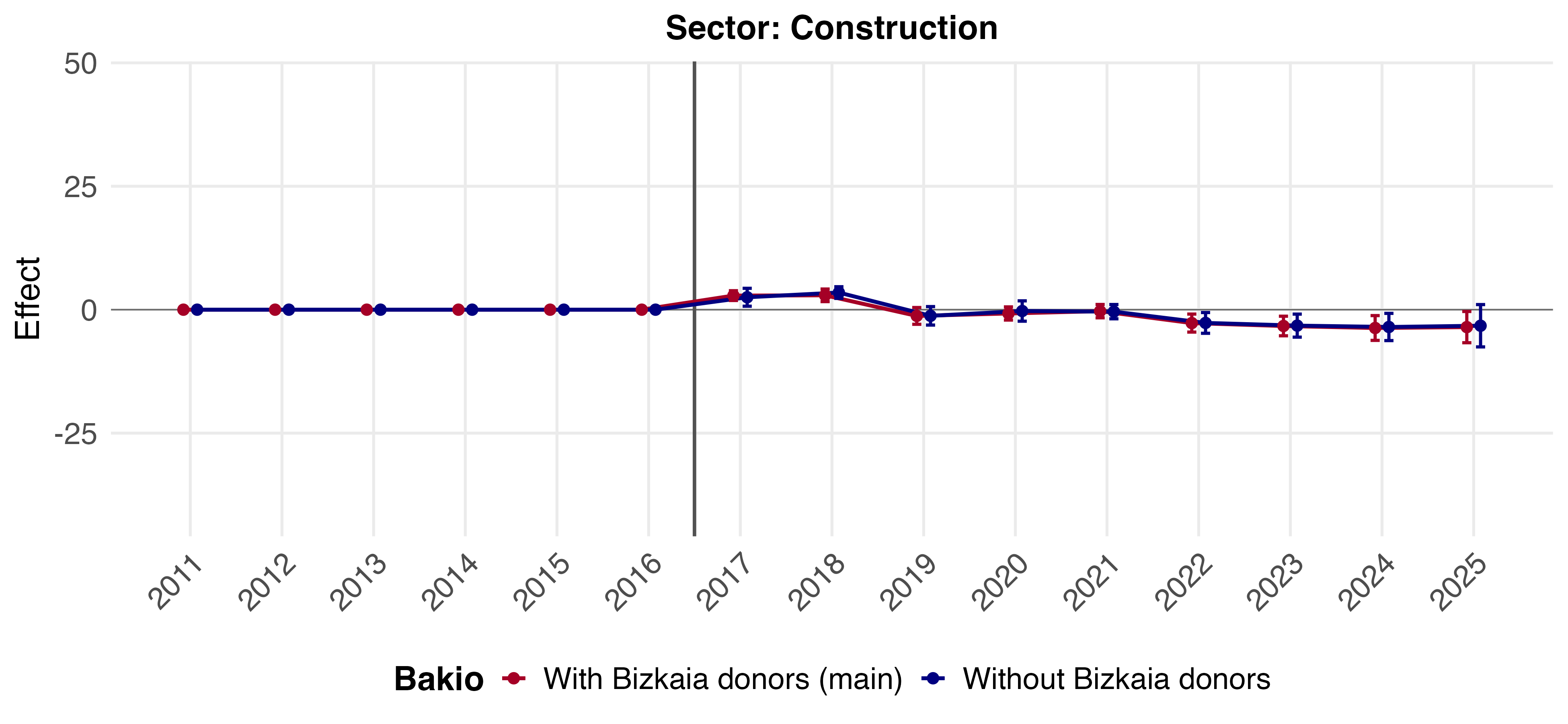} &
    \includegraphics[width=0.3\linewidth]{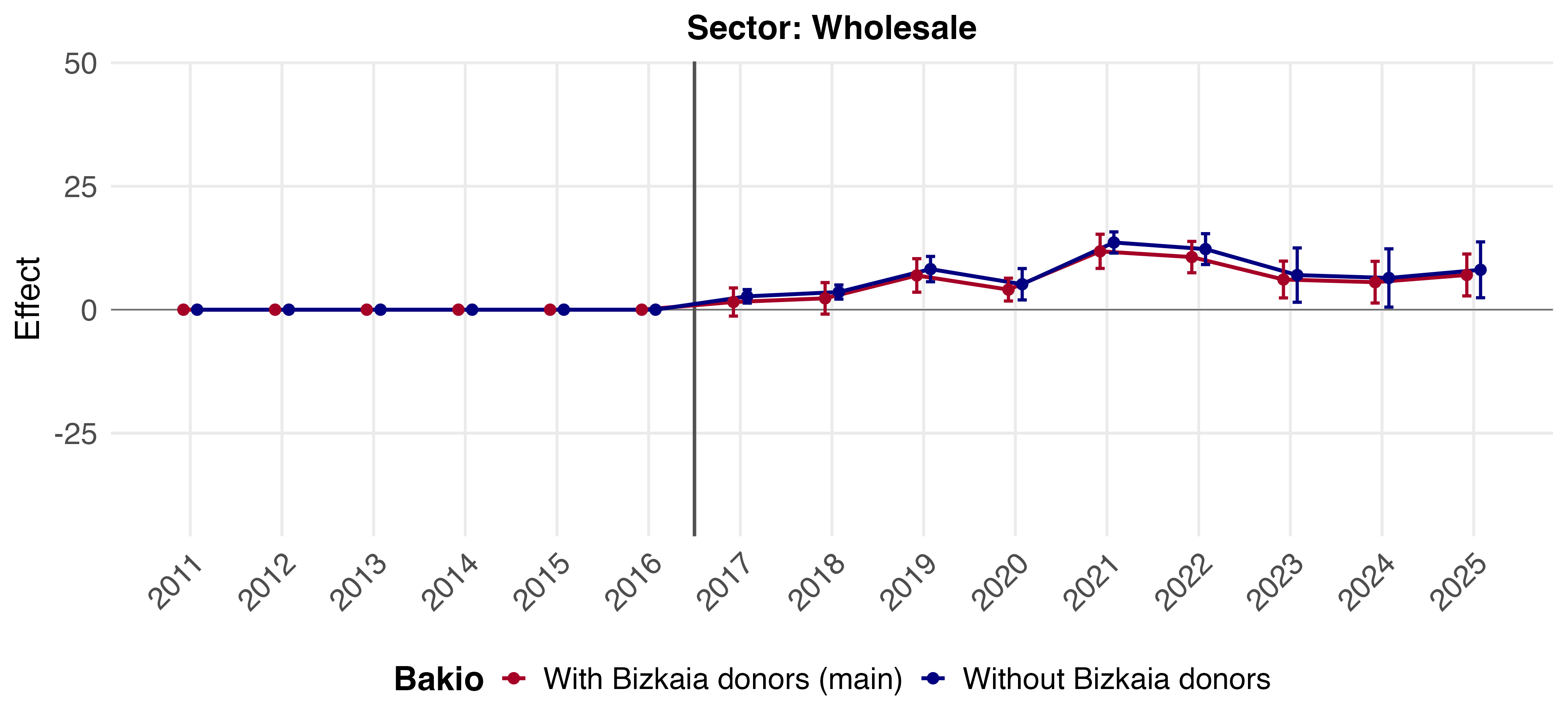} &
    \includegraphics[width=0.3\linewidth]{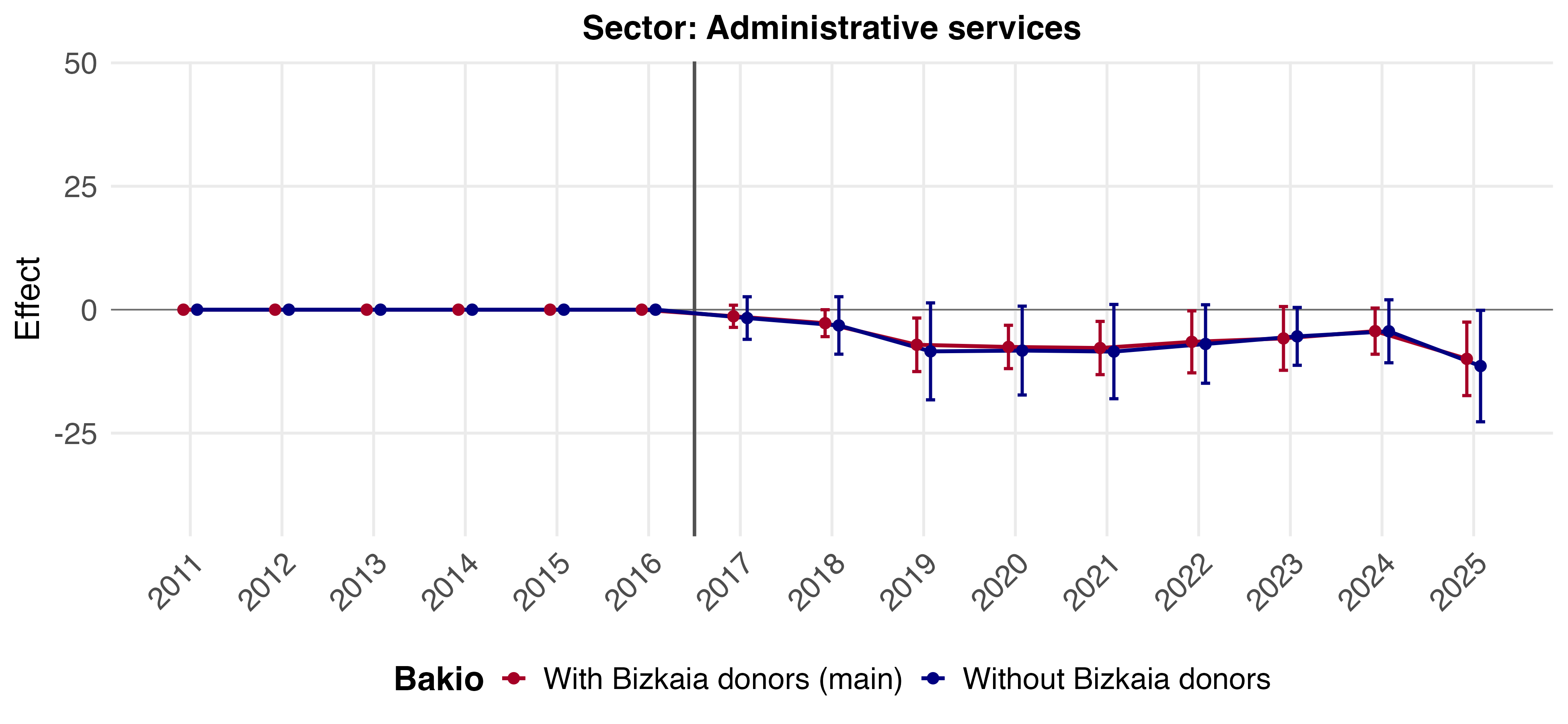} \\

    \includegraphics[width=0.3\linewidth]{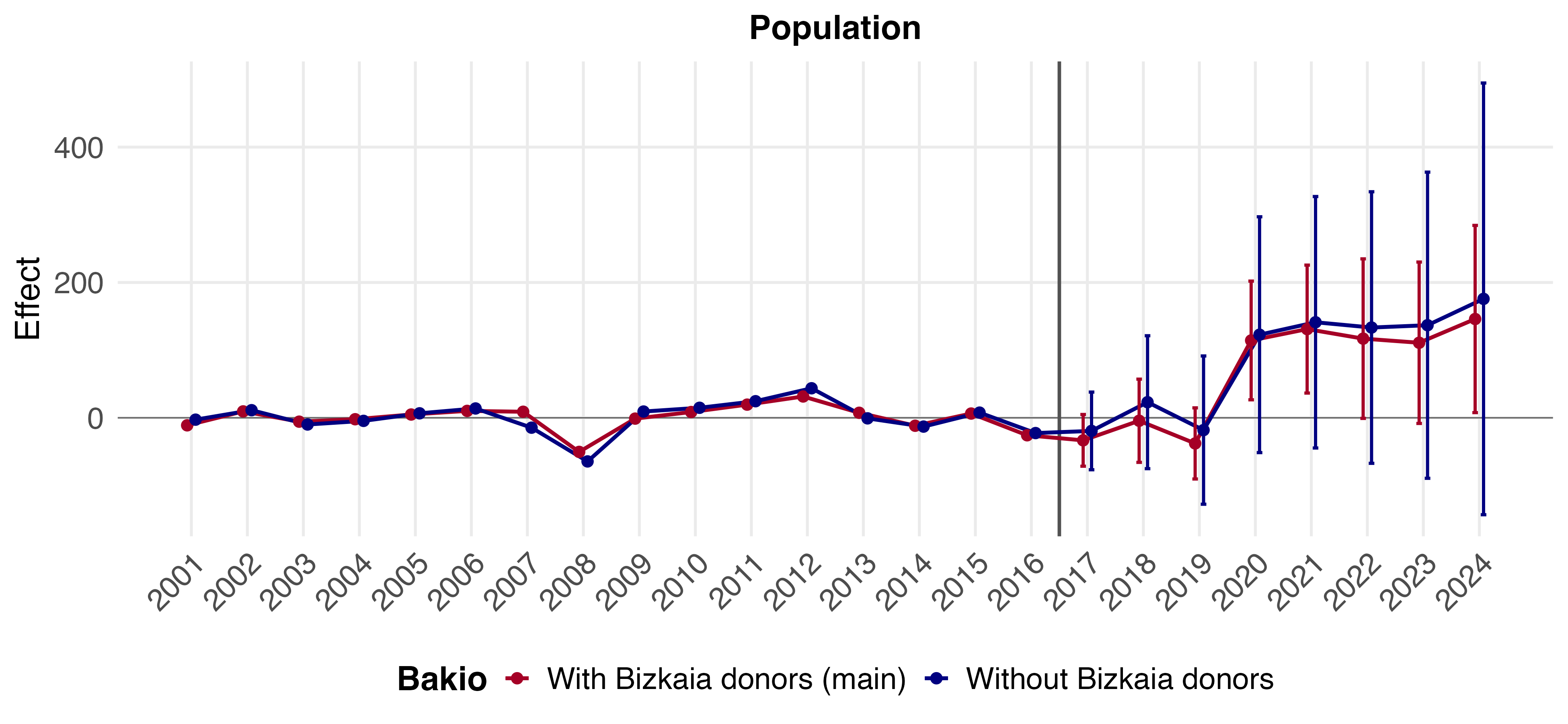} &
    \includegraphics[width=0.3\linewidth]{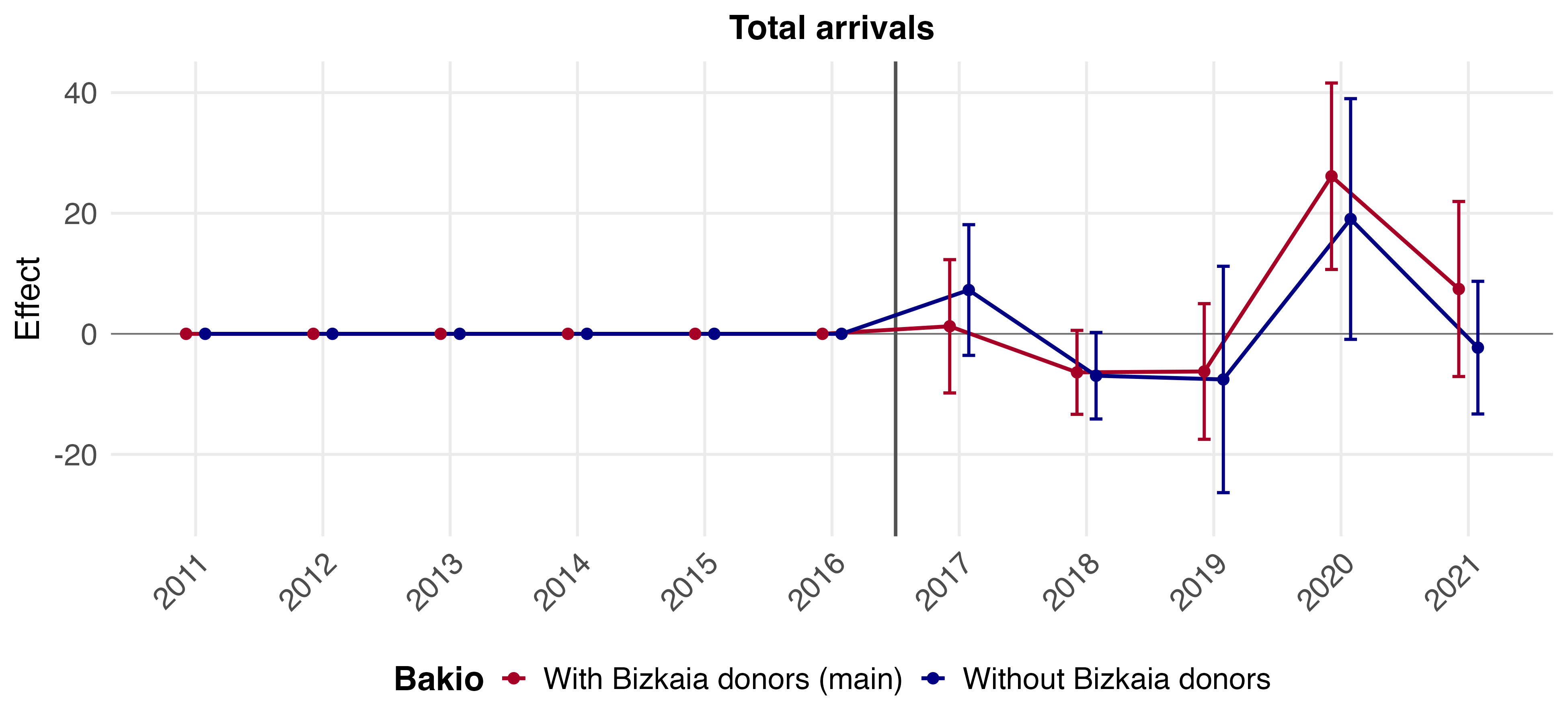} &
    \includegraphics[width=0.3\linewidth]{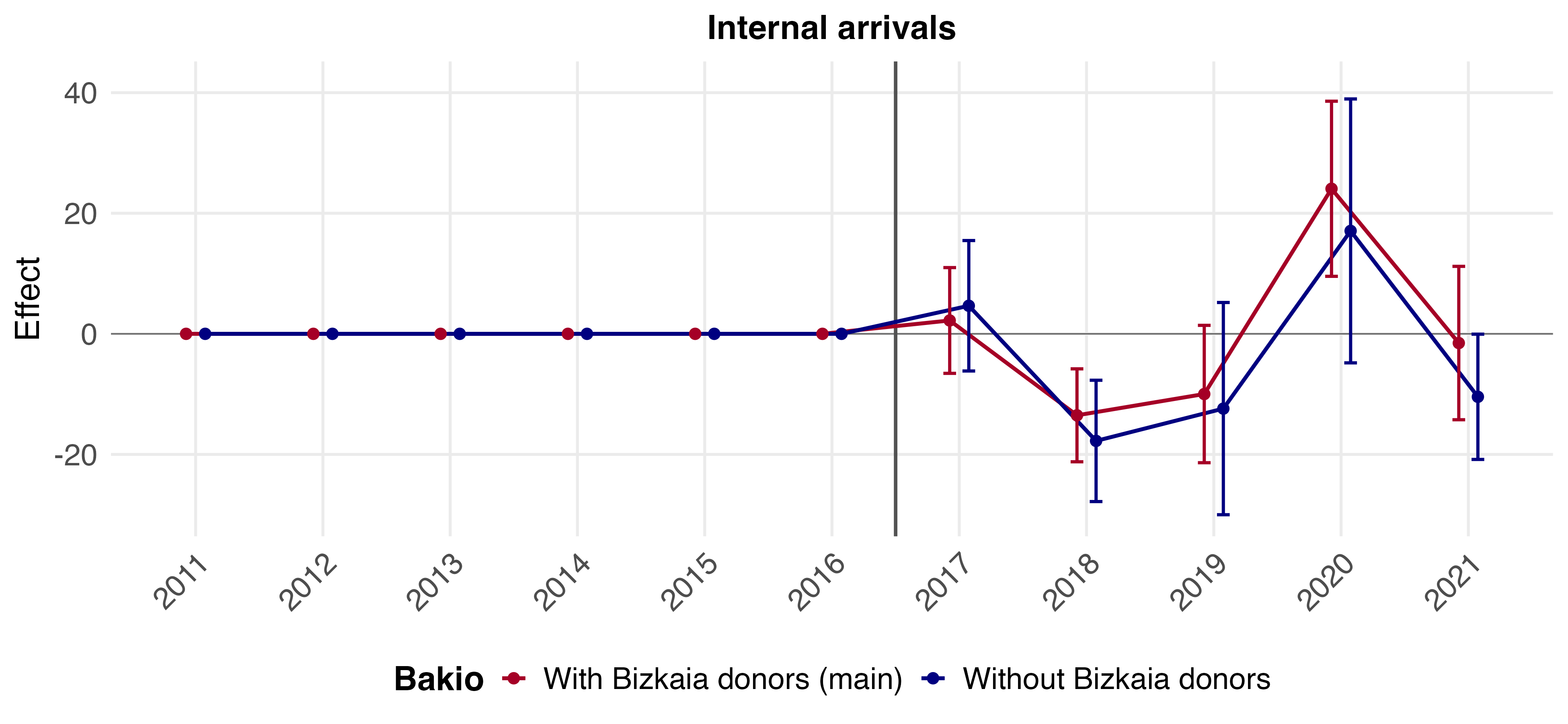} \\

    \includegraphics[width=0.3\linewidth]{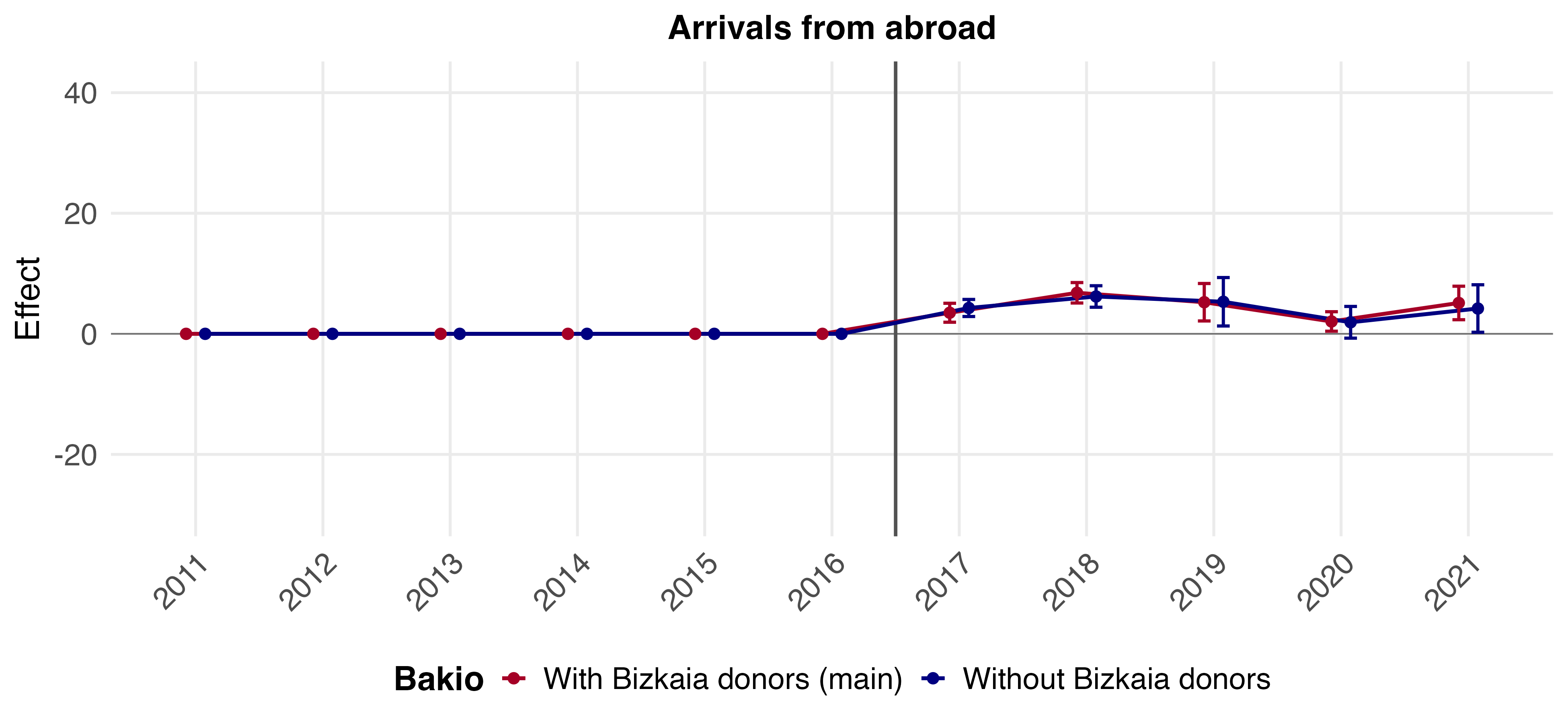} &
    \includegraphics[width=0.3\linewidth]{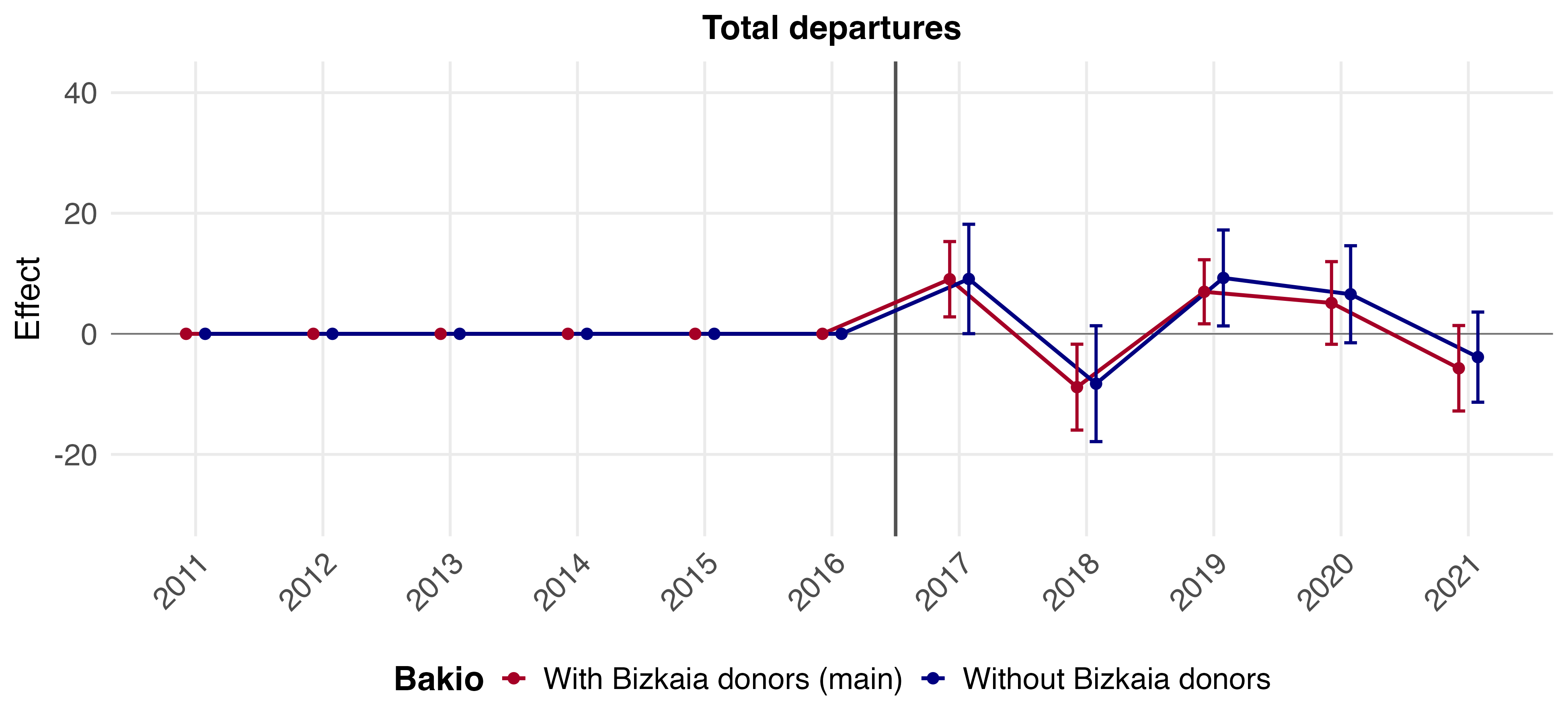} &
    \includegraphics[width=0.3\linewidth]{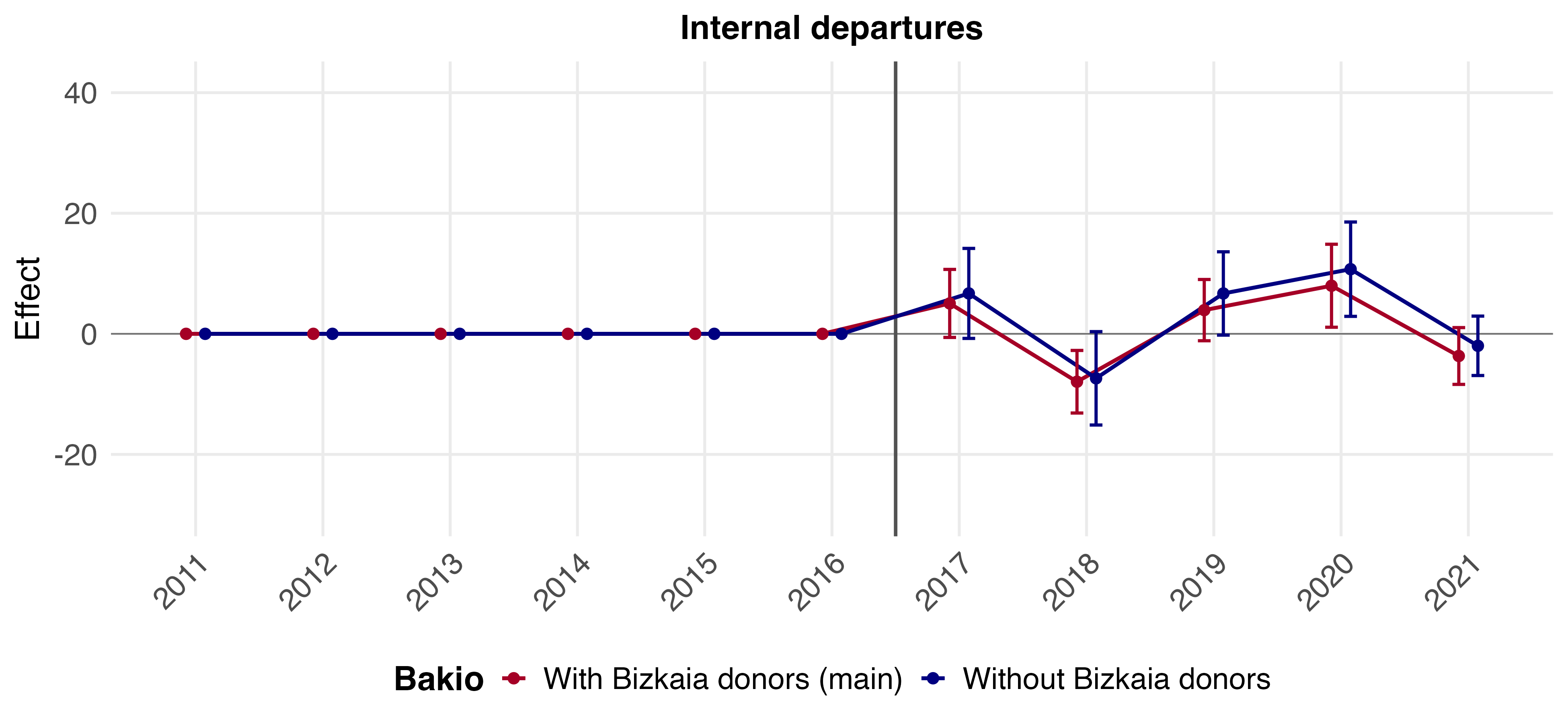} \\

    \includegraphics[width=0.3\linewidth]{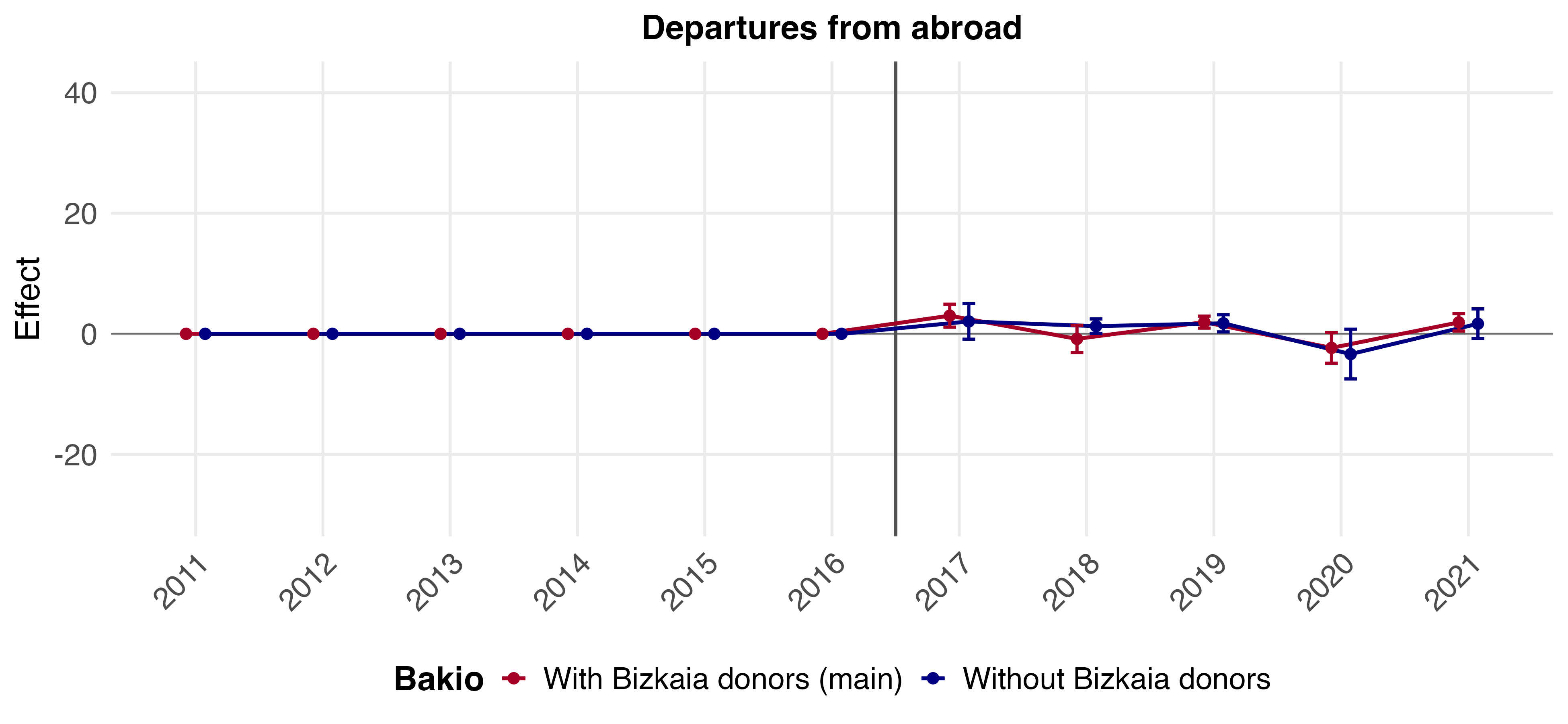} 
    \end{tabular}
    \vspace{-0.2cm}
    \begin{minipage}{\linewidth}
    {\scriptsize \setstretch{1}\textit{Notes:} The plots show the difference between the observed outcome and its synthetic control, together with 95\% confidence intervals. The red line corresponds to the main text's results, that is, the synthetic control results using all the municipalities outside the Bakio's or Gaztelugatxe's counties, that is, all the blue units in the map in Figure \ref{fig:map_groups}. The blue line corresponds to the synthetic control excluding all the municipalities from Bizkaia province, that is, all the blue units in the map in Figure \ref{fig:map_groups_robust}. Estimates are obtained using the Ridge Augmented Synthetic Control estimator \citep{ben-michael_augmented_2021}, estimated separately for each outcome. 95\% confidence interval are displayed for post-treatment periods. }
    \end{minipage}
\end{figure}

One concern is that the shock was not confined to Bakio and the two neighboring counties, Uribe-Butroe and Busturia-Urdaibai. Visitors to San Juan de Gaztelugatxe may also have stayed in, or passed through, other municipalities in the province of Bizkaia. Because these municipalities belong to the donor pool, any such effect would contaminate the synthetic control and bias our estimates. To address this, we drop every municipality in Bizkaia province other than Bakio. This leaves a donor pool made up only of municipalities in Araba and Gipuzkoa, and creates a much larger buffer between Bakio and the closest control municipality. We then re-estimate the synthetic control for Bakio on this restricted sample, keeping the specification, sample period and outcomes of the main analysis unchanged. Map~\ref{fig:map_groups_robust} shows the new control pool, and Figure~\ref{fig:robustness_bizkaia} compares the resulting event studies with those of the main specification. For most outcomes, the two sets of estimates are nearly identical, both in magnitude and in the width of their confidence intervals. For a few outcomes, the point estimates move, but the new effect has the same sign and lies within the 95\% confidence interval of the main estimate.

\subsection{Comparing different estimators' results}\label{app:robust_estimators}
Our main estimates use the ridge-augmented synthetic control (ASC) of \citet{ben-michael_augmented_2021}. To check that the results do not depend on this choice, we re-estimate the effect on Bakio with three alternatives on exactly the same sample: (i) the standard synthetic control (SC) of \citet{abadie_synthetic_2010}, i.e. the same donor weights without the ridge correction; (ii) synthetic difference-in-differences (SDID; \citet{arkhangelsky_synthetic_2021}), which allows a level difference between Bakio and its synthetic control; and (iii) a two-way fixed effects difference-in-differences (TWFE), which compares Bakio with the unweighted average of all donors. The first three estimators build a counterfactual from donors that track Bakio's pre-2017 path. TWFE instead relies on parallel trends with every donor weighted equally, and it serves as a benchmark. ASC and SC give practically the same estimates for every outcome: the pre-treatment fit is already good, so the ridge correction hardly changes anything. SDID yields the same sign and similar magnitudes for most outcomes, and its estimates always fall within the 95\% confidence interval of our preferred specification. There are few exceptions in which the effect is close to zero and not significant, and the SDID estimate flips sign. In these exceptions, the SDID remains close to zero and is not significant. For some outcomes, the SDID estimates are more precisely estimated. One plausible explanation is that the SDID weights pre-treatment periods more aggressively than the ASC. Finally, TWFE is the estimator that differs most from our main results. Where its estimates differ substantially from ASC, it also fails to reproduce Bakio's path before 2017: its pre-treatment coefficients are large and persistent, far from the near-zero pre-treatment gaps of the synthetic-control estimators. This suggests that the parallel-trends assumption underlying TWFE is not plausible in this setting, and that weighting donors is needed to build a credible counterfactual, which is why we use a synthetic-control estimator.

\begin{figure}[H]
    \caption{ATT over all the post-treatment periods, by estimator.}
    \label{fig:model_comparison}
    \centering
    
    \includegraphics[width=0.99\linewidth]{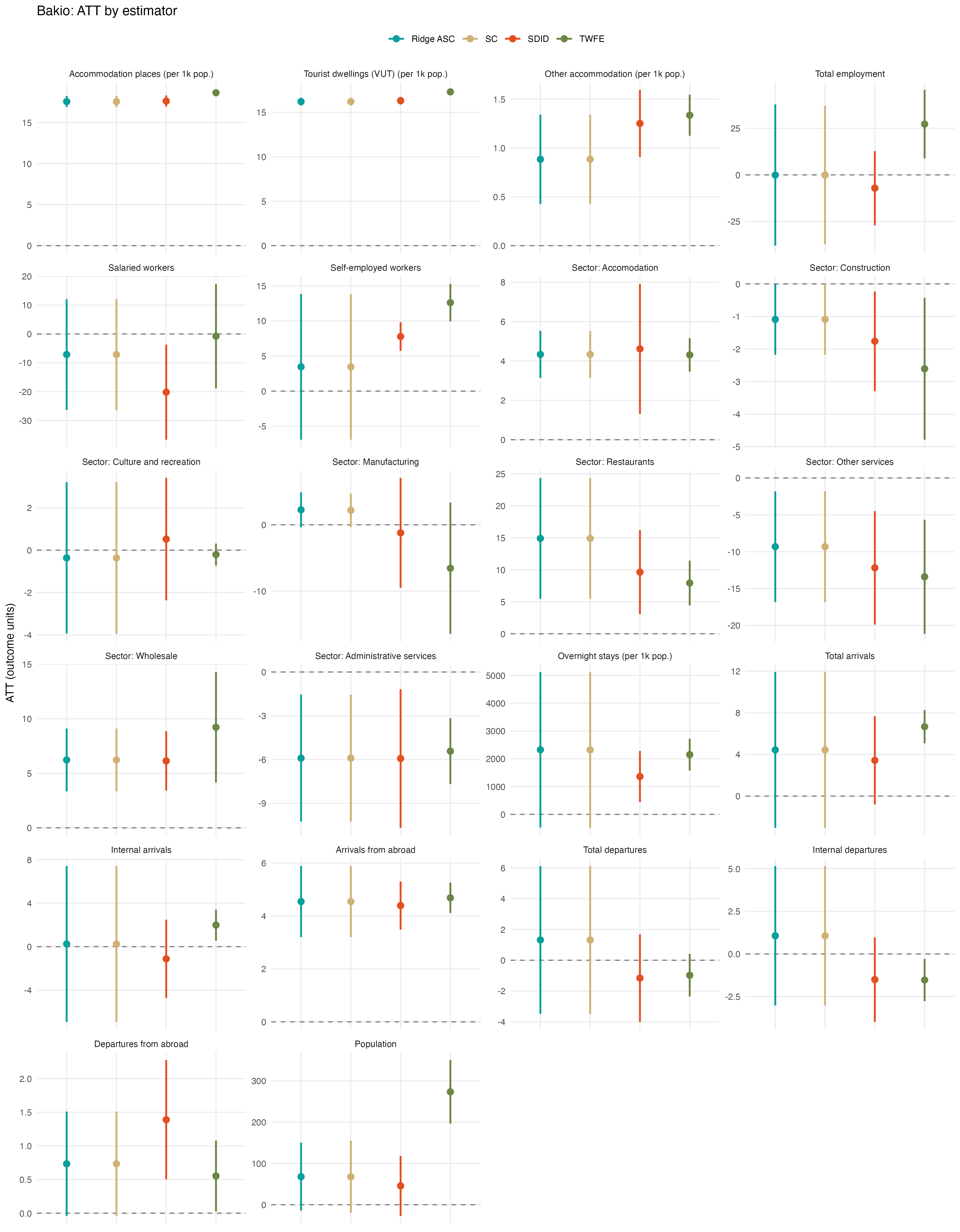} 
    \vspace{-0.2cm}
    \begin{minipage}{\linewidth}
    {\scriptsize \setstretch{1}\textit{Notes:} The plots show the ATT for Bakio, averaged over the post-treatment periods, together with 95\% confidence intervals, for the four candidate estimators: Ridge-Augmented Synthetic Control (Ridge ASC, blue line), Synthetic Control (SC, yellow line), Synthetic Difference-in-Differences (SDID, red line) and Two-Way Fixed Effects (TWFE, green line).}
    \end{minipage}
\end{figure}

\begin{figure}[H]
    \caption{Event study estimates by estimator.}
    \label{fig:model_comparison_es}
    \centering
    
    \includegraphics[width=0.99\linewidth]{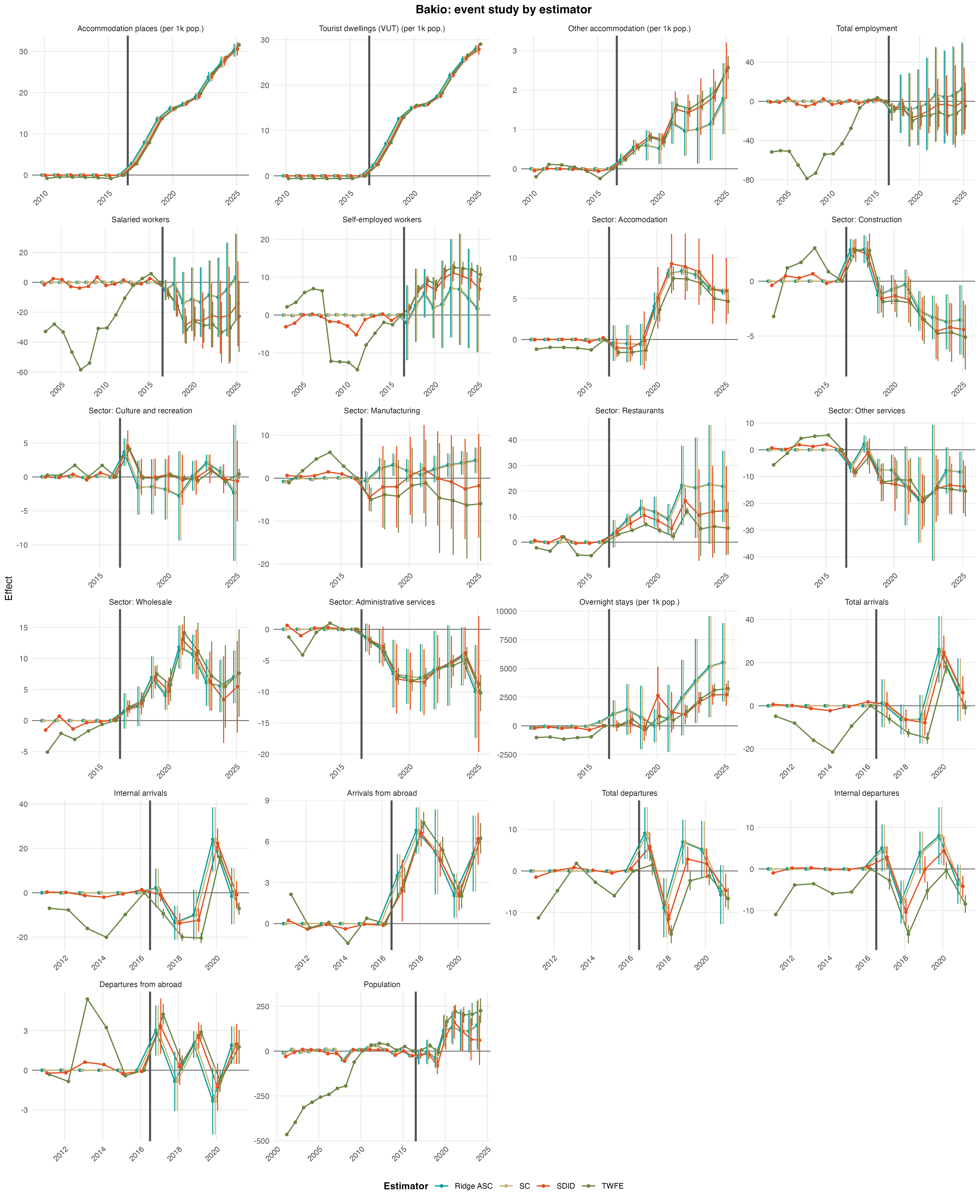} 
    \vspace{-0.2cm}
    \begin{minipage}{\linewidth}
    {\scriptsize \setstretch{1}\textit{Notes:} The plots show the difference between the observed outcome and its estimated counterfactual for the municipality of Bakio, together with 95\% confidence intervals, for the four candidate estimators: Ridge-Augmented Synthetic Control (Ridge ASC, blue line), Synthetic Control (SC, yellow line), Synthetic Difference-in-Differences (SDID, red line) and Two-Way Fixed Effects (TWFE, green line).}
    \end{minipage}
\end{figure}

\clearpage
\renewcommand{\theequation}{B\arabic{equation}}
\renewcommand{\thetable}{B\arabic{table}}
\renewcommand{\thefigure}{B\arabic{figure}}
\renewcommand{\thelemma}{B\arabic{lemma}}
\renewcommand{\theproposition}{B\arabic{proposition}}
\renewcommand{\thecorollary}{B\arabic{corollary}}
\setcounter{equation}{0}
\setcounter{table}{0}
\setcounter{figure}{0}
\setcounter{lemma}{0}
\linespread{1}

\section{Additional Figures}
\label{app:figures}

\subsection{Descriptive Maps}
\begin{figure}[H]
    \caption{Municipalities in the Basque Country by treatment group and population}
    \label{fig:map_population}
    \centering
\includegraphics[width = \linewidth]{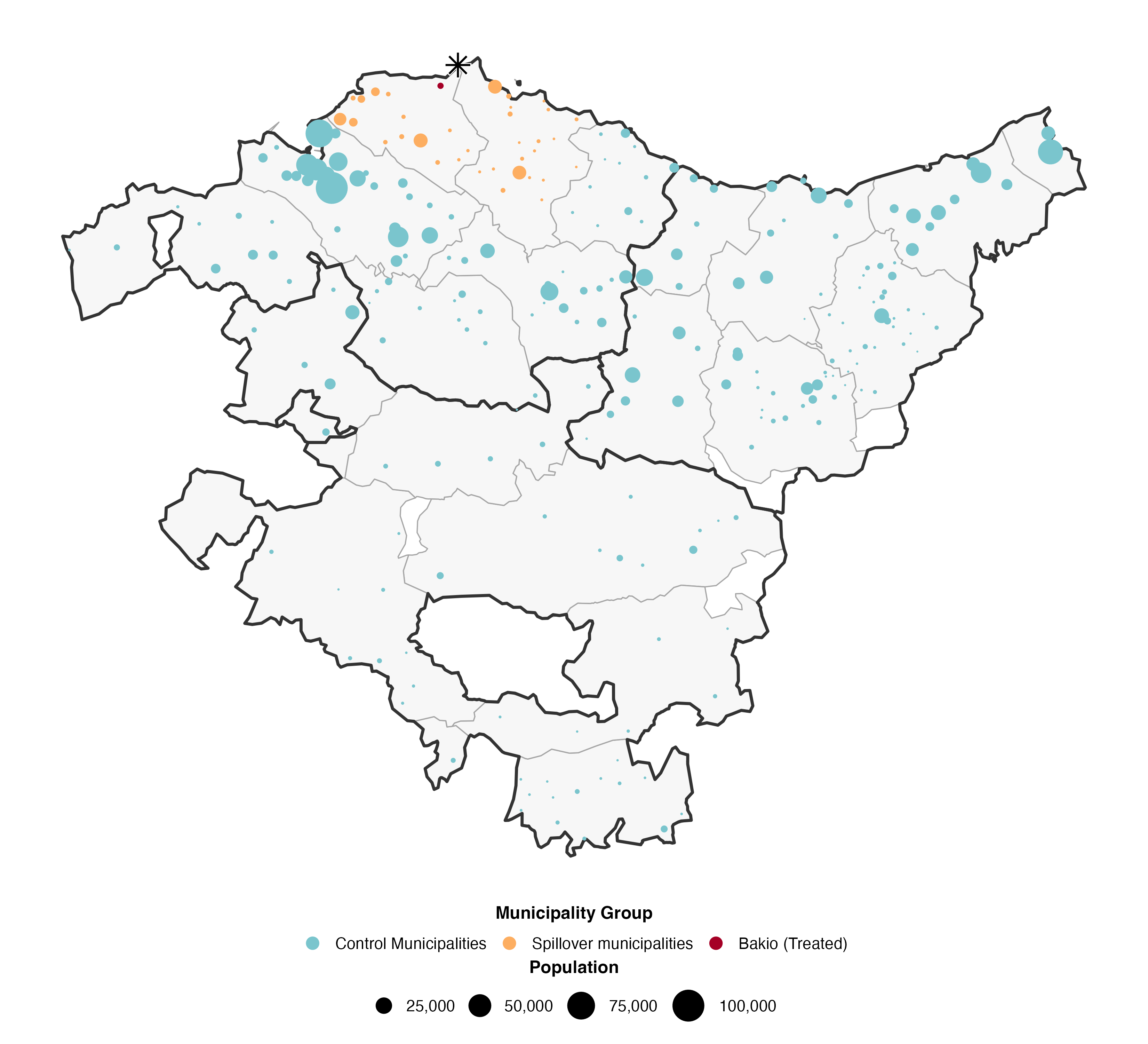}
\vspace{-0.2cm}
\begin{minipage}{\linewidth}
{\scriptsize \setstretch{1}\textit{Notes:} The star marks San Juan de Gaztelugatxe. Red: the treated municipality (Bakio). Orange: municipalities in the same counties as the treated unit and Gaztelugatxe, excluded from the donor pool and used to estimate spillover effects. Blue: the donor pool. The three provincial capitals are excluded from this figure and from the analysis. The size of the circle weighted by the municipality's population. Province capitals are excluded.}
 \end{minipage}
\end{figure}

\begin{figure}[H]
    \caption{Municipalities in the Basque Country by treatment group and distance to Gaztelugatxe}
    \label{fig:map_distance}
    \centering
\includegraphics[width = \linewidth]{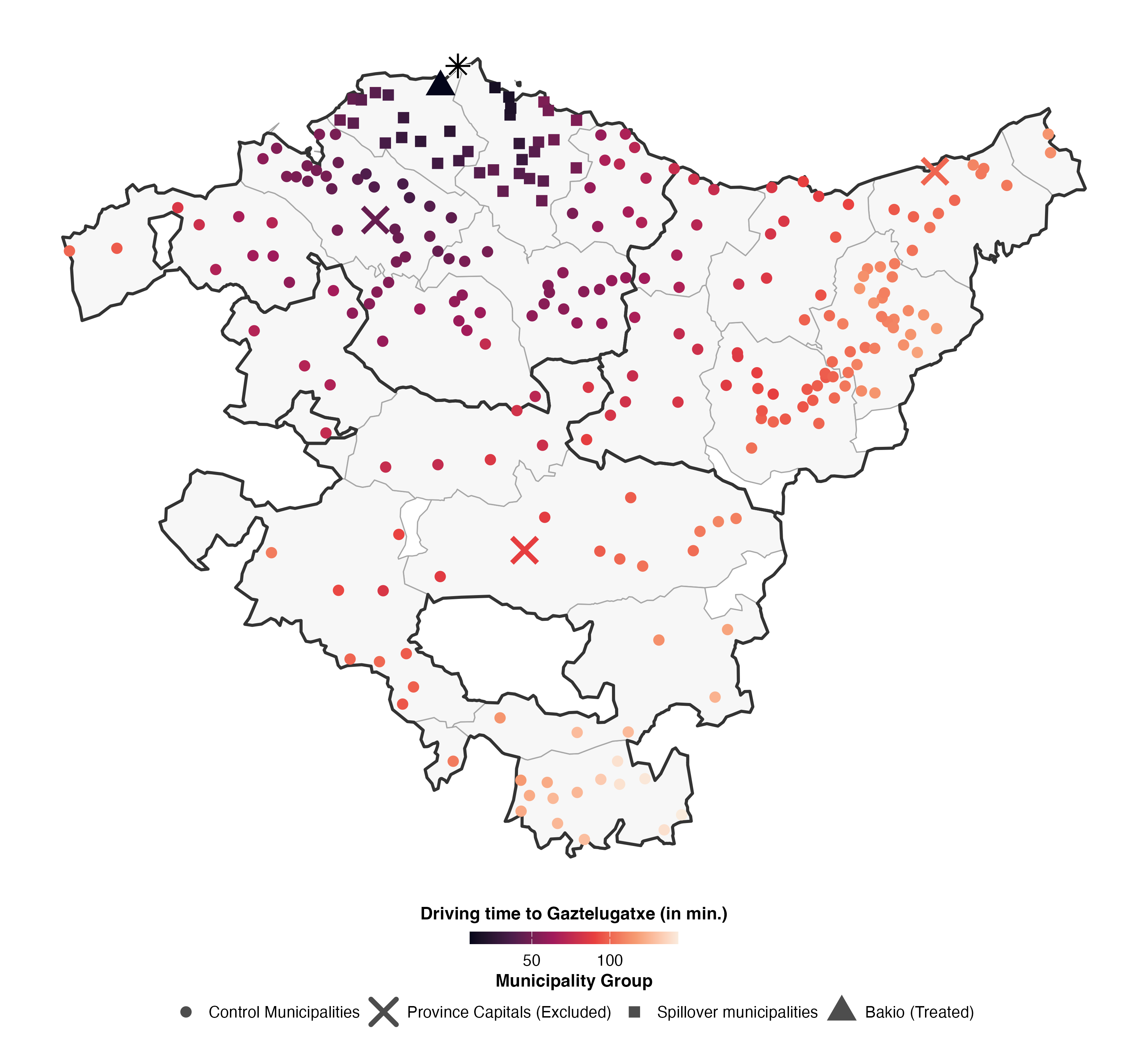}
\vspace{-0.2cm}
\begin{minipage}{\linewidth}
{\scriptsize \setstretch{1}\textit{Notes:} The star marks San Juan de Gaztelugatxe. Triangle: the treated municipality (Bakio). Squares: municipalities in the same counties as the treated unit and Gaztelugatxe, excluded from the donor pool and used to estimate spillover effects. Circles: the donor pool. Cross: The three provincial capitals, which are excluded from the analysis. The municipalities shaded by driving time to San Juan de Gaztelugatxe (darker is closer). Driving times are computed with the OSRM routing engine.}
 \end{minipage}
\end{figure}

\subsection{Observed outcomes and synthetic control trends} \label{app:trends}
This section presents the same results as in Section \ref{sec:results}, but instead of plotting the gap between the observed outcome and the synthetic control in an event-study fashion, we plot the observed and synthetic-control outcomes directly in levels.

\subsubsection{Tourism Development}

\begin{figure}[H]
    \caption{Observed and synthetic control trends for tourist accommodation.}
    \label{fig:tourism_trends}
 \centering
    \begin{tabular}{cc}
    \includegraphics[width = 8 cm]{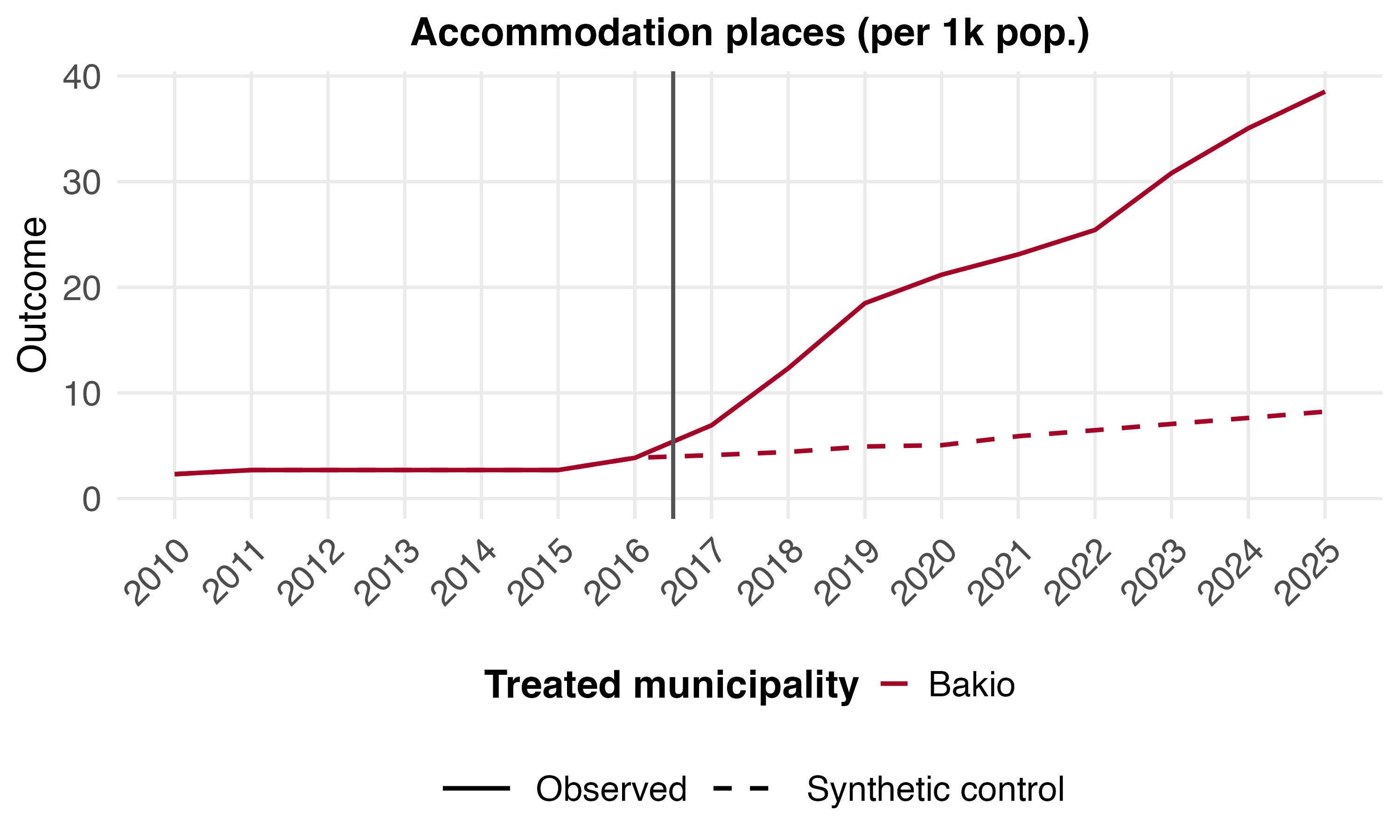} 
    &
    \includegraphics[width = 8 cm]{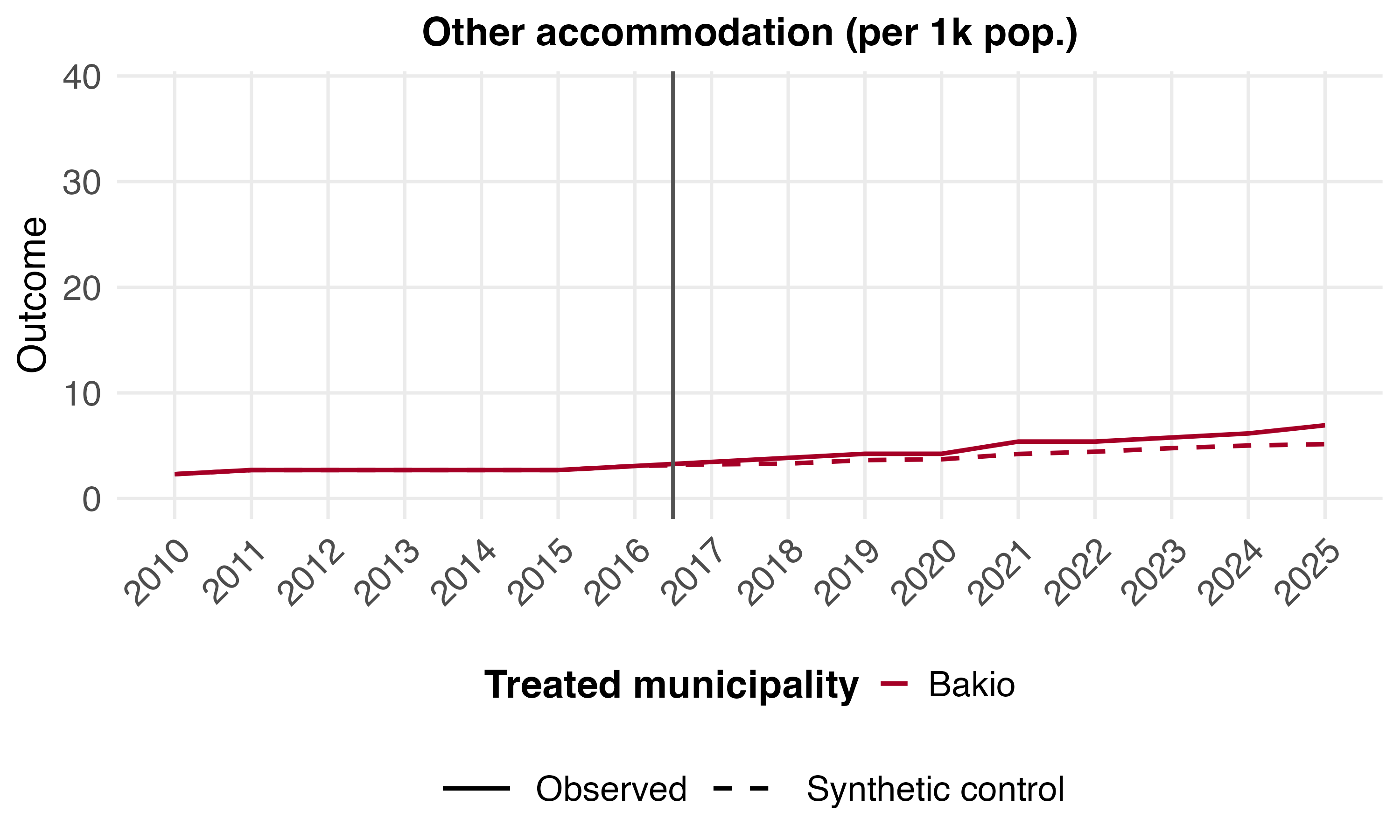}    \\
    \includegraphics[width = 8 cm]{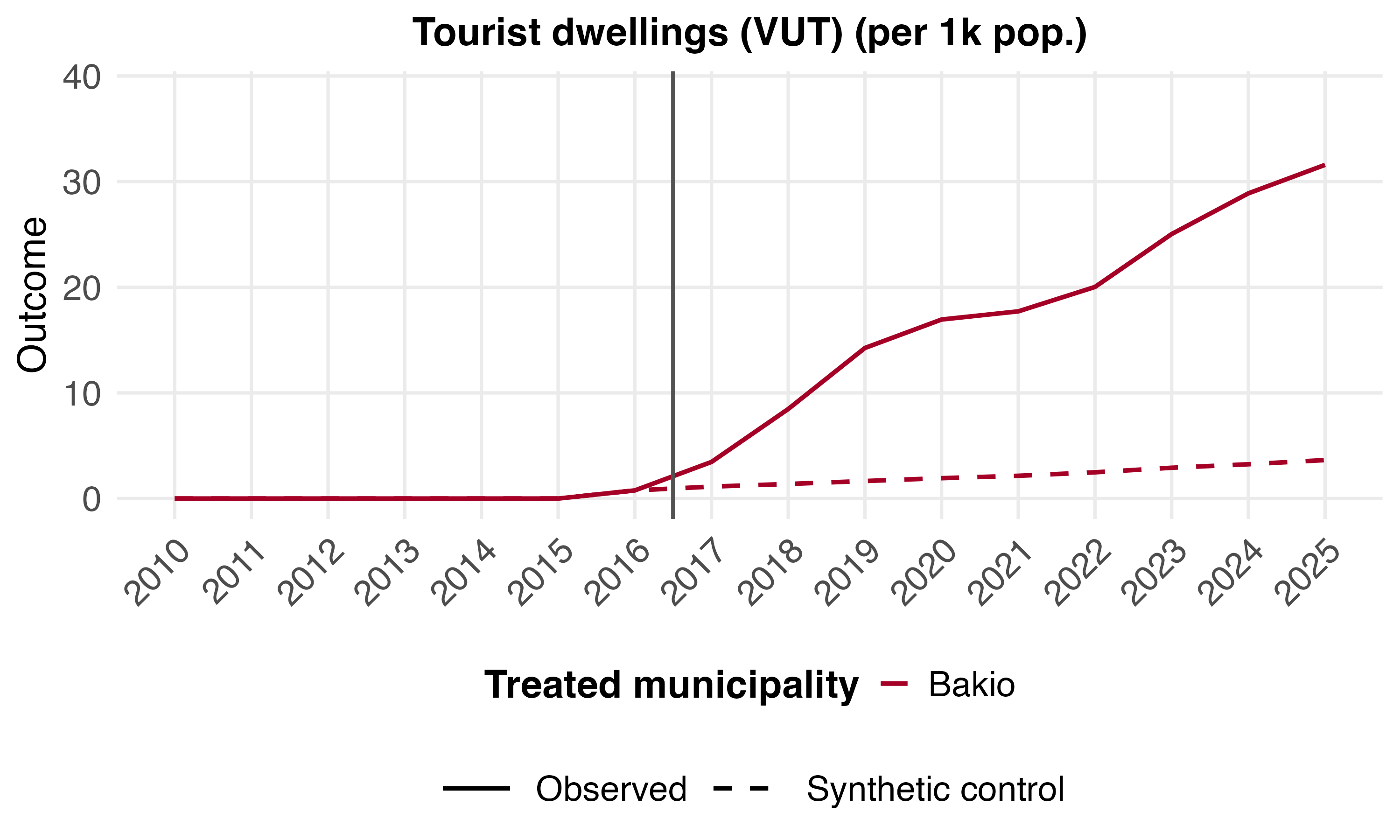} 
    &
    \includegraphics[width = 8 cm]{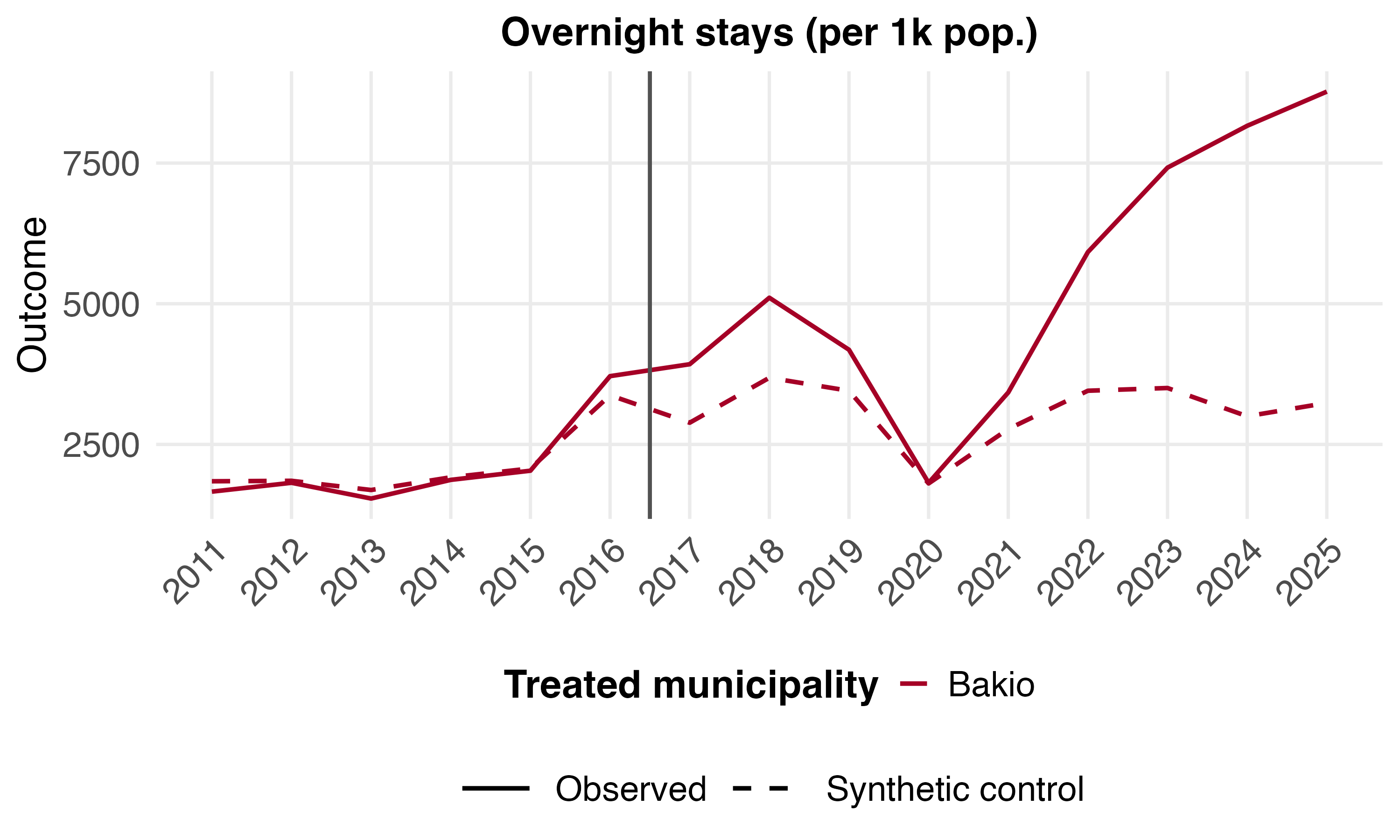} 
\end{tabular}
\vspace{-0.2cm}
\begin{minipage}{\linewidth}
{\scriptsize \setstretch{1}\textit{Notes:} The plots show the time trends of the observed outcome (solid line) and the synthetic control (dashed line). The red line corresponds to Bakio and the blue line to Bermeo. Estimates are obtained using the Ridge Augmented Synthetic Control estimator \citep{ben-michael_augmented_2021}, estimated separately for each outcome and each treated municipality using the same donor pool of Basque municipalities.}
 \end{minipage}
\end{figure}

\subsubsection{Local Labor Markets}

\begin{figure}[H]
    \caption{Observed and synthetic control trends for total employment.}
    \label{fig:labor_trends}
    \centering
\begin{tabular}{cc}
\includegraphics[width = 0.6\linewidth]{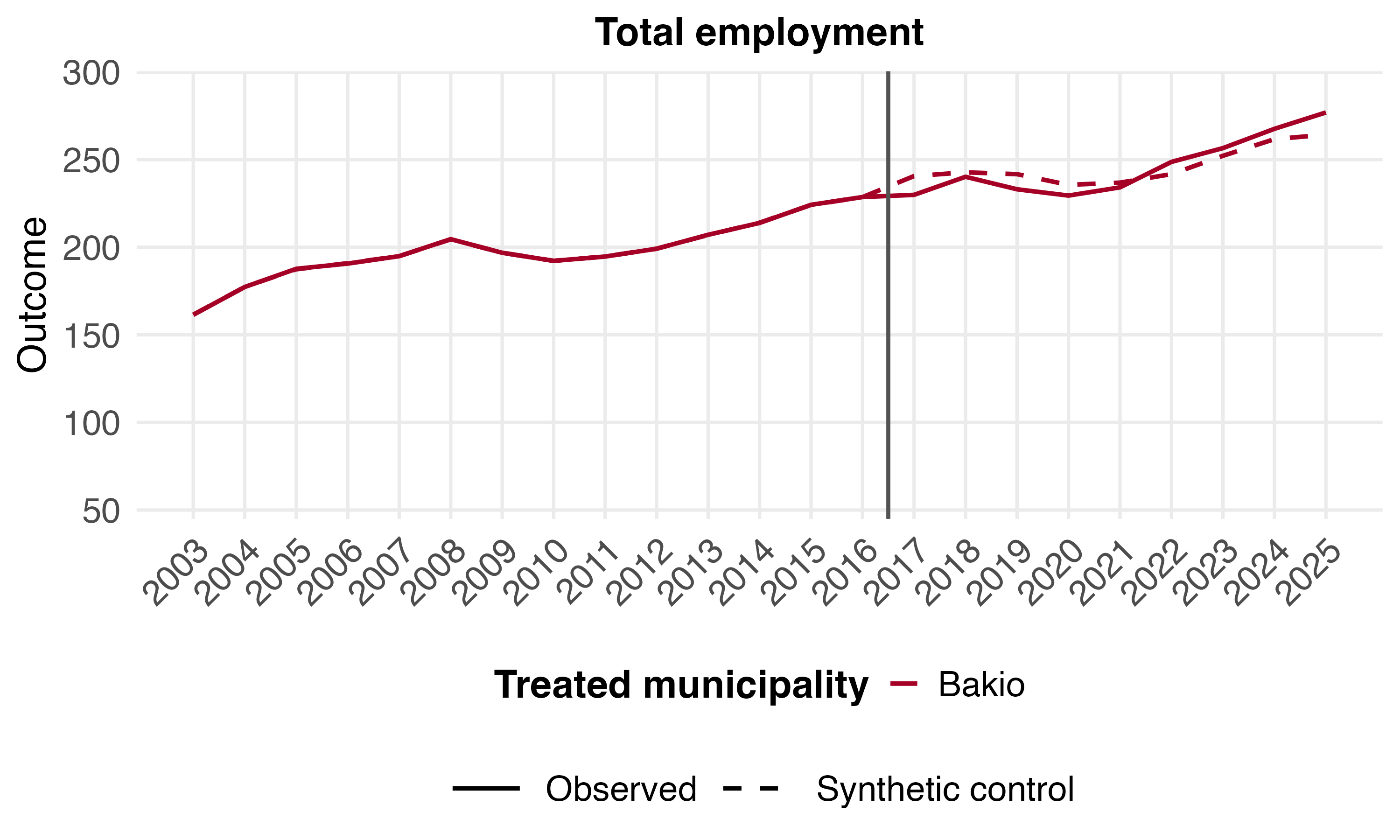}    
\end{tabular}
\vspace{-0.2cm}
\begin{minipage}{\linewidth}
{\scriptsize \setstretch{1}\textit{Notes:} The plots show the time trends of the observed outcome (solid line) and the synthetic control (dashed line). The red line corresponds to Bakio and the blue line to Bermeo. Estimates are obtained using the Ridge Augmented Synthetic Control estimator \citep{ben-michael_augmented_2021}, estimated separately for each outcome and each treated municipality using the same donor pool of Basque municipalities.}
 \end{minipage}
\end{figure}

\begin{figure}[H]
    \caption{Observed and synthetic control trends for regime breakdown.}
    \label{fig:labor_reg_trends}
    \centering
\begin{tabular}{cc}
\includegraphics[width = 8 cm]{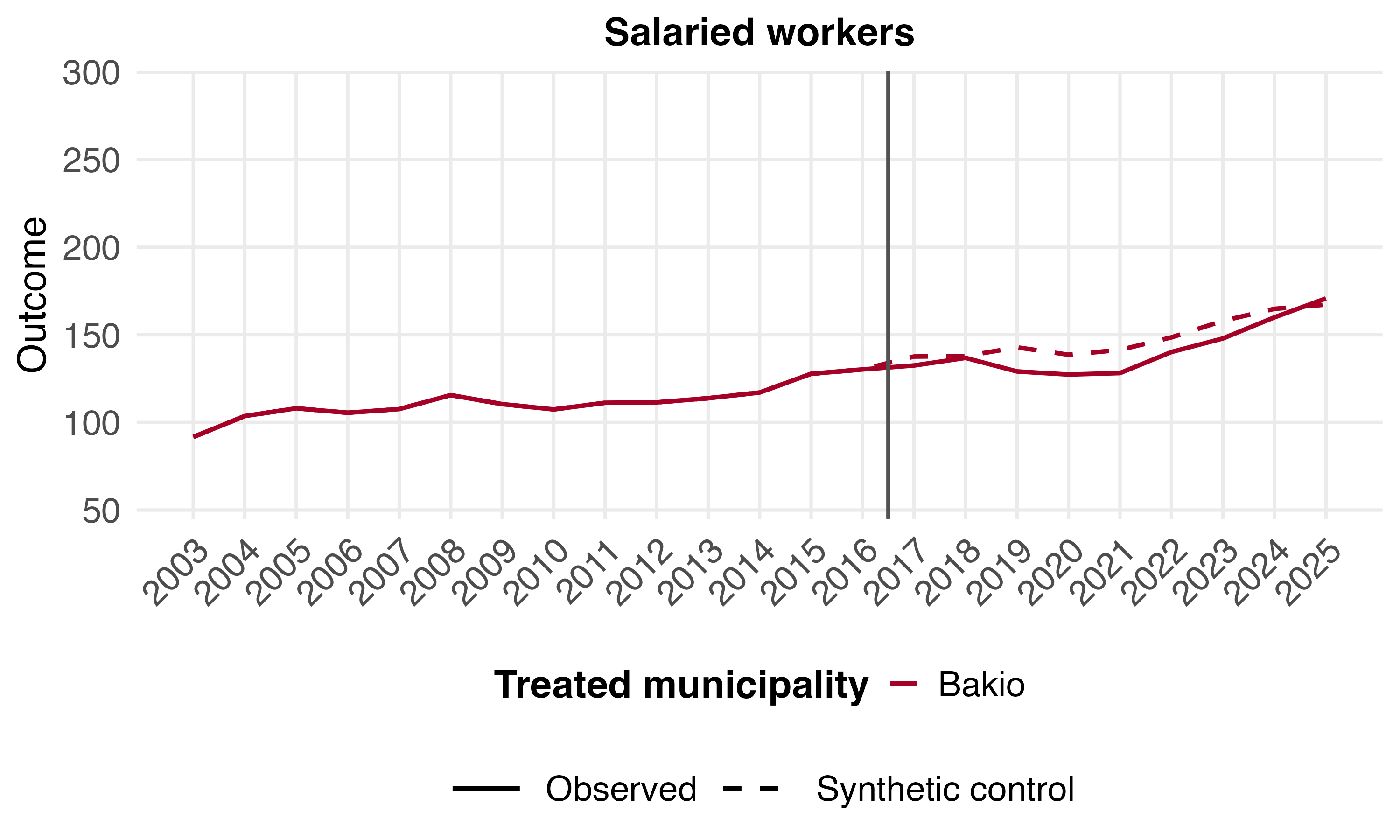}    
&  
\includegraphics[width = 8 cm]{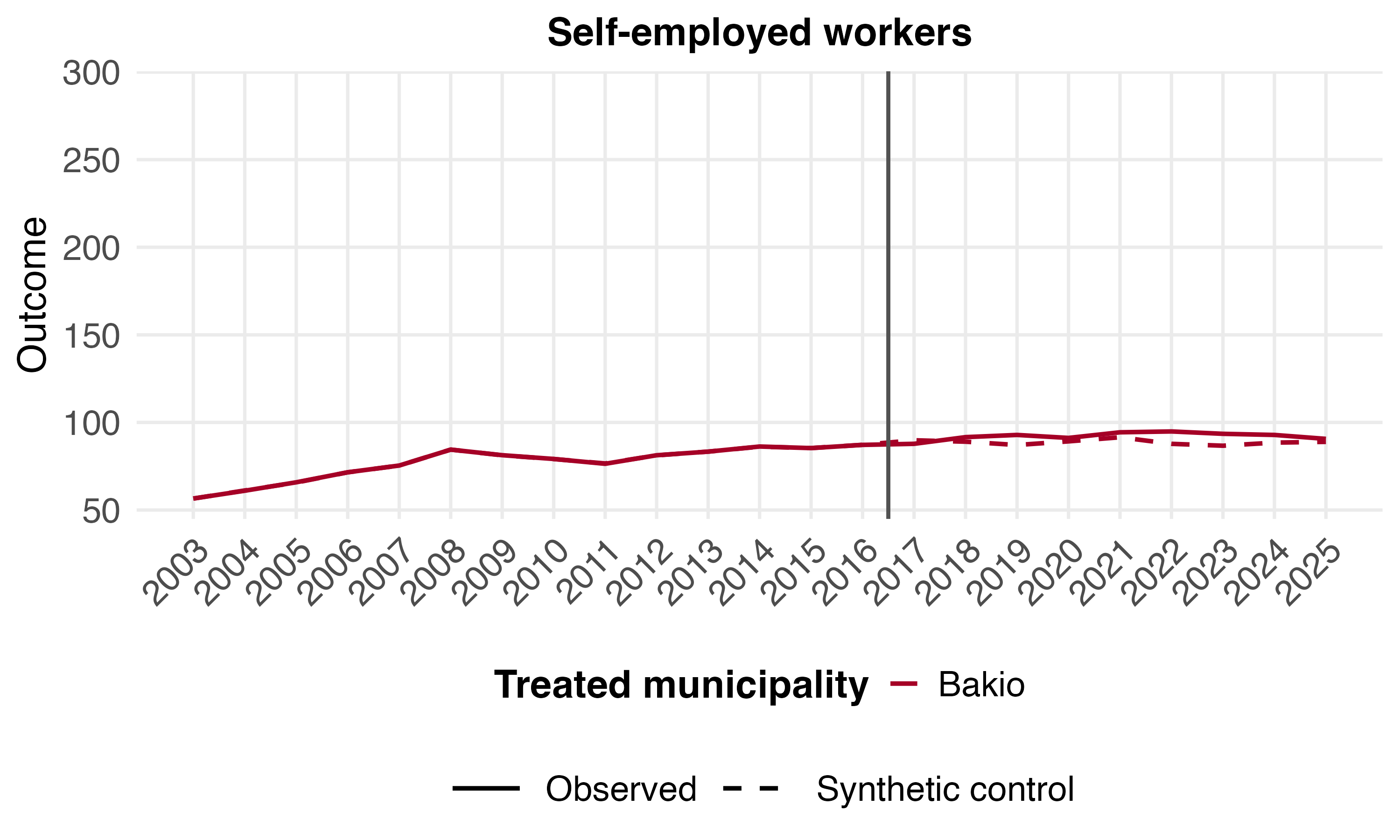} 
\end{tabular}
\vspace{-0.2cm}
\begin{minipage}{\linewidth}
{\scriptsize \setstretch{1}\textit{Notes:} The plots show the time trends of the observed outcome (solid line) and the synthetic control (dashed line). The red line corresponds to Bakio and the blue line to Bermeo. Estimates are obtained using the Ridge Augmented Synthetic Control estimator \citep{ben-michael_augmented_2021}, estimated separately for each outcome and each treated municipality using the same donor pool of Basque municipalities.}
 \end{minipage}
\end{figure}

\begin{figure}[H]
    \caption{Observed and synthetic control trends for sector breakdown.}
    \label{fig:labor_sector_trends}
    \centering
    \begin{tabular}{cc}
    \includegraphics[width=8cm]{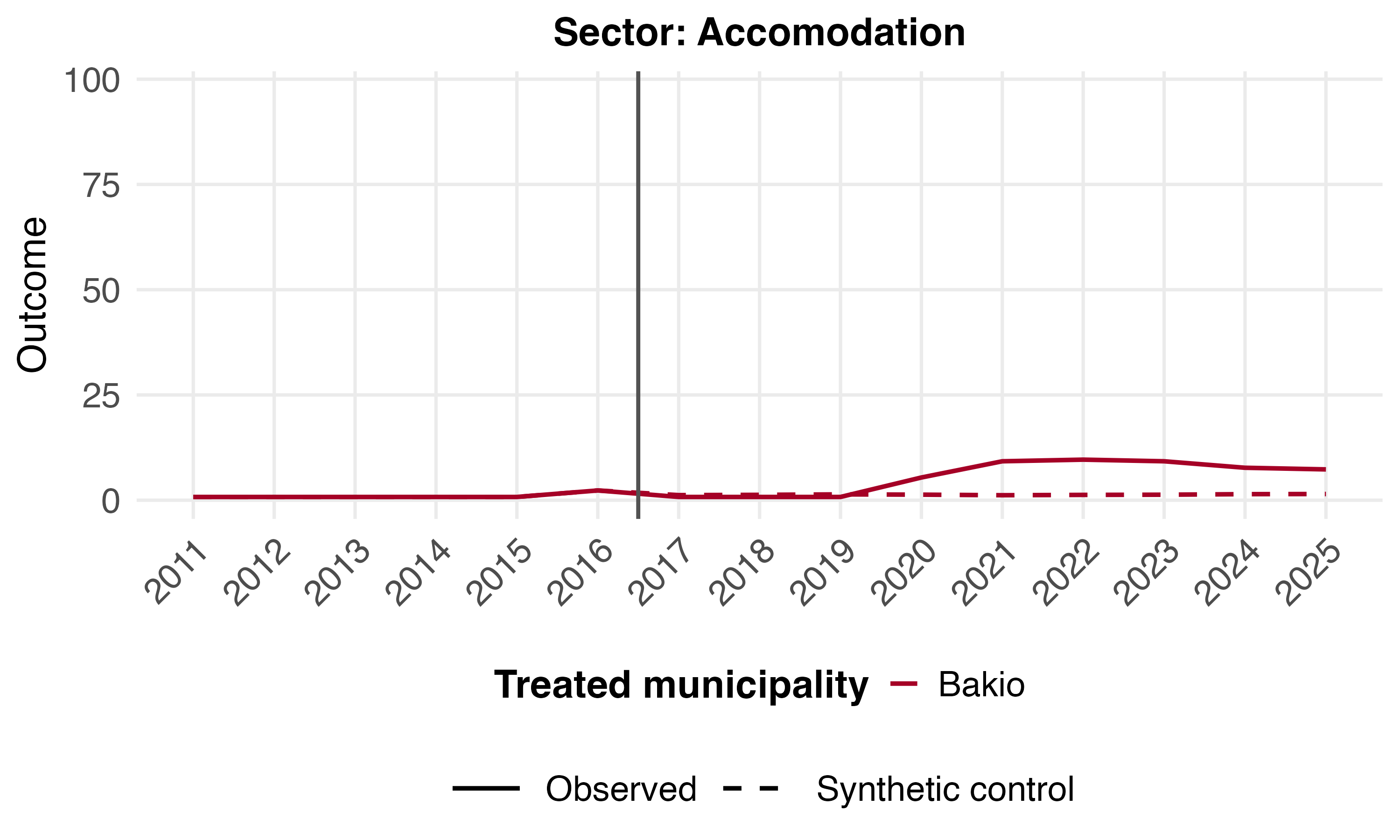} &
    \includegraphics[width=8cm]{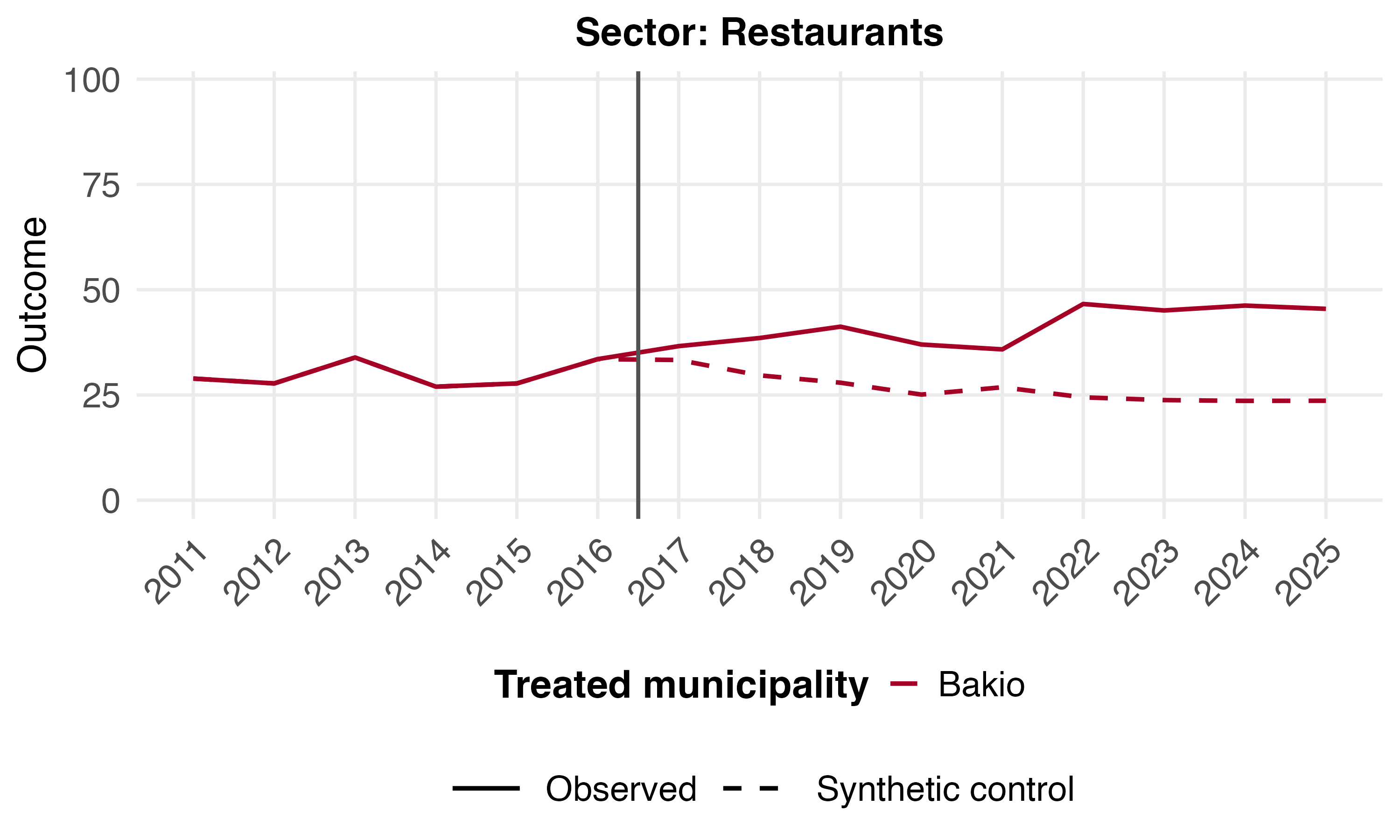} \\
    \includegraphics[width=8cm]{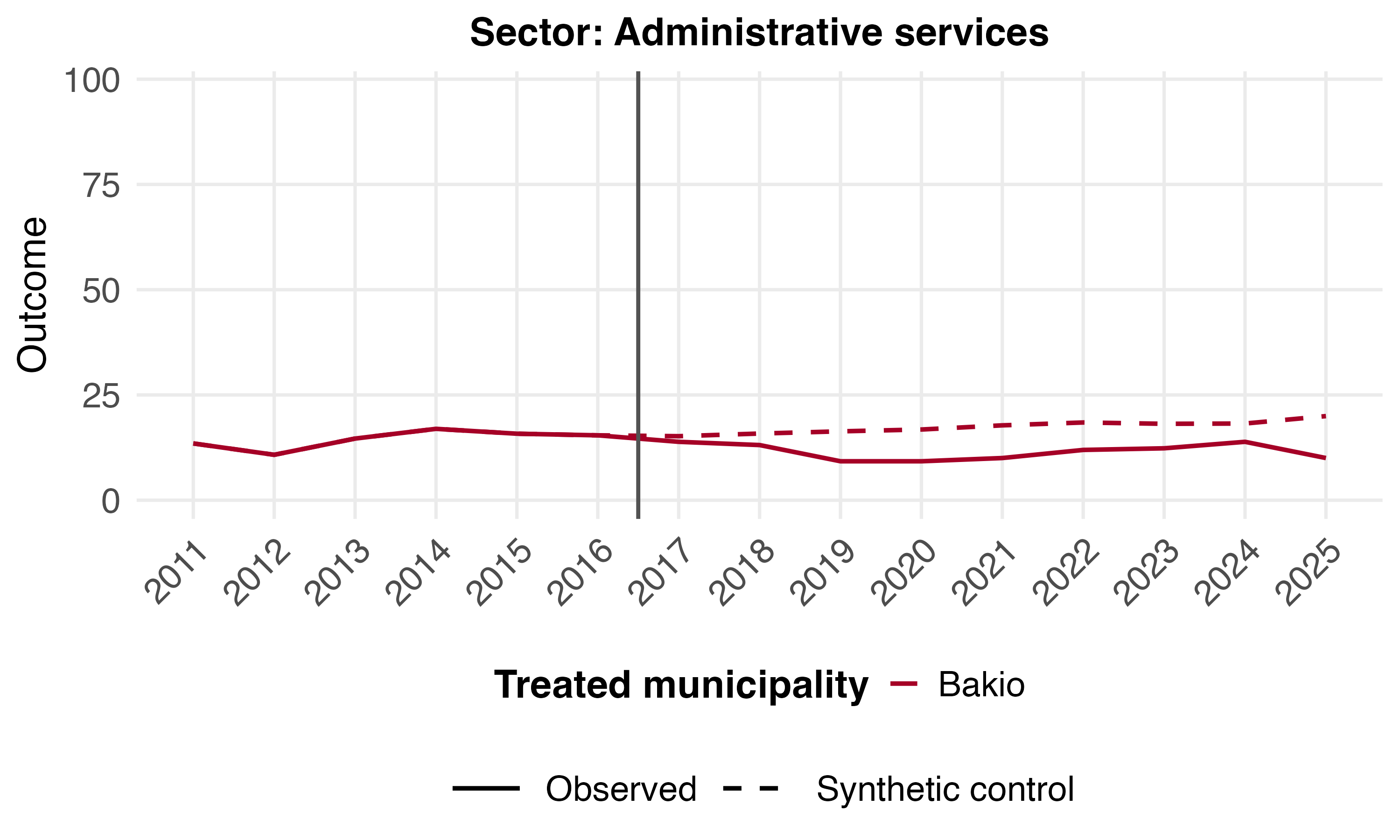} &
    \includegraphics[width=8cm]{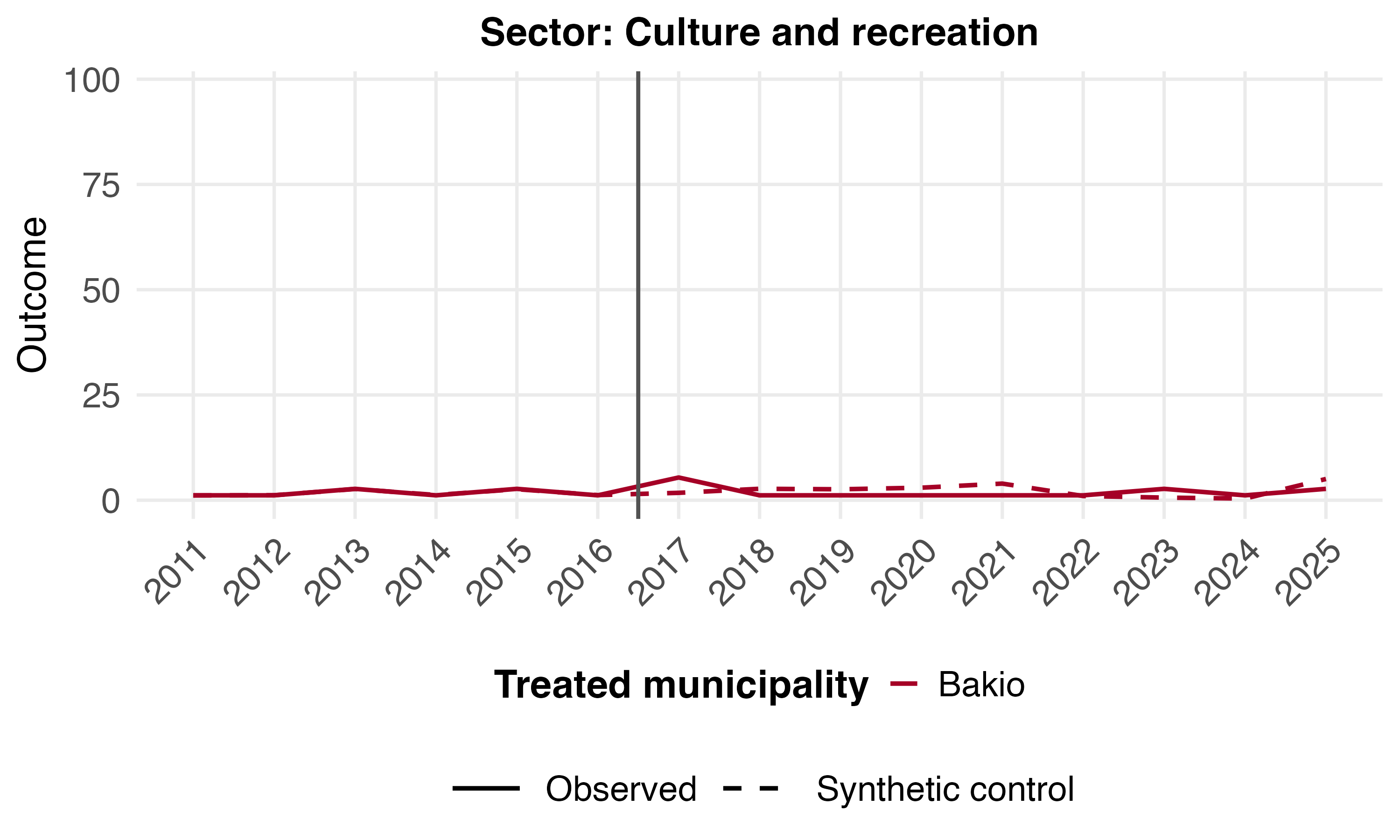} \\
    \includegraphics[width=8cm]{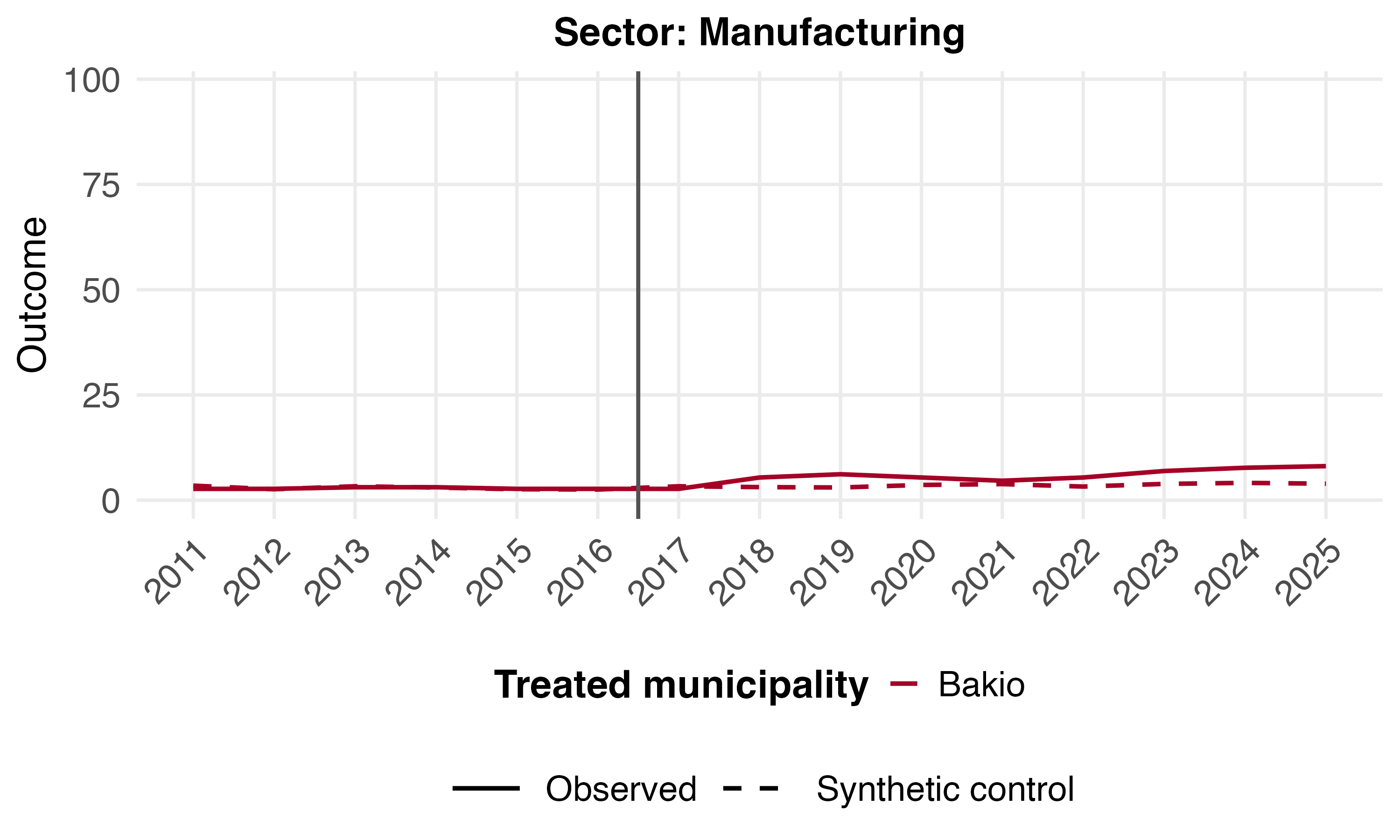} &
    \includegraphics[width=8cm]{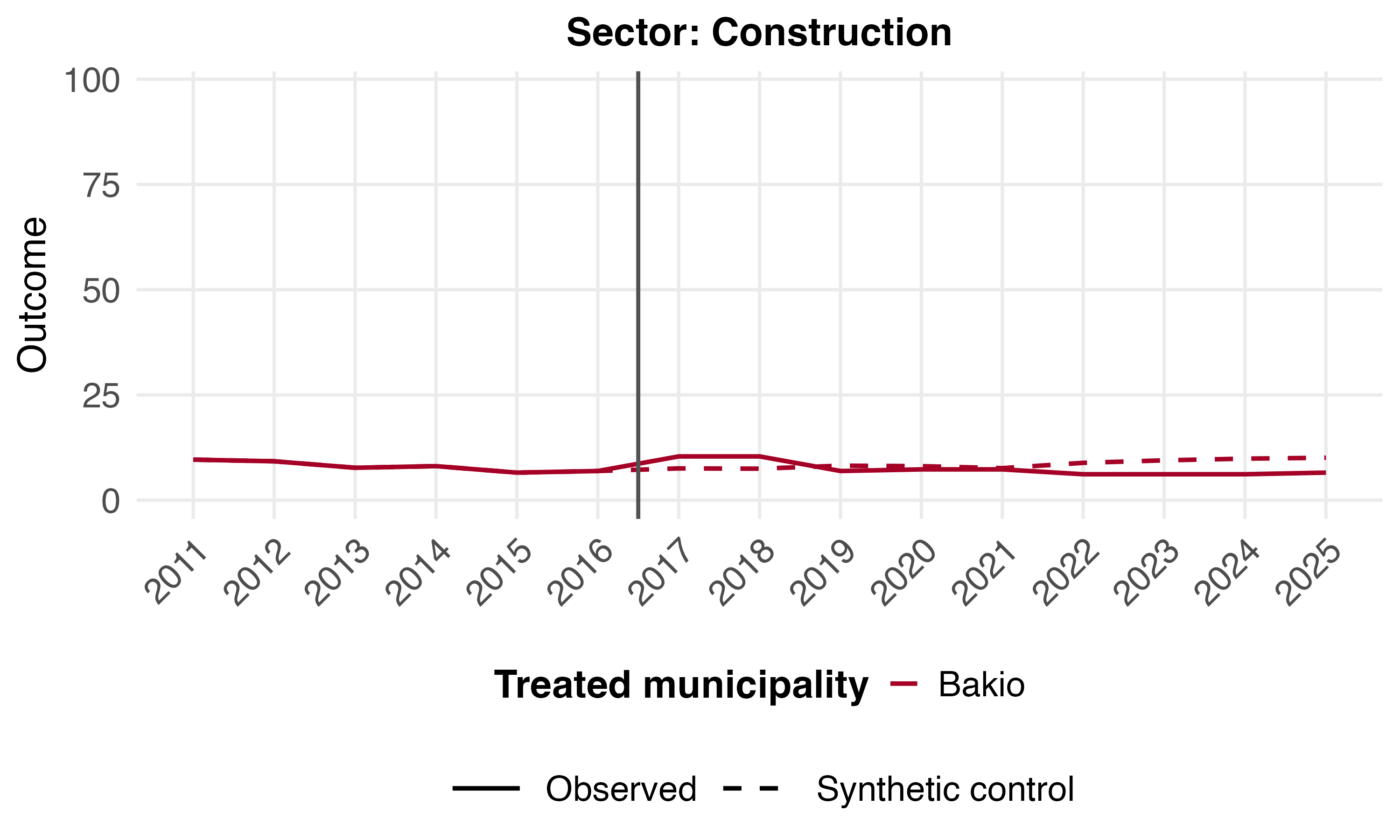} \\
    \includegraphics[width=8cm]{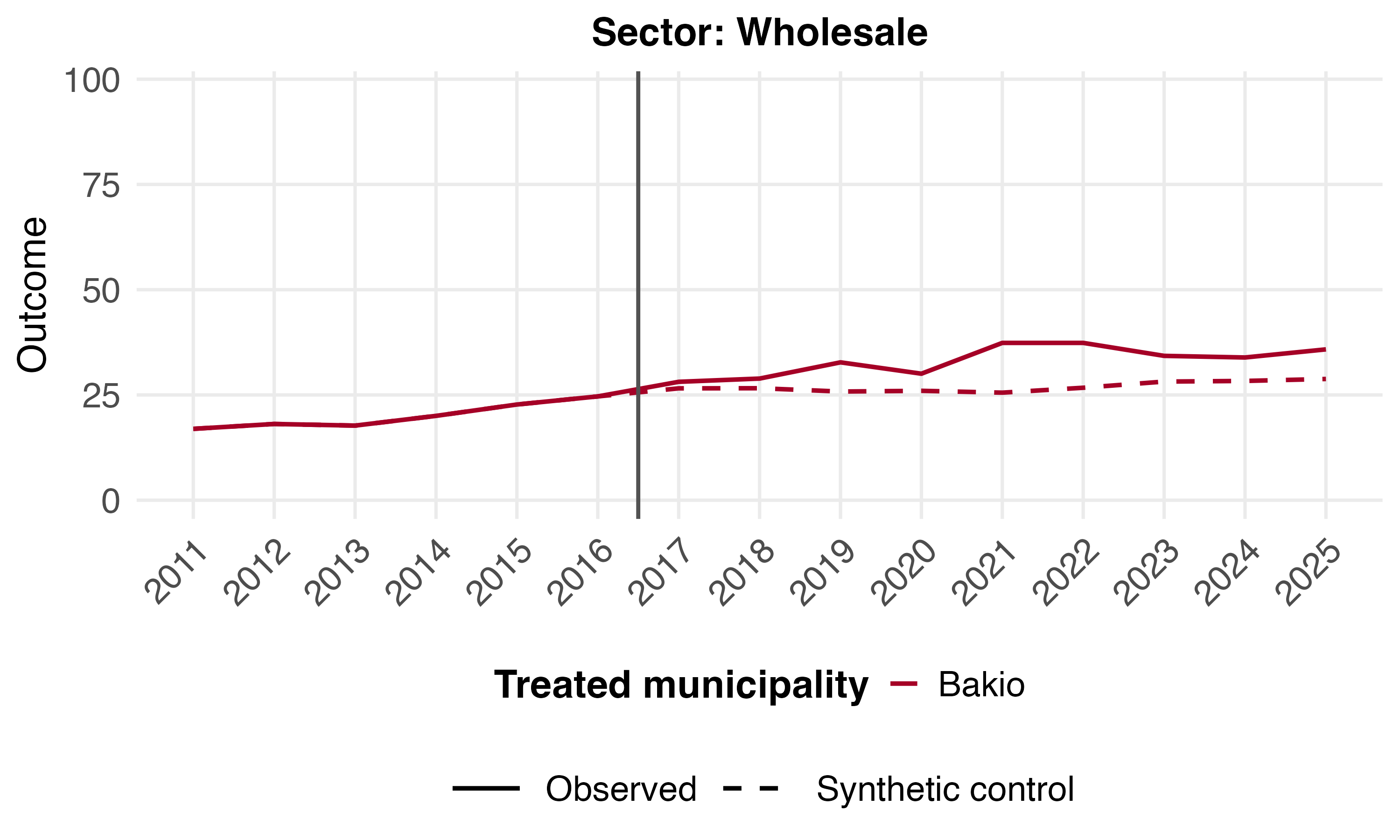} &
    \includegraphics[width=8cm]{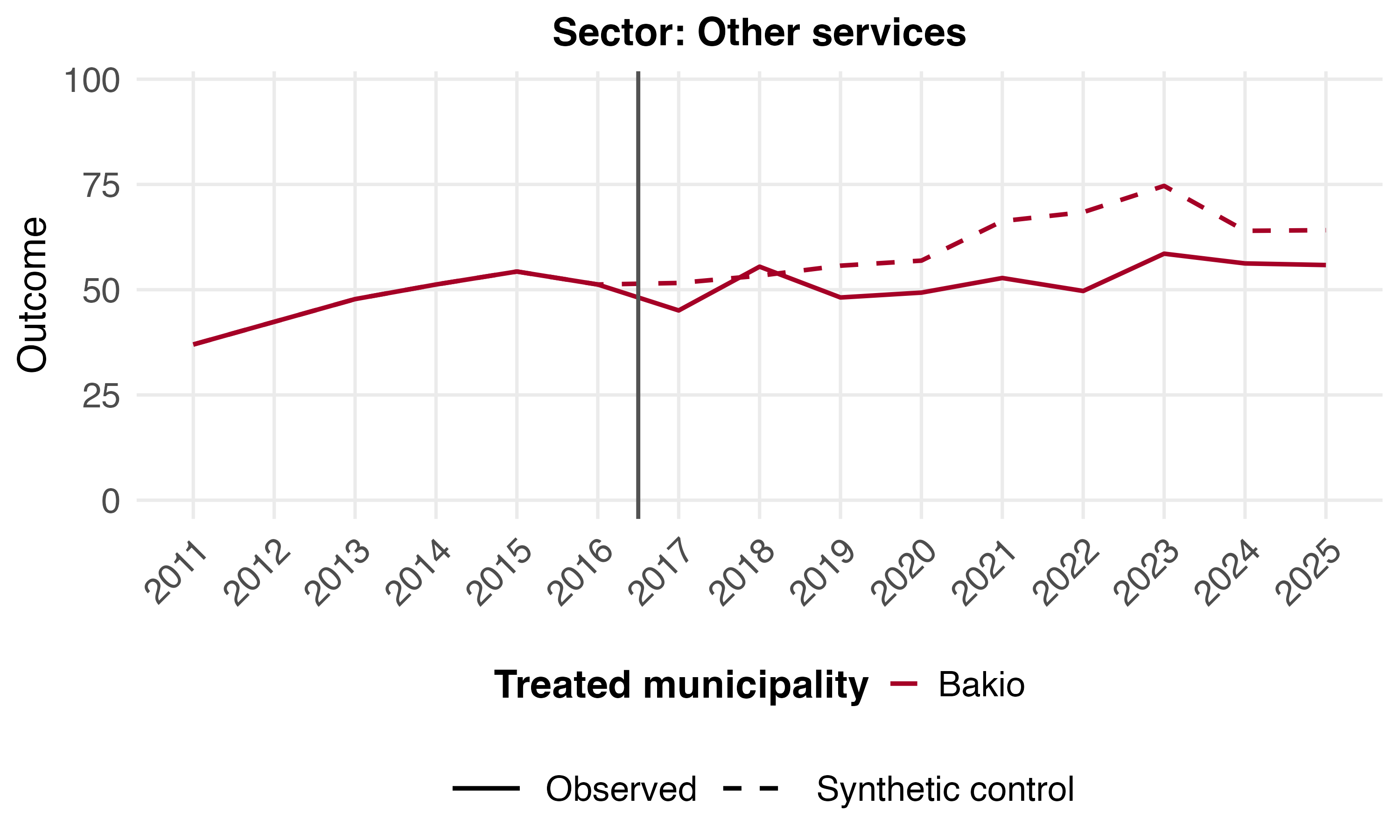} 
    \end{tabular}

    \vspace{-0.2cm}
    \begin{minipage}{\linewidth}
    {\scriptsize \setstretch{1}\textit{Notes:} The plots show the time trends of the observed outcome (solid line) and the synthetic control (dashed line). The red line corresponds to Bakio and the blue line to Bermeo. Estimates are obtained using the Ridge Augmented Synthetic Control estimator \citep{ben-michael_augmented_2021}, estimated separately for each outcome and each treated municipality using the same donor pool of Basque municipalities.}
    \end{minipage}
\end{figure}

\subsubsection{Housing Market}

\begin{figure}[H]
    \ContinuedFloat
    \caption{Observed and synthetic control trends for housing market outcomes.}
    \label{fig:housing_trends}
    \centering
    \begin{tabular}{cc}
    \includegraphics[width=8cm]{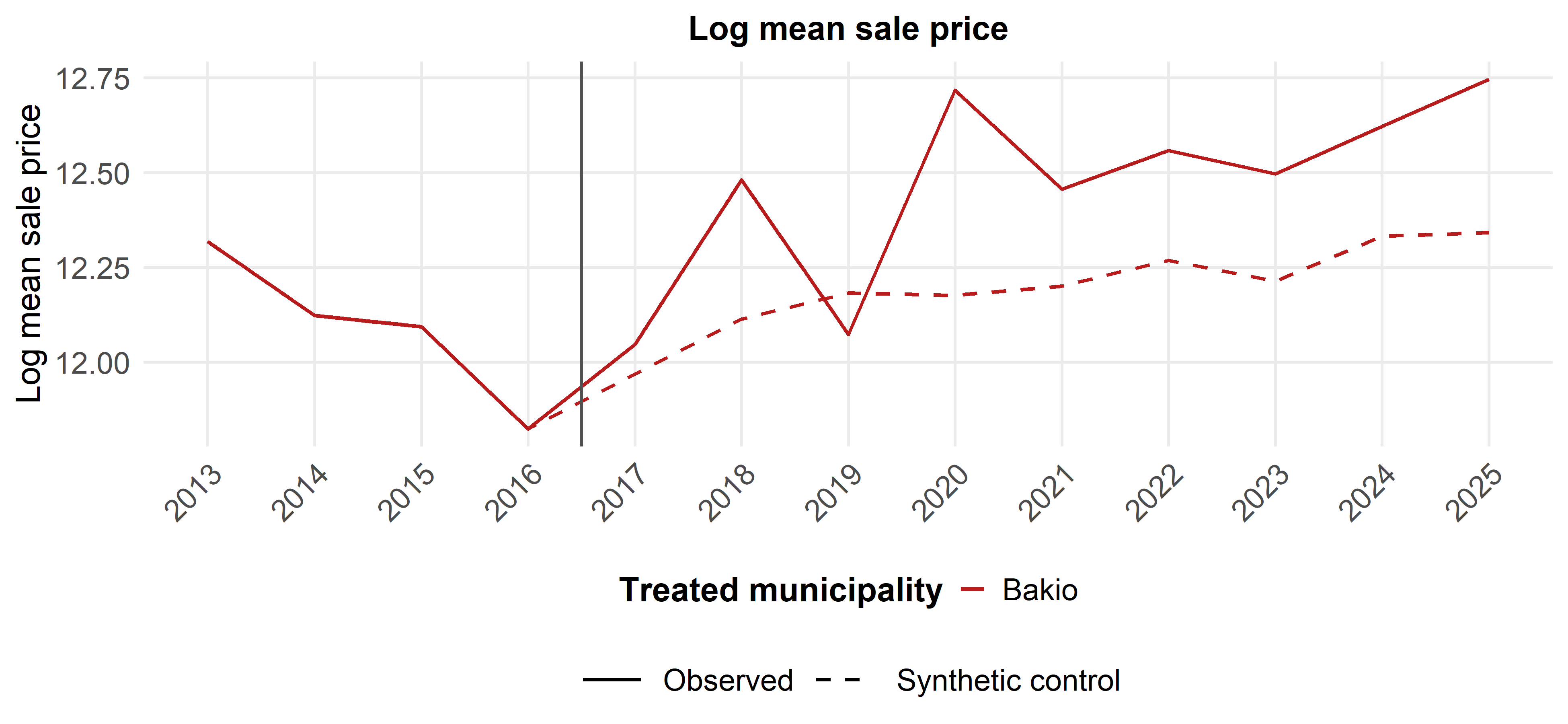} &
    \includegraphics[width=8cm]{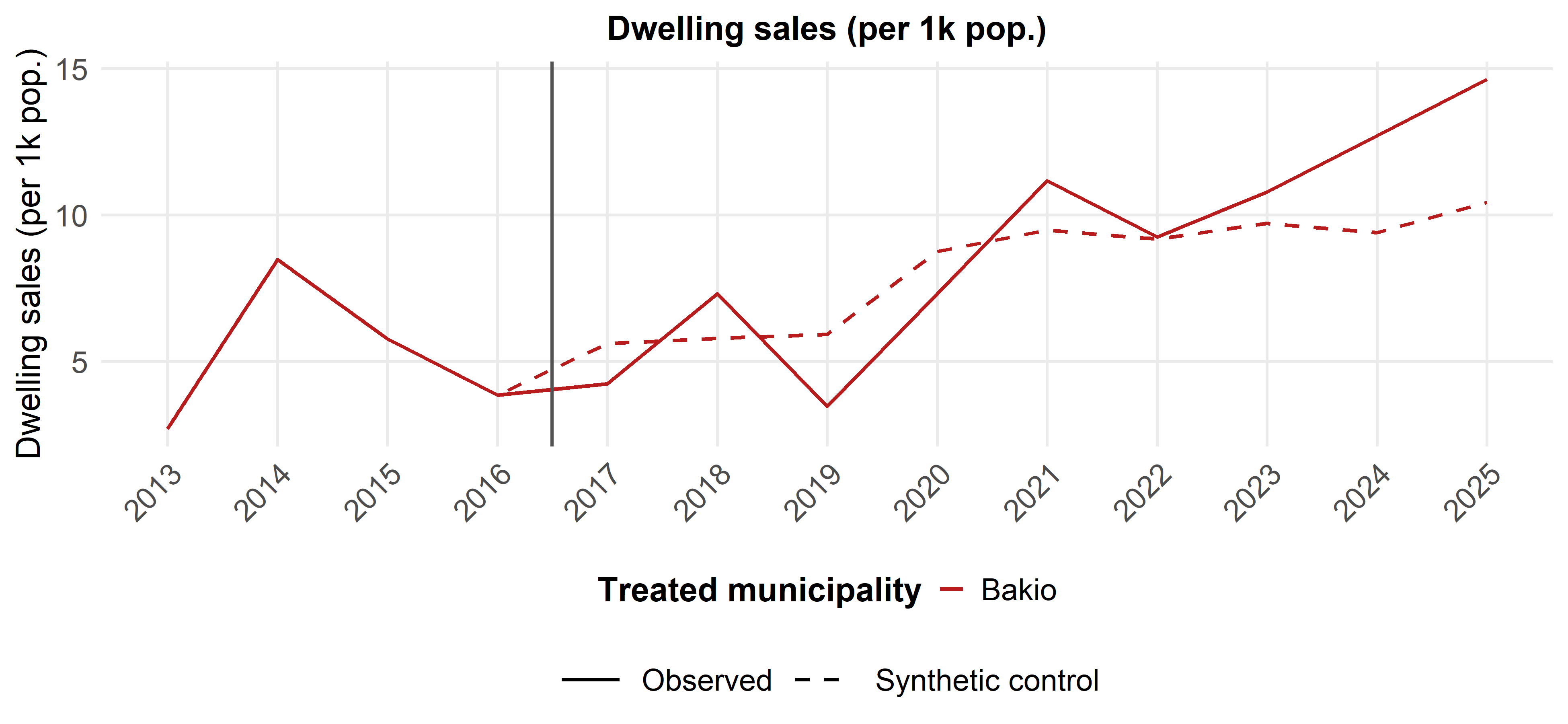} 
    \end{tabular}

    \vspace{-0.2cm}
    \begin{minipage}{\linewidth}
   {\scriptsize \setstretch{1}\textit{Notes:} The plots show the time trends of the observed outcome (solid line) and the synthetic control (dashed line). The red line corresponds to Bakio and the blue line to Bermeo. Estimates are obtained using the Ridge Augmented Synthetic Control estimator \citep{ben-michael_augmented_2021}, estimated separately for each outcome and each treated municipality using the same donor pool of Basque municipalities.}
    \end{minipage}
\end{figure}

\subsubsection{Residential Mobility}

\begin{figure}[H]
    \caption{Observed and synthetic control trends for population.}
    \label{fig:population_trends}
    \centering
\begin{tabular}{cc}
\includegraphics[width = 0.6\linewidth]{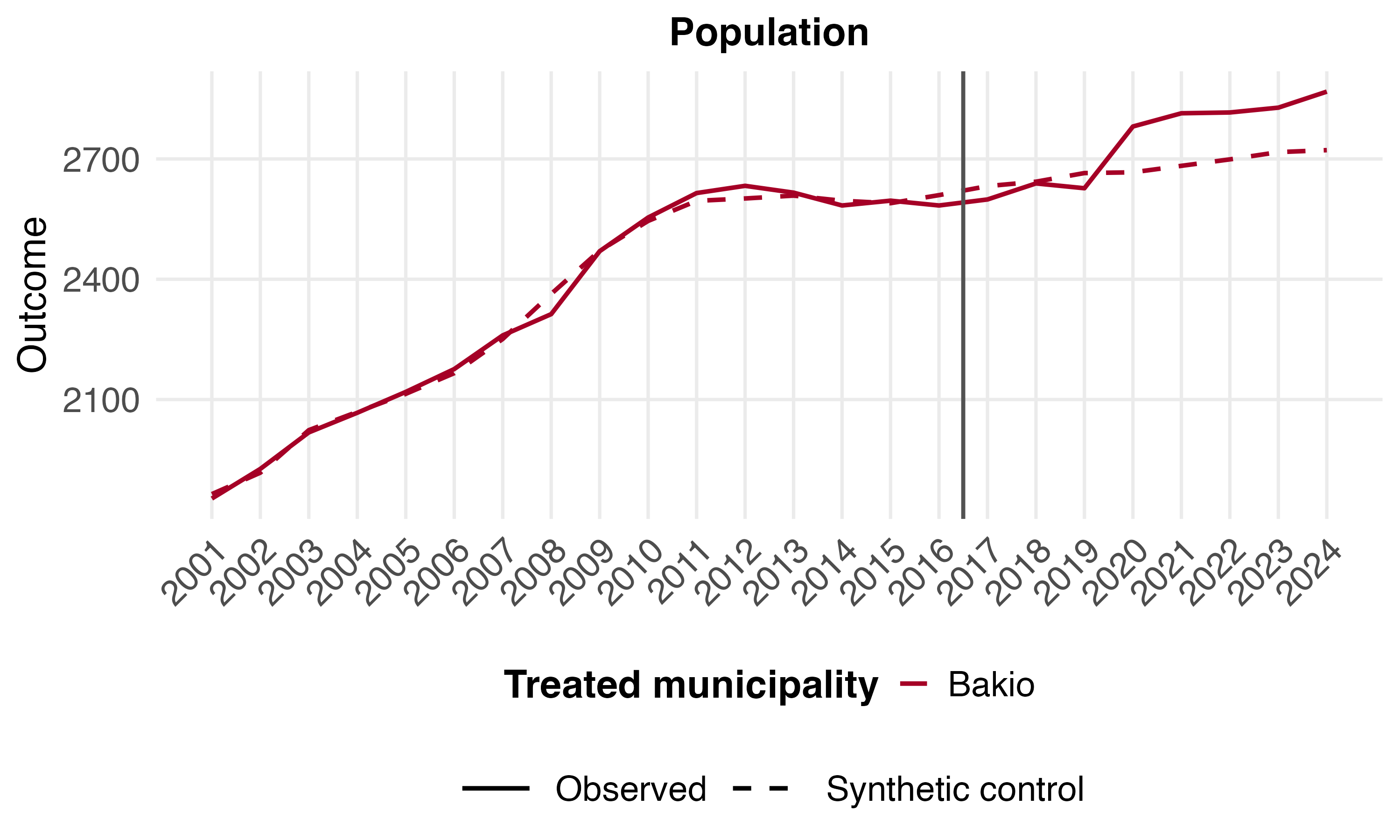}    
\end{tabular}
\vspace{-0.2cm}
\begin{minipage}{\linewidth}
{\scriptsize \setstretch{1}\textit{Notes:} The plots show the time trends of the observed outcome (solid line) and the synthetic control (dashed line). The red line corresponds to Bakio and the blue line to Bermeo. Estimates are obtained using the Ridge Augmented Synthetic Control estimator \citep{ben-michael_augmented_2021}, estimated separately for each outcome and each treated municipality using the same donor pool of Basque municipalities.}
 \end{minipage}
\end{figure}

\begin{figure}[H]
    \caption{Observed and synthetic control trends for residential mobility.}
    \label{fig:mobility_trends}
    \centering
    \setlength{\tabcolsep}{1pt}
    \begin{tabular}{cc}
    \includegraphics[width=0.45\linewidth]{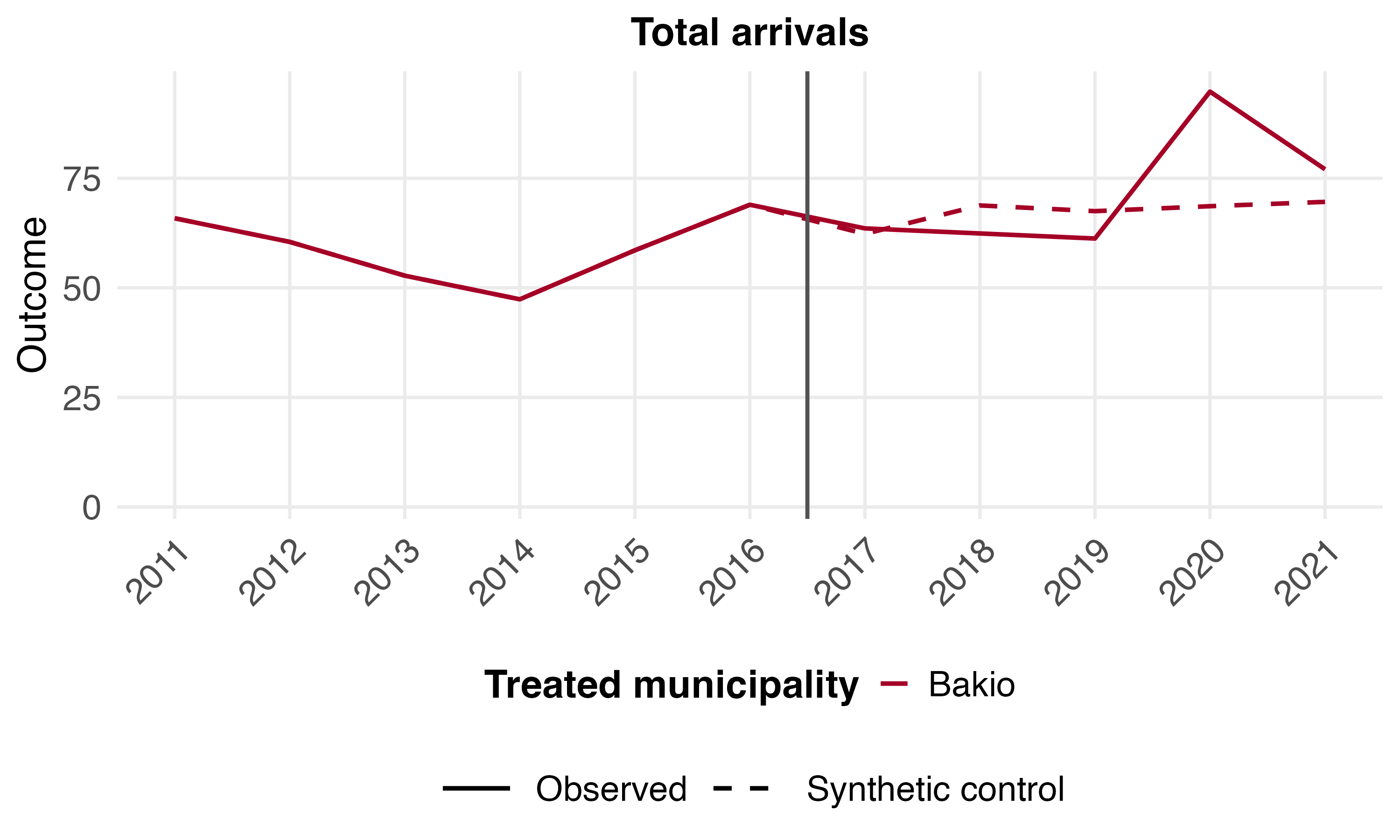} &
    \includegraphics[width=0.45\linewidth]{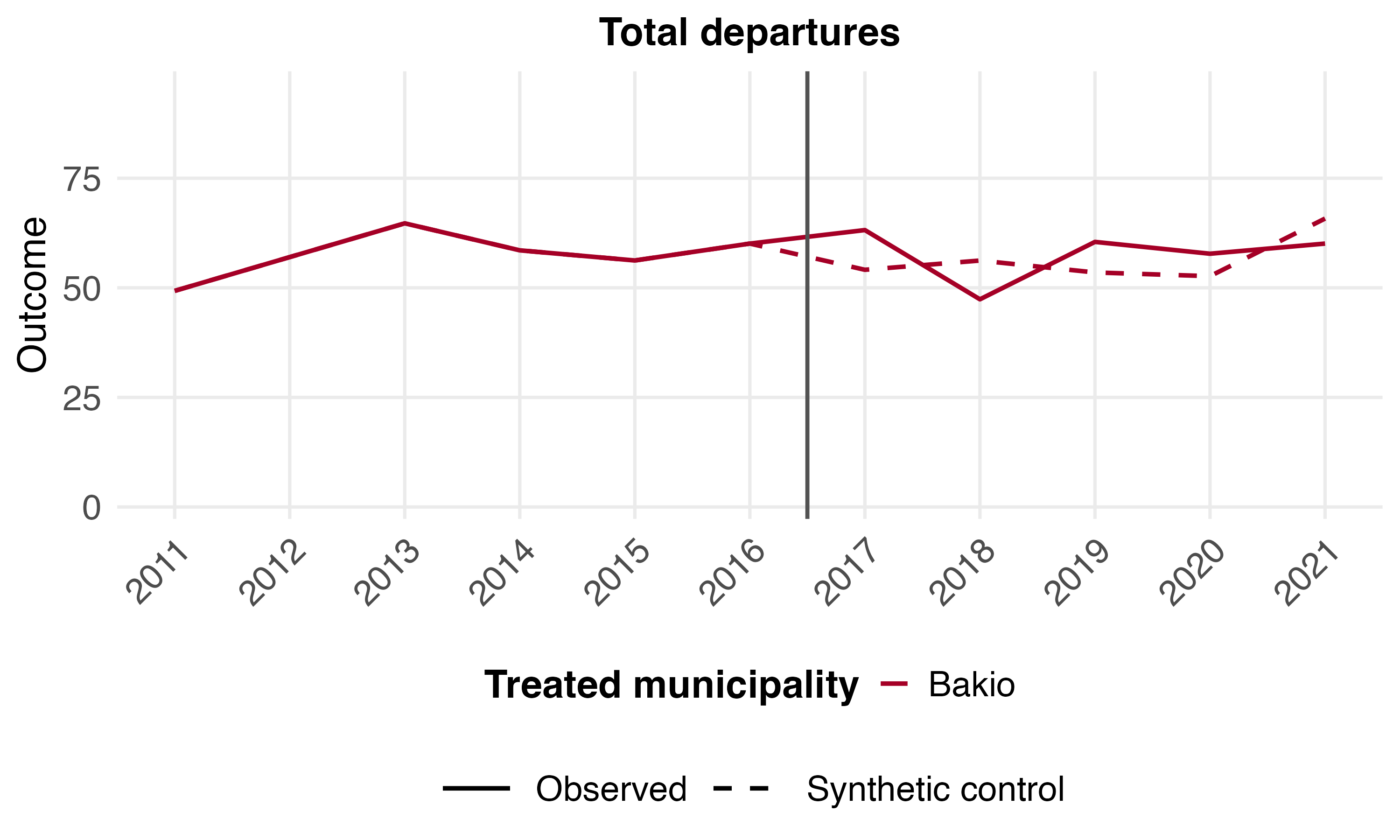} \\
    \includegraphics[width=0.45\linewidth]{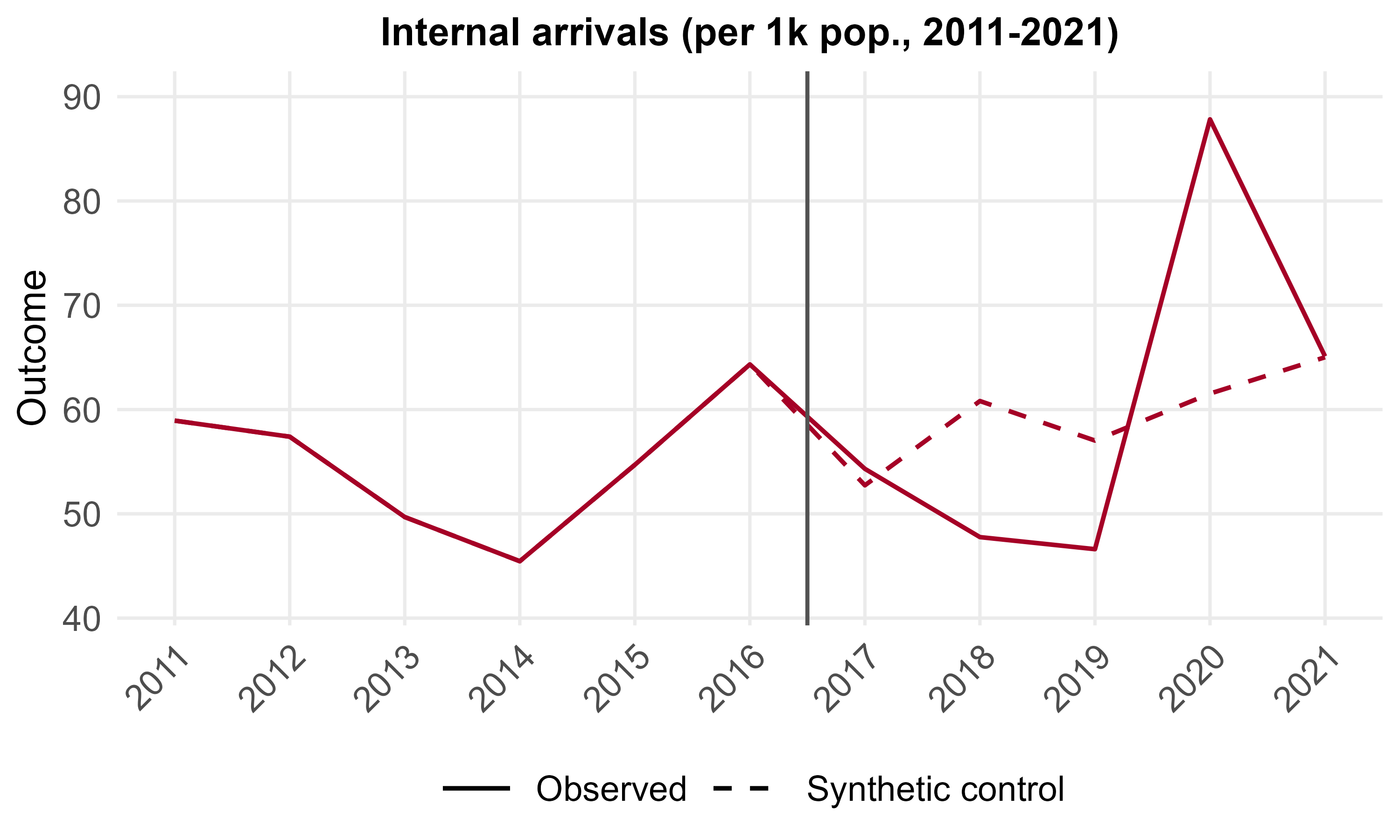} &
    \includegraphics[width=0.45\linewidth]{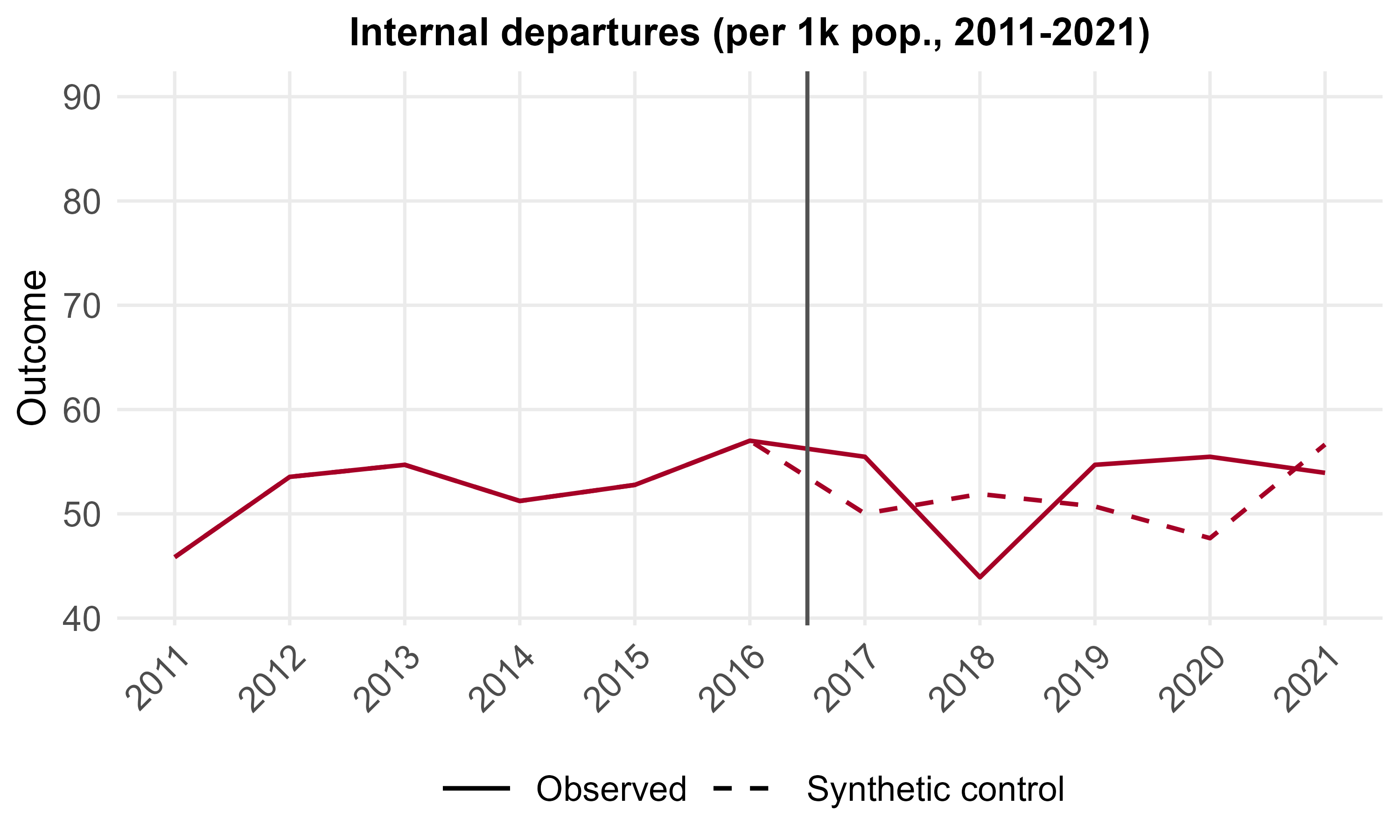} \\
    \includegraphics[width=0.45\linewidth]{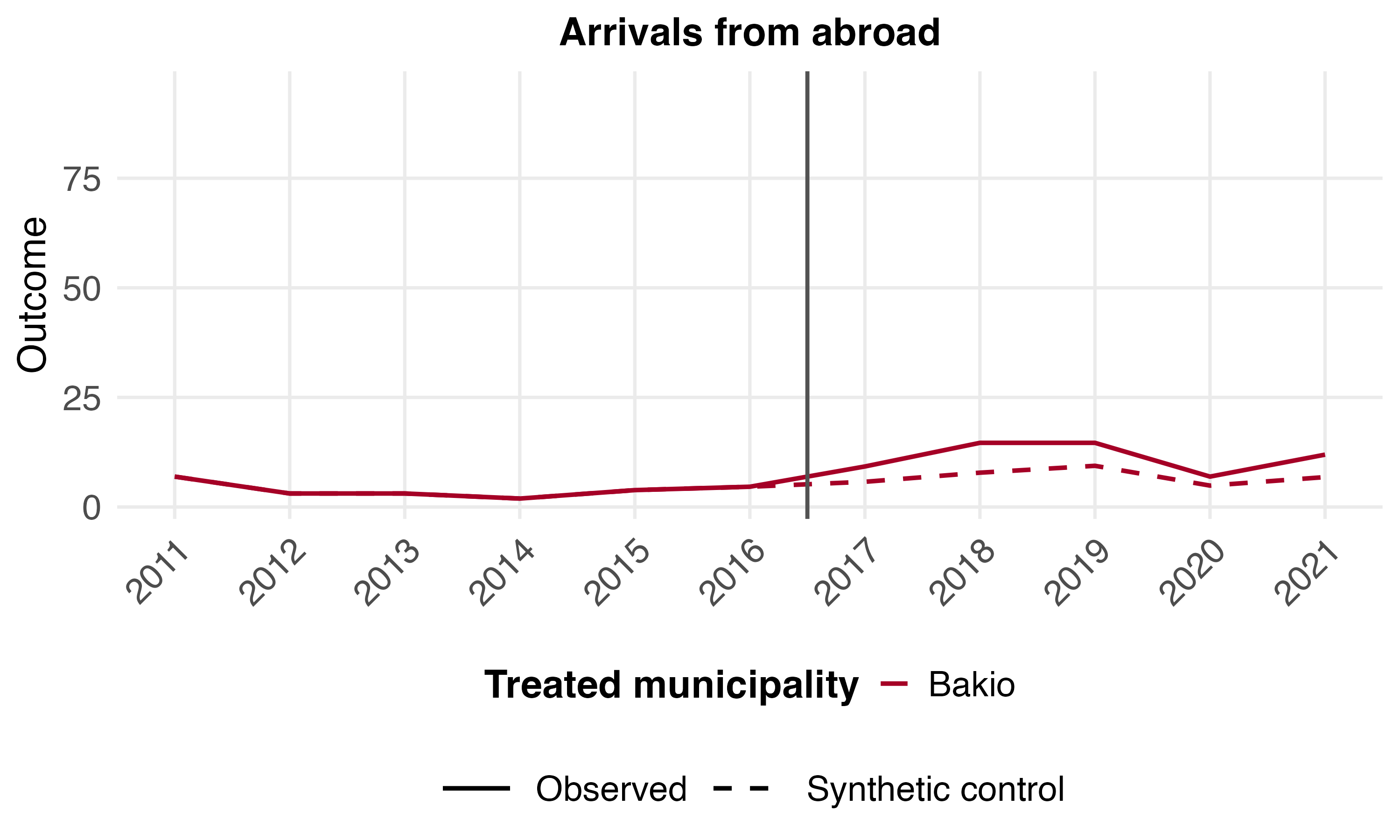} &
    \includegraphics[width=0.45\linewidth]{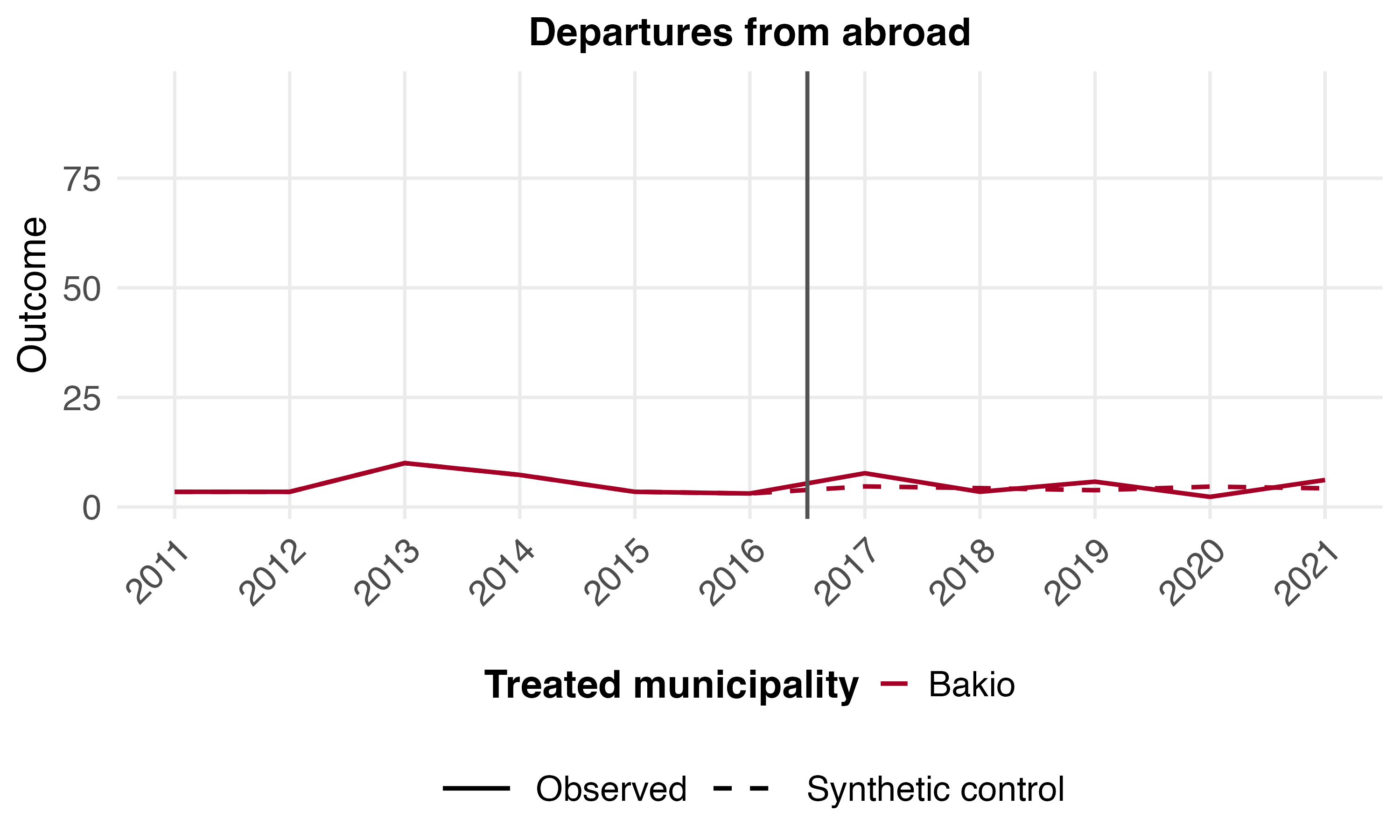}
    \end{tabular}
    \vspace{-0.2cm}
    \begin{minipage}{\linewidth}
    {\scriptsize \setstretch{1}\textit{Notes:} The plots show the time trends of the observed outcome (solid line) and the synthetic control (dashed line) for the treated municipality, Bakio. The left column shows arrivals and the right column departures; rows show all moves (abroad and internal), moves between Spanish municipalities and moves to and from abroad. Outcomes are annual moves per 1,000 residents (2016 population), 2011--2021, from the \textit{Estad\'{i}stica de Variaciones Residenciales} of the Spanish Statistical Office (INE). Estimates are obtained using the Ridge Augmented Synthetic Control estimator \citep{ben-michael_augmented_2021}, estimated separately for each outcome using the same donor pool of Basque municipalities. Within each row, arrivals and departures are drawn on the same vertical scale.}
    \end{minipage}
\end{figure}

\subsection{Synthetic control weights} \label{app:sc_weights}
\begin{figure}[H]
    \caption{Donor weights in the synthetic control of Bakio: tourism outcomes.}
    \label{fig:weights_tourism}
    \centering
    \includegraphics[width=\linewidth]{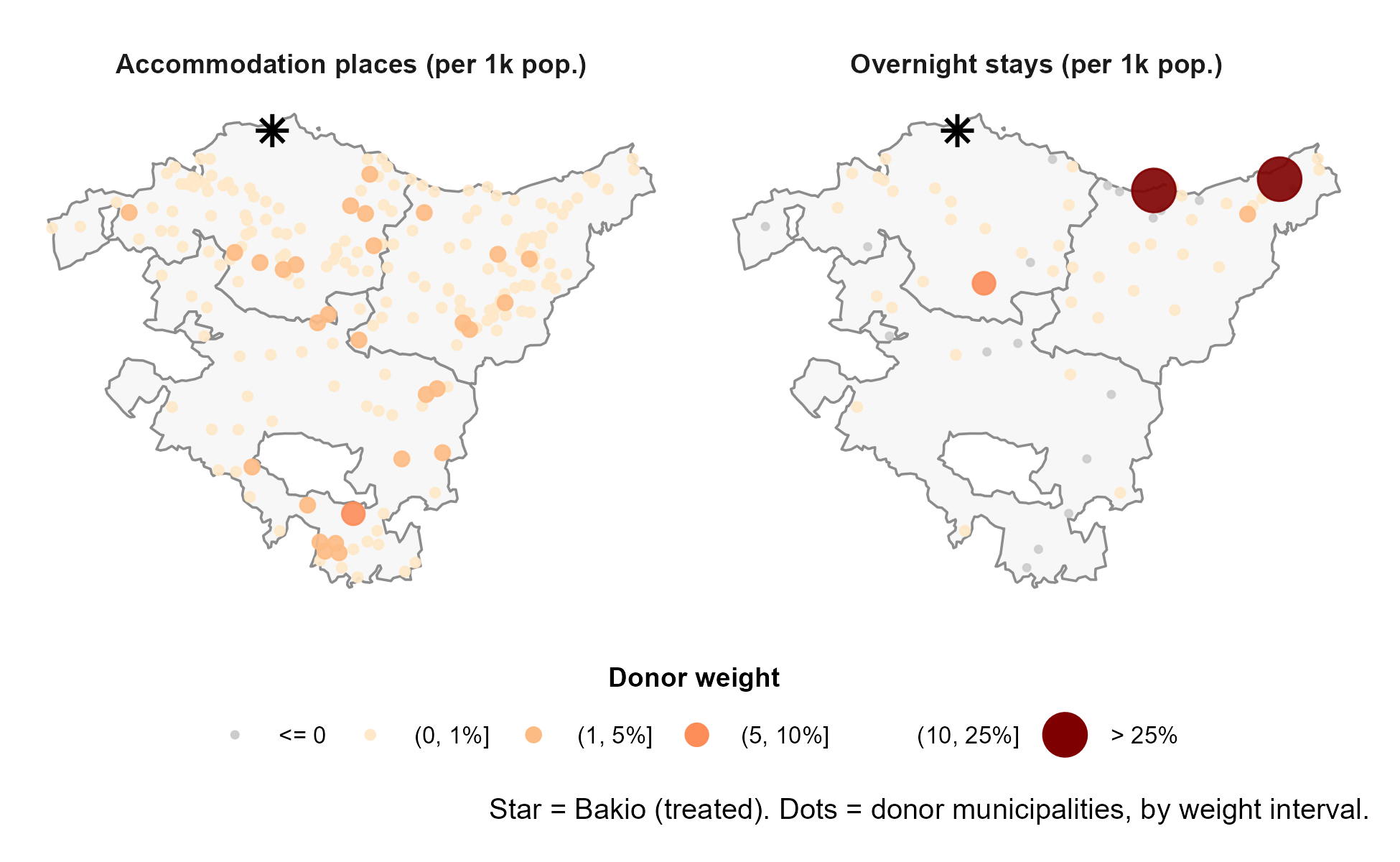}
    \vspace{-0.2cm}
    \begin{minipage}{\linewidth}
    {\scriptsize \setstretch{1}\textit{Notes:} The maps show the weight that each donor municipality receives in the Ridge Augmented Synthetic Control of Bakio \citep{ben-michael_augmented_2021}, estimated separately for each outcome. The star marks Bakio. Dots are donor municipalities, colored and sized by weight interval. The three provincial capitals and the municipalities in the same counties as Bakio and Gaztelugatxe are not in the donor pool. Donors with a weight above 10\% (up to five largest): overnight stays, Lezo (56.6\%) and Zumaia (33.1\%); accommodation places, none.}
    \end{minipage}
\end{figure}

\begin{figure}[H]
    \caption{Donor weights in the synthetic control of Bakio: employment by regime.}
    \label{fig:weights_regime}
    \centering
    \includegraphics[width=\linewidth]{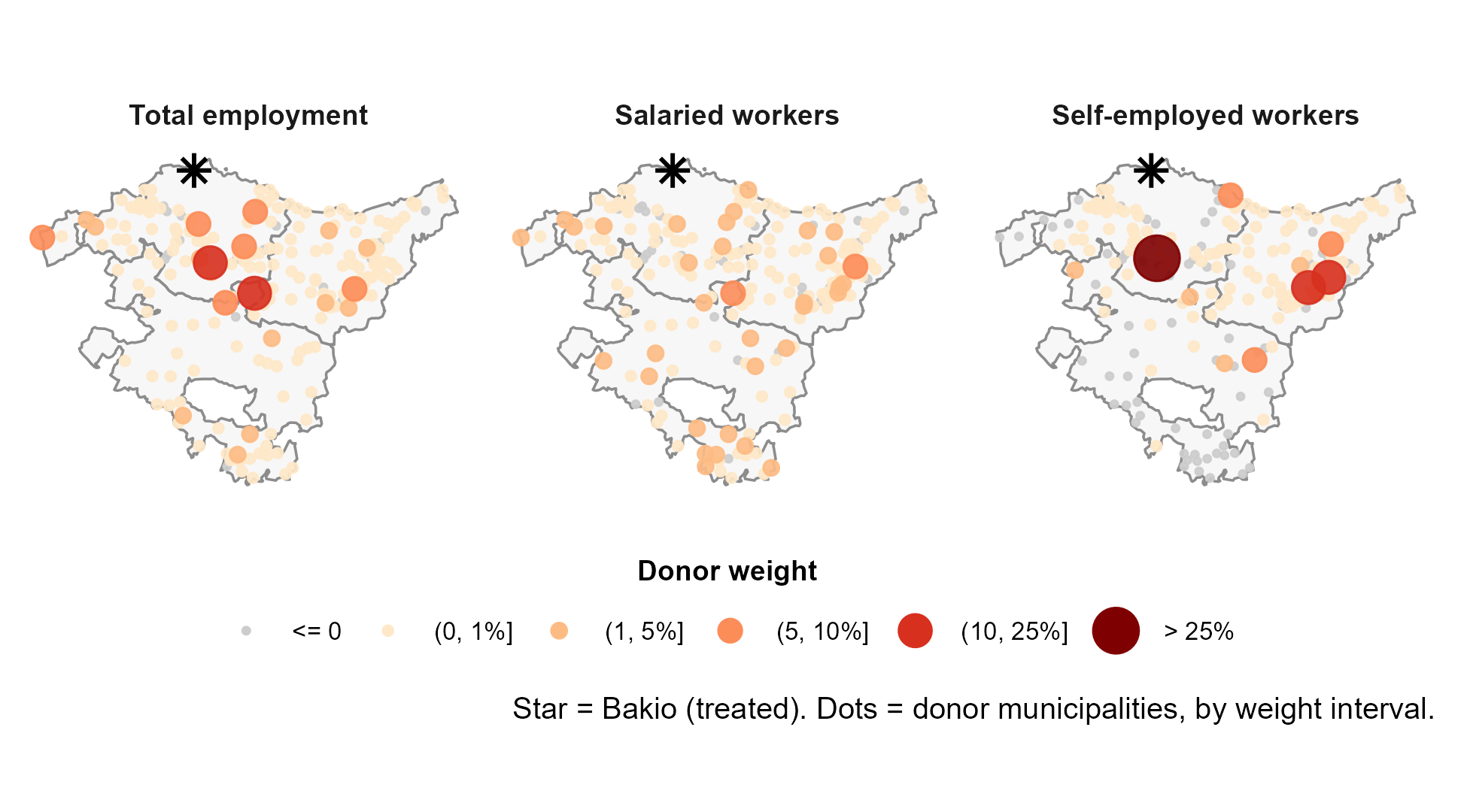}
    \vspace{-0.2cm}
    \begin{minipage}{\linewidth}
    {\scriptsize \setstretch{1}\textit{Notes:} The maps show the weight that each donor municipality receives in the Ridge Augmented Synthetic Control of Bakio \citep{ben-michael_augmented_2021}, estimated separately for each outcome. The star marks Bakio. Dots are donor municipalities, colored and sized by weight interval; the intervals are the same in all panels and figures. The three provincial capitals and the municipalities in the same counties as Bakio and Gaztelugatxe are not in the donor pool. Donors with a weight above 10\% (up to five largest): total employment, Aramaio (14.8\%) and Dima (14.0\%); salaried workers, none; self-employed workers, Arantzazu (26.3\%), Itsasondo (24.7\%) and Altzo (10.8\%).}
    \end{minipage}
\end{figure}

\begin{figure}[H]
    \caption{Donor weights in the synthetic control of Bakio: housing outcomes.}
    \label{fig:weights_housing}
    \centering
    \includegraphics[width=\linewidth]{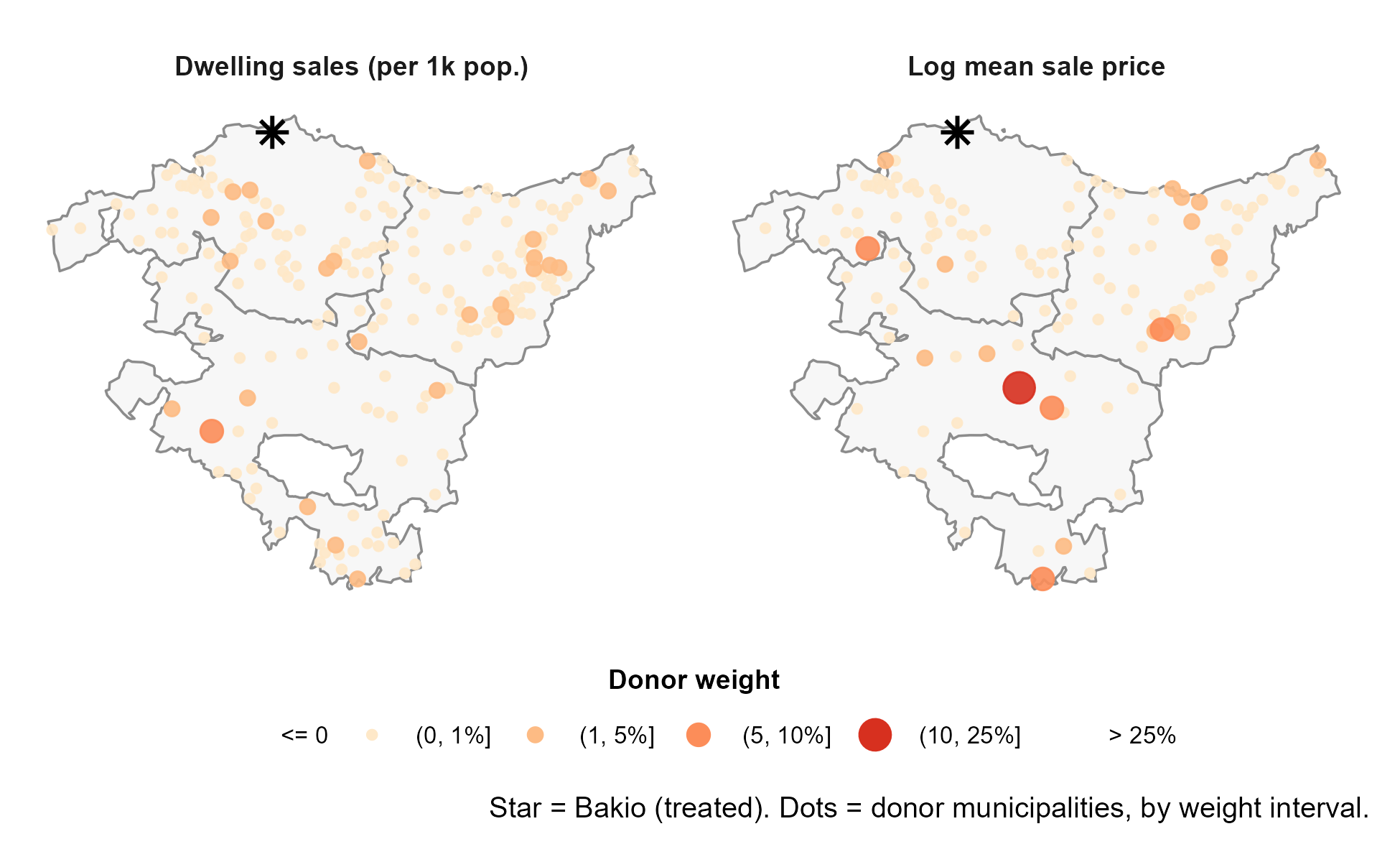}
    \vspace{-0.2cm}
    \begin{minipage}{\linewidth}
    {\scriptsize \setstretch{1}\textit{Notes:} The maps show the weight that each donor municipality receives in the Ridge Augmented Synthetic Control of Bakio \citep{ben-michael_augmented_2021}, estimated separately for each outcome. The star marks Bakio. Dots are donor municipalities, colored and sized by weight interval. The three provincial capitals and the municipalities in the same counties as Bakio and Gaztelugatxe are not in the donor pool. Donors with a weight above 10\% : dwelling sales, none; log mean sale price, Arratzua-Ubarrundia (13.9\%).}
    \end{minipage}
\end{figure}

This section shows which municipalities contribute to the synthetic control of Bakio. Because the Ridge Augmented Synthetic Control is estimated separately for each outcome, each outcome has its own set of donor weights. Each figure maps the weights for a group of outcomes: tourism (Figure \ref{fig:weights_tourism}), employment by regime (Figure \ref{fig:weights_regime}), employment by sector (Figures \ref{fig:weights_sectors1} and \ref{fig:weights_sectors2}) and housing (Figure \ref{fig:weights_housing}). Weights are grouped into the same intervals, and shown with the same colors and dot sizes, in every panel and every figure, so that they can be compared across outcomes. As discussed in Section \ref{sec:strategy}, the Ridge ASC estimator can assign negative weights to some donors; these are grouped with zero-weight donors in the lowest interval ($\leq 0$).

\begin{figure}[H]
    \caption{Donor weights in the synthetic control of Bakio: employment by sector.}
    \label{fig:weights_sectors1}
    \centering
    \includegraphics[width=0.9\linewidth]{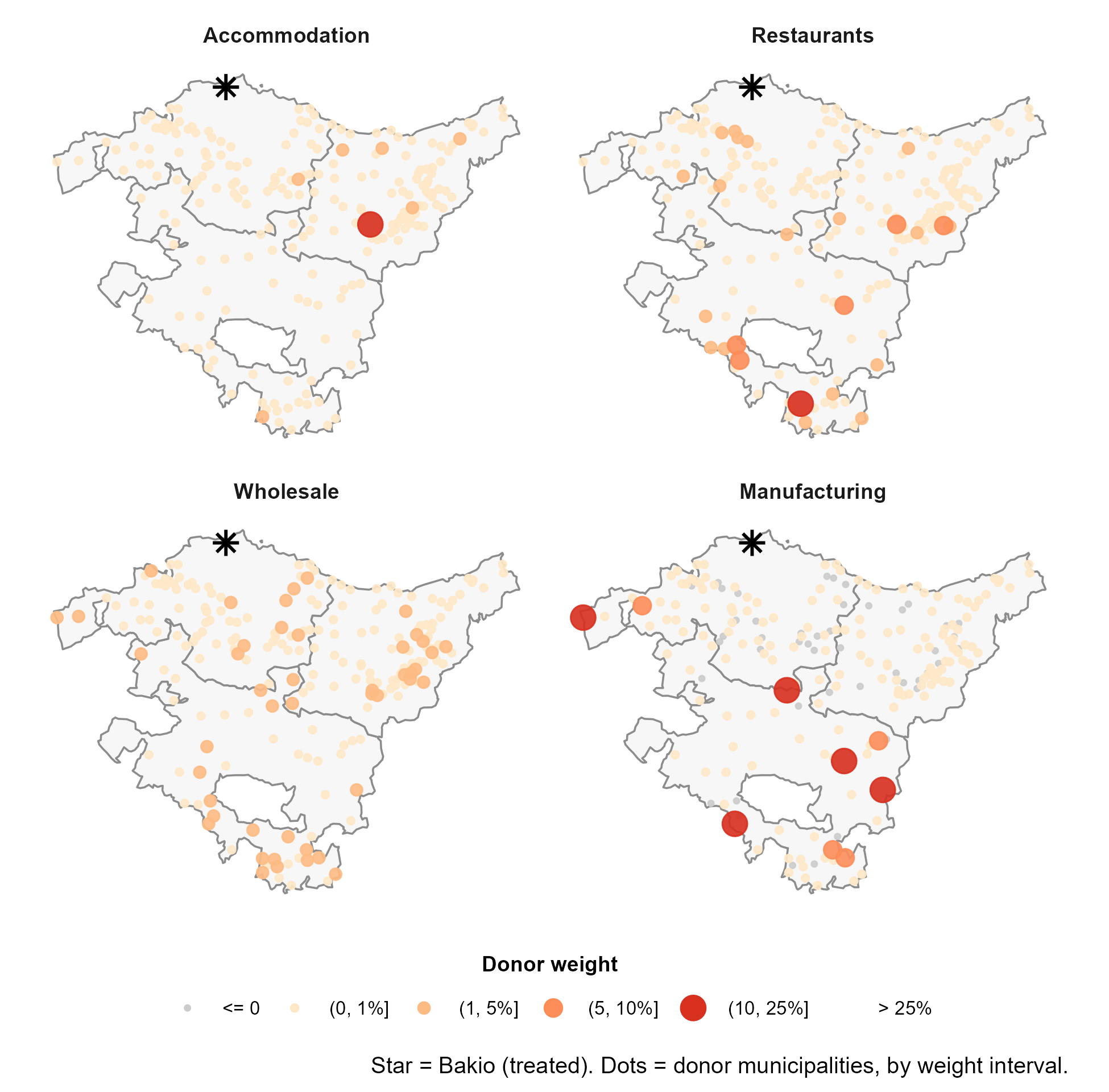}
    \vspace{-0.2cm}
    \begin{minipage}{\linewidth}
    {\scriptsize \setstretch{1}\textit{Notes:} The maps show the weight that each donor municipality receives in the Ridge Augmented Synthetic Control of Bakio \citep{ben-michael_augmented_2021}, estimated separately for each outcome. The star marks Bakio. Dots are donor municipalities, colored and sized by weight interval; the intervals are the same in all panels and figures. The three provincial capitals and the municipalities in the same counties as Bakio and Gaztelugatxe are not in the donor pool. Donors with a weight above 10\% (up to five largest): accommodation, Gabiria (14.7\%); restaurants, Leza (16.1\%); wholesale, none; manufacturing, Lanestosa (17.7\%), Harana/Valle de Arana (15.3\%), Ubide (14.9\%), Zambrana (11.5\%) and Iruraiz-Gauna (11.2\%).}
    \end{minipage}
\end{figure}

\begin{figure}[H]
    \ContinuedFloat
    \caption{Donor weights in the synthetic control of Bakio: employment by sector (continued).}
    \label{fig:weights_sectors2}
    \centering
    \includegraphics[width=0.9\linewidth]{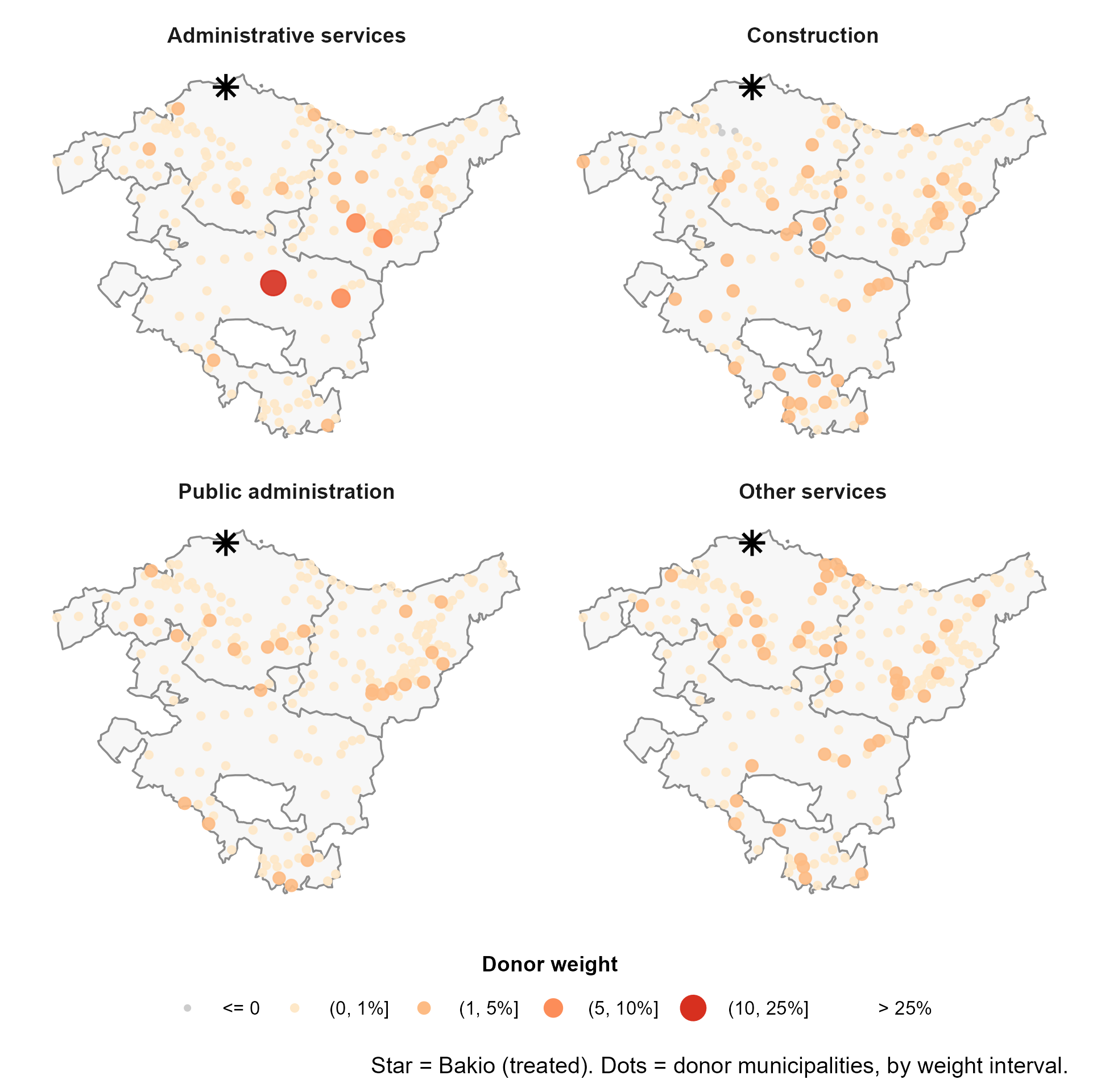}
    \vspace{-0.2cm}
    \begin{minipage}{\linewidth}
    {\scriptsize \setstretch{1}\textit{Notes:} The maps show the weight that each donor municipality receives in the Ridge Augmented Synthetic Control of Bakio \citep{ben-michael_augmented_2021}, estimated separately for each outcome. The star marks Bakio. Dots are donor municipalities, colored and sized by weight interval; the intervals are the same in all panels and figures. The three provincial capitals and the municipalities in the same counties as Bakio and Gaztelugatxe are not in the donor pool. Donors with a weight above 10\% (up to five largest): administrative services, Arratzua-Ubarrundia (13.3\%); construction, public administration and other services, none.}
    \end{minipage}
\end{figure}

\begin{figure}[H]
    \caption{Donor weights in the synthetic control of Bakio: arrivals.}
    \label{fig:weights_arrivals}
    \centering
    \includegraphics[width=\linewidth]{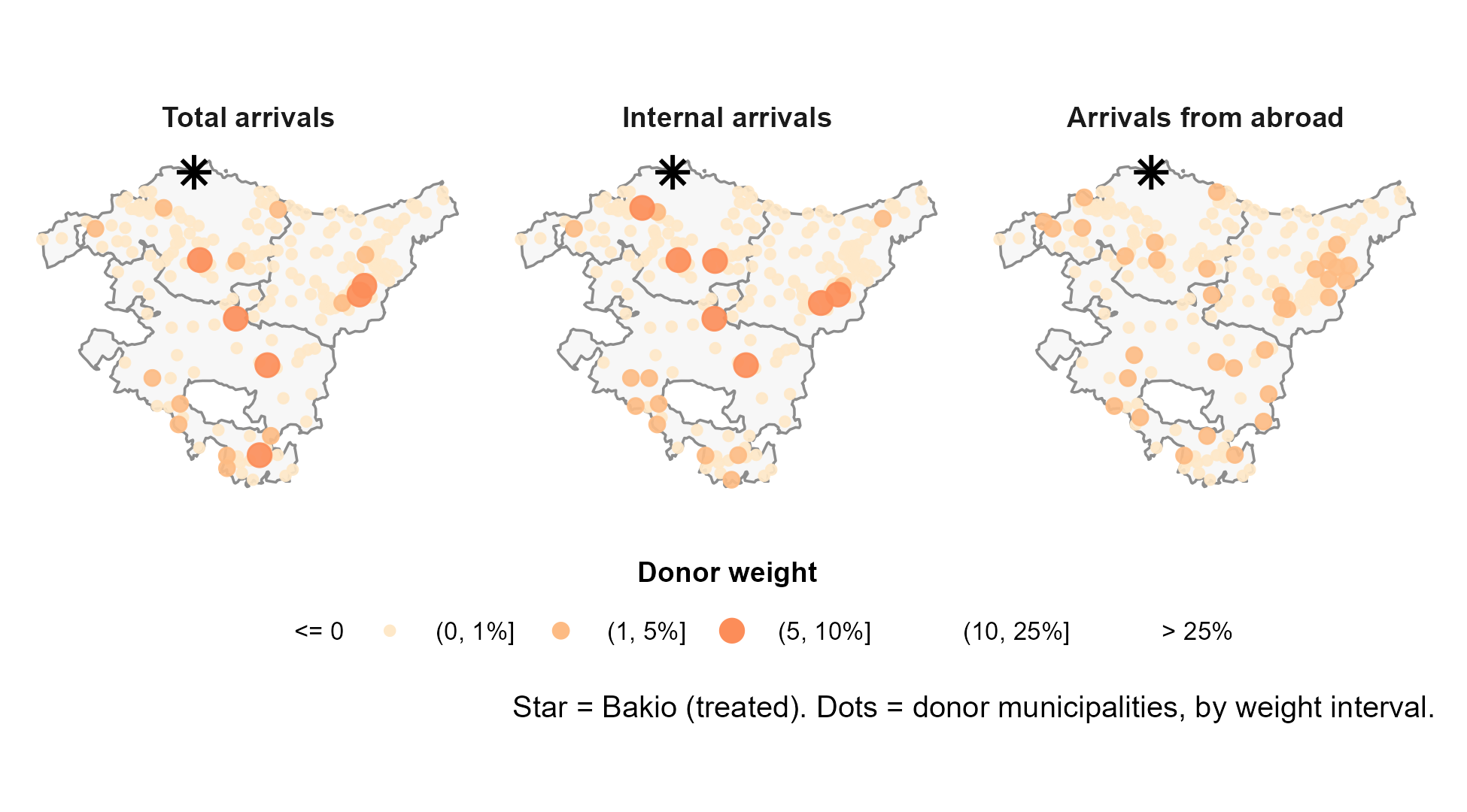}
    \vspace{-0.2cm}
    \begin{minipage}{\linewidth}
    {\scriptsize \setstretch{1}\textit{Notes:} The maps show the weight that each donor municipality receives in the Ridge Augmented Synthetic Control of Bakio \citep{ben-michael_augmented_2021}, estimated separately for each outcome. The star marks Bakio. Dots are donor municipalities, colored and sized by weight interval. The three provincial capitals and the municipalities in the same counties as Bakio and Gaztelugatxe are not in the donor pool. Donors with a weight above 10\% : none for any of the outcomes (the largest single weight is 9.3\%, Legutio, for total and internal arrivals)}
    \end{minipage}
\end{figure}

\begin{figure}[H]
    \caption{Donor weights in the synthetic control of Bakio: departures.}
    \label{fig:weights_departures}
    \centering
    \includegraphics[width=\linewidth]{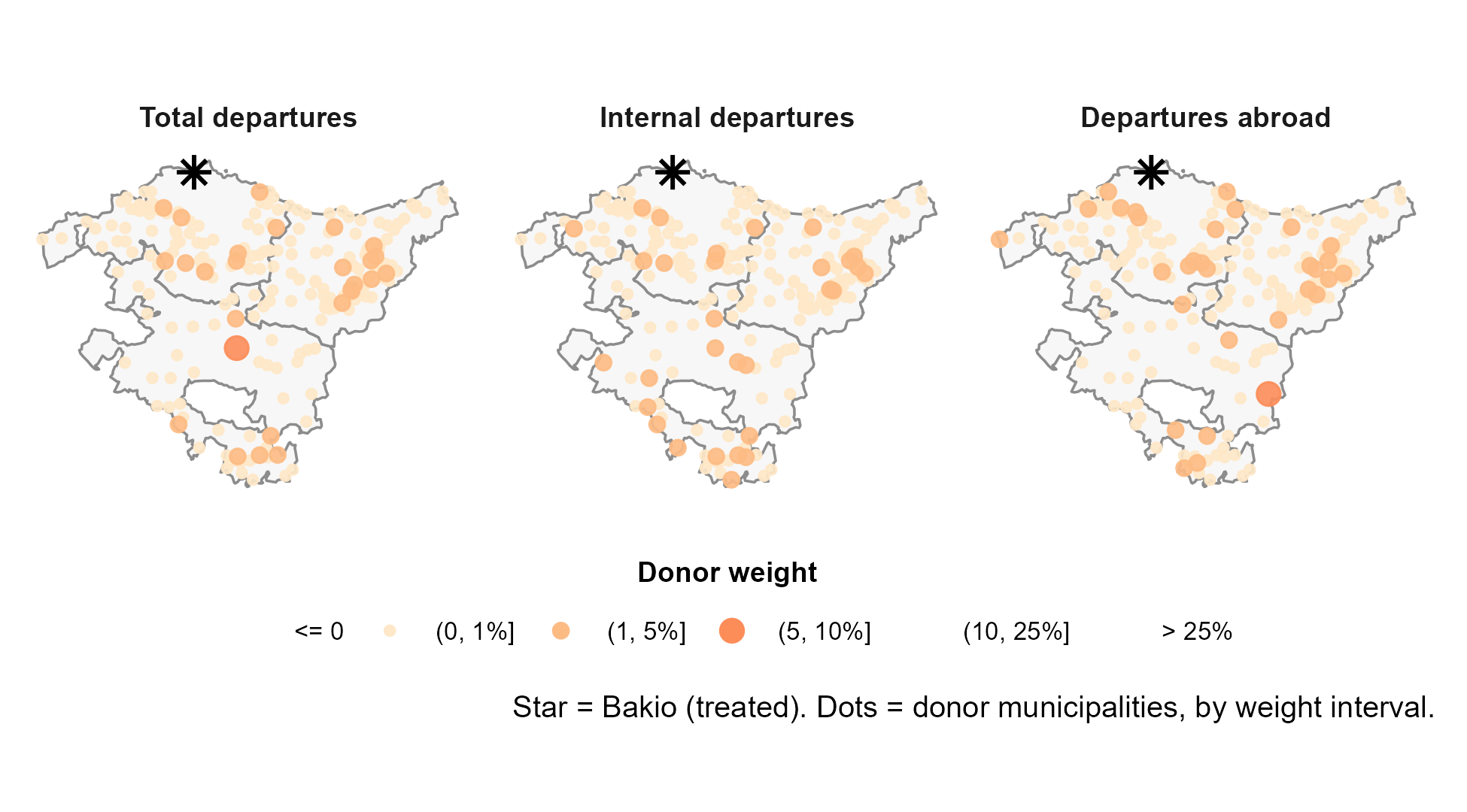}
    \vspace{-0.2cm}
    \begin{minipage}{\linewidth}
    {\scriptsize \setstretch{1}\textit{Notes:} The maps show the weight that each donor municipality receives in the Ridge Augmented Synthetic Control of Bakio \citep{ben-michael_augmented_2021}, estimated separately for each outcome. The star marks Bakio. Dots are donor municipalities, colored and sized by weight interval. The three provincial capitals and the municipalities in the same counties as Bakio and Gaztelugatxe are not in the donor pool. Donors with a weight above 10\% : none.}
    \end{minipage}
\end{figure}

%\end{appendices}

\end{spacing}
\end{document}